\documentclass[12pt,oneside]{book}

\usepackage{amsmath}%
\usepackage{amsfonts}%
\usepackage{amssymb}%
\usepackage{graphicx}
\usepackage{longtable}
\usepackage{url}
\usepackage[all]{xy}
\usepackage{subfig}
\usepackage[noend]{algorithmic}
\usepackage{algorithm}
\usepackage{multirow}
\usepackage{fancyhdr}
\usepackage{booktabs}
\usepackage{setspace}
\usepackage{paralist}
\usepackage{color}
\usepackage{colortbl}
\usepackage{caption}
\usepackage[refpage,noprefix]{nomencl}
\usepackage[pdftex]{hyperref}

\hypersetup{}

\makenomenclature

\definecolor{grey}{rgb}{0.9,0.9,0.9}

\fancypagestyle{plain}{
\fancyhf{}
\fancyhead[R]{\thepage}

}

\newenvironment{proof}[1][Proof]{\textbf{#1.} }{\ \rule{0.5em}{0.5em}}
\newcommand{\qed}{\hfill \ensuremath{\Box}}
\newtheorem{theorem}{Theorem}

\newtheorem{definition}{Definition}

\newcommand{\func}[1]{\ensuremath{\mathit{#1}}}
\newcommand{\var}[1]{\ensuremath{\mathit{#1}}}
\newcommand{\instance}[1]{\texttt{#1}}
\newcommand{\heuristic}[1]{\textsf{#1}}

\DeclareMathOperator*{\argmin}{argmin}

\newcommand{\Cluster}[0]{\func{Cluster}}
\newcommand{\Turn}[0]{\func{Turn}}
\newcommand{\Swap}[0]{\heuristic{Swap}}
\newcommand{\ImprovePath}{\func{ImprovePath}}

\newcommand{\LKtsp}{\heuristic{LK$_\text{tsp}$}}
\newcommand{\LK}{\heuristic{LK}}
\newcommand{\LKb}[3]{\heuristic{B}\ensuremath{_\text{#1}^\text{#2#3}}}
\newcommand{\LKc}[3]{\heuristic{C}\ensuremath{_\text{#1}^\text{#2#3}}}
\newcommand{\LKs}[3]{\heuristic{S}\ensuremath{_\text{#1}^\text{#2#3}}}
\newcommand{\LKe}[2]{\heuristic{G}\ensuremath{_\text{#1}^\text{#2}}}
\newcommand{\opt}[3]{\heuristic{#1-opt}$_\text{#2}^\text{#3}$}
\newcommand{\optshort}[3]{\heuristic{#1o}$_\text{#2}^\text{#3}$}
\newcommand{\twoopt}[2]{\opt{2}{#1}{#2}}
\newcommand{\Insertion}[2]{\heuristic{Ins}$_\text{#1}^\text{#2}$}
\newcommand{\NN}[0]{\heuristic{NN}}

\newcommand{\FO}[1]{\heuristic{FO}$_\text{#1}$}
\newcommand{\CO}{\heuristic{CO}}

\newcommand{\citep}[1]{\cite{#1}}
\newcommand{\citet}[1]{\cite{#1}}

\newcommand{\ExactSolver}{\heuristic{Exact}}
\newcommand{\SD}{\heuristic{SD}}
\newcommand{\GK}{\heuristic{GK}}
\newcommand{\SG}{\heuristic{SG}}
\newcommand{\TSPma}{\heuristic{TSP}}

\newcommand{\ifamily}[1]{{\small\textsf{{#1}}}}

\newcommand{\random}{{\ifamily{Random}}}
\newcommand{\clique}{{\ifamily{Clique}}}
\newcommand{\geometric}{{\ifamily{Geometric}}}
\newcommand{\GP}{{\ifamily{GP}}}
\newcommand{\squareroot}{{\ifamily{SquareRoot}}}
\newcommand{\product}{{\ifamily{Product}}}

\newcommand{\OptPlain}[1]{\mbox{#1-opt}}
\newcommand{\Opt}[1]{\heuristic{\OptPlain{#1}}}

\newcommand{\TwooptPlain}{\OptPlain{2}}
\newcommand{\Twoopt}{\Opt{2}}

\newcommand{\ThreeoptPlain}{\OptPlain{3}}
\newcommand{\Threeopt}{\Opt{3}}

\newcommand{\KoptPlain}{\OptPlain{$k$}}
\newcommand{\Kopt}{\Opt{$k$}}

\newcommand{\OptStarPlain}[1]{\mbox{#1-opt*}}
\newcommand{\OptStar}[1]{\heuristic{\OptStarPlain{#1}}}

\newcommand{\VoptPlain}{\OptPlain{v}}
\newcommand{\Vopt}{\Opt{v}}

\newcommand{\OneDVPlain}{1DV}
\newcommand{\OneDV}{{\heuristic{\OneDVPlain}}}

\newcommand{\TwoDVPlain}{2DV}
\newcommand{\TwoDV}{{\heuristic{\TwoDVPlain}}}

\newcommand{\MDVPlain}{$s$DV}
\newcommand{\MDV}{{\heuristic{\MDVPlain}}}

\newcommand{\VDI}{VDI}

\newcommand{\CombPlain}[2]{#1$_\text{#2}$}
\newcommand{\Comb}[2]{\heuristic{\CombPlain{#1}{#2}}}

\newcommand{\OneDVtwoPlain}{\CombPlain{\OneDVPlain}{2}}
\newcommand{\OneDVthreePlain}{\CombPlain{\OneDVPlain}{3}}
\newcommand{\OneDVVPlain}{\CombPlain{\OneDVPlain}{\text{v}}}

\newcommand{\TwoDVtwoPlain}{\CombPlain{\TwoDVPlain}{2}}
\newcommand{\TwoDVthreePlain}{\CombPlain{\TwoDVPlain}{3}}
\newcommand{\TwoDVVPlain}{\CombPlain{\TwoDVPlain}{\text{v}}}

\newcommand{\MDVtwoPlain}{\CombPlain{\MDVPlain}{2}}
\newcommand{\MDVthreePlain}{\CombPlain{\MDVPlain}{3}}
\newcommand{\MDVVPlain}{\CombPlain{\MDVPlain}{\text{v}}}

\newcommand{\OneDVtwo}{\heuristic{\OneDVtwoPlain}}
\newcommand{\OneDVthree}{\heuristic{\OneDVthreePlain}}
\newcommand{\OneDVV}{\heuristic{\OneDVVPlain}}

\newcommand{\TwoDVtwo}{\heuristic{\TwoDVtwoPlain}}
\newcommand{\TwoDVthree}{\heuristic{\TwoDVthreePlain}}
\newcommand{\TwoDVV}{\heuristic{\TwoDVVPlain}}

\newcommand{\MDVtwo}{\heuristic{\MDVtwoPlain}}
\newcommand{\MDVthree}{\heuristic{\MDVthreePlain}}
\newcommand{\MDVV}{\heuristic{\MDVVPlain}}

\newcommand{\Trivial}{{\heuristic{Trivial}}}
\newcommand{\Greedy}{{\heuristic{Greedy}}}
\newcommand{\MaxRegret}{{\heuristic{Max-Regret}}}
\newcommand{\ROM}{{\heuristic{ROM}}}

\newcommand{\Chain}{{\heuristic{Chain}}}
\newcommand{\Multichain}{{\heuristic{Multichain}}}

\newcommand{\greedy}{{\heuristic{Greedy}}}
\newcommand{\maxregret}{{\heuristic{Max-Regret}}}
\newcommand{\rom}{{\heuristic{ROM}}}
\newcommand{\shiftrom}{{\heuristic{Shift-ROM}}}

\newcommand{\HL}{\heuristic{HL}}

\newcommand{\SA}{\heuristic{SA}}

\newcommand{\LocalSearch}{\mathit{LocalSearch}}
\newcommand{\mutation}{\mathit{mutation}}
\newcommand{\crossover}{\mathit{crossover}}
\newcommand{\selection}{\mathit{selection}}
\newcommand{\code}{\mathit{code}}
\newcommand{\decode}{\mathit{decode}}
\newcommand{\perturb}{\mathit{perturb}}

\begin{document}

\pagenumbering{alph}

\title{
Ph.\ D.\ Thesis\\[5ex]
Design, Evaluation and Analysis of Combinatorial Optimization Heuristic Algorithms
}

\date{}
\author{$\begin{array}{r@{\quad}l}
\text{Author} & \text{Daniil Karapetyan} \\ 
\text{Supervisor} & \text{Prof.\ Gregory Gutin} \\ 
\end{array}$\\[4ex]
Department of Computer Science\\
Royal Holloway College\\
University of London}

\date{July 2010}

\maketitle

\pagenumbering{arabic}
\setcounter{page}{1}

\clearpage
\phantomsection
\addcontentsline{toc}{chapter}{Declaration}
\chapter*{Declaration}

The work presented in this thesis is the result of original research carried out by myself.  This work has not been submitted for any other degree or award in any other university or educational establishment.

\bigskip
\bigskip


\bigskip
\bigskip

\hfill Daniil Karapetyan

\clearpage
\phantomsection
\addcontentsline{toc}{chapter}{Abstract}
\chapter*{Abstract}

Combinatorial optimization is widely applied in a number of areas nowadays. Unfortunately, many combinatorial optimization problems are NP-hard which usually means that they are unsolvable in practice.  However, it is often unnecessary to have an exact solution.  In this case one may use heuristic approach to obtain a near-optimal solution in some reasonable time.

We focus on two combinatorial optimization problems, namely the Generalized Traveling Salesman Problem and the Multidimensional Assignment Problem.  The first problem is an important generalization of the Traveling Salesman Problem; the second one is a generalization of the Assignment Problem for an arbitrary number of dimensions.  Both problems are NP-hard and have hosts of applications.

In this work, we discuss different aspects of heuristics design and evaluation.  A broad spectrum of related subjects, covered in this research, includes test bed generation and analysis, implementation and performance issues, local search neighborhoods and efficient exploration algorithms, metaheuristics design and population sizing in memetic algorithm.

The most important results are obtained in the areas of local search and memetic algorithms for the considered problems.  In both cases we have significantly advanced the existing knowledge on the local search neighborhoods and algorithms by systematizing and improving the previous results.  We have proposed a number of efficient heuristics which dominate the existing algorithms in a wide range of time/quality requirements.  

Several new approaches, introduced in our memetic algorithms, make them the state-of-the-art metaheuristics for the corresponding problems.  Population sizing is one of the most promising among these approaches; it is expected to be applicable to virtually any memetic algorithm.

\clearpage
\phantomsection
\addcontentsline{toc}{chapter}{Dedication}
\null\vspace{\stretch{1}} 
\begin{center}
\emph{In memory of my mother}
\end{center}
\vspace{\stretch{2}}\null

\clearpage
\phantomsection
\addcontentsline{toc}{chapter}{Acknowledgment}
\chapter*{Acknowledgment}

I would like to thank my supervisor Gregory Gutin for his patience, encouragement and advice in all aspects of academic life over the last three years.

I would also like to thank my family and, especially, my grandfather who was continuously inspiring my interest in technical sciences for at least two decades.

This research was supported by the Computer Science Department of the Royal Holloway College and a Thomas Holloway Scholarship.

\clearpage
\phantomsection
\addcontentsline{toc}{chapter}{Contents}
\tableofcontents

\clearpage
\phantomsection
\addcontentsline{toc}{chapter}{List of Tables}
\listoftables

\clearpage
\phantomsection
\addcontentsline{toc}{chapter}{List of Figures}
\listoffigures

\chapter{Introduction}
\label{sec:intro}

\markright{\thechapter.\ Introduction}

Nowadays combinatorial optimization problems arise in many circumstances, and we need to be able to solve these problems efficiently.  Unfortunately, many of these problems are proven to be NP-hard, i.e., it is often impossible to solve the instances in any reasonable time.

However, in practice one usually does not need an exact solution of the problem.  In this case one can use a heuristic algorithm which yields a near-optimal solution in a satisfactory time.  Some of the heuristics, so-called approximation algorithms, guarantee certain solution quality and the polynomial running time.  Unfortunately, this nice theoretical property is usually achieved at the cost of relatively poor performance.  In other words, a simple heuristic is often faster and yields better solutions than an approximation algorithm, though a simple heuristic does not guarantee any quality and in certain cases it yields very bad solutions.

In this research we focus on heuristic algorithms which usually have no guaranteed solution quality.  We are interested in design and selection of the most efficient algorithms for real-world use and, thus, we pay a lot of attention to experimental evaluation.

As a case study we consider two combinatorial optimization problems: the Generalized Traveling Salesman Problem (GTSP) and the Multidimensional Assignment Problem (MAP).  Both problems are known to be NP-hard and each has a host of applications.

Though both GTSP and MAP are very important, the researchers did not pay enough attention to certain areas around these problems.  In particular, the literature lacks any thorough surveys of GTSP or MAP local search; there are just a few metaheuristics for MAP and all of them are designed for only one special case of the problem; there exists no standard test bed for MAP which would include a wide range of instances of different sizes and types.  This work fills many of these and some other gaps, and, moreover, some of the ideas proposed here can be applied to many other optimization problems.

\bigskip

We start from a thorough explanation of the case study problems, their applications and existing solution methods.

Chapter~\ref{sec:heuristics_design} is devoted to some approaches in heuristic design.  It discusses test bed construction, shows several examples on how theoretical tools can help in design of practically efficient heuristics and provides a set of advices on high-performance implementation of an algorithm.

Chapter~\ref{sec:gtsp_ls} introduces a classification of GTSP neighborhoods, proposes several new ones and includes a number of algorithms and improvements which significantly speed up exploration of these neighborhoods both theoretically and practically.  Special attention is paid to adaptation for GTSP of the well-known Lin-Kernighan heuristic, originally designed for the Traveling Salesman Problem.

Chapter~\ref{sec:map_ls}, similar to Chapter~\ref{sec:gtsp_ls}, considers the MAP neighborhoods and local search algorithms.  It splits all the MAP neighborhoods into two classes, generalizes the existing approaches, proposes some new ones and, finally, considers a combined local search which explores neighborhoods of both types together.  An extensive experimental analysis is intended to select the most successful heuristics.

Chapter~\ref{sec:memetic} is devoted to the so-called Memetic Algorithms (MA)\@.  MA is a kind of evolutionary algorithms which applies an improvement procedure to every candidate solution.  Several evolutionary algorithms for GTSP, including MAs, are already presented in the literature.  We propose a new MA which features a powerful local search and an efficient termination criterion.  It also uses some other improvements like variation of the population size according to the instance size.  In our experiments, this algorithm clearly outperforms all GTSP metaheuristics known from the literature with respect to both solution quality and running time.

We develop the idea of adaptive population size and apply it to MAP\@.  In our new MA, we use a time based termination criterion, i.e., the algorithm is given some certain time to proceed.  The population size is selected to exploit the given time with maximum efficiency.  The experimental results provided in Chapter~\ref{sec:memetic} show that the designed algorithm is extremely flexible in solution quality/running time trade-off.  In other words, it works efficiently for a wide range of given times.

Most of the experimental results are presented in detailed tables and placed in Appendix~\ref{sec:tables}.

\bigskip

We have already published many results provided here.  Some aspects of heuristic design are discussed in~\cite{GK_GTSP_Reduction2008,GK_Greedy_Chapter,GK_Some_Theory,GK_MAP_Construction}.  The test bed for MAP was developed in~\cite{GK_MAP_LS_2010,GK_MAP_Construction}.  The Lin-Kernighan heuristic is adapted for GTSP in~\cite{GK_GTSP_LK}.  Other GTSP local searches are presented in~\cite{GK_GTSP_LS_2010}.  A similar discussion of MAP local search can be found in~\cite{GK_MAP_LS_2008_old} and~\cite{GK_MAP_LS_2010}.  The memetic algorithm for GTSP is proposed in~\cite{GK_GTSP_Reduction2008}.  The population sizing and the memetic algorithm for MAP are suggested in~\cite{GK_MAP_MA_2009} and~\cite{GK_Population_Sizing_MAP}.

\section{Generalized Traveling Salesman Problem}
\label{sec:gtsp}

The Generalized Traveling Salesman Problem (GTSP) is an extension of the Traveling Salesman Problem (TSP)\@.  In GTSP, we are given a complete graph $G = (V, E)$, where $V$ is a set of $n$ vertices, and every edge $x \to y \in E$ is assigned a weight $w(x \to y)$.  We are also given a proper partition of $V$ into clusters $C_1, C_2, \ldots, C_m$, i.e., $C_i \cap C_j = \varnothing$ and $\bigcup_i C_i = V$.  A feasible solution, or a \emph{tour}, is a cycle visiting exactly one vertex in every cluster.  The objective is to find the shortest tour.
\nomenclature{GTSP}{stands for the Generalized Traveling Salesman Problem}
\nomenclature{Tour}{is a cycle visiting every cluster of the problem exactly once; sometimes we consider a tour as a set of its edges $T = \{ T_1 \to T_2,\ T_2 \to T_3,\ \ldots,\ T_m \to T_1 \}$}
\nomenclature[V]{$V$ (GTSP)}{is a set of all the vertices in the problem}
\nomenclature{TSP}{stands for the Traveling Salesman Problem}
\nomenclature[n]{$n$ (GTSP)}{is the number of vertices in the problem}
\nomenclature[m]{$m$ (GTSP)}{is the number of clusters in the problem}
\nomenclature[w]{$w(x \to y)$}{is the weight of an edge $x \to y$ in GTSP}

There also exists a variation of the problem where the tour is allowed to visit a cluster more than once, see, e.g.,~\cite{Fischetti1997}.  However, this variation is equivalent if the weights in the graph $G$ satisfy the triangle inequality $w(x \to y) \le w(x \to z \to y)$ for any $x, y, z \in V$.  In what follows, we consider the problem of finding the shortest cycle which visits \emph{exactly} one vertex in each cluster.

If the weight matrix is symmetric, i.e., $w(x \to y) = w(y \to x)$ for any $x \in V$ and $y \in V$, the problem is called \emph{symmetric}.  Otherwise it is an \emph{asymmetric} GTSP.
\nomenclature{Symmetric GTSP}{is a GTSP where $w(x \to y) = w(y \to x)$ for every $x$ and $y$}
\nomenclature{Asymmetric GTSP}{is a GTSP where $w(x \to y) \neq w(y \to x)$ for some $x$ and $y$}

There are many publications on GTSP (see, e.g., the surveys~\cite{Fischetti2002,Gutin2003}) and the problem has many applications in warehouse order picking with multiple stock locations, sequencing computer files, postal routing, airport selection and routing for courier planes and some others, see, e.g., \citep{Fischetti1995,Fischetti1997,Laporte1996,Noon1991} and references there.

The problem is NP-hard, since the \emph{Traveling Salesman Problem} (TSP) is a special case of GTSP when $|C_i| = 1$ for each $i$.  GTSP is trickier than TSP in the following sense: it is an NP-hard problem to find a minimum weight collection of vertex-disjoint cycles such that each cluster has exactly one vertex in the collection (and the claim holds even when each cluster has just two vertices)~\cite{Gutin2003a}.  Compare it with the well-known fact that a minimum weight collection of vertex-disjoint cycles covering the whole vertex set in a weighted complete digraph can be found in polynomial time~\cite{Gutin2007}.

\subsection{Additional Notation}

In what follows we use the following notation:
\begin{itemize}
	\item $s$ is the maximum cluster size.  Obviously $\lceil n / m \rceil \le s \le n - m + 1$.
	\nomenclature[s]{$s$ (GTSP)}{is the size of the largest cluster in an instance $s = \max_{i = 1, 2, \ldots, m} "|C_i"|$}

	\item $\gamma$ is the minimum cluster size.  Obviously $1 \le \gamma \le \lfloor n / m \rfloor$.
	\nomenclature[gamma]{$\gamma$}{is the size of the smallest cluster in a GTSP instance: $\gamma = \min_{i = 1, 2, \ldots, m} "|C_i"|$}

	\item $\Cluster(x)$ is the cluster containing the vertex $x$.
	\nomenclature[Cluster]{$\Cluster(x)$}{is the cluster in a GTSP instance containing the vertex $x$}

	\item $w(x_1 \to x_2 \to \ldots \to x_k)$ is the weight of a path $x_1 \to x_2 \to \ldots \to x_k$, i.e., $w(x_1 \to x_2 \to \ldots \to x_k) = w(x_1 \to x_2) + w(x_2 \to x_3) + \ldots + w(x_{k-1} \to x_k)$.
	\nomenclature[w]{$w(x \to y)$}{is the weight of the path $x_1 \to x_2 \to \ldots \to x_k$ in GTSP}
	
	\item $\displaystyle{w_\text{min}(X \to Y) = \min_{x \in X, y \in Y} w(x \to y)}$ denotes the minimum weight of an edge from a vertex set $X$ to a vertex set $Y$.  If one substitutes a vertex $v$ instead of, e.g., $X$ then we assume that $X = \{ v \}$.	Function $w_\text{max}(X \to Y)$ is defined similarly.
	\nomenclature[wmin]{$w_\text{min}(X \to Y)$}{is the weight of the shortest edge from a vertex set $X$ to a vertex set $Y$ in GTSP}
	\nomenclature[wmax]{$w_\text{max}(X \to Y)$}{is the weight of the longest edge from a vertex set $X$ to a vertex set $Y$ in GTSP}
	
	\item $T_i$ denotes the vertex at the $i$th position in the tour $T$.  We assume that $T_{i+m} = T_i$.
	\nomenclature[Ti]{$T_i$ (GTSP)}{is the vertex at the $i$th position in a tour $T$}

	\item Tour $T$ is also considered as a set of its edges, i.e., $T = \{ T_1 \to T_2,\ T_2 \to T_3,\ \ldots,\ T_{m-1} \to T_m,\ T_m \to T_1 \}$.
	
	\item $\Turn(T, x, y)$ denotes a tour obtained from $T$ by replacing the fragment $T_{x+1} \to T_{x+2} \to \ldots \to T_y$ with $T_y \to T_{y-1} \to \ldots \to T_{x+1}$:
	\begin{multline*}
	\Turn(T, x, y) = T_x \to T_y \to T_{y-1} \to T_{y-2} \to \ldots \to T_{x+1} \\
	\to T_{y+1} \to T_{y+2} \to \ldots \to T_{x-1} \to T_x \,.
	\end{multline*}
	Observe that for the symmetric GTSP the $\Turn(T, x, y)$ tour can be obtained by deleting the edges $T_x \to T_{x+1}$ and $T_y \to T_{y+1}$ and adding the edges $T_x \to T_y$ and $T_{x+1} \to T_{y+1}$:
	$$
	\Turn(T, x, y) = T \setminus \{ T_x \to T_{x+1},\ T_y \to T_{y+1} \} \cup \{ T_x \to T_y,\ T_{x+1} \to T_{y+1} \}
	$$
	and, hence, the weight of the obtained tour is as follows:
	\begin{multline}
	\label{eq:turn_delta}
	w(\Turn(T, x, y)) = w(T) - w(T_x \to T_{x+1}) - w(T_y \to T_{y+1})\\
	+ w(T_x \to T_y) + w(T_{x+1} \to T_{y+1}) \,.
	\end{multline}

	\nomenclature[Turn]{$\Turn(T, x, y)$}{is a tour obtained from a tour $T$ by replacing the fragment $T_{x+1} \to T_{x+2} \to \ldots \to T_y$ with $T_y \to T_{y-1} \to \ldots \to T_{x+1}$}
\end{itemize}

\subsection{Existing Approaches}
\label{sec:gtsp_existing_approaches}

Various approaches to GTSP have been studied.  There are exact algorithms such as branch-and-bound and branch-and-cut algorithms in~\cite{Fischetti1997}. While exact algorithms are very important, they are unreliable with respect to their running time that can easily reach many hours or even days. For example, the well-known TSP solver \textsc{Concorde}~\cite{Applegate2005} can easily solve some TSP instances with
several thousand cities, but it could not solve several asymmetric instances with 316 cities within the time limit of $10^4$~s~\cite{Fischetti1997}.

Several researchers~\citep{Ben-Arieh2003,Laporte1999,Noon1993} proposed transformations of GTSP into TSP\@.  At the first glance, the idea to transform a little-studied problem into a well-known one seems to be natural.  However, this approach has a very limited application.  Indeed, it requires exact solutions of the obtained TSP instances because even a near-optimal solution of such TSP may correspond to an infeasible GTSP solution.  At the same time, the produced TSP instances have quite unusual structure which is hard for the existing TSP solvers.  A more efficient way to solve GTSP exactly is a branch-and-bound algorithm designed by Fischetti et~al.~\citet{Fischetti1997}\@.  This algorithm was able to solve instances with up to 89 clusters.  Two approximation algorithms were proposed in the literature, however, both of them are unsuitable for the general case of the problem, and the guarantied solution quality is unreasonably low for the real-world applications, see~\cite{Bontoux2009} and references therein.

In order to obtain good (but not necessary exact) solutions for larger GTSP instances, one should use heuristic approach.  Several construction heuristics and local searches were discussed in~\cite{Bontoux2009,GK_GTSP_GA_2008,Hu2008,Renaud1998,Snyder2000} and some others.  A number of metaheuristics were proposed in~\cite{Bontoux2009,Huang2005,Pintea2007,Silberholz2007,Snyder2000,Tasgetiren2007,Yang2008}.

\section{Multidimensional Assignment Problem}
\label{sec:map}

The \emph{Multidimensional Assignment Problem} (MAP), abbreviated $s$-AP in the case of $s$ dimensions and also called \emph{(axial) Multi Index Assignment Problem} (MIAP) \citep{Bandelt2004,Pardalos2000}, is a well-known optimization problem.  It is an extension of the \emph{Assignment Problem} (AP), which is exactly the two dimensional case of MAP\@.  While AP can be solved in polynomial time \citep{Kuhn1955}, $s$-AP for every $s \ge 3$ is NP-hard \citep{Garey1979} and inapproximable \citep{Burkard1996}\footnote{Burkard et al.\ show it for a special case of 3-AP and since 3-AP is a special case of $s$-AP the result can be extended to the general MAP.}, i.e., there exists no $k$-approximation algorithm for any fixed $k$.
\nomenclature{MAP}{stands for the Multidimensional Assignment Problem}
\nomenclature[s-AP]{$s$-AP}{stands for the Multidimensional Assignment Problem}

The most studied case of MAP is the case of three dimensions \citep{Aiex2005,Andrijich2001,Balas1991,Crama1992,Huang2006,Spieksma2000} though the problem has a host of applications for higher numbers of dimensions, e.g., in matching information from several sensors (data association problem), which arises in plane tracking \citep{Murphey1998,Pardalos2000b}, computer vision \citep{Veenman2003} and some other applications \citep{Andrijich2001,Bandelt2004,Burkard1999}, in routing in meshes \citep{Bandelt2004}, tracking elementary particles \citep{Pusztaszeri1996}, solving systems of polynomial equations \citep{Bekker2005}, image recognition \citep{Grundel2004}, resource allocation \citep{Grundel2004}, etc.

For a fixed $s \ge 2$, the problem $s$-AP is stated as follows.  Let $X_1 = X_2 = \ldots = X_s = \{1, 2, \ldots, n \}$; we will consider only vectors that belong to the Cartesian product $X = X_1 \times X_2 \times \ldots \times X_s$.  Each vector $e \in X$ is assigned a non-negative weight $w(e)$.  For a vector $e \in X$, the component $e_j$ denotes its $j$th coordinate, i.e., $e_j \in X_j$.  A collection $A$ of $t \le n$ vectors $A^1, A^2, \ldots, A^t$ is a \emph{(feasible) partial assignment} if $A^i_j \neq A^k_j$ holds for each $i \neq k$ and $j \in \{ 1, 2, \ldots, s \}$.  The \emph{weight} of a partial assignment $A$ is $w(A) = \sum_{i=1}^t w(A^i)$.  A partial assignment with $n$ vectors is called \emph{assignment}.  The objective of $s$-AP is to find an assignment of minimal weight.  
\nomenclature[s]{$s$ (MAP)}{is the number of dimensions in the problem}
\nomenclature[n]{$n$ (MAP)}{is the cardinality of the sets $X_i$}
\nomenclature[Xi]{$X_i$ (MAP)}{is a set corresponding to the $i$th dimenstion of the problem (note that $"|X_i"| = n$)}
\nomenclature{AP}{stands for the Assignment Problem}
\nomenclature{Assignment}{is a set of $n$ disjoint vectors}
\nomenclature[Ai]{$A^i$ (MAP)}{is the $i$th vector of the assignment $A$ for $i = 1, 2, \ldots, n$}
\nomenclature[Aji]{$A_j^i$ (MAP)}{is the $j$th component of the $i$th vector of the assignment $A$ for $i = 1, 2, \ldots, n$ and $j = 1, 2, \ldots, s$}
\nomenclature[ej]{$e_j$ (MAP)}{is the $j$th component of the vector $e$ for $j = 1, 2, \ldots, s$}
\nomenclature{Partial assignment}{is a set of $t \le n$ disjoint vectors}
\nomenclature{Vector (MAP)}{is a vector $e$ of size $s$ such that $e_j \in X_j$ for every $j = 1, 2, \ldots, s$}
\nomenclature[w]{$w(e)$ (MAP)}{is the weight assigned to a vector $e$}
\nomenclature[w]{$w(A)$ (MAP)}{is the weight of an assignment $A$: $w(A) = \sum_{i=1}^n w(A^i)$}

\begin{figure}[ht]
\centerline{
\xymatrix@-1pc@R=20pt@C=50pt{
		*++[o][F-]{1} \ar@{.}[ddr] 
	&	*++[o][F-]{1} \ar@{-}[r] 
	&	*++[o][F-]{1}
\\
		*++[o][F-]{2} \ar@{-}[ur] 
	&	*++[o][F-]{2} \ar@{~}[dr] 
	&	*++[o][F-]{2}
\\
		*++[o][F-]{3} \ar@{~}[ur] 
	&	*++[o][F-]{3} \ar@{.}[dr] 
	&	*++[o][F-]{3}
\\
		*++[o][F-]{4} \ar@{-}[r] 
	&	*++[o][F-]{4} \ar@{-}[uur] 
	&	*++[o][F-]{4}
\\
	X_1 & X_2 & X_3
}
}
\caption[An example of a MAP assignment.]{An example of an assignment for a MAP with $s = 3$ and $n = 4$.  This assignment contains the following vectors: (1, 3, 4), (2, 1, 1), (3, 2, 3) and (4, 4, 2).  Note that to simplify the picture we show only a subset of edges for every clique.}
\label{fig:map_example}
\end{figure}
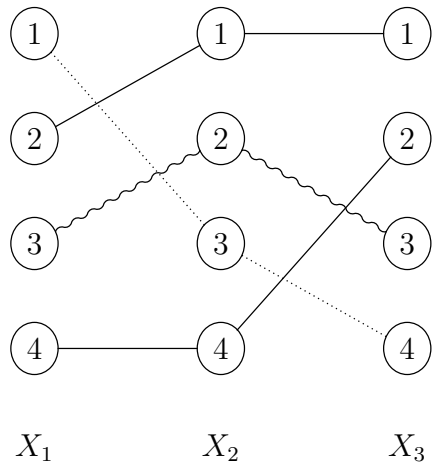

A graph formulation of the problem (see Fig.~\ref{fig:map_example}) is as follows.  Having a complete $s$-partite graph $G$ with parts $X_1$, $X_2$, \ldots, $X_s$, where $|X_i| = n$, find a set of $n$ disjoint cliques in $G$, each of size $s$, of the minimal total weight with every clique $Q$ in $G$ assigned a weight $w(Q)$ (note that in the general case $w(Q)$ is not simply a function of the edges of $Q$).

We also provide a \emph{permutation form} of the assignment which is sometimes very convenient.  Let $\pi_1, \pi_2, \ldots, \pi_s$ be permutations of $X_1, X_2, \ldots, X_s$, respectively.  Then $\pi_1 \pi_2 \ldots \pi_s$ is an assignment of weight $\sum_{i=1}^n w(\pi_1(i) \pi_2(i) \ldots \pi_s(i))$.  It is obvious that some permutation, say the first one, may be fixed without any loss of generality: $\pi_1 = 1_n$, where $1_n$ is the identity permutation of $n$ elements.  Then the objective of the problem is as follows:
$$
\min_{\pi_2, \ldots, \pi_s} \sum_{i=1}^n w(i \pi_2(i) \ldots \pi_s(i))
$$
and it becomes clear that there exist $n!^{s-1}$ feasible assignments and the fastest known algorithm to find an optimal assignment takes $O(n!^{s-2} n^3)$ operations.  Indeed, without loss of generality set $\pi_1 = 1_n$ and for every combination of $\pi_2, \pi_3, \ldots, \pi_{s-1}$ find the optimal $\pi_s$ by solving corresponding AP in $O(n^3)$.

Thereby, MAP is very hard; it has $n^s$ values in the weight matrix, there are $n!^{s-1}$ feasible assignments and the best known algorithm takes $O(n!^{s-2} n^3)$ operations.  Compare it, e.g., with the Travelling Salesman Problem which has only $n^2$ weights, $(n - 1)!$ possible tours and which can be solved in $O(n^2 \cdot 2^n)$ time \citep{Held1962}.

Finally, an integer programming formulation of the problem is as follows.
$$
\min \sum_{i_1 \in X_1, \ldots, i_s \in X_s} w(i_1 \ldots i_s) \cdot x_{i_1 \ldots i_s}
$$
subject to
$$
\sum_{i_2 \in X_2, \ldots, i_s \in X_s} x_{i_1 \ldots i_s} = 1 \qquad {\forall i_1 \in X_1} \,,
$$
$$
\ldots
$$
$$
\sum_{i_1 \in X_1, \ldots, i_{s-1} \in X_{s-1}} x_{i_1 \ldots i_s} = 1 \qquad {\forall i_s \in X_s} \,,
$$
where $x_{i_1 \ldots i_s} \in \{ 0, 1 \}$ for all $i_1$, \ldots, $i_s$ and $|X_1| = \ldots = |X_s| = n$.

The problem described above is called \emph{balanced} \citep{Clemons2004}.  Sometimes MAP is formulated in a more general way if $|X_1| = n_1$, $|X_2| = n_2$, \ldots, $|X_s| = n_s$ and the requirement $n_1 = n_2 = \ldots = n_s$ is omitted.  However, this case can be easily transformed into the balanced problem by complementing the weight matrix to an $n \times n \times \ldots \times n$ matrix with zeros, where $n = \max_i n_i$.

In what follows we assume that the number of dimensions $s$ is a small fixed constant while the size $n$ can be arbitrary large.  This corresponds to the real applications (see above) and also follows from the previous research, see, e.g., \cite{Bekker2005,Pusztaszeri1996,Robertson2001}.

\subsection{Existing Approaches}
\label{sec:map_existing_approaches}

MAP was studied by many researchers.  Several special cases of the problem were intensively studied in the literature (see \cite{Kuroki2007} and references there) but only for a few classes of them polynomial time exact algorithms were found, see, e.g., \cite{Burkard1996a,Burkard1996,Isler2005}.  In many cases MAP remains hard to solve \citep{Burkard1996,Crama1992,Kuroki2007,Spieksma1996}.  For example, if there are three sets of points of size $n$ on a Euclidean plane and the objective is to find $n$ triples, every triple has a point in each set, such that the total circumference or area of the corresponding triangles is minimal, the corresponding 3-APs are still NP-hard \citep{Spieksma1996}.  Apart from proving NP-hardness, researchers studied asymptotic properties of some special instance families~\citep{Grundel2004}.

As regards the solution methods, there exist several exact and approximation algorithms~\citep{Balas1991,Crama1992,Kuroki2007,Pasiliao2005,Pierskalla1968} and a number of heuristics including construction heuristics~\cite{Balas1991,Gutin2008,Oliveira2004}, greedy randomized adaptive search procedures~\cite{Aiex2005,Murphey1998,Oliveira2004,Robertson2001} (including several concurrent implementations, see, e.g.,~\cite{Aiex2005,Oliveira2004}) and a host of local search procedures~\cite{Aiex2005,Balas1991,Bandelt2004,Burkard1996,Clemons2004,Huang2006,Oliveira2004,Robertson2001}.

Two metaheuristics were proposed for MAP in the literature, namely a simulated annealing procedure \citep{Clemons2004} and a memetic algorithm \citep{Huang2006}.
\chapter{Some Aspects of Heuristics Design}
\label{sec:heuristics_design}

Heuristic design still mostly depends on the researcher's skills; there are just a few tools to support the scientist in this process.  In this chapter we show several examples of how such tools can help in heuristic design.

If a heuristic provides no solution quality or running time guarantee, the only choice to evaluate it is to use empirical analysis.  One of the most important aspects of computational experiment design is test bed selection.  In Section~\ref{sec:map_testbed} we discuss MAP test bed design.  

It turns out that there exist no standard test bed for MAP which would cover at least the most natural cases of the problem.  In Section~\ref{sec:map_testbed}, we gather all the instance classes proposed in the literature and systematize them.  We also split all the instances into two classes according to some important properties.  This helps in further experimental analysis.

Unfortunately, there is no way to find the optimal solutions for the instances of the MAP test bed even of a moderate size in any reasonable time.  However, in certain circumstances, it is possible to estimate the optimal solution values.  In Section~\ref{sec:map_random_estimation} we show an example of such estimation for one of the most widely used MAP instances family.

In Section~\ref{sec:gtsp_testbed} we show a successful example of producing a test bed for GTSP from a well-known TSP test bed.

Then we show two examples of improvement of heuristic performance.  Observe that even a small reduction of the problem size can noticeably speed up a powerful solver.  In Section~\ref{sec:gtsp_preprocessing} we propose two algorithms intended to reduce the size of a GTSP instance.  Our experiments show that this preprocessing may successfully reduce the running time of many GTSP algorithms known from the literature.


Another way to improve heuristic performance is to optimize the algorithm with respect to the hardware architecture.  In particular, an extremely important aspect is how the algorithm uses the main memory.  Indeed, computer memory is a complicated subsystem and its performance significantly depends on the way it is used.  It appears that one has to follow just a few simple rules in order to improve virtually any algorithm to make it `friendly' with respect to computer memory.  In Section~\ref{sec:performance_issues} we use three existing and one new construction heuristics for MAP as an example and show how these algorithms can be improved.

We also discuss the questions of selecting the most convenient and efficient data structures on the example of GTSP in Section~\ref{sec:gtsp_data_structures}.

\section{MAP Test Bed}
\label{sec:map_testbed}

The question of selecting proper test bed is one of the most important questions in heuristic experimental evaluation~\cite{Rardin2001}.  While many researchers of MAP focused on instances with random independent weights (\cite{Andrijich2001,Balas1991,Krokhmal2007,Pasiliao2005} and some others) or random instances with predefined solutions~\cite{Clemons2004,Grundel2005}, several more sophisticated models are of greater practical interest~\cite{Bandelt2004,Burkard1996,Crama1992,Frieze1981,Kuroki2007}.  There is also a number of papers which consider real-world and pseudo real-world instances~\cite{Bekker2005,Murphey1998,Pardalos2000b} but we suppose that these instances do not well represent all the instance classes and building a proper benchmark with the real-world instances is a subject for another research.

In this work, we propose splitting all the instance families into two classes: instances with independent weights (Section~\ref{sec:independent}) and instances with decomposable weights (Section~\ref{sec:decomposable}).  Later we will show that the heuristics perform differently on the instances of these classes and, thus, this division is very important for experimental analysis.

\subsection{Instances With Independent Weights}
\label{sec:independent}

One of the most studied classes of instances for MAP is \emph{Random Instance Family}.  In \random{}, the weight assigned to a vector is a random integer value uniformly distributed in the interval $[a, b - 1]$.  \random{} instances were used in~\cite{Aiex2005,Andrijich2001,Balas1991,Pierskalla1968} and some others.  Later, in Section~\ref{sec:map_random_estimation}, we will show that it is possible to estimate the optimal solution value of a large enough \random{} instance.
\nomenclature[random]{\random{} (MAP)}{is an instance family with independent weights, where the weight of a vector is a random number uniformly distributed in a certain range}

Another class of instances with almost independent weights is \emph{GP Instance Family} which contains pseudo-random instances with predefined optimal solutions.  \GP{} instances are generated by an algorithm produced by Grundel and Pardalos~\cite{Grundel2005}.  The generator is naturally designed for $s$-AP for arbitrary large values of $s$ and $n$.  However, the generating algorithm is exponential and, thus, it was impossible to generate any \GP{} instances even of a moderate size.  Nevertheless, this is what we need since finally we have both small (\GP) and large (\random) instances with independent weights with known optimal solutions.
\nomenclature[GP]{\GP}{is a MAP instance family with independent weights with predefined optimal solutions}

\subsection{Instances With Decomposable Weights}
\label{sec:decomposable}

In many cases it is not easy to define a weight for an $s$-tuple of objects but it is possible to define a relation between every pair of objects from different sets.  In this case one should use \emph{decomposable weights}~\cite{Spieksma2000}.  Then the weight of a vector $e$ is defined as follows:
\begin{equation}
\label{eq:decomposable_w}
w(e) = f\left(d^{1, 2}_{e_1, e_2}, d^{1, 3}_{e_1, e_3}, \ldots, d^{s-1,s}_{e_{s-1}, e_s}\right) \,,
\end{equation}
where $d^{i,j}$ is a weight matrix for the sets $X_i$ and $X_j$ and $f$ is some function.

The most straightforward instance family with decomposable weights is \clique{}.  It defines the function $f$ as the sum of all the arguments:
\begin{equation}
\label{eq:clique_w}
w_\text{c}(e) = \sum_{i=1}^{n-1} \sum_{j=i+1}^{n} d^{i,j}_{e_i, e_j} \,.
\end{equation}
The \clique\ instance family was investigated in~\cite{Bandelt2004,Crama1992,Frieze1981} and some others.  It was proven~\cite{Crama1992} that MAP restricted to \clique\ instances remains NP-hard.
\nomenclature[clique]{\clique{}}{is a MAP instance family with decomposable weights}

A special case of \clique{} is \emph{Geometric Instance Family}.  In \geometric{}, each of the sets $X_1$, $X_2$, \ldots, $X_s$ corresponds to a set of points in Euclidean space, and the distance between two points $u \in X_i$ and $v \in X_j$ is defined as Euclidean distance; we consider the two dimensional Euclidean space:
$$
d_\text{g}(u, v) = \sqrt{(u_x - v_x)^2 + (u_y - v_y)^2} \,.
$$
It is proven~\cite{Spieksma1996} that the \geometric{} instances are NP-hard to solve for $s = 3$ and, thus, \geometric{} is NP-hard for every $s \ge 3$.
\nomenclature[geometric]{\geometric{}}{is a MAP instance family with decomposable weights}

We propose a new instance family with decomposable weights, \squareroot{}.  It is a modification of the \clique{} instance family.  Assume we have $s$ radars and $n$ planes and each radar observes all the planes.  The problem is to assign signals which come from different radars to each other.  It is quite natural to define some distance function between each pair of signals from different radars which would correspond to the similarity of these signals.  Then for a set of signals which correspond to one plane the sum of these distances is expected be small and, hence, (\ref{eq:clique_w}) is a good choice.  However, it is not actually correct to minimize the total distance between the signals; one should also ensure that none of these distances is too large.  Note that the same requirement appears in a number of other applications.  We propose a weight function which aims to both small total distance between the assigned signals and small dispersion of these distances:
\begin{equation}
w_\text{sq}(e) = \sqrt{\sum_{i=1}^{n-1} \sum_{j=i+1}^{n} \left(d^{i,j}_{e_i, e_j}\right)^2 } \,.
\end{equation}
Similar approach is used in~\cite{Kuroki2007} though they do not use square root, i.e., a vector weight is just a sum of squares of the edge weights in a clique.  In addition, the edge weights in~\cite{Kuroki2007} are calculated as distances between some nodes in a Euclidean space.
\nomenclature[squareroot]{\squareroot{}}{is a MAP instance family with decomposable weights}

Another special case of the decomposable weights, \product{}, is studied in~\cite{Burkard1996}.  Burkard et al.\ consider 3-AP and define the weight $w(e)$ as $w(e) = a^1_{e_1} \cdot a^2_{e_2} \cdot a^3_{e_3}$, where $a^1$, $a^2$ and $a^3$ are random vectors of positive numbers.  We generalize the \product{} instance family for $s$-AP: $\displaystyle{w_\text{p}(e) = \prod_{i=1}^s a_{e_i}^i}$.
It is easy to show that the \product{} weight function can be represented in the form~(\ref{eq:decomposable_w}).
Note that the minimization problem for the \product{} instances is proven to be NP-hard in case $s = 3$ and, thus, it is NP-hard for every $s \ge 3$.
\nomenclature[product]{\product{}}{is a MAP instance family with decomposable weights}

\subsection{Additional Details}
\label{sec:map_testbed_technical_detail}

We include the following instances in our test bed:
\begin{itemize}
    \item \random{} instances where each weight was randomly chosen in $\{ 1, 2, \ldots, 100 \}$, i.e., $a = 1$ and $b = 101$.  We will show in Section~\ref{sec:independent} that the weights of the optimal solutions of all the considered \random\ instances are very likely to be $an = n$.

    \item \GP{} instances with predefined optimal solutions.

    \item \clique\ and \squareroot\ instances, where the weight of each edge in the graph was randomly selected from $\{ 1, 2, \ldots, 100 \}$.  Instead of the optimal solution value we use the best known solution value.

    \item \geometric\ instances, where both coordinates of every point were randomly selected from $\{ 1, 2, \ldots, 100 \}$.  The distances between the points are calculated precisely while the weight of a vector is rounded to the nearest integer.  Instead of an optimal solution value we use the best known solution value.

    \item \product\ instances, where every value $a_i^j$ was randomly selected from $\{ 1, 2, \ldots, 10 \}$.  Instead of an optimal solution value we use the best known solution value.
\end{itemize}

An instance name consists of three parts: the number $s$ of dimensions, the type of the instance (`gp' for \GP{}, `r' for \random{}, `cq' for \clique{}, `g' for \geometric, `p' for \product\ and 'sr' for \squareroot), and the size $n$ of the instance.  For example, \texttt{5r40} means a five dimensional \random\ instance of size 40.  For every combination of instance size and type we generated 10 instances, using the number $\var{seed} = s + n + i$ as a seed of the random number sequences, where $i$ is an index of the instance of this type and size, $i \in \{ 1, 2, \ldots, 10 \}$.  Thereby, every experiment is conducted for 10 different instances of some fixed type and size, i.e., every number reported in the tables below is an average for 10 runs, one for each of the 10 instances.  This smooths out the experimental results.

\section{Probabilistic Analysis of Test Beds}
\label{sec:map_random_estimation}

In experimental analysis of heuristics, it is important to know the optimal solutions for test bed instances.  However, it is not always possible to solve every instance to optimality.  Indeed, heuristic approach is usually applied when exact algorithms fail to solve the problem in any reasonable time.

If the optimal solution cannot be obtained for an instance, one can use a lower or an upper bound instead or apply probabilistic analysis of the instance if it is generated randomly.  In this section we apply the latter approach to the MAP \random{} instance family.  

Recall that in \random{} instances, the weight assigned to a vector is an independent random uniformly distributed integer in the interval $[a, b - 1]$.  Let us estimate the average solution value for \random.  In fact, we prove that it is very likely that every large enough \random{} instance has an assignment of weight $an$, i.e., a minimal possible assignment (observe that a minimal assignment includes $n$ vectors of weight $a$).

Let $\alpha$ be the number of assignments of weight $na$ and let $c = b - a$.  We would like to have an upper bound on the probability $\Pr(\alpha = 0)$.  Such an upper bound is given in the following theorem whose proof is based on the Extended Jansen Inequality given in Theorem 8.1.2 of~\cite{Alon2000}.

\begin{theorem}
\label{th:pr}
For values of $n$ such that $n\ge 3$ and
\begin{equation}
\label{eq:cbound}
\left( \frac{n-1}{e} \right)^{s-1} \ge c \cdot 2^{\frac{1}{n-1}},
\end{equation}
we have $\Pr(\alpha = 0) \le e^{-\frac{1}{2 \sigma}}$, where
$\sigma = \sum\limits_{k = 1}^{n - 2} \frac{\binom{n}{k} \cdot c^k}{\left[n \cdot (n - 1) \cdots (n - k + 1) \right]^{s - 1}}$.
\end{theorem}
\proof Let $[n] = \{ 1, 2, \ldots, n \}$.  Let $t$ be the number of
all feasible assignments and let $A$ be an arbitrary assignment
consisting of vectors $e^1, e^2, \ldots, e^n$ such that $e^i_1 = i$
for each $i \in [n]$.  There are $n!$ possibilities to choose the
$j$th coordinate of all vectors in $A$ for each $j = 2, 3, \ldots,
n$ and, thus, $t = (n!)^{s-1}$.

Let $R$ be the set of vectors in $X$ of weight $a$ and let $\{ A_1,
A_2, \ldots, A_t \}$ be the set of all assignments.  Let $B_i$ be
the event $\{ A_i \subset R\}$ for each $i \in [t]$.  Let $\mu =
\sum_{i = 1}^t \Pr(B_i)$ and $\Delta = \sum_{i \sim j} \Pr(B_i \cap
B_j)$, where $i \sim j$ if $i \neq j$ and $A_i \cap A_j \neq
\varnothing$ and the sum for $\Delta$ is taken over all ordered pairs
$(B_i, B_j)$ with $i \sim j$.

By the Extended Jansen Inequality,
\begin{equation}
\label{eq:EJI}
\Pr(\alpha = 0) \le e^{-\frac{\mu^2}{2 \Delta}}
\end{equation}
provided $\Delta \ge \mu.$  We will compute $\mu$ and estimate
$\Delta$ to apply (\ref{eq:EJI}) and to show $\Delta \ge \mu$. It is
easy to see that $\mu = \frac{t}{c^n}$.

Now we will estimate $\Delta$.  Let $A_i \cap A_j = K$, $k = |K|$ and $i \neq j$.  Thus, we have
$$
\Pr(B_i \cap B_j) = \Pr(K \subset R) \cdot \Pr(A_i \setminus K \subset R) \cdot \Pr(A_j \setminus K \subset R) = \frac{1}{c^k} \left( \frac{1}{c^{n - k}} \right)^2 = \frac{1}{c^{2n - k}}.
$$

Let $(f^1, f^2, \ldots, f^n)$ be an assignment with $f^i_1 = i$ for
every $i \in [n]$ and consider the following two sets of assignments.
Let
$$
P(k) = \{ (e^1, e^2, \ldots, e^n):\ \forall i \in [n]\ (e^i_1=i) \mbox{
and } \forall j\in [k]\ (e^j=f^j) \}
$$
and let $Q(n-k) = \{(e^1, e^2, \ldots, e^n):\ \forall i\in [n]\ (e^i_1=i) \mbox{
and } \forall j\in [n-k]\ (e^{k+j}\neq f^{k+j}) \}$.  Let $h(n,k)=|P(k)\cap Q(n-k)|$. Clearly, $h(n,k) \le |P(k)|=\big((n-k)!\big)^{s-1}$. Observe that
$$h(n,k)\ge |P(k)|-(n-k)|P(k+1)|=L(n,k,s),$$ where $L(n,k,s) = \big((n-k)!\big)^{s-1} - (n-k) \cdot \big((n-k-1)!\big)^{s-1}$.

Let $g(n, k)$ be the number of ordered pairs $(A_i, A_j)$ such that
$|A_i \cap A_j| = k$. Observe that $g(n,k) = t \cdot \binom{n}{k}
\cdot h(n,k)$ and, thus, $t \cdot {\binom{n}{k}} \cdot L(n, k, s)
\le g(n, k) \le t \cdot \binom{n}{k} \cdot \big((n-k)!\big)^{s-1}$.

Observe that $\Delta = \sum\limits_{k = 1}^{n - 2} \sum\limits_{|A_i
\cap A_j| = k} \Pr(B_i \cap B_j) = \sum\limits_{k = 1}^{n - 2} g(n,
k) \cdot c^{k - 2n}$.  Thus,
\begin{equation}
\label{eq:alpha=0}
\frac{(n!)^{s-1}}{c^{2n}} \cdot \sum_{k=1}^{n-2}{\binom{n}{k}} \cdot c^k \cdot L(n, k, s)
\le \Delta
\le \frac{(n!)^{s-1}}{c^{2n}} \sum_{k = 1}^{n - 2}{\binom{n}{k}} \cdot c^k \cdot \big((n-k)!\big)^{s-1}
\end{equation}

Now  $\Pr(\alpha = 0) \le e^{-\frac{1}{2 \sigma}}$ follows from
(\ref{eq:EJI}) by substituting $\mu$ with $\frac{(n!)^{s-1}}{c^n}$
and $\Delta$ with its upper bound in (\ref{eq:alpha=0}).  It remains
to prove that $\Delta\ge \mu.$  Since $n \ge 3$, $L(n, 1, s) \ge
\frac{1}{2} \big((n-1)!\big)^{s-1}$.  By the lower bound for
$\Delta$ in (\ref{eq:alpha=0}), we have $\Delta \ge
\frac{(n!)^{s-1}}{c^{2n-1}} \cdot L(n,1,k)$.  Therefore,
$\frac{\Delta}{\mu} \ge \frac{0.5 ((n-1)!)^{s-1}}{c^{n-1}}$.  Now
using the inequality $(n-1)!>(\frac{n-1}{e})^{n-1}$, we conclude
that $\frac{\Delta}{\mu} \ge 1$ provided (\ref{eq:cbound})
holds.\qed

\smallskip

Useful results can also be obtained from (11) in~\cite{Grundel2004} that is an upper bound for the average optimal solution.  Grundel, Oliveira and Pardalos~\cite{Grundel2004} consider the same instance family except the weights of the vectors are real numbers uniformly distributed in the interval $[a, b]$. However the results from~\cite{Grundel2004} can be extended to our discrete case.  Let $w'(e)$ be a real weight of the vector $e$ in a continuous instance.  Consider a discrete instance with $w(e) = \lfloor w'(e) \rfloor$ (if $w'(e) = b$, set $w(e) = b - 1$).  Note that the weight $w(e)$ is a uniformly distributed integer in the interval $[a, b - 1]$.  The optimal assignment weight of this instance is not larger than the optimal assignment weight of the continuous instance and, thus, the upper bound for the average optimal solution for the discrete case is correct.

In fact, the upper bound $\bar{z}^*_u$ (see~\cite{Grundel2004}) for the average optimal solution is not really accurate.  For example, $\bar{z}^*_u \approx an + 6.9$ for $s = 3$, $n = 100$ and $b - a = 100$, and $\bar{z}^*_u \approx an + 3.6$ for $s = 3$, $n = 200$ and $b - a = 100$.  It gives a better approximation for larger values of $s$, e.g., $\bar{z}^*_u \approx an + 1.0$ for $s = 4$, $n = 40$ and $b - a = 100$, but Theorem~\ref{th:pr} provides stronger results ($\Pr(\alpha > 0) \approx 1.000$ in the latter case).

The following table gives the probabilities for $\Pr(\alpha > 0)$ for various values of $s$ and $n$:
\begin{center}
\begin{tabular*}{1\textwidth}{@{\extracolsep{\fill}} *5{c}}
\toprule
$s = 4$
    & $s = 5$
    & $s = 6$
    & $s = 7$ \\

\cmidrule(r){1-4}

\begin{tabular}[t]{r @{\quad} l}
$n$ & $\Pr(\alpha > 0)$ \\
\cmidrule(){1-2}
15 & 0.575 \\
20 & 0.823 \\
25 & 0.943 \\
30 & 0.986 \\
35 & 0.997 \\
40 & 1.000 \\
\end{tabular}
&
\begin{tabular}[t]{r @{\quad} l}
$n$ & $\Pr(\alpha > 0)$ \\
\cmidrule(){1-2}
10 & 0.991 \\
11 & 0.998 \\
12 & 1.000 \\
\end{tabular}
&
\begin{tabular}[t]{r @{\quad} l}
$n$ & $\Pr(\alpha > 0)$ \\
\cmidrule(){1-2}
8 & 1.000 \\
\end{tabular}
&
\begin{tabular}[t]{r @{\quad} l}
$n$ & $\Pr(\alpha > 0)$ \\
\cmidrule(){1-2}
7 & 1.000 \\
\end{tabular} \\
\bottomrule
\end{tabular*}
\end{center}

\section{GTSP Test Bed}
\label{sec:gtsp_testbed}

There is a standard test bed which was used in most of the recent literature on GTSP, see, e.g.,~\citep{Silberholz2007,Snyder2000,Tasgetiren2007,Tasgetiren2010}\@.  It is produced from a well-known TSP test bed called TSPLIB~\citep{TSPLIB}\@.  TSPLIB contains both symmetric and asymmetric instances of different sizes and types, some pseudo-random and some real-world ones.

The procedure of generating a GTSP instance from a TSP instance was proposed by Fischetti, Salazar, and Toth~\citep{Fischetti1997}.  It is applicable to both symmetric and asymmetric TSP instances and produces symmetric and asymmetric GTSP instances, respectively.  The number of vertices $n$ in the produced GTSP instance is the same as in the original TSP instance; the number $m$ of clusters is $m = \lceil n / 5 \rceil$.  The clusters are `localized', i.e., the procedure attempts to group close vertices into the same clusters.

TSPLIB includes instances from as few as 14 to as many as 85900 vertices.  In our experiments, we usually consider instances with $10 \le m \le 217$.  The same test bed is used in a number of papers, see, e.g.,~\citep{Bontoux2009,GK_GTSP_GA_2008,GK_GTSP_LK,Silberholz2007}.  In other papers the bounds are even smaller.

Every instance name consists of three parts: `$m$ $t$ $n$', where $m$ is the number of clusters, $t$ is the type of the original TSP instance (see~\citep{TSPLIB} for details) and $n$ is the number of vertices.

Observe that the optimal solutions are known only for certain instances with up to 89 clusters~\citep{Fischetti1997}.  For the rest of the test bed we use the best known solutions obtained in our experiments and from the literature~\cite{Bontoux2009,Silberholz2007}.

\section{Preprocessing}
\label{sec:gtsp_preprocessing}

Preprocessing is a procedure of an instance simplification.  It is used to reduce the computation time of a solver.  There are several examples of such approaches in integer and linear programming (e.g.,~\cite{Gutman2007, Savelsbergh1994}) as well as for the Vehicle Routing Problem~\cite{Laporte1992}.  In some cases preprocessing plays the key role in an algorithm (see, e.g.,~\cite{Fourer1994}).  Next we propose two efficient preprocessing procedures for GTSP\@.

\subsection{GTSP Reduction Algorithms}

An important feature of GTSP is that a feasible tour does not visit every vertex of the problem and, thus, GTSP may contain vertices that a priori cannot be included in the optimal tour and, hence, may be removed in advance.

\begin{definition} 
\label{definition:excess}
Let $C$ be a cluster, $|C| > 1$. We say that a vertex $r \in C$ is \emph{redundant} if, for each pair $x$ and $y$ of vertices from distinct clusters different from $C$, there exists $r' \in C \setminus \{ r \}$ such that $w(x \to r' \to y) \leq w(x \to r \to y)$.
\end{definition}

Testing this condition for every vertex takes approximately $O(n^3 s)$ operations.  In some cases it is possible to significantly reduce the preprocessing time for symmetric instances.

Let us fix some vertex $r \in C$.  For every $r' \in C$ and every $x \notin C$ calculate the value $\Delta_{x}^{r, r'} = w(x \to r) - w(x \to r')$.  Observe that a vertex $r$ is redundant if there is no pair of vertices $x, y \notin C$ from different clusters such that $\Delta_x^{r,r'} + \Delta_y^{r,r'} < 0$ for every $r'$, i.e., $r$ is redundant if for every $x, y \notin C$, $\Cluster(x) \neq \Cluster(y)$, there exists $r' \in C \setminus \{ r \}$ such that $\Delta_{x}^{r, r'} + \Delta_{y}^{r, r'} \geq 0$.  That is due to $\Delta_{x}^{r, r'} + \Delta_{y}^{r, r'} = w(x \to r) - w(x \to r') + w(y \to r) - w(y \to r') = w(x \to r \to y) - w(x \to r' \to y)$.

There is a way to accelerate the algorithm.  If
$$
\min_{x \notin Z} \max_{r' \in C} \Delta_{x}^{r, r'} + \min_{x \in Z} \max_{r' \in C} \Delta_{x}^{r, r'} < 0
$$
for some cluster $Z$, then we know immediately that $r$ cannot be reduced.  We can use an equivalent condition:
$$
\min_{x \in \bigcup_{j < i} V_j} \max_{r' \in C} \Delta_{x}^{r, r'}
    + \min_{x \in V_i} \max_{r' \in C} \Delta_{x}^{r, r'} < 0
$$
This condition can be tested during the $\Delta$ values calculation by accumulating the value of $\displaystyle{\min_{x \in \bigcup_{j < i} V_j} \max_{r' \in C} \Delta_{x}^{r, r'}}$.

Removing a redundant vertex may cause a previously irredundant vertex to become redundant.  Thus, it is useful to check redundancy of vertices in cyclic order until we see that, in the last cycle, no vertices are found to be redundant.  However, in the worst case, that would lead to the total number of the redundancy tests to be $\Theta(n^2)$.  Our computational experience has shown that almost all redundant vertices are found in two cycles.  Hence, we never conduct the redundancy test more than twice for a vertex.

\bigskip

Similar to the vertices, it is sometimes possible to say in advance that a certain edge cannot be included in the optimal tour.

\begin{definition} 
\label{definition:redundantedge}
Let $u$ and $v$ be a pair of vertices from distinct clusters $U$ and $C$ respectively.  Then the edge $u \to v$ is \emph{redundant} if for each vertex $x \in V \setminus (U \cup C)$ there exists $v' \in C \setminus \{ v \}$ such that $w(u \to v' \to x) \leq w(u \to v \to x)$.
\end{definition}

We propose the following algorithm for the edge reduction.  Given a
vertex $v \in C$, $|C| > 1$, we detect redundant edges incident with
$v$ using the following procedure:
\begin{enumerate}
    \item Select an arbitrary vertex $v'' \in C \setminus \{ v \}$.

    \item Set $P_x = \Delta_{x}^{v, v''}$ for each vertex $x \in V \setminus C$.

    \item Sort the array $P$ in non-decreasing order.

    \item For each cluster $U \neq C$ and for each vertex $u \in U$ do the following:
        \begin{enumerate}
            \item $\delta = \Delta_u^{v, v''}$.
            \item \label{EdgeRedundantCheck} For each item $\Delta_{x}^{v, v''}$ of the array $P$ such that $\Delta_x^{v, v''} + \delta < 0$ check: if $x \notin U$ and $\Delta_x^{v, v'} + \Delta_{u}^{v, v'} < 0$ for every $v' \in C \setminus \{ v, v'' \}$, the edge $u \to v$ is immediately not redundant, continue with the next $u$.
            \item Edge $u \to v$ is redundant, set $w(u \to v) = \infty$.
        \end{enumerate}
\end{enumerate}

To prove that the above edge reduction algorithm works correctly, let us fix some edge $u \to v$, $u \in U$, $v \in C$, $U \neq C$.  The algorithm declares this edge redundant if the following condition holds for each $x \notin C$ (see \ref{EdgeRedundantCheck}):
\begin{align*} \label{EdgeRedundantCondition}
    \Delta_{x}^{v, v''} + \Delta_{u}^{v, v''} \geq 0 & \qquad \text{or}\\
    \Delta_{x}^{v, v'} + \Delta_{u}^{v, v'} \geq 0 & \qquad \text{for some } v' \in C \setminus \{ v, v'' \} \,.
\end{align*}
This is equivalent to
$$
    \Delta_{x}^{v, v'} + \Delta_{u}^{v, v'} \geq 0 \qquad \text{for some } v' \in C \setminus \{ v \} \,.
$$

So the algorithm declares the edge $u \to v$ redundant if for each $x \in V \setminus C \setminus U$ there exists $v' \in C \setminus \{ v \}$ such that $\Delta_x^{v, v'} + \Delta_u^{v, v'} \geq 0$:
\begin{align*}
    & w(x \to v) - w(x, v') + w(u \to v) - w(u \to v') \geq 0 \text{ and, hence,} \\
    & w(u \to v \to x) \geq w(u \to v' \to x) \,.
\end{align*}

\bigskip

The edge reduction procedure is executed exactly once for every vertex $v$ such
that $|\Cluster(v)| > 1$.  The whole algorithm takes
$O(n^3 s)$ operations.

\subsection{GTSP Reduction Experimental Evaluation}

We have tested three reduction algorithms: the Vertex Reduction Algorithm, the Edge Reduction Algorithm, and the Combined Algorithm which first applies the Vertex Reduction and then the Edge Reduction.

The columns of Table~\ref{tab:gtsp_reduction_results} are as follows:
\begin{itemize}
    \item \emph{Instance} is the instance name.  Note that the suffix number in the name is the number of vertices before any preprocessing.
    \item $R_v$ is the number of vertices detected as redundant.
    \item $R_e$ is the number of edges detected as redundant.  For the Combined Algorithm, $R_e$ shows the number of redundant edges in the instances already reduced by the Vertex Reduction.
    \item $T$ is preprocessing time, in seconds.
\end{itemize}

All the algorithms are implemented in C++; the evaluation platform is based on an AMD Athlon 64 X2 Core Dual processor (3~GHz frequency).

The results of the experiments (Table~\ref{tab:gtsp_reduction_results}) show that the preprocessing time for the Vertex Reduction is negligible (less than 50~ms) for all the instances up to \texttt{212u1060}, i.e., for almost all \texttt{TSPLIB}-based GTSP instances used in the literature. The average percentage of detected redundant vertices for these instances is 14\%, and it is 11\% for all considered instances.  The experimental complexity of the Vertex Reduction algorithm is about $O(n^{2.4})$.

The Edge Reduction is more time-consuming than the Vertex Reduction.  The running time is negligible for all instances up to \texttt{115rat575}.  Note that in most of the GTSP literature, only instances with $m \le 89$ are considered.  The average per cent of the detected redundant edges for these instances is about 27\%, and it is 21\% for all the instances in Table~\ref{tab:gtsp_reduction_results}.  The experimental complexity of the Edge Reduction algorithm is $O(n^{2.6})$.

\subsection{Influence of GTSP Reduction on Solvers}

Certainly, one can doubt the usefulness of our reduction algorithms since they may not necessarily decrease the running time of GTSP solvers.  Therefore, we have experimentally checked if the reductions are beneficial for several powerful GTSP solvers (obviously, preprocessing is useless in combination with a fast solver since preprocessing may take more time than the solver itself):
\begin{itemize}
    \item An exact algorithm (\ExactSolver{}) based on a transformation of GTSP to TSP~\cite{Ben-Arieh2003}; the algorithm from~\cite{Fischetti2002} was not available.  The algorithm that we use converts a GTSP instance with $n$ vertices to a TSP instance with $3 n$ vertices in the polynomial time, solves the obtained TSP using the \textsc{Concorde} solver~\cite{Applegate2005}, and then converts the obtained TSP solution to GTSP solution also in the polynomial time.

    \item A memetic algorithm from~\cite{Snyder2000} (\SD{}).

    \item A memetic algorithm from~\cite{Silberholz2007} (\SG{}).

    \item A modified version of our memetic algorithm (see Section~\ref{sec:gtsp_ma}) (\GK{}).
\end{itemize}


For each stochastic algorithm, every test was repeated ten times and an average was used.  The columns of the tables not described above are as follows:
\begin{itemize}
    \item $T_0$ is the original solution time.
    \item $B$ is the time benefit, i.e., $\dfrac{T_0 - T_\text{pr}}{T_0} \cdot 100\%$, where $T_\text{pr}$ is the solution time after preprocessed; this includes the preprocessing time.
\end{itemize}


The experiments (see Tables~\ref{tab:reduction_exact},~\ref{tab:reduction_sd},~\ref{tab:reduction_sg} and~\ref{tab:reduction_gk}) show that the Vertex Reduction, the Edge Reduction and the Combined Reduction Techniques significantly reduce the running time of the \ExactSolver{} and \SD{} solvers.  However, the Edge Reduction (and because of that the Combined Reduction Technique) is not that successful for \SG{} (Table~\ref{tab:reduction_sg}) and the original version of \GK{}\@.  That is because not every algorithm processes infinite or enormous edges well.  

We have adjusted our solver \GK{} to work better with preprocessed instances.  The details of the modified version can be found in Section~\ref{sec:gtsp_ma_modified}.  The modified algorithm does not reproduce exactly the results of the original \GK{} heuristic; it produces slightly better solutions at the cost of slightly larger running times.  However, one can see (Table~\ref{tab:reduction_gk}) that all the Reduction Algorithms proposed in this section influence the modified \GK{} algorithm positively.

Different reductions have different degree of success for different solvers.  The Edge Reduction is more efficient than the Vertex Reduction for \SD{}; in other cases the Vertex Reduction is more successful.  For every solver except \SG{} the Combined Technique is preferred to single reductions.

\bigskip

Preprocessing is called to reduce the solution time.  On the other hand, there is no guarantee that the outcome of the preprocessing will be noticeable.  Thus, it is important to ensure that, at least, preprocessing is significantly faster than the solver.

Four GTSP solvers are considered in this section.  The first solver, \ExactSolver{}, is exponential and, thus, it is clear that its time complexity is larger than the one of the reduction algorithms.  The time complexities of the other three solvers were estimated experimentally.  The experimental complexity of \SD{} is about $\Theta(n^3)$ and it is about $\Theta(n^{3.5})$ for \GK{} and \SG{}\@.

Since the time complexity of every considered solver is higher than the time complexity of preprocessing, we can conclude that preprocessing remains relatively fast for an arbitrary large instance.

Note that the solution quality of the considered solvers was not affected by the reductions, on average.

\subsection{MAP Preprocessing}

We do not discuss any preprocessing for MAP\@.  Observe that the Vertex Reduction proposed above for GTSP preserves the structure of the problem and, hence, any ordinary GTSP solver may be applied to a reduced instance.  However, it is not the case for the Edge Reduction (see above) since the yielded instances may contain infinite edges.

It is unlikely that a MAP preprocessing can reduce the values of $s$ or $n$.  One can rather think of removing certain vectors from the $X$ set, however, this would change the problem structure and, hence, complicate the solvers.

\section{Implementation Performance Issues}
\label{sec:performance_issues}

It may seem that implementation is a technical question which is not worth discussion because its influence on the algorithm's performance is negligible and, moreover, platform-dependent.  In this section we will show that this assumption is sometimes very wrong.  It turns out that two formally equal implementations of some algorithm may have very different running times in certain circumstances.  In particular, we will show that some simple transformations of an algorithm may be crucial with respect to efficiency of processor cache usage.  Note that this discussion does not involve any specifics of particular CPUs and is relevant to all the computers produced in at least last 30 years.

For a case study we need some algorithms which deal with large amounts of data.  For this purpose we selected MAP construction heuristics.  Recall that MAP instance is defined by an $s$-dimensional matrix of size $n$, i.e., it requires $n^s$ values.  Construction heuristics are very quick and, hence, we are able to consider very large instances such that the weight matrices exceed the size of processor cache.

In Section~\ref{sec:map_construction_heuristics} we describe all MAP construction heuristics known from the literature and propose a new one, \shiftrom{}\@.  In Section~\ref{sec:performance_notes} we discuss the efficiency of computer memory in certain circumstances and provide several simple rules to improve performance of an algorithm implementation.  Then, in Section~\ref{sec:map_construction_heuristics_improved}, we show how these rules can be applied to the MAP construction heuristics.  We do not provide any experimental analysis here; one can refer to~\cite{GK_MAP_Construction} for details.  We only declare here that the refinements proposed below speed up each of the considered heuristics in, roughly speaking, 2 to 5 times.

\subsection{MAP Construction Heuristics}
\label{sec:map_construction_heuristics}

\subsubsection{Greedy}
\label{sec:map_greedy}

The \greedy{} heuristic starts with an empty partial assignment $A = \varnothing$.  On each of $n$ iterations \greedy{} finds a vector $e \in X$ of minimum weight, such that $A \cup \{ e \}$ is a feasible partial assignment, and adds it to $A$.

The time complexity of \greedy{} heuristic is $O(n^s + (n-1)^s + \ldots + 2^s + 1) = O(n^{s+1})$ (if the \greedy{} algorithm is implemented via sorting of all the vectors according to their weights, the algorithm complexity is $O(n^s \cdot \log{n^s})$ however this implementation is inefficient, see Section~\ref{sec:map_greedy_improved}).

\subsubsection{Max-Regret}
\label{sec:map_maxregret}

The \maxregret{} heuristic was first introduced in~\cite{Balas1991} for 3-AP and its modifications for s-AP were considered in~\cite{Bekker2005}.

\maxregret{} proceeds as follows.  Initialize a partial assignment $A = \varnothing$.  Set $V_d = \{ 1, 2, \ldots, n \}$ for each $1 \le d \le s$.  For each dimension $d$ and each coordinate value $v \in V_d$ consider every vector $e \in X'$ such that $e_d = v$, where $X' \subset X$ is the set of `available' vectors, i.e., $A \cup \{ e \}$ is a feasible partial assignment if and only if $e \in X'$.  Find two vectors ${e^1}_\text{min}$ and ${e^2}_\text{min}$ in the considered subset $Y_{d, v} = \{ e \in X' :\ e_d = v \}$ such that $\displaystyle{{e^1}_\text{min} = \argmin_{e \in Y_{d, v}} w(e)}$, and $\displaystyle{{e^2}_\text{min} = \argmin_{e \in Y_{d, v} \setminus \{ {e^1}_\text{min} \}} w(e)}$.  Select the pair $(d, v)$ that corresponds to the maximum difference $w({e^2}_\text{min}) - w({e^1}_\text{min})$ and add the vector ${e^1}_\text{min}$ for the selected $(d, v)$ to $A$.



The time complexity of \maxregret{} is $O(s \cdot n^s + s \cdot (n-1)^s + \ldots + s \cdot 2^s + s) = O(s \cdot n^{s+1})$.

\subsubsection{ROM}
\label{sec:map_rom}

The \emph{Recursive Opt Matching} (\rom{}) is introduced
in~\cite{Gutin2008} as a heuristic of large domination number (see~\cite{Gutin2008} for definitions and results in domination analysis).  \rom{} proceeds as follows.  Initialize $A$ with a trivial assignment: $A^i = (i, i, \ldots, i)$.  On each $j$th iteration of the heuristic, $j = 1, 2, \ldots, s - 1$, calculate an $n \times n$ matrix $\displaystyle{M_{i, v} = \sum_{e \in Y(j, i, v)} w(e)}$, where $Y(j, i, v)$ is a set of all vectors $e \in X$ such that the first $j$ coordinates of the vector $e$ are equal to the first $j$ coordinates of the vector $A^i$ and the $(j+1)$th coordinate of $e$ is $v$: $Y(j, i, v) = \{ e \in X :\ e_k = A^i_k, 1 \le k \le j \text{ and } e_{j+1} = v \}$.  Let permutation $\pi$ be a solution of the 2-AP for the matrix $M$.  Set $A^i_{j+1} = \pi(i)$ for each $1 \le i \le n$.

The time complexity of \rom{} heuristic is $O((n^s + n^3) + (n^{s-1} + n^3) + \ldots + (n^2 + n^3)) = O(n^s + s n^3)$.

\subsubsection{Shift-ROM}
\label{sec:map_shiftrom}

A disadvantage of the \rom{} heuristic is that it is not symmetric
with respect to the dimensions.  For example, if the vector weights
do not depend significantly on the last coordinate then the
algorithm is likely to work badly. \shiftrom{} is intended to solve
this problem by trying \rom{} for different permutations of the
instance dimensions. However, we do not wish to try all $s!$
possible dimension permutations as that would increase the running
time of the algorithm quite significantly.  We apply only
$s$ permutations: $(X_1 X_2 \ldots X_s)$, $(X_s X_1 X_2
\ldots X_{s-1})$, $(X_{s-1} X_s X_1 X_2 \ldots X_{s-2})$, \ldots,
$(X_2 X_3 \ldots X_{s} X_1)$.

In other words, on each run \shiftrom{} applies \rom{} to the problem; upon completion, it renumbers the dimensions for the next run in the following way: $X_1 \gets X_2,\ X_2 \gets X_3,\ \ldots,\ X_{s-1} \gets X_s,\ X_s \gets X_1$.  After $s$ runs, the best solution is selected.

The time complexity of \shiftrom{} heuristic is $O((n^s + s n^3) \cdot s) = O(s n^s + s^2 n^3)$.

\subsubsection{Time Complexity Comparison}
\label{sec:timecomplexity}

Now we can gather all the information about the time complexity of the considered heuristics.  The following table shows the time complexity of each of the heuristics for different values of $s$:
\bigskip

\begin{tabular}{lllll}
\toprule
    & \greedy
    & \maxregret
    & \rom
    & \shiftrom
\\ \cmidrule(){1-5}

Arbitrary $s$
    & $O(n^{s+1})$
    & $O(s n^{s+1})$
    & $O(n^s + s n^3)$
    & $O(s n^s + s^2 n^3)$
\\

Fixed $s = 3$
    & $O(n^4)$
    & $O(n^4)$
    & $O(n^3)$
    & $O(n^3)$
\\

Fixed $s \ge 4$
    & $O(n^{s+1})$
    & $O(n^{s+1})$
    & $O(n^s)$
    & $O(n^s)$
\\ 
\bottomrule
\end{tabular}

\subsection{Performance Notes}
\label{sec:performance_notes}

In a standard computer model it is assumed that all the operations take approximately the same time.  However, it is not true since the architecture of a modern computer is complex.  We will use a more sophisticated model in our further discussion.  The idea is to differentiate fast and slow memory access operations.

The weight matrix of a MAP instance is normally stored in the Random Access Memory (RAM) of a computer.  RAM's capacity is large enough even for very large instances, e.g., nowadays RAM of an average desktop PC is able to hold a weight matrix for 3-AP with $n = 750$, i.e., $4.2 \cdot 10^8$ weights\footnote{Here and further we assume that every weight is represented with a 4 byte integer.  The calculations are provided for 2 Gb of RAM.}.  RAM is a fast storage; one can load gigabytes of data from RAM in one second.  However, RAM has a comparatively high latency, i.e., it takes a lot of time for the processor to access even a small portion of data in RAM.  Processor cache is intended to minimize the time spent by the processor for waiting for RAM response.

The processor cache exploits two heuristics: firstly, if some data was recently used then there is a high probability that it will be used again soon, and, secondly, the data is usually used successively, i.e., if some portion of data is used now then it is likely that the successive portion of data will be used soon.  As an example, consider an in place vector multiplication algorithm: on every iteration the algorithm loads a value from the memory, multiplies it and saves the result at the same memory position.  So, the algorithm accesses every portion of data twice and the data is accessed successively, i.e., the algorithm accesses the first element of the vector, then it accesses the second element, the third one, etc.


Processor cache\footnote{We provide a simplified overview of cache; for detailed information, see, e.g.,~\cite{Bailey2006b}.} is a temporary data storage, relatively small and fast, usually located on the same chip as the processor.  It contains several \emph{cache lines} of the same size; each cache line holds a copy of some fragment of the data stored in RAM\@.  Each time the processor needs to access some data in RAM it checks whether this data is already presented in the cache.  If this is the case, it accesses this data in the cache instead.  Otherwise that if a `miss' is detected, the processor suspends, some cache line is freed and a new portion of data is loaded from RAM to cache.  Then the processor resumes and accesses the data in the cache as normally.  Note that in case of a `miss' the system loads the whole cache line that is currently 64 bytes on most of the modern computers~\cite{AMD2005} and this size tends to grow with the development of computer architecture.  Thus, if a program accesses some value in the memory several times in a short period of time it is very likely that this data will be loaded from RAM just once and then will be stored in the cache so the access time will be minimal.  Moreover, if some value is accessed and, thus, loaded from RAM to the processor cache, it is likely that the next value is also loaded since the cache line is large enough to store several values.

With respect to MAP heuristics, there are two key rules for improving the memory subsystem performance:
\begin{enumerate}
\item The successive access to the weight matrix (scan), i.e., access to the matrix in the order of its alignment in the memory,
is strongly preferred (we use the row-major order~\cite{Knuth1997}
for weight matrix in our implementations of the algorithms). Note
that if an algorithm accesses, e.g., every second weight in the
matrix and does it in the proper order, the real complexity of this
scan with respect to the memory subsystem is the same as the complexity of a
full scan since loading of one value causes loading of several
neighbor values.

    \item One should minimize the number of the weight matrix scans as much as possible.  Even a partial matrix scan is likely to access much more data than the processor cache is able to store, i.e., the data will be loaded from RAM all over again for every scan.
\end{enumerate}

Following these rules may significantly improve the running time of the heuristics.  In our experiments, the benefit of following these rules was a speed-up of roughly speaking 2 to 5 times.

\subsection{MAP Construction Heuristics Improvement}
\label{sec:map_construction_heuristics_improved}

\subsubsection{Greedy Heuristic Optimization}
\label{sec:map_greedy_improved}

A common implementation of the greedy approach for a combinatorial optimization problem involves sorting of all the weights in the problem.  In case of MAP this approach is inefficient since we actually need only $n$ vectors from the set of size $n^s$.  Another natural implementation of the \greedy{} heuristic is to scan all available vectors and to choose the lightest one on each iteration but it is very unfriendly with respect to the memory subsystem: it performs $n$ scans of the weight matrix.

We propose a combination of these approaches; our algorithm proceeds
as follows.  Let $A = \varnothing$ be a partial assignment and $B$ an
array of vectors.  While $|A| < n$, i.e., $A$ is not a full
assignment, the following is repeated.  We scan the weight matrix to
fill the array $B$ with $k$ vectors corresponding to $k$ minimal weights
in non-decreasing order: if the weight of the current vector is less
than the largest weight in $B$ then we insert the current vector to
$B$ in the appropriate position and, if necessary, remove the last
element of $B$.  Then, for each vector $e \in B$, starting from the
lightest, we check whether $A \cup \{ e \}$ is a feasible partial
assignment and, if so, add $e$ to $A$.  Note, that during the second
and further cycles we scan not the whole weight matrix but only a
subset $X' \subset X$ of the vectors that can be included into the
partial assignment $A$ with the feasibility preservation: $A \cup \{
x \}$ is a partial assignment for any $x \in X'$.  The size of the
array $B$ is calculated as $k = \min \{ 64, |X'| \}$ in our implementation.  The constant 64 is obtained empirically.

The algorithm is especially efficient on the first iterations, i.e., in the hardest part of its work, while the most of the vectors are feasible.  However, there exists a bad case for this heuristic.  Assume that the weight matrix contains a lot of vectors of the minimal weight $w_\text{min}$.  Then the array $B$ will be filled with vectors of the weight $w_\text{min}$ at the beginning of the scan and, thus, it will contain a lot of similar vectors (recall that the weight matrix is stored in the row-major order and only the last coordinates are varied at the beginning of the scan, so all the vectors processed at the beginning of the scan are likely to have the same first coordinates).  As a result, selecting the first of these vectors will cause infeasibility for the other vectors in $B$.  We use an additional heuristic to decrease the running time of the \greedy{} algorithm for such instances.  Let $w_\text{min}$ be the minimum possible weight: $w_\text{min} = \min_{e \in X'} w(e)$ (sometimes this value is known like for \random{} instance family it is 1, see Section~\ref{sec:map_testbed}).  If it occurs during the matrix scan that all the vectors in $B$ have the weight $w_\text{min}$, i.e., $w(B_i) = w_\text{min}$ for every $1 \le i \le k$, then the rest of the scan can be skipped because there is certainly no vector lighter than $B_k$.  Moreover, it is safe to update $w_\text{min}$ with $w(B_k)$ every time before the next matrix scan.

\subsubsection{Max-Regret Heuristic Optimization}
\label{sec:map_maxregret_improved}

The \maxregret{} heuristic naturally requires $O(n^2 s)$ weight
matrix partial scans.  Each of these scans fixes one coordinate and,
thus, every available vector $e \in X'$ (see
Subsection~\ref{sec:map_greedy_improved}) is accessed $s$ times during each
iteration, and this access is very inefficient when the last
coordinate is fixed (recall that the weight matrix is stored in a
row-major order and, thus, if the last coordinate is fixed then the
algorithm accesses every $n$th value in the memory, i.e., the access
is very non-successive and one can assume that this scan will load
the whole weight matrix from RAM to cache).  In our more detailed
computer model (see Section~\ref{sec:performance_notes}), the time complexity of the non-optimized
\maxregret{} is $O((s - 1) \cdot n^{s+1} + n^{s+2})$.

We propose another way to implement \maxregret{}.  Let us scan the whole set $X'$ of available vectors on each iteration.  Let $L$ be an $n \times s$ matrix of the lightest vector pairs: $L_{i, j}^1$ and $L_{i, j}^2$ are the lightest vectors when the $j$th coordinate is fixed as $i$, and $w(L_{i, j}^1) \le w(L_{i, j}^2)$.
To fill the matrix $L$ we do the following: for every vector $e \in X'$ and for every coordinate $1 \le d \le s$ check: if $w(e) < w(L_{e_d, d}^1)$, set $L_{e_d, d}^2 = L_{e_d, d}^1$ and $L_{e_d, d}^1 = e$.  Otherwise if $w(e) < L_{e_d, d}^2$, set $L_{e_d, d}^2 = e$.  Thus, we update the $L_{e_d, d}$ item of the matrix with the current $e$ if $w(e)$ is small enough.  Having the matrix $L$, we
can easily find the coordinate $d$ and the fixed value $v$ such that $w(L_{v, d}^2) - w(L_{v, d}^1)$ is maximized.  The vector $L_{v, d}^1$ is added to the solution and the next iteration of the algorithm is executed.

The proposed algorithm performs just $n$ partial scans of the weight matrix.  The matrix $L$ is usually small enough to fit in the processor cache, so the access to $L$ is fast.  Thus, the time complexity of the optimized \maxregret{} in our more detailed computer model is $O(n^{s+1})$.

\subsubsection{ROM Heuristic Optimization}
\label{sec:map_rom_improved}

The \rom{} heuristic can be implemented in a very friendly way with
respect to the memory access.  On the first iteration it fixes
the first two coordinates ($n^2$ combinations) and enumerates all
vectors with these fixed coordinates.  Thus, it scans the whole
weight matrix successively.  On the next iteration it fixes three
coordinates ($n^2$ combinations as the second coordinate depends on
the first one), and enumerates all vectors with these fixed
coordinates.  Thus, it scans $n^2$ solid $n^{s-3}$-size fragments of
the weight matrix; further iterations are similar.  As a result, the
time complexity of \rom{} in our more detailed computer model is the
same as in a simple one: $O(n^s + s n^3)$.

\subsubsection{Shift-ROM Heuristic Optimization}
\label{sec:map_shiftrom_improved}

The \shiftrom{} heuristic is an extension of \rom{}; it simply runs \rom{} $s$ times, starting it from different dimensions.  However, not every run of \rom{} is efficient when it is a part of \shiftrom{}\@.  Let us consider the case when the first iteration of \rom{} fixes the last two coordinates.  For each of the $n^2$ combinations of the last two coordinate values, the heuristic scans the whole weight matrix with the step $n^2$ between the accessed weights, i.e., the distance between the successively accessed weights in the memory is $n^2$ elements, which is very inefficient.  A similar situation occurs when the first and the last dimensions are fixed.

To avoid this disadvantage, we propose the following algorithm.  Let $M^d$ be an $n \times n$ matrix for every $1 \le d \le s$.  Initialize $M^d_{i, j} = 0$ for every $1 \le d \le s$ and $1 \le i, j \le n$.  For each vector $e \in X$ and for each $1 \le d \le s$ set $M^d_{e_d, e_{d+1}} = M^d_{e_d, e_{d+1}} + w(e)$ (here we assume that $e_{s+1} = e_{1}$).  Now the matrices $M^d$ can be used for the first iteration of every \rom{} run.

When applying this technique, only one full matrix scan is needed for the heuristic and this scan is successive.  There are several other inefficient iterations like fixing of the last three coordinates but their influence on the algorithm's performance is negligible.

\section{Data Structures}
\label{sec:gtsp_data_structures}

In some cases data structure plays the key role in an algorithm's theoretical efficiency (see, e.g.,~\cite{Johnson2002} and references there).  In other cases it does not change the theoretical time complexity of an algorithm but it is still worth a separate discussion.  Below we consider several data structures for GTSP algorithms.

\subsection{GTSP Tour Storage}

It is a non-trivial question how one should store a GTSP solution.  The most common approach is to store a sequence of vertices in the visiting order.  It was used in \citet{Silberholz2007,Tasgetiren2010} and many others.  The advantages of this method are simplicity, compactness (it requires only an integer array of size $m$) and quickness of the weight calculation.  The disadvantages are difficulty in some tour modifications (observe that moving one vertex requires up to $m$ operations) and absence of a trivial way to check the tour correctness.  Sliding along the tour is easy in this representation but requires some additional checks.

Another tour representation, random-key, was used in \citet{Snyder2000}.  It represents the tour as a sequence of real numbers $x_1, x_2, \ldots, x_m$; the $i$th number $x_i$ corresponds to the $i$th cluster $C_i$ of the problem.  The integer part $\lfloor x_i \rfloor$ of the number is the vertex index within the cluster $C_i$ and the fractional part $x_i - \lfloor x_i \rfloor$ determines the position of the cluster in the tour---the clusters are ordered according to these fractional parts, in ascending order.  The main advantage of random-key tours is that almost any sequence of numbers represent a correct tour; one only needs to ensure that $1 \le \lfloor x_i \rfloor \le |C_i|$ for every $i$.  It is also relatively easy to implement some modifications of the tour.  The disadvantages are difficulty in sliding along the tour and the high cost of the tour weight calculation.

We propose a new tour representation which is base on double-linked lists.  In particular, we store three integer arrays of size $m$: \var{prev}, \var{next} and \var{vertices}, where $\var{prev}_i$ is the cluster preceding the cluster $C_i$ in the tour, $\var{next}_i$ is the cluster succeeding the cluster $C_i$ in the tour, and $\var{vertices}_i$ is the vertex within cluster $C_i$.  There are several important advantages of this representation.  Unlike other approaches, it naturally represents the cycle which simplifies the algorithms.  Consider, e.g., a typical local search implementation (Algorithm~\ref{alg:ls_implementation}):
\begin{algorithm}[ht]
\caption{A typical implementation of a local search based on the double-linked list tour representation.  In this example the algorithm preforms as few iterations as possible to ensure that the tour is a local minimum.}
\label{alg:ls_implementation}
\begin{algorithmic}
\STATE Set the current cluster $X \gets 1$.
\STATE Set the counter $t \gets m$.
\WHILE {$t > 0$}
	\IF {there exist some improvements for the current cluster $X$}
		\STATE Update the tour accordingly.
		\STATE Update the counter $t \gets m$.
	\ELSE
		\STATE Decrease the counter $t \gets t - 1$.
	\ENDIF
	\STATE Move to the next cluster $X \gets \var{next}_X$.
\ENDWHILE
\end{algorithmic}
\end{algorithm}
the algorithm smoothly slides along the tour until no improvement is found for exactly one loop.  Observe that one does not need the concept of position when using this tour representation; it is possible to use cluster index instead.  In this context the procedure of tour rotation becomes meaningless; one can simply consider any cluster as the first cluster in the tour.  Moreover, it allows one to find a certain cluster in $O(1)$ time; we use it, e.g., to start the \CO{} calculations from the smallest cluster with no extra effort.  Our representation clearly splits the cluster order and the vertex selection; note that some algorithms do not require the information on the vertex selection while some others do not modify the cluster order.  It is useful that linked lists allow quick removing and inserting elements.  To turn the tour backwards, one only has to swap \var{prev} and \var{next}.  Observe that this tour representation is deterministic, i.e., each GTSP tour has exactly one representation in this form.  If the problem is symmetric, every tour $(\var{prev}, \var{next}, \var{vertices})$ has exactly one clone $(\var{next}, \var{prev}, \var{vertices})$.  

The main disadvantage of this representation is that it takes three times more space than the sequence of vertices.  In fact, implementation of many algorithms do not require backward links.  In this case one can avoid using the \var{prev} array and hence use only two $m$-elements arrays.  When necessary, one can quickly restore the \var{prev} array according to \var{next}.

Note that a similar tour representation was used in~\cite{Tasgetiren2007}.

\subsection{GTSP Weights Storage}

Another important decision is how to store the weights of a GTSP instance.  There are two obvious solutions of this problem:
\begin{enumerate}
	\item Store a two dimensional matrix $M$ of size $n \times n$ as follows: $M_{i,j} = w(V_i \to V_j)$.  Note that this data structure stores $\sum_{i=1}^m |C_i|^2$ redundant weights.

	\item Store $m (m - 1)$ matrices, one matrix $M^{X,Y}$ of size $|X| \times |Y|$ per every pair of distinct clusters $X$ and $Y$.
\end{enumerate}

If we have a pair of vertices and we need to get the weight between them, it is obviously better to use the first approach.  However, if we need to use many weights between two clusters (consider, e.g., calculation of the smallest weight between clusters $X$ and $Y$: $w_\text{min}(X \to Y)$), the second approach is preferable.  Indeed, in the first approach we have to use something like $M_{X_i, Y_j}$, i.e., look for the absolute index of every vertex in $X$ and $Y$.  In the second approach we just find the matrix $M^{X,Y}$ and then use it like this: $M^{X,Y}_{i,j}$.  Observe that the second approach provides a sequential access to the weight matrix which is very friendly with respect to the computer architecture, see Section~\ref{sec:performance_notes}.

Our experimental analysis shows that the second approach improves the performance of some algorithms approximately twice.  However, it is not efficient for some other algorithms which behave as a TSP heuristic, i.e., consider only one vertex in every cluster.  We decided to use both approaches in our implementations, i.e., to store the weights in a single matrix and, in addition, to store a matrix for every pair of clusters.

\section{Conclusion}

Several aspects of optimization heuristic design and analysis are discussed in this chapter.  A lot of attention is paid to the questions of test bed selection.  Observe that a typical heuristic does not provide any solution quality guarantee and, hence, experimental evaluation is vastly important.  

We consider two examples of test bed generation.  For MAP, we systematized the existing instance families.  For one of these instance families we have successfully applied probabilistic analysis in order to estimate the exact solution values.  Note that for most of instances of this type our estimation is really precise.

There exist several speed-up approaches applicable to virtually any optimization heuristic.  One of these approaches in preprocessing.  Observe that almost any algorithm hugely depends on the input size.  Hence, even a small decrease of the instance size may significantly reduce the running time of a heuristic.  We show an example of GTSP preprocessing which removes some vertices and/or edges from an instance if they may not be included in the optimal solution.  Our experiments confirm the success of this technique.

At last, we discuss some aspects related to implementation details of an algorithm.  In turns out that a simple transformation of an algorithm may significantly speed it up.  We provide an example of a very successful optimization of MAP construction heuristics.  In addition, we discuss the efficiency of several data structures.  We show that selecting a proper data structure may often improve and simplify an algorithm.
\newcommand{\TourCluster}[2]{\ensuremath{\mathcal{#1}_{#2}}}
\newcommand{\T}[1]{\TourCluster{T}{#1}}

\chapter{Local Search Algorithms for GTSP}
\label{sec:gtsp_ls}

While GTSP is a very important combinatorial optimization problem and is well-studied in many aspects, researches still did not pay enough attention to GTSP specific local search and mostly use simple TSP heuristics with basic adaptations for GTSP\@.  This section aims at thorough and deep investigation of the neighborhoods specific for GTSP and algorithms that can explore these neighborhoods quickly.  

We formalize the procedure of adaptation of a TSP neighborhood for GTSP and propose efficient algorithms to explore the obtained neighborhoods.  We also generalize all other existing and some new GTSP neighborhoods.  Apart from these theoretical results, we also provide the results of a thorough experimental analysis to compare the proposed algorithms implementations and find out which neighborhoods are the most efficient in practice.

Note that some neighborhoods were used in \cite{Snyder2000,Silberholz2007,Tasgetiren2007}, but they were not systematized or analyzed in detail.

We introduce a classification of GTSP neighborhoods.  We divide all the neighborhoods into three classes:
\begin{enumerate}
	\item Cluster Optimization neighborhoods are the neighborhoods which preserve the cluster order in the tour.  This class is discussed in Section~\ref{sec:gtsp_cluster_optimization}.
	\item TSP neighborhoods are the neighborhoods produced from the TSP ones.  They usually perform some global rearrangements in the cluster order.  In Section~\ref{sec:gtsp_ls_adaptation} we show that there exist several ways to adapt a TSP neighborhood for GTSP and propose a number of improvements to make these adaptations fast.  We thoroughly investigate possible adaptations of the state-of-the-art TSP Lin-Kernighan heuristic in Section~\ref{sec:gtsp_lin_kernighan}.
	\item Fragment Optimization neighborhoods include only tours which are different from the original one in at most some small tour fragment.  Neighborhoods of this type were not widely used before.  In Section~\ref{sec:gtsp_fragment_optimization}, we propose two efficient algorithms for these neighborhoods.
\end{enumerate}

In order to compare the efficiency of different neighborhoods and implementations, a series of experiments is conducted in Section~\ref{sec:gtsp_ls_experiments}.

\section{Cluster Optimization}
\label{sec:gtsp_cluster_optimization}

In this section we discuss GTSP neighborhoods which preserve the order of clusters in the tour.  In other words, these neighborhoods may only vary the vertices within certain clusters.  The virtually smallest neighborhood of this type is 
$$
N_\text{L}(T, i) = \{ T_1 \to T_2 \to \ldots \to T_{i-1} \to T'_i \to T_{i+1} \to T_{i+2} \to T_m \to T_1 :\ T'_i \in \Cluster(T_i) \} \,.
$$
Its size is $|N_\text{L}(T, i)| = |\Cluster(T_i)|$ and it takes $O(s)$ operations to explore it.  One can extend it for two or more clusters: $N_\text{L}(T, I)$, where $I$ is a set of cluster indices.  The size of such neighborhood $|N_\text{L}(T, I)| = \prod_{i \in I} |\Cluster(T_i)|$.  Observe that while the set $I$ contains no neighbor indices, i.e., if $i \in I$ then $i - 1, i + 1 \notin I$, it takes only $O(|I|s)$ operations to explore it.  If $I = \{ i, i + 1 \}$, the neighborhood $N_\text{L}(T, I)$ changes its structure.  Now it takes $O(s^2)$ operations to explore it.  One may assume that, if $I = \{ i, i + 1, \ldots, i + k - 1 \}$, the time complexity of the local search is $O(s^k)$.  However, we will show that it remains quadratic for any fixed $k < m$.

Consider the case when $k = m$, i.e., when the vertices are optimized in all the clusters of the tour.  This is the most powerful neighborhood of this type and we call it \emph{Cluster Optimization}.  In fact, there is an exact algorithm \CO{} that finds the optimal vertex selection for the whole solution in $O(n \gamma s)$ time.  In other words, given a fixed cluster order, it finds the best cycle through these clusters.
\nomenclature[CO]{\CO}{stands for the Cluster Optimization algorithm}

\CO{} was introduced by Fischetti, Salazar-Gonz{\'a}lez and Toth~\citep{Fischetti1997} (see its detailed description also in~\citep{Fischetti2002}) and used in~\citep{Hu2008,Pintea2007,Renaud1998} and others.  It is based on the shortest path algorithm for acyclic digraphs (see, e.g.,~\citep{Bang-Jensen2000}).

Let $T = T_1 \to T_2 \to \ldots \to T_m \to T_1$ be the given tour and $\T{i} = \Cluster(T_i)$ for every $i$.  The algorithm builds a layered network $G_\text{CO} = (V_\text{CO}, E_\text{CO})$, where $V_\text{CO} = V \cup \T{1}'$ is the set of the GTSP instance vertices extended by a copy $\T{1}'$ of the cluster $\T{1}$, and $E_\text{CO}$ is a set of edges in the digraph $G_\text{CO}$.  An edge $x \to y \in E_\text{CO}$ exists if there exists $i$ such that $x \in \T{i}$ and $y \in \T{i+1}$ (assume $\T{m+1} = \T{1}'$).  The weight of the edge $x \to y$ is $w(x \to y)$.  For each vertex $v_1 \in \T{1}$ and its copy $v_1' \in \T{1}'$, the algorithm finds the shortest $(v_1, v_1')$-path in $G_\text{CO}$.  It selects the shortest $(v_1, v_1')$-path which represents the best vertex selection within the given cluster sequence.  A formal procedure based on the dynamic programming approach is presented in Algorithm~\ref{alg:co}.
\begin{algorithm}[ht]
\caption{Cluster Optimization.  Basic implementation.}
\label{alg:co}
\begin{algorithmic}
\REQUIRE Tour $T = T_1 \to T_2 \to \ldots \to T_m \to T_1$, where $|\Cluster(T_1)| = \gamma$.
\STATE Let $\T{i} = \Cluster(T_i)$ for every $i$.
\FORALL {$r \in \T{1}$ and $v \in \T{2}$}
	\STATE Set $p_{r, v} \gets (r \to v)$.
\ENDFOR
\FOR {$i \gets 3, 4, \ldots, m$}
	\FORALL {$r \in \T{1}$ and $v \in \T{i}$}
		\STATE Set $p_{r, v} \gets p_{r, u} + (u \to v)$, where $u \in \T{i-1}$ is selected to minimize $w\big(p_{r, u} + (u \to v)\big)$.
	\ENDFOR
\ENDFOR
\RETURN $p_{r, v} + (v \to r)$, where $r \in \T{1}$ and $v \in \T{m}$ are selected to minimize $w\big(p_{r, v} + (v \to r)\big)$.
\end{algorithmic}
\end{algorithm}
Note that there is no need to repeat the search several times since it finds the local minimum after the first run.

\subsection{Cluster Optimization Refinements}
\label{sec:gtsp_co_refinements}

Several improvements can noticeably reduce the running time of \CO{}.

Observe (see Algorithm~\ref{alg:co}) that the algorithm's time complexity grows linearly with the size of the cluster $\mathcal{T}_1$.  Thus, before applying \CO{}, we rotate the solution such that $|\mathcal{T}_1| = \gamma$.  Hence, the time complexity of the algorithm is $O(n \gamma s)$.  Moreover, in some applications one can assume that $\gamma \in O(1)$ which changes the time complexity to $O(n s)$.

This improvement was widely used in the literature.

\bigskip

Since the running time of the algorithm significantly depends on the size $\gamma$ of the first cluster, it is worth checking whether if we can reduce its size.  Some attempts to reduce the cluster sizes in GTSP were proposed in Section~\ref{sec:gtsp_preprocessing}.  The idea was to remove a vertex $r \in R$ if for every $v \in V$ and $u \in U$ there exists some $r' \in R \setminus \{ r \}$ such that $w(v \to r' \to u) \le w(v \to r \to u)$, where $R$, $U$ and $V$ are arbitrary distinct clusters.  In our case the reduction can be significantly more efficient.  Indeed, we do not need to consider all the combinations of $R$, $U$ and $V$.  Let $R = \mathcal{T}_1$.  Then the clusters $U$ and $V$ are fixed to $U = \mathcal{T}_m$ and $V = \mathcal{T}_2$.
	
A straightforward reduction algorithm would take $O(s^2 \gamma^2)$ operations.  We propose Algorithm~\ref{alg:reduce_cluster} which reduces the size of cluster $\mathcal{T}_1$ in $O(s^2 \gamma)$ time.
\begin{algorithm}[ht]
\caption{Reduction of a cluster in a tour.}
\label{alg:reduce_cluster}

\begin{algorithmic}
\REQUIRE Tour $T = T_1 \to T_2 \to \ldots \to T_m \to T_1$, where $|\Cluster(T_1)| = \gamma$.
\STATE Let $U = \Cluster(T_m)$, $R = \Cluster(T_1)$ and $V = \Cluster(T_2)$.
\FORALL {$u \in U$ and $v \in V$}
	\STATE Find the shortest distance $l_{u,v} \gets \min_{r \in R} w(u \to r \to v)$.
	\STATE Find the number $c_{u,v}$ of paths $u \to r \to v$ such that $w(u \to r \to v) = l_{u,v}$, i.e., $c_{u,v} \gets \big| \{ r : r \in R \text{ and } w(u \to r \to v) = l_{u,v} \} \big|$.
\ENDFOR

\FORALL {$r \in R$}
	\FORALL {$u \in U$ and $v \in V$}
		\IF {$w(u \to r \to v) = l_{u,v}$ and $c_{u,v} = 1$}
			\STATE Go to the next $r$.
		\ENDIF
	\ENDFOR
	\FORALL {$u \in U$ and $v \in V$}
		\IF {$w(u \to r \to v) = l_{u,v}$}
			\STATE Update $c_{u,v} \gets c_{u,v} - 1$.
		\ENDIF
	\ENDFOR
	\STATE Remove $r$ from $R$.
\ENDFOR
\end{algorithmic}
\end{algorithm}
One can try to reduce the size of every cluster but this will likely only slow down the \CO{} algorithm.  We apply this reduction only to the smallest cluster $\mathcal{T}_1 = \Cluster(T_1)$.  

Note that this reduction is valid only for a certain cluster order and, hence, it is suitable only for a local search of the Cluster Optimization class.  This means that the cluster $\T{1}$ should be restored after the run of \CO{}.
	
\bigskip
	
Observe that Algorithm~\ref{alg:co} goes sequentially along the tour.  However, there are many other ways to calculate the shortest paths in a layered network using the dynamic programming approach.  In particular, one can interpret an arbitrary dynamic programming algorithm for the shortest paths as in Algorithm~\ref{lab:co_general}.
\begin{algorithm}[ht]
\caption{Calculation of the shortest paths in a layered network.}
\label{lab:co_general}

\begin{algorithmic}
\REQUIRE Network layers $\T{1}$, $\T{2}$, \ldots, $\T{m}$.
\FOR {$i \gets 1, 2, \ldots, m - 2$}
	\STATE Find the shortest paths from $\T{X_i - 1}$ to $\T{X_i + 1}$.
	\STATE Remove the layer $\T{X_i}$ and set the weights between $\T{X_i - 1}$ and $\T{X_i + 1}$ to the calculated shortest paths.  Renumber the layers.
\ENDFOR
\end{algorithmic}
\end{algorithm}
Here $X$ is a sequence of $m - 2$ numbers, $1 < X_i \le m - i$.  It defines the behavior of the algorithm: on the $i$th iteration the algorithm removes the cluster $\T{X_i}$ from the sequence by calculating the shortest paths from $\T{X_i - 1}$ to $\T{X_i + 1}$.

Let us calculate the number of times the \CO{} algorithm takes a weight between two vertices.  This number adequately reflects the running time of the algorithm.

In general, the dynamic programming algorithm takes
\begin{equation}
\label{eq:co_optimal_order}
t_\text{optimal} = 2 \cdot \left[ |\T{1}| |\T{m}| + \sum_{i = 1}^{m-2} |\T{x_i}| |\T{y_i}| |\T{z_i}| \right] \text{ weight operations,}
\end{equation}
where the ordered lists $x$, $y$ and $z$ correspond to $X$.  Without loss of generality, let $x_i < y_i < z_i$.

The sequential algorithm always removes the second cluster in the sequence ($X_i = 2$ for every $i$), i.e., the number of weight operations required for the algorithm is as follows:
\begin{equation}
\label{eq:co_sequential}
t_\text{seq} = 2 \cdot \left[ |\T{1}| |\T{m}| + \sum_{i = 2}^{m-1} |\T{1}| |\T{i}| |\T{i+1}| \right] \,.
\end{equation}

Consider the following example.  Let $m$ be odd, $|\T{i}| = z > 1$ for every $i = 2, 4, 6, \ldots, m - 1$ and $|\T{i}| = 1$ for every $i = 1, 3, 5, \ldots, m$.  According to (\ref{eq:co_sequential}), the sequential algorithm takes $2 \cdot (m - 2) \cdot z + 2$ weight operations.  Consider a different algorithm which first removes all the clusters $\T{2}$, $\T{4}$, \ldots, $\T{m-1}$; it requires only $(m - 1) \cdot z + (m - 3) + 2$ weight operations.  Hence, the ratio is:
$$
\lim_{m \to \infty}{\lim_{z \to \infty}{\frac{2 \cdot (m - 2) \cdot z + 2}{(m - 1) \cdot z + (m - 3) + 2}}} = \lim_{m \to \infty}{2 \cdot \frac{m - 2}{m - 1}} = 2 \,.
$$
Note that the time ratio between the sequential calculation and the improved one can be significant in practice.  Even for the modest values of $m = 7$ and $z = 7$ in this example the ratio is $1.5$.

A natural question is how much it is possible to speed up the sequential algorithm by changing the calculation order.
\begin{theorem}
Let the first layer in a layered network be the smallest one.  Then the sequential (see Algorithm~\ref{alg:co}) calculation of the shortest paths in this network is up to 2 times slower than the optimal dynamic programming algorithm, and this bound is sharp.
\end{theorem}
\proof Let $\T{1}$, $\T{2}$, \ldots, $\T{m}$ be the layers of the network.  Then Algorithm~\ref{lab:co_general} allows one to find all the shortest paths from every vertex in $\T{1}$ to every vertex in $\T{m}$.  Having these paths, one can find the shortest cycle in $O(s^2)$ operations.

Observe that, whatever is $X'$, the distances between the layers $\T{i}$ and $\T{i+1}$ are used in the algorithm exactly once.  In other words, (\ref{eq:co_optimal_order}) contains exactly one term which includes $|\T{i}| |\T{i+1}|$.  Note that a term in (\ref{eq:co_optimal_order}) may be either $|\T{i-1}| |\T{i}| |\T{i+1}|$ for some $i$, or $|\T{i}| |\T{i+1}| |\T{j}|$, where $j \notin \{ i - 1, i + 2 \}$, or $|\T{i}| |\T{j}| |\T{k}|$, where $j \notin \{ i - 1, i + 1 \}$ and $k \notin \{ i - 1, i + 1, j - 1, j + 1 \}$.

Let us match every term $|\T{1}| |\T{i}| |\T{i+1}|$ in (\ref{eq:co_sequential}) to the term $|\T{x_j}| |\T{y_j}| |\T{z_j}|$ in (\ref{eq:co_optimal_order}), where either $x_j = i$ and $y_j = i + 1$ or $y_j = i$ and $z_j = i + 1$.  Observe that this term exists and it is the only term containing $|\T{i}| |\T{i+1}|$.  Indeed, the distances between the clusters $\T{i}$ and $\T{i+1}$ are used exactly once in the dynamic programming algorithm.

Obviously, every term $|\T{x_i}| |\T{y_i}| |\T{z_i}|$ in (\ref{eq:co_optimal_order}) may be matched to at most two terms $|\T{1}| |\T{x_i}| |\T{y_i}|$ and $|\T{1}| |\T{y_i}| |\T{z_i}|$ in (\ref{eq:co_sequential}).  Now observe that, since $|\T{1}| \le |\T{j}|$ for any $j$,
$$
|\T{j}| |\T{i}| |\T{i+1}| \ge |\T{1}| |\T{i}| |\T{i+1}| \text{ for any $i$ and $j$}
$$
and, thus, 
$$
2 \cdot |\T{x_i}| |\T{y_i}| |\T{z_i}| \ge |\T{1}| |\T{x_i}| |\T{y_i}| + |\T{1}| |\T{y_i}| |\T{z_i}| \,.
$$
Hence, $2 \cdot t_\text{optimal} \ge t_\text{seq}$.
\qed

However, it would take too long to find the optimal sequence of calculations, and, thus, we propose a simple heuristic.  On every iteration, it looks one step ahead; if the following condition is met:
\begin{equation}
\label{eq:co_order_condition}
|\T{1}| |\T{2}| |\T{3}| + |\T{1}| |\T{3}| |\T{4}|	> |\T{2}| |\T{3}| |\T{4}| + |\T{1}| |\T{2}| |\T{4}| \,,
\end{equation}
it removes the cluster $\T{3}$ before removing $\T{2}$, see Algorithm~\ref{alg:co_improved_order}.
\begin{algorithm}[ht]
\caption{Cluster Optimization with an improved order of calculations.}
\label{alg:co_improved_order}

\begin{algorithmic}
\REQUIRE Tour $T = T_1 \to T_2 \to \ldots \to T_m \to T_1$, where $|\Cluster(T_1)| = \gamma$.
\STATE Let $\mathcal{T}_i = \Cluster(T_i)$ for every $i$.
\FOR {$i \gets 2, 3, \ldots, k - 1$}
	\IF {$i < k - 1$ and $|\T{1}| |\T{i}| |\T{i+1}| + |\T{1}| |\T{i+1}| |\T{i+2}|	> |\T{i}| |\T{i+1}| |\T{i+2}| + |\T{1}| |\T{i}| |\T{i+2}|$}
		\STATE Calculate the shortest paths from $\T{i}$ to $\T{i+2}$.
		\STATE Calculate the shortest paths from $\T{1}$ to $\T{i+2}$.
		\STATE Set the weights between $\T{1}$ and $\T{4}$ to the calculated values.
		\STATE Set $i \gets i + 1$.
	\ELSE
		\STATE Calculate the shortest paths from $\T{1}$ to $\T{i+1}$.
		\STATE Set the weights between $\T{1}$ and $\T{i+1}$ to the calculated values.
	\ENDIF
\ENDFOR
\end{algorithmic}
\end{algorithm}

\section{TSP Neighborhoods Adaptation}
\label{sec:gtsp_tsp_neighborhoods}

GTSP is an extension of TSP and, hence, it is natural to use TSP neighborhoods for GTSP\@.  In this section we discuss different ways to adapt a TSP neighborhood.  These approaches are later applied to the most efficient TSP neighborhoods.

In order to use a TSP neighborhood for GTSP, one may propose splitting GTSP into two problems~\citep{Renaud1998}: solving the TSP instance induced by the given tour to find the cluster order and then applying \CO{} algorithm to it (see Section~\ref{sec:gtsp_cluster_optimization}).  We will show now that this approach is generally poor with regards to solution quality.  Let $N_\text{TSP}(T)$ be a set of tours which can be obtained from the tour $T$ by reordering the vertices in $T$.  Observe that one has to solve a TSP instance induced by $T$ to find the best tour in $N_\text{TSP}(T)$.  Let $N_\text{CO}(T)$ be the neighborhood of the \CO{} local search (see Section~\ref{sec:gtsp_cluster_optimization}).  Recall that the size of $N_\text{CO}(T)$ neighborhood is $|N_\text{CO}(T)| = \prod_{i=1}^m |C_i| \in O(s^m)$ but it can be explored in the polynomial time.

The following theorem shows that splitting GTSP into two problems (search in $N_\text{TSP}(T)$ and then search in $N_\text{CO}(T)$) does not guarantee any solution quality.

\begin{figure}[ht]
\centering  
\subfloat[The instance $G$ and the optimal GTSP tour $T_\text{opt}$.]{
\label{fig:gtsp_problem}
\xymatrix@R=2.5em@C=2.5em@L=0.1em{
	&	*++[o][F-]{2}		\ar@{-}[r]^1 
										\ar@{-}[rd]_(0.7)1
	&	*++[o][F-]{3}		\ar@{-}[rd]^0 
										\ar@{-}@<0.5\linethickness>[rd] 
										\ar@{-}@<-0.5\linethickness>[rd]
\\
		*++[o][F-]{1}		\ar@{-}[ru]^1 
										\ar@{-}[r]_1
										\ar@{-}@<0.5\linethickness>[r]
										\ar@{-}@<-0.5\linethickness>[r]
	&	*++[o][F-]{2'}	\ar@{-}[ru]_(0.7)1 
										\ar@{-}[r]_0
										\ar@{-}@<0.5\linethickness>[r]
										\ar@{-}@<-0.5\linethickness>[r]
	&	*++[o][F-]{4}		\ar@{-}[r]_0 
										\ar@{-}[u]^0
										\ar@{-}@<0.5\linethickness>[u]
										\ar@{-}@<-0.5\linethickness>[u]
	&	*++[o][F-]{5}		\ar@/^2em/@{-}[lll]_0
										\ar@/^2em/@{-}@<0.5\linethickness>[lll]
										\ar@/^2em/@{-}@<-0.5\linethickness>[lll]
}}
\qquad
\subfloat[A local minimum $T$ which is the worst possible GTSP tour.]{
\label{fig:worst_tour}
\xymatrix@R=2.5em@C=2.5em@L=0.1em{
	&	*++[o][F-]{2}		\ar@{-}[r]^1
	&	*++[o][F-]{3}		\ar@{-}[d]^0
\\
		*++[o][F-]{1}		\ar@{-}[ru]^1 
										\ar@{--}[r]_1
	&	*++[o][F-]{2'}	\ar@{--}[ru]_(0.7)1
	&	*++[o][F-]{4}		\ar@{-}[r]_0
	&	*++[o][F-]{5}		\ar@/^2em/@{-}[lll]_0
}}

\caption{An example of a local minimum in both $N_\text{TSP}(T)$ and $N_\text{CO}(T)$ which is a longest possible GTSP tour.}

\label{fig:tsp_co_local_minimum}
\end{figure}
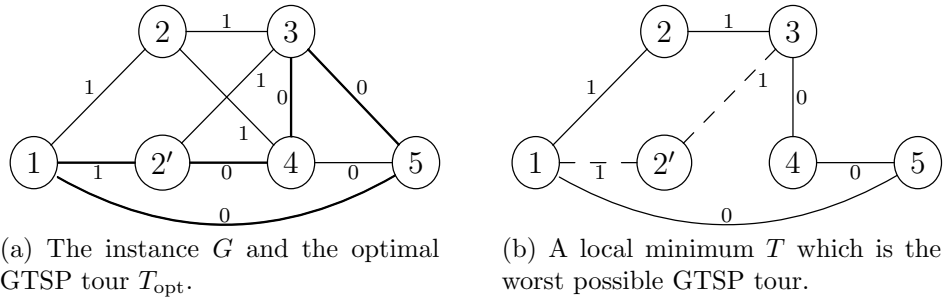

\begin{theorem}
\label{th:tsp_co_local_minimum}
The best tour among $N_\text{CO}(T) \cup N_\text{TSP}(T)$ can be a longest GTSP tour different from a shortest one.
\end{theorem}
\proof
Consider the GTSP instance $G$ in Figure~\ref{fig:gtsp_problem}.  It is a symmetric GTSP containing 5 clusters $\{ 1 \}$, $\{ 2, 2' \}$, $\{ 3 \}$, $\{ 4 \}$ and $\{ 5 \}$.  The weights of the edges not displayed in the graph are as follows: $w(1 \to 3) = w(1 \to 4) = 0$ and $w(2 \to 5) = w(2' \to 5) = 1$.

Observe that the tour $T = 1 \to 2 \to 3 \to 4 \to 5 \to 1$, shown in Figure~\ref{fig:worst_tour}, is a local minimum in both $N_\text{CO}(T)$ and $N_\text{TSP}(T)$.  The dashed line shows the second solution in $N_\text{CO}(T)$ but it gives the same objective value.  It is also clear that $T$ is a local minimum in $N_\text{TSP}(T)$.  Indeed, all the edges incident to the vertex 2 are of weight 1, and, hence, any tour through the vertex 2 is at least of weight 2.

The tour $T$ is in fact a longest tour in $G$\@.  Observe that all nonzero edges in $G$ are incident to vertices 2 and $2'$.  Since only one of these vertices can be visited by a tour, at most two nonzero edges can be included into a tour.  Hence, the weight of the worst tour in $G$ is 2.

However, there exists a better GTSP tour $T_\text{opt} = 1 \to 2' \to 4 \to 3 \to 5 \to 1$ of weight 1, see Figure~\ref{fig:gtsp_problem}.

\qed

In fact, TSP and GTSP behave quite differently during optimization.  Observe that there exists no way to find out quickly if some modification of the cluster order improves the tour.  Indeed, choosing wrong vertices within clusters may lead to an arbitrary large increase of the tour weight.  And since a replacement of a vertex within one cluster may require a replacement of vertices in the neighbor clusters, any local change influences the whole tour in general case.

\subsection{Original TSP Neighborhoods}

In order to continue this discussion, let us briefly list the most known TSP neighborhoods.  Here we assume that $m$ is the number of vertices in the TSP instance.

\begin{description}
	\item[$k$-opt] is the most general TSP neighborhood\footnote{We use the `$k$-opt' notation for both the neighborhood and the local search and specify explicitly, if necessary, what is meant in every particular case.  Note that in some literature this neighborhood is called $k$-exchange.}.  It includes all the tours which can be obtained from the given one by removing $k$ edges and inserting $k$ new edges.  Obviously any tour can be obtained from a given one by an $m$-opt move.
	\nomenclature[k-opt]{$k$-opt}{is a neighborhood and a local search used for many combinatorial optimization problems and, in particular, for TSP, GTSP and MAP; on every move it removes $k$ edges and inserts $k$ new edges such that the feasibility of the solution is preserved}
	
	\item[Insertion] includes all the tours which can be obtained from the given one by removing a vertex from the tour and inserting it at some other position.  It can be represented as a special case of 3-opt.
	\nomenclature{Insertion}{is a neighborhood and a local search for TSP and GTSP; on every move it tries to remove a vertex from the tour and insert it at a different position}

	\item[Or-opt] heuristic is an extension of Insertion.  First, it tries to insert every fragment of three vertices to every feasible position in the tour; then it does the same for every fragment of two vertices and finally it performs as simple Insertion.  Or-opt neighborhood can be represented as a special case of 3-opt.
	
	\item[Swap (also known as Exchange)] includes all the tours which can be obtained from the given one by swapping two vertices in the tour.  It can be represented as a special case of 4-opt.
	\nomenclature{Swap}{is a neighborhood and a local search for TSP and GTSP; on every move it tries to swap two vertices in the tour}
		
	\item[Lin-Kernighan] is a sophisticated heuristic which does not have any certain neighborhood; it explores some areas of $k$-opt neighborhood without fixing $k$.
	\nomenclature{Lin-Kernighan}{is a sophisticated TSP heuristic considered as an extension of the $k$-opt local search, where $k$ is not fixed}
\end{description}

For more information on these and some other TSP local searches, see, e.g., \cite{Johnson2002,Johnson2002a}.

\subsection{Adaptation of TSP local search for GTSP}
\label{sec:gtsp_ls_adaptation}

A typical local search with a neighborhood $N(T)$ performs as Algorithm~\ref{alg:typical_ls}.
\begin{algorithm}[ht]
\caption{Typical local search with neighborhood $N(T)$.}
\label{alg:typical_ls}
\begin{algorithmic}
\REQUIRE The original solution $T$.
\FORALL {$T' \in N(T)$}
	\IF {$w(T') < w(T)$}
		\STATE $T \gets T'$.
		\STATE Run the whole algorithm again.
	\ENDIF
\ENDFOR
\RETURN $T$.
\end{algorithmic}
\end{algorithm}
Let $N_1(T) \subseteq N_\text{TSP}(T)$ be a neighborhood of some TSP local search $\func{LS}_1(T)$.  Let $N_2(T) \subseteq N_\text{CO}(T)$ be a neighborhood of the Cluster Optimization class and $\func{LS}_2$ be corresponding local search, see Section~\ref{sec:gtsp_cluster_optimization}.  Then one can think of the following two adaptations of a TSP local search for GTSP:
\begin{enumerate}[(i)]
	\item \label{item:coinsidetsp} Enumerate all candidates $T' \in N_1(T)$.  For every candidate $T'$ run $T' \gets \func{LS}_2(T')$ to optimize it in $N_2(T')$.
	\item \label{item:tspinsideco} Enumerate all candidates $T' \in N_2(T)$.  For every candidate $T'$ run $T' \gets \func{LS}_1(T')$ to optimize it in $N_1(T')$.
\end{enumerate}

Observe that the TSP neighborhood $N_1(T)$ is normally much harder to explore than the cluster optimization neighborhood $N_2(T)$.  Consider, e.g., $N_1(T) = N_\text{TSP}(T)$ and $N_2(T) = N_\text{CO}(T)$.  Then both options yield an optimal GTSP solution but Option~(\ref{item:coinsidetsp}) requires $O(n \gamma s \cdot m!)$ operations while Option~(\ref{item:tspinsideco}) requires $O(s^m \cdot m!)$ operations.

Moreover, many practical applications of GTSP have some localization of clusters, i.e., $|w(x \to y_1) - w(x \to y_2)| \ll \max \{ w(x \to y_1), w(x \to y_2) \}$ on average, where $\Cluster(y_1) = \Cluster(y_2) \neq \Cluster(x)$.  Hence, the dependency of the $N_2(T)$ landscape on the cluster order is higher than the dependency of the $N_1(T)$ landscape on the vertex selection.  Hence, Option~(\ref{item:coinsidetsp}) is preferable.

Option~(\ref{item:tspinsideco}) was used in~\citet{Hu2008}.  Note that using $N_2(T) = N_\text{CO}(T)$ would lead to a non-polynomial algorithm; the cluster optimization neighborhood $N_2(T)$ they use includes only the tours which differ from $T$ in exactly one vertex.  For every $T' \in N_2(T)$, the Chained Lin-Kernighan heuristic is applied.  This results in $n$ runs of the Chained Lin-Kernighan heuristic which makes the heuristic unreasonably slow while the vertex selection is given a very little freedom.

\bigskip

Option~(\ref{item:coinsidetsp}) may be improved as in Algorithm~\ref{alg:co_inside_tsp_improved}.
\begin{algorithm}[ht]
\caption{Improved adaptation of a TSP neighborhood for GTSP according to Option~(\ref{item:coinsidetsp}).}
\label{alg:co_inside_tsp_improved}
\begin{algorithmic}
\REQUIRE The original tour $T$.
\FORALL {$T' \in N_1(T)$}
	\STATE $T' \gets \func{QuickImprove}(T')$.
	\IF {$w(T') < w(T)$}
		\STATE $T \gets \func{SlowImprove}(T')$.
		\STATE Run the whole algorithm again.
	\ENDIF
\ENDFOR
\RETURN $T$.
\end{algorithmic}
\end{algorithm}
Here $\func{QuickImprove}(T)$ and $\func{SlowImprove}(T)$ are some tour improvement heuristics of the Cluster Otimization class.  Formally, these heuristics should meet the following requirements: 
\begin{itemize}
	\item $\func{QuickImprove}(T), \func{SlowImprove}(T) \in N_\text{CO}(T)$ for any tour $T$;
	\item $w(\func{QuickImprove}(T)) \le w(T)$ and $w(\func{SlowImprove}(T)) \le w(T)$ for any tour $T$.
\end{itemize}
\func{QuickImprove} is applied to every candidate $T'$ before its evaluation.  \func{SlowImprove} is only applied to successful candidates in order to further improve them.  One can think of the following implementations of \func{QuickImprove} and \func{SlowImprove}: 
\begin{itemize}
	\item Trivial $I(T)$ which leaves the solution without any change: $I(T) = T$.
	\item Global cluster optimization $\mathit{CO}(T)$ which applies the \CO{} algorithm to the given solution.  The time complexity is $O(n \gamma s)$.
	\item Local cluster optimization $L(T) = L(T, I)$, see Section~\ref{sec:gtsp_cluster_optimization}.  It updates the vertices only within clusters $i \in I$, affected by the latest solution change.  E.g., if a tour $x_1 \to x_2 \to x_3 \to x_4 \to x_1$ was changed to $x_1 \to x_3 \to x_2 \to x_4 \to x_1$, we can use $L(T, \{ 2, 3 \})$ which will yield the best solution among $x_1 \to x'_3 \to x'_2 \to x_4 \to x_1$, where $x'_2 \in \Cluster(x_2)$ and $x'_3 \in \Cluster(x_3)$.  The time complexity of $L(T)$ is $O(s)$ or $O(s^2)$ if the number of affected clusters is fixed.
\end{itemize}

There are five meaningful combinations of \func{QuickImprove} and \func{SlowImprove}:
\begin{description}
	\item[Basic] $\func{QuickImprove}(T) = I(T)$ and $\func{SlowImprove}(T) = I(T)$.  This actually yields the original TSP local search applied to the TSP instance induced by the GTSP tour $T$.
	\nomenclature{Basic Adaptation}{is an adaptation of a TSP neighborhood for GTSP where no cluster optimization is applied}

	\item[Basic with CO] $\func{QuickImprove}(T) = I(T)$ and $\func{SlowImprove}(T) = \func{CO}(T)$, i.e., the algorithm explores the original TSP neighborhood but every time an improvement $T'$ is found, it is optimized in $N_\text{CO}(T')$.  One can also consider $\func{SlowImprove}(T) = L(T)$, but it has no practical interest.  Indeed, \func{SlowImprove} is used quite rarely and so its influence on the total running time is negligible.  At the same time, $\func{CO}(T)$ is much more powerful than $L(T)$ with respect to solution quality.
	\nomenclature{Basic with CO Adaptation}{is an adaptation of a TSP neighborhood for GTSP such that \CO{} is applied to every successful candidate}
	
	\item[Local] $\func{QuickImprove}(T) = L(T)$ and $\func{SlowImprove}(T) = I(T)$, i.e., every candidate $T' \in N_1(T)$ is improved locally before it is compared to the original solution.
	\nomenclature{Local Adaptation}{is an adaptation of a TSP neighborhood for GTSP such that some local cluster optimization is applied to every candidate solution}

	\item[Local with CO] $\func{QuickImprove}(T) = L(T)$ and $\func{SlowImprove}(T) = \func{CO}(T)$, which is the same as \emph{Local} but in addition it optimizes every improvement $T'$ globally in $N_\text{CO}(T')$.
	\nomenclature{Local with CO Adaptation}{is an adaptation of a TSP neighborhood for GTSP such that some local cluster optimization is applied to every candidate solution and then \CO{} is applied to every successful candidate}

	\item[Global] $\func{QuickImprove}(T) = \func{CO}(T)$ and $\func{SlowImprove}(T) = I(T)$, i.e., every candidate $T' \in N_1(T)$ is optimized globally in $N_\text{CO}(T')$ before it is compared to the original solution $T$.
	\nomenclature{Global Adaptation}{is the most powerful adaptation of a TSP neighborhood for GTSP such that \CO{} is applied to every candidate solution}
\end{description}

For a local search \func{LS} we use $\func{LS}_\text{B}$, $\func{LS}_\text{B}^\text{co}$, $\func{LS}_\text{L}$, $\func{LS}_\text{L}^\text{co}$ and $\func{LS}_\text{G}$ to denote the Basic, Basic with CO, Local, Local with CO and Global adaptations of \func{LS}, respectively.

Some of these adaptations were applied in the literature.  For example, the heuristics G2 and G3~\citep{Renaud1998} are actually Global adaptations of 2-opt and 3-opt TSP heuristics, respectively.  An enhanced implementation of the Global 2-opt adaptation is proposed in~\citet{Hu2008}; asymptotically, it is faster than the naive implementation by factor 3.  Local adaptations of 2-opt and some other neighborhoods were used in~\citet{Fischetti1997,GK_GTSP_GA_2008,Silberholz2007,Snyder2000,Tasgetiren2007}.  Some Basic adaptations were used in~\citet{Bontoux2009,GK_GTSP_GA_2008,Silberholz2007,Snyder2000}.

\subsection{Global Adaptation}
\label{sec:gtsp_global}

The most powerful adaptation of a TSP local search for GTSP is the Global adaptation.  It applies \CO{} to every candidate tour before it is evaluated.  In other words, if $N_1(T) \subseteq N_\text{TSP}(T)$ is the original TSP neighborhood, than the adapted neighborhood $N(T)$ is as follows:
$$
N(T) = \bigcup_{T' \in N_1(T)} N_\text{CO}(T') \,.
$$
Observe that, apart from other adaptations (see Section~\ref{sec:gtsp_ls_adaptation}), the Global one turns a polynomial TSP neighborhood into a very large neighborhood, i.e., into a neighborhood of the exponential size which can be explored in polynomial time.  Indeed, $N_\text{CO}(T_1) \cap N_\text{CO}(T_2) = \varnothing$ if the tours $T_1$ and $T_2$ have different cluster order.  Hence, the size of $N(T)$ is exactly 
$$
|N(T)| = |N_1(T)| \cdot \prod_{i=1}^m{|C_i|} \in O(|N_1(T)| \cdot s^m) \,.
$$

A straightforward exploration of the $N(T)$ neighborhood takes $O(n \gamma s \cdot |N_1(T)|)$.  This or slightly improved approach was applied in \citet{Renaud1998} and \citet{Hu2008}.

We propose a new technique which is $n \gamma / s^2$ times faster than a naive adaptation.  Apart from general discussion of this approach, we also provide an example of its application and introduce several efficient speed-up heuristics.

The main idea is to generate the candidates $T' \in N_1(T)$ in a certain order such that previously calculated shortest paths could be reused.  Observe that any TSP local search is a special case of $k$-opt.  Indeed, any transformation of a TSP tour may be represented as a $k$-opt move, subject to a sufficiently large value of $k$.

Let $\func{k\text{-opt}}(T, \alpha, \beta)$ be a tour obtained from $T$ by removing the edges $\alpha$ and adding the edges $\beta$, where $\alpha$ and $\beta$ are edge sets, $|\alpha| = |\beta| = k$.  We need to group all the candidates $T' \in N_1(T)$ as follows:
\begin{itemize}
	\item Let $T^1$, $T^2$, \ldots, $T^l$ be a group of candidates and $T^i = \func{k\text{-opt}}(T, \alpha^i, \beta^i)$.  The value of $k$ should be the same for all the candidates in the group.
	
	\item Let $\alpha = \bigcap_i{\alpha^i}$ and let $\alpha'^i = \alpha^i \setminus \alpha$.  Similarly, $\beta = \bigcap_i{\beta^i}$.
	
	\item Let $Q = T \setminus \alpha \cup \beta$, i.e., $Q$ is a set of paths and/or cycles produced from $T$ by removing the edges $\alpha$ and adding the edges $\beta$.
	
	\item Removing the edges $\alpha'^i$ from $Q$ yields a number of paths, let us say $P^i_1, P^i_2, \ldots, P^i_{k - |\beta|}$.  Our requirement is that every of these paths has at least one fixed end:
$$
\func{beginning}(P^i_x) = \func{beginning}(P^j_x) \text{ for every $i$ and $j$ or }
$$
$$
\func{end}(P^i_x) = \func{end}(P^j_x) \text{ for every $i$ and $j$}
$$
for every path index $x = 1, 2, \ldots, {k - |\beta|}$.

	\item In order to achieve an $m \cdot \gamma / s$ times speed-up, each group should contain at least $l \in \Theta(m)$ candidates, and the number of edges in every $\alpha^i$ should be fixed: $k - |\alpha| \in O(1)$.
\end{itemize}

If the declared above conditions are met, the Global adaptation may be implemented as in Algorithm~\ref{alg:global_adaptation}.
\begin{algorithm}[!ht]
\caption{General implementation of the Global adaptation of a TSP local search.}
\label{alg:global_adaptation}

\begin{algorithmic}
\REQUIRE Tour $T$.
\REQUIRE A group of candidates $T^1, T^2, \ldots, T^l$ such that $T^i = \func{k\text{-opt}}(T, \alpha^i, \beta^i)$.
\STATE Let $\alpha = \bigcap_i{\alpha^i}$ and $\beta = \bigcap_i{\beta^i}$. Let $\alpha'^i = \alpha^i \setminus \alpha$.
\STATE Let $Q = T \setminus \alpha \cup \beta$.  Let $Q \setminus \alpha^i = \{ P^i_1, P^i_2, \ldots, P^i_{k - |\beta|}\}$.  Note that the paths $P^i_j$ meet the conditions above.
\FOR {$j = 1, 2, \ldots, k - |\beta|$}
	\STATE Calculate all the shortest paths through the cluster sequences corresponding to $P^1_j$, $P^2_j$, \ldots, $P^l_j$.  Since one of the ends of all of these paths is fixed, this should take only $O(ms^3)$ operations.
\ENDFOR
\FOR {$i = 1, 2, \ldots, l$}
	\STATE Construct a layered network as follows:
	\begin{compactitem}
		\item Each layer $2j - 1$ corresponds to the beginning cluster of the path $P^i_j$;
		\item Each layer $2j$ corresponds to the end cluster of the path $P^i_j$;
		\item The weights of the edges between the layers $2j - 1$ and $2j$ corresponds to the shortest paths in $P^i_j$;
		\item The weights of the edges between the layers $2j$ and $2j + 1$ are the weights between corresponding clusters.
		\item Layer $l + 1$ is a copy of the layer 1 and the weights between the layers 1 and $l + 1$ are the weights between corresponding clusters.
	\end{compactitem}

	\STATE Find the shortest cycle $C$ in the constructed layered network using the \CO{} algorithm.  It will take only $O((k - |\beta|) \cdot s^3) = O(s^3)$ operations (recall that $k - |\beta| \in O(1)$.
	
	\IF {$w(C) < w(T)$}
		\STATE $T \gets C$.
		\STATE Restart the algorithm.
	\ENDIF
\ENDFOR
\RETURN $T$.
\end{algorithmic}
\end{algorithm}

Observe that these results can be easily used for the assymmetric case.  Indeed, even if orientation of some path in the candidate tour does not conincide with orientation of this path in the original tour, one can calculate the shortest paths within this fragment in the backward direction.

\subsubsection{Example Implementation}
\label{sec:gtsp_two_opt_global_adaptation}

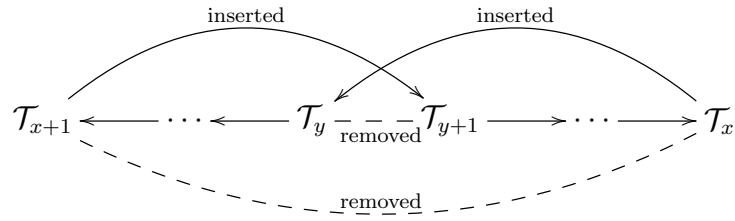
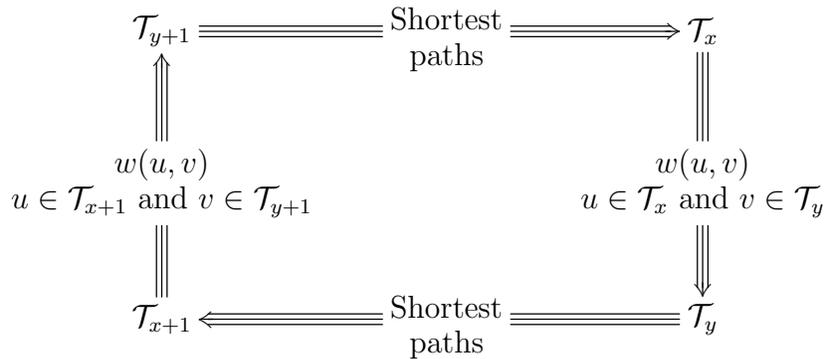
\begin{figure}[ht]
\centering
\subfloat[The clusters $\mathcal{T}_{x+1}$ and $\mathcal{T}_x$ are fixed while $\mathcal{T}_y$ and $\mathcal{T}_{y+1}$ `slide' from left to right.]
{
	\label{fig:two_opt_global_tour}
\xymatrix@R=3em@C=2.5em@L=0.2em{
		*+{\mathcal{T}_{x+1}} \ar@{<-}[r]
	&	*+{\cdots} \ar@{<-}[r]
	&	*+{\mathcal{T}_y} \ar@/^3em/@{<-}[rrr]^{\text{inserted}} \ar@{--}[r]_{\text{removed}}
	&	*+{\mathcal{T}_{y+1}} \ar@/_3em/@{<-}[lll]_{\text{inserted}}
	&	*+{\cdots} \ar@{<-}[l]
	&	*+{\mathcal{T}_x} \ar@{<-}[l] \ar@/^3em/@{--}[lllll]_{\text{removed}}
}
}
\\[2em]
\subfloat[Having all the shortest paths from $\mathcal{T}_y$ to $\mathcal{T}_{x+1}$, and from $\mathcal{T}_{y+1}$ to $\mathcal{T}_x$, one can construct this layered network and apply \CO{} to it in order to find the shortest cycle in the whole rearranged tour.]
{
	\label{fig:two_opt_global_network}
\xymatrix@R=2em@C=2em@L=0.5em{
		*+{\mathcal{T}_{y+1}} \ar@3{-}[r]
	&	\txt{Shortest\\paths} \ar@3{->}[r]
	&	*+{\mathcal{T}_x} \ar@3{-}[d]
\\
		\txt{$w(u, v)$\\$u \in \mathcal{T}_{x+1}$ and $v \in \mathcal{T}_{y+1}$} \ar@3{->}[u]
	&
	&	\txt{$w(u, v)$\\$u \in \mathcal{T}_x$ and $v \in \mathcal{T}_y$} \ar@3{->}[d]
\\
		*+{\mathcal{T}_{x+1}} \ar@3{-}[u]
	&	\txt{Shortest\\paths} \ar@3{->}[l]
	&*+{\mathcal{T}_y} \ar@3{-}[l]
}
}
\caption{Global adaptation of the TSP 2-opt heuristic.}
\label{fig:two_opt_global}
\end{figure}

Let us consider the 2-opt TSP neighborhood and its Global adaptation.  Algorithm~\ref{alg:two_opt_enumerate} enumerates all the candidates in $N_\text{2-opt}(T)$.
\begin{algorithm}[ht]
	\caption{Enumeration of all the candidates in the TSP 2-opt neighborhood.}
	\label{alg:two_opt_enumerate}
\begin{algorithmic}
\REQUIRE The original solution $T$.
\FOR {$x = 1, 2, \ldots, m - 2$}
	\FOR {$y = x + 2, x + 3, \ldots, \min\{m, x + m - 2\}$}
		\STATE List the candidate $\Turn(T, x, y)$ (see Section~\ref{sec:intro}).
	\ENDFOR
\ENDFOR
\end{algorithmic}
\end{algorithm}
Observe that all the candidates which share the same value of $x$ meet the conditions above (see Section~\ref{sec:gtsp_global}).  Indeed, for each $x$ there exist $\Theta(m)$ candidates such that the set $\alpha^i = \{ T_x \to T_{x+1}, T_{y(i)} \to T_{y(i)+1} \}$ and the set $\beta^i = \{ T_x \to T_{y(i)}, T_{x+1} \to T_{y(i)+1} \}$ (see Figure~\ref{fig:two_opt_global_tour}).
We get $\alpha = \{ T_x \to T_{x+1} \}$ and $\beta = \varnothing$.  Hence, $Q$ is a path obtained from $T$ by removing the edge $T_x \to T_{x+1}$.  Removing the edge $\alpha'^i = \{ T_{y(i)} \to T_{y(i)+1} \}$ splits $Q$ into two paths $T_{x+1} \to \ldots \to T_{y(i)}$ and $T_{y(i)+1} \to \ldots \to T_x$.  Observe that the first of these paths has a fixed beginning, and the second of these paths has a fixed end.

The algorithm exploring the neighborhood for some fixed $x$ is presented in Algorithm~\ref{alg:two_opt_global}.
\begin{algorithm}[ht]
	\caption{Global adaptation of the 2-opt heuristic.}
	\label{alg:two_opt_global}
\begin{algorithmic}
\REQUIRE The original tour $T$.
\STATE Let $\mathcal{T}_i = \Cluster(T_i)$.
\FOR {$x = 1, 2, \ldots, m - 2$}
	\STATE Calculate the shortest paths along the tour $T$ from every vertex in $\mathcal{T}_{x+1}$ to every vertex in $\mathcal{T}_{y}$ and from every vertex in $\mathcal{T}_x$ to every vertex in $\mathcal{T}_{y + 1}$ for every $y = x + 2, x + 3, \ldots, \min\{m, x + m - 2\}$.
	\FOR {$y = x + 2, x + 3, \ldots, \min\{m, x + m - 2\}$}
		\STATE Construct a layered network as in Figure~\ref{fig:two_opt_global_network}.
		\STATE Apply \CO{} to this layered network to get the shortest cycle $C$.
		\IF {$w(C) < w(T)$}
			\STATE Replace $T$ with $C$.
			\STATE Restart the whole algorithm.
		\ENDIF
	\ENDFOR
\ENDFOR
\end{algorithmic}
\end{algorithm}
Compare the time complexity of the naive exploration of $N_\text{2-opt}(T)$, which is $O(m^2 n \gamma s)$, with our adaptation, which takes only $O(m n s^2)$ operations.  If $s / \gamma \ll m$, which is a very natural assumption, our implementation is significantly faster than the naive one.

\subsection{Global Adaptation Refinements}
\label{sec:gtsp_global_refinements}

In certain cases it is possible to significantly speed up the Global adaptation algorithm proposed in Section~\ref{sec:gtsp_global}.  Consider the \twoopt{G}{} implementation described in Section~\ref{sec:gtsp_two_opt_global_adaptation}.  For a fixed $x$, its time complexity is $O(n s^2)$.  A more accurate estimation of the number of operations required for every value of $x$ is as follows (it consists of calculating the shortest paths, see Figure~\ref{fig:two_opt_global_tour}, and finding the shortest cycles, see Figure~\ref{fig:two_opt_global_network}):
\begin{multline}
\label{eq:two_opt_time_estimation}
t(x) = \left( |\mathcal{T}_{x+1}| \cdot \sum_y{|\mathcal{T}_{y-1}| |\mathcal{T}_y|} \right)
+ \left( |\mathcal{T}_x| \cdot \sum_y{|\mathcal{T}_{y+2}| |\mathcal{T}_{y+1}|} \right) \\
+ \sum_y{\min\Big\{|\mathcal{T}_x| |\mathcal{T}_{x+1}| \cdot \big(|\mathcal{T}_y| + |\mathcal{T}_{y+1}|\big),\ |\mathcal{T}_y| |\mathcal{T}_{y+1}| \cdot \big(|\mathcal{T}_x| + |\mathcal{T}_{x+1}|\big)\Big\}} \,.
\end{multline}
The minimum in (\ref{eq:two_opt_time_estimation}) is taken because our \CO{} implementation finds the optimal calculations order if the number of layers in the network is 4 (see Section~\ref{sec:gtsp_co_refinements}).
  
Recall that an important property of \CO{} is that its running time hugely depends on the size of the first cluster, and, hence, we choose the smallest cluster as the first one.  Similarly, our 2-opt adaptation depends on the clusters \T{x}, \T{x+1}, \T{y} and \T{y+1}, but, unfortunately, selection of these clusters does not depend on us.  However, we can introduce a supporting cluster, i.e., break the tour at some extra position, and choose this cluster arbitrarily.

\begin{figure}[ht]
\centerline{
\xymatrix@R=2em@C=1em@L=0.5em{
		*+{\T{y+1}} \ar@3{-}[r]
	&	\txt{Shortest\\paths} \ar@3{->}[r]
	&	*+{\T{1}} \ar@3{-}[r]
	&	\txt{Shortest\\paths} \ar@3{->}[r]
	&	*+{\T{x}} \ar@3{-}[d]
\\
		\txt{$w(u, v)$\\$u \in \mathcal{T}_{x+1}$ and $v \in \mathcal{T}_{y+1}$} \ar@3{->}[u]
	&
	&
	&
	&	\txt{$w(u, v)$\\$u \in \mathcal{T}_x$ and $v \in \mathcal{T}_y$} \ar@3{->}[d]
\\
		*+{\T{x+1}} \ar@3{-}[u]
	&
	&	\txt{Shortest\\paths} \ar@3{->}[ll]
	&
	&*+{\T{y}} \ar@3{-}[ll]
}
}
	\caption{Global adaptation of the TSP 2-opt heuristic with a supporting cluster.}
	\label{fig:two_opt_supporting_cluster}
\end{figure}
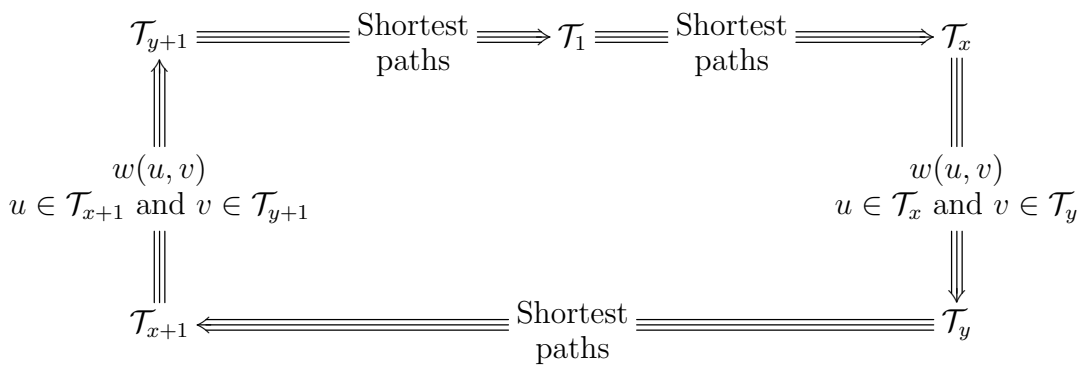

Without loss of generality, assume that $\T{1}$ is the smallest cluster in the problem: $|\T{1}| = \gamma$.  If any of \T{x}, \T{x+1}, \T{y} or \T{y+1} coincide with \T{1}, then the search of the shortest cycle is already quick.  Otherwise let us include \T{1} in the layered network (see Figure~\ref{fig:two_opt_supporting_cluster}).
Now it takes only $O(\gamma s^2)$ operations to find the shortest cycle in the rearranged tour.

Observe that $\T{1}$ always belongs to the fragment \T{y+1}, \T{y+2}, \ldots, \T{x}.  Hence, if $x \neq 1$ and $y + 1 \neq 1$ (i.e., $y \neq m$), then, instead of the shortest paths from \T{y+1} to \T{x}, we need the shortest paths from \T{y+1} to \T{1} and from \T{1} to \T{x}, see Figure~\ref{fig:two_opt_supporting_cluster}.  Whatever is the calculations scheme, it takes only $O(n \gamma s)$ operations to calculate all the shortest paths between \T{y+1} and \T{x} for a fixed $x$.  Moreover, these shortest paths can be reused for different values of $x$, i.e., they should be calculated once for the whole procedure.  One only has to update some of these paths when a tour improvement is found.

It is more difficult to speed up the calculations of the shortest paths from \T{y} to \T{x+1}.  Indeed, there is no cluster in this fragment which could be used as a supporting one for all the values of $x$ and $y$.  Our workaround is as follows.  Consider a problem of finding the shortest paths from every $u \in X_1$ to every $v \in X_i$, where $i = 2, 3, \ldots, k$ in a sequence $X_1$, $X_2$, \ldots, $X_k$ of clusters.  A straightforward approach to this problem is presented in Algorithm~\ref{alg:shortestpaths_straightforward}.
\begin{algorithm}[ht]
\caption{Straightforward calculation of the shortest paths.}
\label{alg:shortestpaths_straightforward}

\begin{algorithmic}
\REQUIRE Sequence of clusters $X_1$, $X_2$, \ldots, $X_k$.

\FORALL {$u \in X_1$ and $v \in X_2$}
	\STATE $l(u \to v) \gets w(u \to v)$.
\ENDFOR

\FOR {$i \gets 3, 4, \ldots, k$}
	\FORALL {$u \in X_1$ and $v \in X_i$}
		\STATE $l(u \to v) \gets \min_{t \in X_{i-1}} \{ l(u \to t) + w(t \to v) \}$.
	\ENDFOR
\ENDFOR
\end{algorithmic}
\end{algorithm}
To proceed, it takes approximately 
$
\displaystyle{
|X_1| \cdot \sum_{i=3}^k |X_{i-1}| |X_i|
}
$
operations.  Since its running time significantly depends on the size of the cluster $X_1$, we can do the following.  Let $|X_j| < |X_1|$ for some $j$.  Then we can calculate all the shortest paths $l_{u, v}$ from every $u \in X_1$ to every $t \in X_i$, $i = 2, 3, \ldots, j$, and then for the rest of the cluster sequence calculate the shortest paths $l_{t, v}$ from every $t \in X_j$ to every $X_i$, where $i = j + 1, j + 2, \ldots, k$.  This will take approximately 
$$
|X_1| \cdot \sum_{i=3}^j |X_{i-1}| |X_i| + |X_j| \cdot \sum_{i=j+2}^k |X_{i-1}| |X_i| \text{ operations.}
$$
Hence, introducing the supporting cluster $X_j$ will save
$$
|X_1| |X_j| |X_{j+1}| + (|X_1| - |X_j|) \cdot \sum_{i=j+2}^k |X_{i-1}| |X_i| \text{ operations.}
$$
However, the refined algorithm does not yield the shortest paths from $u \in X_1$ to $v \in X_k$.  It introduces a supporting cluster, which means that one should do some additional calculations in order to use the obtained results.  In particular, the supporting cluster slows down the \CO{} algorithm which is applied after each local search move to get the shortest cycle.

There is no quick way to decide if introducing a supporting cluster is beneficial, but we can do some estimation.  Let us compare two algorithms: with and without the supporting cluster $X_j$.  Observe that these algorithms behave equally for $i = 2, 3, \ldots, j$ and, thus, we are interested only in $i = j + 1, j + 2, \ldots, k$.  

Algorithm~\ref{alg:global_adaptation} consists of two parts: calculation of the shortest paths through the tour fragments and calculation of the shortest cycles in the rearranged tours.  The second part applies the \CO{} procedure to small layered networks in order to find the shortest cycle through the whole tour.  Recall that \CO{} depends on the size of the smallest layer in the layered network but we can guarantee that the constructed layered network contains a layer of size $\gamma$ (see above).

Let $X_1$, $X_2$, \ldots, $X_k$ correspond to \T{y}, \T{y+1}, \ldots, \T{x+1}.  Without a supporting cluster, the procedure requires
$$
t_\text{pure} = \left(|X_1| \cdot \sum_{i=j+1}^k |X_{i-1}| |X_i| \right) + \gamma \cdot |X_1| \cdot \sum_{i=j+1}^k |X_i| \text{ operations.}
$$
If the supporting cluster exists, it requires
$$
t_\text{sup} = \left(|X_j| \cdot \sum_{i=j+2}^k |X_{i-1}| |X_i| \right) + \gamma \cdot |X_j| \cdot \sum_{i=j+1}^k \big( |X_1| + |X_i| \big) \text{ operations.}
$$
In order to get some meaningful estimation, let us use the expected value of $|\T{i}| = n / m$\footnote{We assume that cluster sizes are distributed uniformly and independantly.}.  Then we can replace $\displaystyle{\sum_{i=j+1}^k |X_i|}$ with $\displaystyle{(k - j) \cdot \frac{n}{m}}$, and 
$\displaystyle{\sum_{i=j+1}^k |X_{i-1}| |X_i|}$ with $\displaystyle{(k - j) \cdot \left(\frac{n}{m}\right)^2}$.  Finally we get:
$$
t_\text{pure} = |X_1| (k - j) (n / m)^2 + \gamma |X_1| (k - j) (n / m) \text{ and}
$$
$$
t_\text{sup} = |X_j| (k - j) (n / m)^2 - |X_j| (n / m)^2 + \gamma |X_j| (k - j) (|X_1| + n / m) \,.
$$
Now we can find the ratio $t_\text{sup} / t_\text{pure}$ to determine when it is beneficial to introduce the supporting cluster:
\begin{multline}
\frac{t_\text{sup}}{t_\text{pure}} = \frac{
|X_j| (k - j) (n / m)^2 - |X_j| (n / m)^2 + \gamma |X_j| (k - j) (|X_1| + n / m)
}{
|X_1| (k - j) (n / m)^2 + \gamma |X_1| (k - j) (n / m)
} \\
= \frac{|X_j|}{|X_1|} \cdot \frac{
(n / m)^2 - (n / m)^2 / (k - j) + \gamma (|X_1| + n / m)
}{
(n / m)^2 + \gamma (n / m)
} \,.
\end{multline}
We are interested in the case when $t_\text{sup} / t_\text{pure} < 1$.  Hence, we claim
\begin{equation}
\label{eq:supporting_cluster_size}
|X_j| < |X_1| \cdot \frac{
(n / m)^2 + \gamma (n / m)
}{
(n / m)^2 - (n / m)^2 / (k - j) + \gamma (|X_1| + n / m)
} \,.
\end{equation}
Let us substitute some reasonable values to (\ref{eq:supporting_cluster_size}).  Let $\gamma = 1$, $k - j \gg 1$, $|X_1| = n / m$ and $n / m = 5$.  Then $|X_j| < 4.29$, i.e., any $X_j$ which is smaller than $X_i$ is beneficial.  For $|X_1| = 2 \cdot n / m$ we get $|X_j| < 7.5$, i.e., if the the first cluster is large, any cluster even of an average size is suitable as a supporting one.

Observe that the estimation (\ref{eq:supporting_cluster_size}) was obtained in the assumption that we need to apply \CO{} for every possible local search move.  In fact, one can avoid some of these calculations by using a lower bound.  Hence, even if the condition (\ref{eq:supporting_cluster_size}) is not met, it is probable that introducing a supporting cluster is still beneficial.  Observe also that, having a supporting cluster $X_j$, it is always beneficial to replace it with a new one $X_t$, $t > j$, if $|X_t| < |X_j|$.  In our implementations we decided to avoid the check (\ref{eq:supporting_cluster_size}) and introduce a supporting cluster as soon as a smaller cluster is found, see Algorithm~\ref{alg:supporting_clusters}.
\begin{algorithm}[ht]
\caption{Calculation of the shortest paths with supporting clusters.}
\label{alg:supporting_clusters}

\begin{algorithmic}
\REQUIRE Sequence of clusters $X_1$, $X_2$, \ldots, $X_k$.

\FORALL {$u \in X_1$ and $v \in X_2$}
	\STATE $l_{u, v} \gets w(u \to v)$.
\ENDFOR

\STATE Initialize the supporting cluster $Z \gets X_1$.

\FOR {$j \gets 3, 4, \ldots, k$}
	\IF {$|X_{j-1}| < |Z|$}
		\IF {$Z \neq X_1$}
			\FORALL {$u \in X_1$ and $v \in X_{j-1}$}
				\STATE $l_{u, v} \gets \min_{z \in Z} \{ l_{u, z} + l_{z, v} \}$.
			\ENDFOR
		\ENDIF
		\STATE Update the supporting cluster $Z \gets X_{j-1}$.
		\FORALL {$u \in Z$ and $v \in X_j$}
			\STATE $l_{u, v} \gets w(u \to v)$.
		\ENDFOR
	\ELSE
		\FOR {every $u \in Z$ and every $v \in X_j$}
			\STATE $l_{u, v} \gets \min_{p \in X_{j-1}} \{ l_{u, p} + w(p \to v) \}$.
		\ENDFOR
	\ENDIF
\ENDFOR
\end{algorithmic}
\end{algorithm}

\bigskip

In the proposed adaptation, we need to calculate the shortest cycle on every iteration, and it takes $O(f s^3)$ time, where $f$ is the number of the fragments to be rearranged.  Having a lower bound for the shortest cycle, one can omit some of these calculations.

Assume that the rearranged tour $T$ consists of $k$ fragments $P^1$, $P^2$, \ldots, $P^k$ such that $\func{end}(P^i)$ is connected to $\func{beginning}(P^{i+1})$ and $\func{end}(P^k)$ is connected to $\func{beginning}(P^1)$, where $\func{beginning}(P^i)$ ($\func{end}(P^i)$) is the first (the last) cluster in $P^i$.  Let $p^i$ be the shortest path through the cluster sequence $P^i$.  Then it is clear that the lower bound for the shortest cycle in this sequence of clusters is 
$$
\func{CO}(T) \ge \sum_{i=1}^k \left[w(p^i) + w_\text{min}\left(\func{beginning}(P^i) \to \func{end}(P^{i+1})\right)\right] \,,
$$
where $w_\text{min}(X \to Y)$ is the weight of the shortest edge from cluster $X$ to cluster $Y$ and $P^{k+1} = P^1$.

It would take too much time to calculate the shortest paths $p^i$ on every iteration.  Instead, we propose a lower bound for $w(p^i)$ according to Theorem~\ref{th:shortest_path_lower_bound}.

\begin{theorem}
\label{th:shortest_path_lower_bound}
For the shortest path from an arbitrary vertex in $\mathcal{T}_a$ to an arbitrary vertex in $\mathcal{T}_b$ in a layered network $\mathcal{T}_1 \cup \mathcal{T}_2 \cup \ldots \cup \mathcal{T}_m$ we have:
\begin{multline}
\label{eq:shortest_path_lower_bound}
w_\text{min}(\mathcal{T}_a \to \mathcal{T}_{a+1} \to \ldots \to \mathcal{T}_b) \ge w(T_a \to T_{a+1} \to \ldots \to T_b)\\
- w_\text{max}(T_a \to \mathcal{T}_{a+1}) - w_\text{max}(\mathcal{T}_{b-1} \to T_b)\\
+ w_\text{min}(\mathcal{T}_a \to \mathcal{T}_{a+1}) + w_\text{min}(\mathcal{T}_{b-1} \to \mathcal{T}_b) \,,
\end{multline}
where $T_1 \to T_2 \to \ldots \to T_m \to T_1$ is the shortest cycle through all the layers of the network.
\end{theorem}
\proof
Observe that $T_a \to T_{a+1} \to \ldots \to T_b$ is the shortest path from $T_a$ to $T_b$ through the layers $\mathcal{T}_{a+1}$, $\mathcal{T}_{a+1}$, \ldots, $\mathcal{T}_{b-1}$.  Indeed, if there would exist a shorter path, the condition that $T_1 \to T_2 \to \ldots \to T_m \to T_1$ would be violated.

Now assume that there exists some path $T'_a \to T'_{a+1} \to \ldots \to T'_b$ which is shorter than the lower bound provided in~(\ref{eq:shortest_path_lower_bound}):
\begin{multline*}
w(T'_a \to T'_{a+1} \to \ldots \to T'_b) < w(T_a \to T_{a+1} \to \ldots \to T_b)\\
- w_\text{max}(T_a \to \mathcal{T}_{a+1}) - w_\text{max}(\mathcal{T}_{b-1} \to T_b)\\
+ w_\text{min}(\mathcal{T}_a \to \mathcal{T}_{a+1}) + w_\text{min}(\mathcal{T}_{b-1} \to \mathcal{T}_b) \,.
\end{multline*}
Observe that 
$$
w(T_a \to T'_{a+1}) - w_\text{max}(T_a \to \mathcal{T}_{a+1}) \le w(T'_a \to T'_{a+1}) - w_\text{min}(\mathcal{T}_a \to \mathcal{T}_{a+1}) \text{ and}
$$
$$
w(T'_{b-1} \to T_b) - w_\text{max}(\mathcal{T}_{b-1} \to T_b) \le w(T'_{b-1} \to T'_b) - w_\text{min}(\mathcal{T}_{b-1} \to \mathcal{T}_b) \,.
$$
Since
$
w(T'_a \to T'_{a+1} \to \ldots \to T'_b)
= w(T'_a \to T'_{a+1}) + w(T'_{a+1} \to T'_{a+2} \to \ldots \to T'_{b-1}) + w(T'_{b-1} \to T'_b)
$, 
we get:
\begin{multline*}
w(T_a \to T'_{a+1}) - w_\text{max}(T_a \to \mathcal{T}_{a+1}) + w_\text{min}(\mathcal{T}_a \to \mathcal{T}_{a+1})\\
+ w(T'_{a+1} \to T'_{a+2} \to \ldots \to T'_{b-1})\\
+ w(T'_{b-1} \to T_b) - w_\text{max}(\mathcal{T}_{b-1} \to T_b) + w_\text{min}(\mathcal{T}_{b-1} \to \mathcal{T}_b)\\
< w(T_a \to T_{a+1} \to \ldots \to T_b)
- w_\text{max}(T_a \to \mathcal{T}_{a+1}) - w_\text{max}(\mathcal{T}_{b-1} \to T_b)\\
+ w_\text{min}(\mathcal{T}_a \to \mathcal{T}_{a+1}) + w_\text{min}(\mathcal{T}_{b-1} \to \mathcal{T}_b) \,.
\end{multline*}
Hence:
\begin{multline*}
w(T_a \to T'_{a+1}) + w(T'_{a+1} \to T'_{a+2} \to \ldots \to T'_{b-1}) + w(T'_{b-1} \to T_b)\\
< w(T_a \to T_{a+1} \to \ldots \to T_b) \text{ or}
\end{multline*}
$$
w(T_a \to T'_{a+1} \to T'_{a+2} \to \ldots \to T'_{b-1} \to T_b)
< w(T_a \to T_{a+1} \to \ldots \to T_b) \,.
$$
The latter means that the path $T_a \to T'_{a+1} \to T'_{a+2} \to \ldots \to T'_{b-1} \to T_b$ is shorter than $T_a \to T_{a+1} \to \ldots \to T_b$ but this contradicts with the fact that $T_a \to T_{a+1} \to \ldots \to T_b$ is the shortest path from $T_a$ to $T_b$.  Hence, our assumption is wrong. \qed

Observe that, having precalculated $w_\text{min}(X \to Y)$ for every pair of clusters $X$ and $Y$ and $w_\text{max}(x \to Y)$ and $w_\text{max}(Y \to x)$ for every pair of vertex $x$ and cluster $Y$, it takes only $O(1)$ time to obtain a lower bound according to Theorem~\ref{th:shortest_path_lower_bound}.  One also has to apply \CO{} once but the time required for this call is usually negligible.

Our experiments have shown that the usa of the lower bound speeds up the 2-opt Global adaptation three times.  The lower bound works better for  large instances since only a fixed number of edges in the lower bound are calculated imprecisely while the rest of it corresponds to exact shortest paths.

Another lower bound is even more efficient if one needs the shortest path through all the clusters in a broken cycle.
\begin{theorem}
\label{th:shortest_path_lower_bound_cycle}
For the shortest path from an arbitrary vertex in $\mathcal{T}_1$ to an arbitrary vertex in $\mathcal{T}_m$ in a layered network $\mathcal{T}_1 \cup \mathcal{T}_2 \cup \ldots \cup \mathcal{T}_m$ we have:
\begin{multline*}
w_\text{min}(\mathcal{T}_1 \to \mathcal{T}_2 \to \ldots \to \mathcal{T}_m) \ge w(T_1 \to T_2 \to \ldots T_m \to T_1) - w_\text{max}(\mathcal{T}_m \to \mathcal{T}_1) \,,
\end{multline*}
where $T_1 \to T_2 \to \ldots \to T_m \to T_1$ is the shortest cycle through all the layers of the network.
\end{theorem}
\proof
Assume there exists a path $T'_1 \to T'_2 \to \ldots \to T'_m$ such that
$$
w(T'_1 \to T'_2 \to \ldots \to T'_m) < w(T_1 \to T_2 \to \ldots T_m \to T_1) - w_\text{max}(\mathcal{T}_m) \,.
$$
Close up this path with the edge $T'_m \to T'_1$.  Observe that the weight of the obtained cycle is
\begin{multline*}
w(T'_1 \to T'_2 \to \ldots \to T'_m \to T'_1) < w(T_1 \to T_2 \to \ldots T_m \to T_1) \\
+ w(T'_m \to T'_1) - w_\text{max}(\mathcal{T}_m \to \mathcal{T}_1) \,.
\end{multline*}
However, this contradicts with the fact that $T_1 \to T_2 \to \ldots \to T_m \to T_1$ is the shortest path.  Hence, our assumption is wrong.
\qed

\subsection{\texorpdfstring{$k$-opt}{k-opt}}
\label{sec:k_opt}

$k$-opt neighborhood is widely used for TSP and some other combinatorial optimization problems, see~\citep{Fischetti1997,GK_MAP_LS_2010,GK_GTSP_GA_2008,Snyder2000}.  It was shown to be very efficient in TSP \cite{Helsgaun2009}.  In general, $N_\text{$k$-opt}(T)$ contains all the solutions which can be obtained from $T$ by selecting $k$ elements in $T$ and then replacing them with $k$ new elements such that the feasibility of the solution is preserved.  In TSP or GTSP, $k$-opt means replacing $k$ existing edges in the solution with $k$ new edges.

The time complexity of $k$-opt increases exponentially with the growth of $k$.  In practice only 2-opt and 3-opt are used for TSP~\citep{Helsgaun2000,Lin1965} with rare exceptions~\citep{Helsgaun2009}.  We do not consider $k$-opt for $k > 3$.

\subsection{2-opt}
\label{sec:gtsp_two_opt}

For $k = 2$ and for a fixed pair of edges $T_x \to T_{x+1}$, $T_y \to T_{y+1}$ there are only two options for every 2-opt move, i.e., to replace these edges either with $T_x \to T_y$ and $T_{x+1} \to T_{y+1}$ or with $T_{y+1} \to T_{x+1}$ and $T_y \to T_x$.  However, for the symmetric case both options are identical and it takes only $O(1)$ operations to evaluate a 2-opt move, see~(\ref{eq:turn_delta}).  Hence, it takes $O(m^2)$ operations to explore the whole neighborhood $N_\text{2-opt}(T)$ in the symmetric case.

We consider two algorithms to explore the 2-opt neighborhood, namely \emph{simple} and \emph{advanced}.  The simple one tries all feasible pairs of $x$ and $y$ with $y > x$.  An efficient approach is used to avoid repetitions, see Algorithm~\ref{alg:two_opt_basic}.
\begin{algorithm}[th]
\caption{Basic 2-opt implementation with an efficient algorithm of avoiding repetitions (symmetric case).}
\label{alg:two_opt_basic}
\begin{algorithmic}
\REQUIRE Tour $T = T_1 \to T_2 \to \ldots \to T_m \to T_1$.
\STATE Initialize $b(T_i) \gets \text{true}$ for every $i = 1, 2, \ldots, m$.
\REPEAT
	\STATE Initialize $\var{optimal} \gets \text{true}$.
	\FOR {$x \gets 1, 2, \ldots, m - 2$}
		\STATE Initialize $\delta \gets 0$.
		\FOR {$y \gets x + 2, x + 3, \ldots, \min\{ m, x + m - 2 \}$}
			\IF {$b(T_x) = \text{false}$ and $b(T_y) = \text{false}$}
				\STATE Go to the next $y$.
			\ENDIF
			\STATE $\Delta \gets w(T_x \to T_y) + w(T_{x+1} \to T_{y+1}) - w(T_x \to T_{x+1}) - w(T_y \to T_{y+1})$.
			\IF {$\Delta < 0$}
				\STATE Replace the edges $T_x \to T_{x+1}$ and $T_y \to T_{y+1}$ in $T$ with the edges $T_x \to T_y$ and $T_{x+1} \to T_{y+1}$.
				\STATE `Invalidate' vertices: $b(T_i) = \text{true}$ for every $i = x, x + 1, \ldots, y$.
				\STATE Set $\var{optimal} \gets \text{false}$.
				\STATE Start the inner loop from scratch, i.e., $y \gets x + 2$.
			\ENDIF
		\ENDFOR
	\ENDFOR
\UNTIL {$\var{optimal} = \text{true}$}
\end{algorithmic}
\end{algorithm}
In particular, the algorithm stores a flag $b(T_i)$ for every vertex $T_i$.  This flag shows if the edge which starts from $T_i$ was changed since the last check.  Observe that a move of $\Turn(T, x, y)$ is redundant if both edges $T_x \to T_{x+1}$ and $T_y \to T_{y+1}$ stay unchanged since the last check of $\Turn(T, x, y)$.

The advanced algorithm is only suitable for symmetric problems.  It considers all the values $x \in \{ 1, 2, \ldots, m \}$ and for every $x$ it takes all feasible $y$ such that $w(T_x \to T_y) < w(T_x \to T_{x+1})$ or $w(T_{x+1} \to T_{y+1}) < w(T_x \to T_{x+1})$.  Note that every pair of edges can be considered twice in this approach.  Hence, if a pair of edges was not considered at all, then both $w(T_x \to T_y) \ge w(T_x \to T_{x+1})$ and $w(T_{x+1} \to T_{y+1}) \ge w(T_y \to T_{y+1})$ which cannot be an improving move.

For every vertex $v$ precalculate a list $l(v)$ of vertices $l(v)_1, l(v)_2, \ldots$ and order them according to the distance $w(v, l(v)_i)$.  Now for some $x$ one should only consider the first entries of $l(T_x)$ for the vertex $T_y$ or the first entries of $l(T_{x+1})$ for $T_{y+1}$.  For details see~\citep{Johnson2002}.

\bigskip

For the asymmetric problem one standalone move $\Turn(T, x, y)$ of \twoopt{}{} requires $O(m)$ operations.  There are two options to reconnect the fragments and each of the options requires one of these fragments to be inverted.  However, it is still possible to explore the whole neighborhood $N_\text{2-opt}(T)$ in $O(m^2)$.  For this purpose the \twoopt{}{} moves should be carried out in a certain sequence, see Algorithm~\ref{alg:two_opt_asymmetric}.
\begin{algorithm}[ht]
\caption{Basic 2-opt implementation for asymmetric problem.}
\label{alg:two_opt_asymmetric}
\begin{algorithmic}
\REQUIRE Tour $T = T_1 \to T_2 \to \ldots \to T_m \to T_1$.
\FOR {$x \gets 1, 2, \ldots, m - 2$}
	\STATE Initialize $\delta \gets 0$.
	\FOR {$y \gets x + 2, x + 3, \ldots, \min\{ m, x + m - 2 \}$}
		\STATE Update $\delta \gets \delta + w(T_{y-1} \to T_y) - w(T_y \to T_{y - 1})$.
		\STATE $\Delta \gets w(T_x \to T_y) + w(T_{x+1} \to T_{y+1}) - w(T_x \to T_{x+1}) - w(T_y \to T_{y+1}) - \delta$.
		\IF {$\Delta < 0$}
			\STATE The tour $T \setminus \{ T_x \to T_{x+1}, T_y \to T_{y+1} \} \cup \{ T_x \to T_y, T_{x+1} \to T_{y+1} \}$ is an improvement over $T$.
		\ENDIF
	\ENDFOR
\ENDFOR
\end{algorithmic}
\end{algorithm}
On every iteration, the variable $\delta$ stores the weight difference caused by flipping the fragment $T_{x+1} \to T_{x+2} \ldots \to T_y$, i.e., 
$$
\delta = w(T_{x+1} \to T_{x+2} \ldots \to T_y) - w(T_y \to T_{y-1} \ldots \to T_{x+1}) \,.
$$
In order to consider the moves $\Turn(T, x, y)$ where $x > y$, inverse the given tour $T = T_m \to T_{m-1} \to \ldots \to T_1 \to T_m$ and apply the procedure again.

Observe that this algorithm's complexity is $O(m^2)$.

\bigskip

Our Local adaptation of 2-opt (\twoopt{L}{}) is based on Algorithm~\ref{alg:two_opt_basic}.  For every pair of $x$ and $y$ it finds the shortest paths $T_{x-1} \to T'_x \to T'_y \to T_{y-1}$ and $T_{x+2} \to T'_{x+1} \to T'_{y+1} \to T_y$, where $T'_i \in \Cluster(T_i)$ for $i \in \{ x, x + 1, y, y + 1 \}$.  The time complexity of the local adaptation of 2-opt is $O(m n s)$.

\bigskip

Our Global adaptation of 2-opt exploits all the approaches proposed in Section~\ref{sec:gtsp_global}.  Some further discussion of the \twoopt{G}{} implementation performance can be found in Section~\ref{sec:gtsp_ls_experiments_implementations}.

Note that \twoopt{G}{} is naturally suitable for both symmetric and asymmetric problems.  However, in order to explore the whole neighborhood for an asymmetric problem, one has to apply the procedure, then inverse the tour and apply the procedure again.

\subsection{3-opt}
\label{sec:three_opt}

Let us remove the edges $T_x \to T_{x+1}$, $T_y \to T_{y+1}$ and $T_z \to T_{z+1}$ from a tour $T$.  Then there exist eight options to non-trivially rearrange the obtained fragments in order to obtain a feasible tour which is not in $N_\text{2-opt}(T)$.  However, we limit ourselves to only one of these options, which does not turn any of the tour fragments.  Note that all the other options can be replaced with sequences of two non-independent 2-opt moves~\citep{Rego2006}: $\Turn(\Turn(T, x, y), x, z)$ or $\Turn(\Turn(T, x, y), y, z)$.

We implemented all the adaptations (see Section~\ref{sec:gtsp_ls_adaptation}) of the 3-opt neighborhood and found out that the obtained algorithms are rather slow than powerful.  However, it is worth noting that the Global adaptation for 3-opt can be implemented quite efficiently.  Indeed, it takes $O(n^2 s)$ time to find the shortest paths from every vertex $u$ to every vertex $v \notin \Cluster(u)$ along the tour.  Then for every triple $x$, $y$ and $z$ one can find the shortest cycle through $\mathcal{T}_x \to \mathcal{T}_{y+1} \to \mathcal{T}_{y+2} \to \ldots \to \mathcal{T}_z \to \mathcal{T}_{x+1} \to \mathcal{T}_{x+2} \to \ldots \to \mathcal{T}_y \to \mathcal{T}_{z+1} \to \mathcal{T}_{z+2} \to \ldots \to \mathcal{T}_x$ using the the \CO{} algorithm.  Hence, the time complexity of the algorithm is $O(m^2 n s^2)$.  Using a supporting cluster, one can reduce it to $O(m^2 n \gamma s + n^2 s)$ operations.  Finally, one can apply the lower bound for the shortest cycle (see Theorem~\ref{th:shortest_path_lower_bound}) which will significantly speed up the algorithm.

\subsection{Insertion}
\label{sec:gtsp_insertion}

The \emph{Insertion} TSP neighborhood includes all the solutions which can be obtained from the given one by removing a vertex and inserting it into some other position.  Observe that $N_\text{ins}(T) \subset N_\text{3-opt}(T)$ (consider 3-opt where one of the fragments consist of exactly one vertex).  The size of the insertion neighborhood is $|N_\text{ins}(T)| = m(m - 2)$.

We implement all the adaptations (see Section~\ref{sec:gtsp_ls_adaptation}) for Insertion (\Insertion{}{}).  As a quick improvement (\func{QuickImprove}) for the local adaptations \Insertion{L}{} and \Insertion{L}{co}, we optimize the vertex within the inserted cluster.  For a lower bound in the Global adaptation (\Insertion{G}{}) we use the results of Theorem~\ref{th:shortest_path_lower_bound_cycle}.

Some of these adaptations were already used in the literature.  For example, \Insertion{L}{} was used in \citet{Snyder2000} (though it is called there \emph{Swap}) and in \citet{Renaud1998} (\emph{G-opt} heuristic).  The \emph{Move} heuristic in \citet{Bontoux2009} is \Insertion{G}{}.  However, in~\citep{Bontoux2009} the neighborhood is explored with a heuristic algorithm which does not guarantee that it finds a local minimum.

\subsection{Swap}
\label{sec:gtsp_swap}

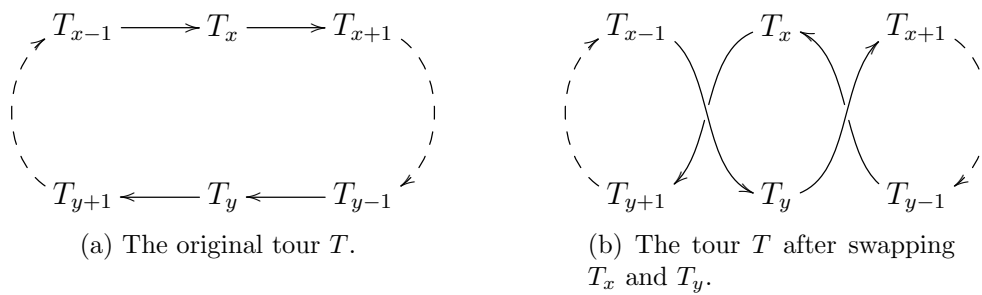
\begin{figure}[ht]
\centering
\subfloat[The original tour $T$.]
{
\xymatrix@R=4em@C=2.5em@L=0.5em{
		*+{T_{x-1}} \ar@{->}[r]
	&	*+{T_x} \ar@{->}[r]
	&	*+{T_{x+1}} \ar@(r,r)@{-->}[d]
\\
		*+{T_{y+1}} \ar@(l,l)@{-->}[u] 
	&	*+{T_y} \ar@{->}[l]
	&	*+{T_{y-1}} \ar@{->}[l]
}
}
\qquad\qquad\qquad
\subfloat[The tour $T$ after swapping $T_x$ and $T_y$.]
{
\xymatrix@R=4em@C=2.5em@L=0.5em{
		*+{T_{x-1}} \ar@(r,l)@{->}[rd]
	&	*+{T_x} \ar@(l,r)@{->}[ld]|\hole
	&	*+{T_{x+1}} \ar@(r,r)@{-->}[d]
\\
		*+{T_{y+1}} \ar@(l,l)@{-->}[u] 
	&	*+{T_y} \ar@(r,l)@{->}[ru]
	&	*+{T_{y-1}} \ar@(l,r)@{->}[lu]|\hole
}
}
\caption{A TSP Swap move.}
\label{fig:swap_move}
\end{figure}

The \emph{Swap} TSP neighborhood $N_\text{swap}(T)$ contains all the solutions obtained from $T$ by swapping two vertices in it, see Figure~\ref{fig:swap_move}.  Observe that $|N_\text{swap}(T)| = m (m - 1)$.

An improtant message is that \Swap{} does not work well for near-optimal solutions.  Indeed, a \Swap{} move can be replaced with a sequence of two \Insertion{}{} or \twoopt{}{} moves.  Moreover, the following theorem proves that a \twoopt{}{} local minimum is also a \Swap{} local minimum for symmetric TSP.

\begin{theorem}
\label{th:swap}
Let $T$ be a local minimum in $N_\text{2-opt}(T)$.  Then $T$ is also a local minimum in $N_\text{swap}(T)$ if the problem is symmetric.
\end{theorem}
\proof
Assume that the tour $T$ is a local minimum in $N_\text{2-opt}(T)$ but it is not a local minimum in $N_\text{swap}(T)$.  Then there exist some $x$ and $y$ such that $w(T') < w(T)$, where $T'$ is obtained from $T$ by swapping $T_x$ and $T_y$ (see Figure~\ref{fig:swap_move}):
\begin{multline}
\label{eq:swap_improves}
w(T_{x-1} \to T_y \to T_{x+1}) + w(T_{y-1} \to T_x \to T_{y+1}) \\
< w(T_{x-1} \to T_x \to T_{x+1}) + w(T_{y-1} \to T_y \to T_{y+1}) \,.
\end{multline}

Let us consider two tours, $A = \Turn(T, x - 1, y)$ and $B = \Turn(T, x, y - 1)$.  (Without loss of generality, one may assume that $x < y$.)  According to (\ref{eq:turn_delta}),
$$
w(A) = w(T) + w(T_{x-1} \to T_y) + w(T_x \to T_{y+1}) - w(T_{x-1} \to T_x) - w(T_y \to T_{y+1}) \text{ and}
$$
$$
w(B) = w(T) + w(T_x \to T_{y-1}) + w(T_{x+1} \to T_y) - w(T_x \to T_{x+1}) - w(T_{y-1} \to T_y) \,.
$$

If $T$ is a local minimum in $N_\text{2-opt}(T)$, then both $w(A) - w(T)$ and $w(B) - w(T)$ are non-negative and their sum is also nonnegative.  Recall that we consider a symmetric problem and observe that
\begin{multline*}
[w(A) - w(T)] + [w(B) - w(T)] \\
= \big[ w(T_{x-1} \to T_y) + w(T_x \to T_{y+1}) - w(T_{x-1} \to T_x) - w(T_y \to T_{y+1}) \big] \\
+ \big[ w(T_x \to T_{y-1}) + w(T_{x+1} \to T_y) - w(T_x \to T_{x+1}) - w(T_{y-1} \to T_y) \big] \\
 = \big[ w(T_{x-1} \to T_y \to T_{x+1}) + w(T_{y-1} \to T_x \to T_{y+1} \big] \\
 - \big[ w(T_{x-1} \to T_x \to T_{x+1}) + w(T_{y-1} \to T_y \to T_{y+1}) \big]
\end{multline*}
However, according to (\ref{eq:swap_improves}) this expression is negative and, hence, the assumption is wrong and the tour $T$ is a local minimum in $N_\text{2-opt}(T)$.
\qed

Note that this effect was also obtained empirically in \citet{GK_GTSP_GA_2008}.

\begin{figure}[ht]
\centerline{
\xymatrix@-1pc@R=5pt@C=25pt{
	&	
	& *++[o][F-]{2} \ar@{-}[rd]
	& *++[o][F-]{3} 
	& *++[o][F-]{4} \ar@(r,u)@{-}[rrddd] 
\\
	&
	&
	&	*++[o][F-]{7'} \ar@{-}[ru]
\\
\\
	  *++[o][F-]{1} \ar@(u,l)@{-}[rruuu] \ar@(d,l)@{-}[rrddd]
	&
	&
	&
	&
	&
	&	*++[o][F-]{5}
\\
\\
	&
	&
	&	*++[o][F-]{3'} \ar@{-}[rd]
\\
	&	
	& *++[o][F-]{8} \ar@{-}[ru]
	& *++[o][F-]{7} 
	& *++[o][F-]{6} \ar@(r,d)@{-}[rruuu]
}
}
\caption{An example of a local minimum in $N_\text{2-opt G}(T)$ which is not a local minimum in $N_\text{swap L}(T)$.}
\label{fig:swap_example}
\end{figure}
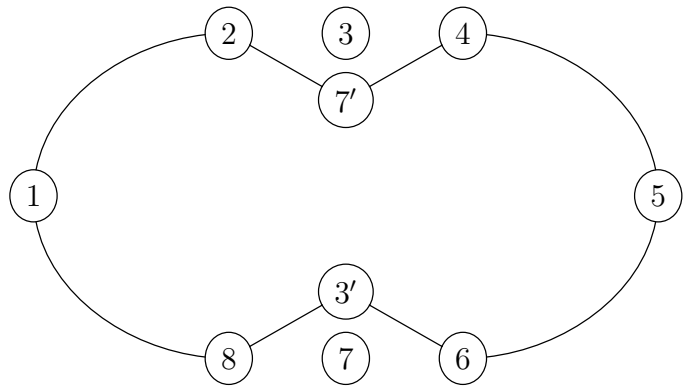

Till now we considered only the TSP Swap neighborhood.  Obviously this result can be extended to the Basic adaptation but it is unclear if it holds for the Local and Global adaptations.
\begin{theorem}
The result of Theorem~\ref{th:swap} does not hold for the Local or Global adaptations of Swap, i.e., a local minimum in $N_\text{2-opt G}(T)$ is not necessarily a local minimum in $N_\text{swap L}(L)$ even if the problem is planar with Euclidean distances.
\end{theorem}
\proof
We will show an example of GTSP tour $T$ which is a local minimum in $N_\text{2-opt L}(T)$ but not a local minimum in $N_\text{swap L}(T)$.  Consider an example on Figure~\ref{fig:swap_example}.
It is a planar GTSP with Euclidean distances and 8 clusters: $\{ 1 \}$, $\{ 2 \}$, $\{ 3, 3' \}$, $\{ 4 \}$, $\{ 5 \}$, $\{ 6 \}$, $\{ 7, 7' \}$ and $\{ 8 \}$.  The initial tour $T$ is $1 \to 2 \to 7' \to 4 \to 5 \to 6 \to 3' \to 8 \to 1$.  Observe that swapping $3'$ and $7'$ together with optimizing the swapped vertices (i.e., replacing $3'$ and $7'$ with $3$ and $7$, respectively) produces the optimal tour $1 \to 2 \to 3 \to 4 \to 5 \to 6 \to 7 \to 8 \to 1$.  At the same time, obviously no adaptation of \twoopt{}{} is able to improve $T$ because whatever is the vertex selection, any \twoopt{}{} move will yield a tour with two intersecting (and, hence, long) edges.
\qed

\section{Lin-Kernighan}
\label{sec:gtsp_lin_kernighan}

Lin-Kernighan heuristic is known to be one of the most successful heuristics for the Traveling Salesman Problem (TSP)\@.  Its efficiency is also proven in application to some other problems.  However, apart from a few naive attempts, it was never applied to the GTSP.

Because of the complexity of the original Lin-Kernighan heuristic, this adaptation is not as straightforward as other adaptations discussed in Section~\ref{sec:gtsp_tsp_neighborhoods}.  At first (see Section~\ref{sec:lin_kernighan}), we provide an easy-to-understand description of a simplified TSP Lin-Kernighan heuristic.  In Section~\ref{sec:lk_adaptation}, we propose several adaptations, both trivial and complicated ones, and analyze them empirically in Section~\ref{sec:gtsp_ls_experiments}.

Since Lin-Kernighan is designed for the symmetric problem, we do not consider asymmetric GTSP in this research.  However, the Global adaptation of Lin-Kernighan, as it was noted in Section~\ref{sec:gtsp_global}, naturally suits both symmetric and asymmetric cases.

Note that a naive adaptation of Lin-Kernighan for GTSP was already proposed in~\cite{Hu2008}; their algorithm constructed a set of TSP instances and solved all of them with the TSP Lin-Kernighan heuristic.  Bontoux et al.~\cite{Bontoux2009} apply the original TSP Lin-Kernighan heuristic to the TSP tours induced by the GTSP tours.  However, it was shown in Section~\ref{sec:gtsp_ls_adaptation} that both of these approaches are relatively weak.

\subsection{TSP Lin-Kernighan Heuristic}
\label{sec:lin_kernighan}

In this section we describe the TSP Lin-Kernighan heuristic (\LKtsp)\@.  It is a simplified version of the original algorithm~\citep{Lin1973}.  Note that~\citep{Lin1973} was published almost 40 years ago, when modest computer resources, obviously, influenced the algorithm design, hiding the main idea behind the technical details.  Also note that, back then, the `goto' operator was widely used; this affects the original algorithm description.  In contrast, our interpretation of the algorithm is easy to understand and implement.

\LKtsp{} is a generalization of the so-called $k$-opt local search.  The $k$-opt neighborhood $N_\text{$k$-opt}(S)$ includes all the TSP tours which can be obtained by removing $k$ edges from the original tour $S$ and adding $k$ different edges such that the resulting tour is feasible.  Observe that exploring the whole $N_\text{$k$-opt}(S)$ takes $O(n^k)$ operations and, thus, with a few exceptions, only 2-opt and rarely 3-opt are used in practice~\citep{Johnson2002,Rego2006}.

As well as $k$-opt, \LKtsp{} also tries to remove and insert edges in the tour but it explores only some parts of the neighborhood that deem to be the most promising.  Consider removing an edge from a tour; this produces a path.  Rearrange the path to minimize its weight.  To close up the tour we only need to add one edge.  Since we did not consider this edge during the path optimization, it is likely that its weight is neither minimized nor maximized.  Hence, the weight of the whole tour is probably reduced together with the weight of the path.  Here is a general scheme of \LKtsp{}:
\begin{enumerate}
	\item Let $T$ be the original tour.
	\item \label{item:nextedge} For every edge $e \to b \in T$ do the following:
	\begin{enumerate}
		\item Let $P = b \to \ldots \to e$ be the path obtained from $T$ by removing the edge $e \to b$.
		\item Rearrange $P$ to minimize its weight.  Every time an improvement is found during this optimization, try to close up the path $P$\@.  If it leads to a tour shorter than $T$, save this tour as $T$ and start the whole procedure again.
		\item If no tour improvement was found, continue to the next edge (Step~\ref{item:nextedge}).
	\end{enumerate}
\end{enumerate}

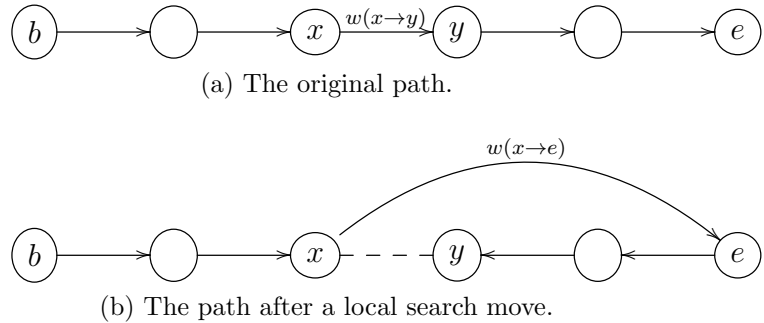
\begin{figure}[ht]
\centering  

\subfloat[The original path.]{
\label{fig:original_path}
\xymatrix@R=3em@C=3em@L=0.1em{
	&	*++[o][F-]{b} \ar@{->}[r]
	&	*++[o][F-]{\phantom{1}} \ar@{->}[r]
	&	*++[o][F-]{x} \ar@{->}[r]^{w(x \to y)}
	&	*++[o][F-]{y} \ar@{->}[r]
	&	*++[o][F-]{\phantom{1}} \ar@{->}[r]
	&	*++[o][F-]{e}
}}
\\[1em]
\subfloat[The path after a local search move.]{
\label{fig:rejoined_path}
\xymatrix@R=3em@C=3em@L=0.1em{
	&	*++[o][F-]{b} \ar@{->}[r]
	&	*++[o][F-]{\phantom{1}} \ar@{->}[r]
	&	*++[o][F-]{x}	\ar@{--}[r]
	&	*++[o][F-]{y} \ar@{<-}[r]
	&	*++[o][F-]{\phantom{1}} \ar@{<-}[r]
	&	*++[o][F-]{e} \ar@/_3em/@{<-}[lll]_{w(x \to e)}
}}

\caption{An example of a local search move for a path improvement.  The weight of the path is reduced by $w(x \to y) - w(x \to e)$.}

\label{fig:path_local_search}
\end{figure}

In order to reduce the weight of the path, a local search is used as follows.  On every move, it tries to break up the path into two parts, invert one of these parts, and then rejoin them (see Figure~\ref{fig:path_local_search}).  In particular, the algorithm tries every edge $x \to y$ to find one which maximize the gain $g = w(x \to y) - w(e \to x)$.  If the maximum $g$ is positive, the move is an improvement of the path and it is accepted.  Note that every move of this local search takes $O(n)$ operations.

Observe that this algorithm tries only the best improvement and skips the other ones.  A natural improvement of the heuristic would be to use a backtracking mechanism to try all the improvements.  However, this would slow down the algorithm too much.  A compromise is to use the backtracking only for the first $\alpha$ moves.  This approach is implemented in a recursive function $\ImprovePath(P, \var{depth}, R)$, see Algorithm~\ref{alg:improvepath_scheme}.

\begin{algorithm}[ht]
\caption{$\ImprovePath{}(P, \var{depth}, R)$ recursive algorithm (\LKtsp{} version).  The function either terminates after an improved tour is found or finishes normally with no profit.}
\label{alg:improvepath_scheme}
\begin{algorithmic}
\REQUIRE Path $P = b \to \ldots \to e$, recursion depth \var{depth} and a set of restricted vertices $R$.
\IF {$\var{depth} < \alpha$}
	\FOR {every edge $x \to y \in P$ such that $x \notin R$}
		\STATE Calculate $g = w(x \to y) - w(e \to x)$ (see Figure~\ref{fig:rejoined_path}).
		\IF {$g > 0$}
			\IF {the tour $b \to \ldots \to x \to e \to \ldots \to y \to b$ is an improvement over the original one}
				\STATE Accept the produced tour and \textbf{terminate}.
			\ELSE
				\STATE $\ImprovePath(b \to \ldots \to x \to e \to \ldots \to y, \var{depth} + 1, R \cup \{ x \})$.
			\ENDIF
		\ENDIF
	\ENDFOR
\ELSE
	\STATE Find the edge $x \to y$ which maximizes $g = w(x \to y) - w(e \to x)$.
	\IF {$g > 0$}
		\IF {the tour $b \to \ldots \to x \to e \to \ldots \to y \to b$ is an improvement over the original one}
			\STATE Accept the produced tour and \textbf{terminate}.
		\ELSE
			\RETURN $\ImprovePath{}(b \to \ldots \to x \to e \to \ldots \to y, \var{depth} + 1, R \cup \{ x \})$.
		\ENDIF
	\ENDIF
\ENDIF
\end{algorithmic}
\end{algorithm}


Execution of $\ImprovePath(P, 1, \varnothing)$ takes $O(n^\alpha \cdot \var{depth}_{\max})$ operations, where $\var{depth}_{\max}$ is the maximum depth of recursion achieved during the run.  Hence, one should use only small values of backtracking depth $\alpha$.

\bigskip

The presented above algorithm is a simplified Lin-Kernighan heuristic.  Here is a list of major differences between the described algorithm and the original one.
\begin{enumerate}
	\item The original heuristic does not accept the first found tour improvement.  It records it and continues the optimization of the path in the hope of finding a better tour improvement.  Note that it was reported in~\citet{Helsgaun2000} that this complicates the algorithm but does not really improve its quality.
	
	\item The original heuristic does not try all the $n$ options when optimizing a path.  It considers only the five shortest edges $x \to e$ in the nondecreasing order.  This hugely reduces the running time and helps to find the best rather than the first improvement on the backtracking stage.  However, this speed-up heuristic is known to be a weak point of the original implementation~\citep{Helsgaun2000,Johnson2002}.  Indeed, even if the edge $x \to y$ is long, the algorithm does not try to break it if the edge $x \to e$ is not in the list of five shortest edges to $e$.	 
	
	Note that looking for the closest vertices or clusters may be meaningless in the application to GTSP\@.  In our implementation, every edge $x \to y$ is considered.

	\item The original heuristic does not allow deleting the previously added edges or adding the previously deleted edges.  It was noted~\citep{Helsgaun2000,Johnson2002} that either of these restrictions is enough to prevent an infinite loop.  In our implementation a previously deleted edge is allowed to be added again but every edge can be deleted only once.  Our implementation also prevents some other moves, however, the experimental evaluation shows that this does not affect the performance of the heuristic.
	
	\item The original heuristic also considers some more sophisticated moves to produce a path from the tour.
	
	\item The original heuristic is, in fact, embedded into a metaheuristic which runs the optimization several times.  There are several tricks related to the metaheuristic which are inapplicable to a single run.
\end{enumerate}

\subsection{\texorpdfstring{Adaptation of \LKtsp{}}{Adaptation of LK}}
\label{sec:lk_adaptation}

In this section we present our adaptation \LK{} of \LKtsp{} for GTSP\@.  A pseudo-code of the whole heuristic is presented in Algorithm~\ref{alg:lk_main}.
\begin{algorithm}[!ht]
\caption{LK general implementation}
\label{alg:lk_main}

\begin{algorithmic}
\REQUIRE Original tour $T$.
\STATE Initialize the number of idle iterations $i \gets 0$.
\WHILE {$i < m$}
	\STATE Cyclically select the next edge $e \to b \in T$.
	\STATE Let $P_o = b \to \ldots \to e$ be the path obtained from $T$ by removing the edge $e \to b$.
	\STATE Run $T' \gets \func{ImprovePath}(P_o, 1, \varnothing)$ (see below).
	\IF {$w(T') < w(T)$}
		\STATE Set $T = \func{ImproveTour}(T')$.
		\STATE Reset the number of idle iterations $i \gets 0$.
	\ELSE
		\STATE Increase the number of idle iterations $i \gets i + 1$.
	\ENDIF
\ENDWHILE
\end{algorithmic}

\rule{\linewidth}{\heavyrulewidth}
\textbf{Procedure} $\func{ImprovePath}(P, \var{depth}, R)$\\
\rule[1em]{\linewidth}{\linethickness}

\vspace{-12pt}
\begin{algorithmic}
\REQUIRE Path $P = b \to \ldots \to e$, recursion depth $\var{depth}$ and the set of restricted vertices $R$.
\IF {$\var{depth} \ge \alpha$}
	\STATE Find the edge $x \to y \in P$, $x \neq b$, $x \notin R$ such that it maximizes the path gain $\mathit{Gain}(P, x \to y)$.
\ELSE
	\STATE Repeat the rest of the procedure for every edge $x \to y \in P$, $x \neq b$, $x \notin R$.
\ENDIF \bigskip

\STATE Conduct the local search move: $P \gets \func{RearrangePath}(P, x \to y)$.
\IF {$\func{GainIsAcceptable}(P, x \to y)$}
	\STATE Replace the edge $x \to y$ with $x \to e$ in $P$.
	
	\STATE $T' = \func{CloseUp}(P)$.
	\IF {$w(T') \ge w(T)$}
		\STATE Run $T' \gets \func{ImprovePath}(P, \var{depth} + 1, R \cup \{ x \})$.
	\ENDIF
	
	\IF {$w(T') < w(T)$}
		\RETURN $T'$.
	\ELSE
		\STATE Restore the path $P$.
	\ENDIF
\ENDIF
\RETURN $T$.
\end{algorithmic}
\end{algorithm}
Some of its details are encapsulated into the following functions:
\begin{itemize}
	\item $\func{Gain}(P, x \to y)$ is intended to calculate the gain of breaking the path $P$ at the edge $x \to y$.
	
	\item $\func{RearrangePath}(P, x \to y)$ removes the edge $x \to y$ from the path $P$ and adds the edge $x \to e$, where $P = b \to \ldots \to x \to y \to \ldots \to e$, see Figure~\ref{fig:path_local_search}.  Together with \func{CloseUp}, it includes an implementation of $\func{QuickImprove}(T)$ (see Section~\ref{sec:gtsp_ls_adaptation}), so \func{RearrangePath} may also apply some cluster optimization.
	
	\item $\func{GainIsAcceptable}(P, x \to y)$ determines if the gain of breaking the path $P$ at the edge $x \to y$ is worth some further effort.
	
	\item $\func{CloseUp}(P)$ adds an edge to the path $P$ to produce a feasible tour.  Together with \func{RearrangePath}, it includes an implementation of $\func{QuickImprove}(T)$ (see Section~\ref{sec:gtsp_ls_adaptation}), so \func{CloseUp} may also apply some cluster optimization.
	
	\item $\func{ImproveTour}(T)$ is a tour improvement function.  It is an analogue to $\func{SlowImprove}(T)$ (see Section~\ref{sec:gtsp_ls_adaptation}).
\end{itemize}

These functions are the key points in the adaptation of \LKtsp{} for GTSP\@.  They determine the behavior of the heuristic.  In Sections~\ref{sec:gtsp_lk_basic}, \ref{sec:gtsp_lk_closest_shortest} and~\ref{sec:gtsp_lk_global} we describe different implementations of these functions.

\subsection{Basic Variation}
\label{sec:gtsp_lk_basic}

\LKb{}{}{} is the Basic adaptation of \LKtsp{}\@.  It defines the functions \func{Gain}, \func{RearrangePath}, \func{CloseUp} and \func{ImproveTour} as follows:
$$
\func{Gain}_\text{B}(b \to \ldots \to e, x \to y) = w(x \to y) - w(e \to x) \,,
$$
$$
\func{RearrangePath}_\text{B}(b \to \ldots \to x \to y \to \ldots \to e, x \to y) = b \to \ldots \to x \to e \to \ldots \to y \,,
$$
$$
\func{CloseUp}_\text{B}(b \to \ldots \to e) = b \to \ldots \to e \to b \,,
$$
and $\func{ImproveTour}_\text{B}(T)$ is trivial.  However, we also consider the Basic with CO adaptations \LKb{}{}{co} which applies \CO{} every time an improvement is found: $\func{ImproveTour}(T) = \func{CO}(T)$.

The implementation of $\func{GainIsAcceptable}(G, P)$ will be discussed in Section~\ref{sec:gain}.

\subsection{Closest and Shortest Variations}
\label{sec:gtsp_lk_closest_shortest}

\heuristic{Closest} and \heuristic{shortest} variations (denoted as \LKc{}{}{} and \LKs{}{}{}, respectively) are two Local adaptations of \LKtsp{}, i.e., $\func{QuickImprove}(T) = L(T)$ and $\func{SlowImprove}(T) = I(T)$.  In other words, some local cluster optimization is applied to every candidate during the path optimization.  

\begin{figure}[ht]
\centerline{
\noindent\xymatrix@R=2.5em@C=2.5em@L=0.1em{
		*++[o][F-]{b} \ar@{->}[r]
	&	*++[o][F-]{\phantom{1}} \ar@{->}[r]
	&	*++[o][F-]{p} \ar@{->}[r]
	&	*++[o][F-]{x}	\ar@{--}[r]
	&	*++[o][F-]{y} \ar@{<-}[r]
	&	*++[o][F-]{\phantom{1}} \ar@{<-}[r]
	&	*++[o][F-]{r} \ar@{<-}[r]
	&	*++[o][F-]{e} \ar@/_3em/@{<-}[llll]_{w(x, e)}
}}
\caption{Path optimization adaptations.}
\label{fig:path_vary}
\end{figure}
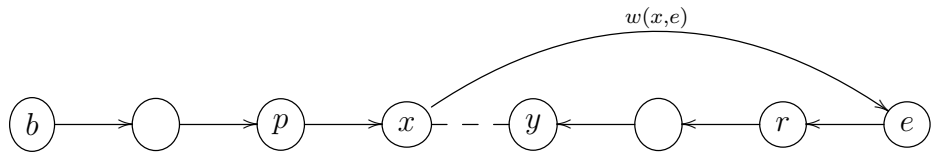

Consider an iteration of the path improvement heuristic \func{ImprovePath}.  Let the path $P = b \to \ldots \to x \to y \to \ldots \to e$ be broken at the edge $x \to y$ (see Figure~\ref{fig:path_vary}).  Then, to calculate $\func{Gain}(P, x \to y)$ in \LKc{}{}{}, we replace $x \in X$ with $x' \in X$ such that the edge $x \to e$ is minimized:
\begin{multline*}
\func{Gain}_\text{C}(b \to \ldots \to p \to x \to y \to \ldots \to e, x \to y) \\
= w(p \to x \to y) - w(p \to x' \to e) \,, 
\end{multline*}
where $x' \in \Cluster(x)$ is chosen to minimize $w(x' \to e)$.

In \LKs{}{}{}, we update both $x$ and $e$ such that the path $p \to x \to e \to r$ is minimized:
\begin{multline*}
\func{Gain}_\text{S}(b \to \ldots \to p \to x \to y \to \ldots \to r \to e, x \to y) = \\
w(p \to x \to y) + w(r \to e) - w(p \to x' \to e' \to r) \,,
\end{multline*}
where $x' \in \Cluster(x)$ and $e' \in \Cluster(e)$ are chosen to minimize $w(p \to x' \to e' \to r)$.

Observe that the most time-consuming part of \LK{} is the path optimization.  In case of the \LKs{}{}{} variation, the bottleneck is the gain evaluation function which takes $O(s^2)$ operations.  In order to reduce the number of gain evaluations in \LKs{}{}{}, we do not consider some edges $x \to y$.  In particular, we assume that the gain $\func{Gain}(b \to \ldots \to e, x \to y)$ is never larger than $w_{\min}(X, Y) - w_{\min}(X, E)$, where $X = \Cluster(x)$, $Y = \Cluster(y)$, $E = \Cluster(e)$ and $w_{\min}(A, B)$ is the weight of the shortest edge from cluster $A$ to cluster $B$:
$$
w_{\min}(A, B) = \min_{a \in A, b \in B} w(a \to b) \,.
$$
Obviously, all the values $w_{\min}(A, B)$ are precalculated.  Note that this speed-up heuristic is used only when $\var{depth} \ge \alpha$, see Algorithm~\ref{alg:lk_main}.

One can hardly speed up the $\func{Gain}$ function in \LKb{}{}{} or \LKc{}{}{}.

The \func{RearrangePath} function does some further cluster optimization in the \LKc{}{}{} variation:
\begin{multline*}
\func{RearrangePath}_\text{C}(b \to \ldots \to p \to x \to y \to \ldots \to e,\ x \to y) \\
= b \to \ldots \to p \to x' \to e \to \ldots \to y \,,
\end{multline*}
where $x' \in \Cluster(x)$ is chosen to minimize the weight $w(p \to x' \to e)$.  

In \LKs{}{}{} it just repeats the optimization performed for the \func{Gain} evaluation:
\begin{multline*}
\func{RearrangePath}_\text{S}(b \to \ldots \to p \to x \to y \to \ldots \to r \to e,\ x \to y) \\
= b \to \ldots \to p \to x' \to e' \to r \to \ldots \to y \,, 
\end{multline*}
where $x' \in \Cluster(x)$ and $e' \in \Cluster(e)$ are chosen to minimize $w(p \to x' \to e' \to r)$.

Every time we want to close up the path, both \LKc{}{}{} and \LKs{}{}{} try all the combinations of the end vertices to minimize the weight of the loop: 
\begin{multline*}
\func{CloseUp}_\text{C, S}(b \to p \to \ldots \to q \to e) = 
	b' \to p \to \ldots \to q \to e' \to b' :\ \\
	b' \in \Cluster(b), e' \in \Cluster(e) \text{ and } w(q \to e' \to b' \to p) \text{ is minimized} \,.
\end{multline*}

We also implemented Local with CO adaptations \LKc{}{}{co} and \LKs{}{}{co} such that \CO{} is applied every time a tour improvement is found: $\func{ImproveTour}(T) = \func{CO}(T)$.

\subsection{Global Variation}
\label{sec:gtsp_lk_global}

Finally we propose a Global adaptation \LKe{}{}.  For every cluster ordering under consideration it finds the shortest path from the first to the last cluster (via all clusters in that order).  After closing up the path it always applies \CO{}.  However, it explores the neighborhood much faster than a naive implementation would do.

The $\func{Gain}$ function for \LKe{}{} is defined as follows:
\begin{multline*}
\func{Gain}_\text{E}(b \to \ldots \to x \to y \to \ldots \to e,\ x \to y) \\
= w_\text{min}(b \to \ldots \to x \to e \to \ldots \to y) - w_\text{min}(b \to \ldots x \to y \to \ldots \to e) \,,
\end{multline*}
where $w_\text{min}(P)$ is the weight of the shortest path through the corresponding clusters: 
$$
w_\text{min}(x_1 \to x_2 \to \ldots \to x_m) = \min_{x'_i \in \Cluster(x_i), i = 1, \ldots, m} w(x'_1 \to x'_2 \to \ldots \to x'_m) \,.
$$
Note that $\func{ImprovePath}$ runs this function sequentially for every $x \to y \in P$.  Observe that this algorithm meets the conditions declared in Section~\ref{sec:gtsp_global}.  Let $X_i = \Cluster(x_i)$ for every $i$.  The improved implementation calculates the shortest paths from $X_1$ to every $v \in X_i$ through $X_2$, $X_3$, \ldots, $X_{i-1}$, where $i = 1, 2, \ldots, m$, in $O(n s)$ operations\footnote{Indeed, we can add a new cluster $X_0 = \{ x_0 \}$ and set $w(x_0 \to x'_1) = 0$ for every $x'_1 \in X_1$.  Then the shortest path from $x_0$ to $v$ coincide with the shortest paths from $X_1$ to $v$.}.  Then it finds the shortest paths from every $e \in X_m$ to every $q \in X_i$ through $X_{m-1}$, $X_{m-1}$, \ldots, $X_{i+1}$ for every $i = 1, 2, \ldots, m$ in $O(n s^2)$ operations.  Having all these shortest paths, it takes $O(s^2)$ to find $\func{Gain}_\text{E}$ using the algorithm for the shortest path in a layered network.  Finally we get $O(n s^2)$ time complexity which is usually much faster than the naive exploration in $O(n m \gamma$ operations.

Observe that the most time consuming part of the algorithm is calculation of the shortest paths from $e \in X_m$ to $q \in X_i$.  However, we apply our improved technique (see Section~\ref{sec:gtsp_global_refinements}) which significantly speeds up the algorithm (in our experiments this speed-up heuristic decreased the running time of the \LKe{}{} algorithm by 30\% to 50\%).


The \func{RearrangePath} function for \LKe{}{} replaces the edge $x \to y$ with $x \to e$ and optimizes the vertices in the path:
\begin{multline*}
\func{RearrangePath}_\text{E}(b \to \ldots \to x \to y \to \ldots \to e) = b' \to \ldots \to x' \to y' \to \ldots \to e' \,, \\
\text{where all the vertices are selected to minimize the weight of the resulting path.}
\end{multline*}
The \func{CloseUp} function for \LKe{}{} simply applies \CO{} to the tour:
$$
\func{CloseUp}_\text{E}(b \to \ldots \to e) = \func{CO}(b \to \ldots \to e \to b) \,.
$$

Observe that, unlike other adaptations or the original \LKtsp{} heuristic, \LKe{}{} is naturally suitable for asymmetric instances.

\subsection{Gain Function}
\label{sec:gain}

Gain is a measure of the path improvement.  It is used to find the best path improvement and to decide whether this improvement should be accepted.  To decide this, we use boolean function $\func{GainIsAcceptable}(P, x \to y)$.  This function greatly influences the performance of the whole algorithm.  We propose four different implementations of $\func{GainIsAcceptable}(P, x \to y)$ to find the most efficient ones:
\begin{enumerate}
	\item \label{item:gain1} $\func{GainIsAcceptable}(P,\ x \to y) = w(P) < w(P_o)$, i.e., the function accepts any changes while the path is shorter than the original one.
	
	\item \label{item:gain2} $\func{GainIsAcceptable}(P,\ x \to y) = w(P) + \dfrac{w(T)}{m} < w(T)$, i.e., it is assumed that an edge of an average weight $\dfrac{w(T)}{m}$ will close up the path.
	
	\item \label{item:gain3} $\func{GainIsAcceptable}(P,\ x \to y) = w(P) + w(x \to y) < w(T)$, i.e., the last removed edge is `restored' for the gain evaluation.  Note that the weight of the edge $x \to y$ cannot be obtained correctly in \LKe{}{}.  Instead of $w(x \to y)$ we use the weight $w_{\min}(X, Y)$ of the shortest edge between clusters $X = \Cluster(x)$ and $Y = \Cluster(y)$.

	\item \label{item:gain4} $\func{GainIsAcceptable}(P,\ x \to y) = w(P) < w(T)$, i.e., the obtained path should be shorter than the original tour.  In other words, the weight of the `close up edge' is assumed to be 0.  Unlike the first three implementations, this one is optimistic and, hence, yields deeper search trees.  This takes more time but also improves the solution quality.

	\item \label{item:gain5} $\func{GainIsAcceptable}(P,\ x \to y) = w(P) + \dfrac{w(T)}{2m} < w(T)$, i.e., it is assumed that an edge of a half of an average weight will close up the path.  It is a mixture of Options~\ref{item:gain2} and~\ref{item:gain4}.
\end{enumerate}

\section{Fragment Optimization}
\label{sec:gtsp_fragment_optimization}

All the adaptations of the TSP local searches, discussed in Section~\ref{sec:gtsp_tsp_neighborhoods}, are intended to improve the structure of the whole tour.  In this section we discuss local improvements, i.e., local search that optimizes only a small fragment of a tour on every iteration.

One can think of many kinds of fragment optimization, but we focus only on the most powerful option, i.e., we only consider a neighborhood containing all possible rearrangements in a given fragment.
\nomenclature{Fragment Optimization}{is a GTSP neighborhood and a local search that optimizes small fragments of the tour}

Consider a tour $T = T_1 \to T_2 \to \ldots \to T_m \to T_1$.  Let $a = T_1$, $b = T_{k + 2}$, $\Omega_i = \Cluster(T_{i+1})$ for $i = 1, 2, \ldots, k$ and $\Omega = \{ \Omega_1, \Omega_2, \ldots, \Omega_k \}$.

Let $\func{FO}(a, b, \Omega)$ be the set of all paths from the vertex $a$ to the vertex $b$ through all the clusters in $\Omega$ being taken in an arbitrary order.  Obviously, $|\func{FO}(a, b, \Omega)| \in O(k! s^k)$.

Using the routine for finding the shortest paths in a layered network, one can find the best path among $\func{FO}(a, b, \Omega)$ in $O(k! \cdot (k-1) s^2)$ operations.  In this work we propose two algorithms $\mathcal{F}_1$ and $\mathcal{F}_2$ which find the best path in $\func{FO}(a, b, \Omega)$ in $O(s^2 k!)$ and $O(s^2 k^2 2^k)$ time, respectively.  The objective is to find the best permutation $\pi$ of the clusters $\Omega_1, \Omega_2, \ldots, \Omega_k$ and the best vertex selection within these clusters.

The first algorithm $\mathcal{F}_1$ proceeds as follows.  Assume that the first $f$ elements in the permutation $\pi$ are fixed.  Then one only needs to find a permutation of the rest $k - f$ elements in $\pi$.  Let $\Omega_{\pi(f)} = \{ x_1, x_2, \ldots, x_c \}$ be the last fixed cluster.  Let $l_1, l_2, \ldots, l_c$ be the weights of the shortest paths from $a$ to the corresponding vertices in $\Omega_{\pi(f)}$ through $\Omega_{\pi(1)}$, $\Omega_{\pi(2)}$, \ldots, $\Omega_{\pi(f - 1)}$ visited in exactly this order.  Try to fix $\Omega_{\pi(f+1)}$ to every of the unused clusters $\Omega \setminus \{ \Omega_{\pi(1)}, \Omega_{\pi(2)}, \ldots, \Omega_{\pi(f)} \}$.  For each $\Omega_{\pi(f+1)} = \{ x'_1, x'_2, \ldots, x'_{c'} \}$ it is easy to calculate the weights of the shortest paths $l'_1, l'_2, \ldots, l'_{c'}$ from $a$ to $\Omega_{\pi(f+1)}$: set $l'_i = \min_j \{l_j + w(x_j, x'_i)\}$.  Now we have $f+1$ fixed elements and the values $l'_1, l'_2, \ldots, l'_{c'}$.  Apply the procedure recursively.  Once all the clusters are fixed, i.e., $f = k$, calculate the total fragment weight $\min_j \{l_j + w(x_j, b)\}$ and if it is shorter than the best one found before then save the solution.  To start the algorithm, apply the procedure for $c = 1$, $x_1 = a$ and $l_1 = 0$.

This procedure takes $O(s^2 k!)$ operations, i.e., it is $O(k)$ times faster than a straightforward algorithm.

\bigskip

The second algorithm $\mathcal{F}_2$ is preferable for large values of $k$.  It is a dynamic programming algorithm which combines the ideas of the Held and Karp's TSP algorithm~\citep{Papadimitriou1982} with the ideas of the layer network shortest path algorithm.  Let $\Delta \subset \Omega$ be a subset of clusters.  We want to find all the shortest paths $p_x^\Delta$ from $a$ to $x \notin \bigcup_{\Omega_i \in \Delta} \Omega_i$ via all the clusters $\Delta$.  Observe that $p_x^\varnothing = w(a \to x)$.  Assume that we know all the shortest paths $p_y^{\Delta \setminus \{ Y \}}$ from $a$ to $y$ through $\Delta \setminus \{ Y \}$ for every $y \in \bigcup_{\Omega_i \in \Delta} \Omega_i$, where $Y = \Cluster(y)$.  Then 
$$
p_x^\Delta = \min_{y \in \bigcup_{\Omega_i \in \Delta} \Omega_i} \left\{ p_y^{\Delta \setminus \{ Y \}} + w(y \to x) \right\} \,.
$$
In other words, having the required information, one can find the shortest path from $a$ to $x$ via clusters $\Delta$ in $O(|\Delta| \cdot s)$ operations.  Observe that for $\Delta = \Omega$ and $x = b$ the algorithm finds the shortest path from $a$ to $b$ via all the clusters in the fragment.

There are $\binom{k}{|\Delta|}$ possible subsets of clusters $\Delta$ of the given size and for every subset there are $O((k - |\Delta|) \cdot s)$ vertices $x$.  It takes $O(|\Delta| \cdot s)$ operations to find each of these shortest paths.  The whole procedure takes 
$$
O\left(\sum_{|\Delta| = 1}^k \binom{k}{|\Delta|} \cdot (k - |\Delta|) \cdot s \cdot |\Delta| \cdot s\right) = O(s^2 k^2 2^k)
$$

Hence, for small values of $k$ the first algorithm $\mathcal{F}_1$ is preferable while for large fragments the second algorithm is significantly faster.

\bigskip

The corresponding $N_\text{$k$-FO}(T)$ neighborhood includes all the tours which can be obtained from $T$ by reordering any $k$ consequent vertices and varying these vertices in the corresponding clusters.  Let $\Phi_i^k(T)$ be the set of all tours obtained from $T$ by rearranging and varying vertices $T_{i+1}$, $T_{i+2}$, \ldots, $T_{i+k}$.  Then $N_\text{$k$-FO}(T) = \bigcup_{i=1}^m \Phi_i^k(T)$, and to explore this neighborhood we have to run our $\mathcal{F}_r$, $r \in \{ 1, 2 \}$, algorithm $m$ times.  Observe that $|\Phi_i^k(T) \cap \Phi_j^k(T)| \gg 1$ for some $i$ and $j$ and, hence, our algorithm explores some of the candidates in $N_\text{$k$-FO}(T)$ more than once.  It is a natural question if avoiding multiple evaluation of these candidates can save any noticeable time.

Let $A_i^k(T) = \{ T' \in \Phi_i^k(T) :\ T'_{i+1} \neq T_{i+1} \}$.  We assume that $k \le m / 2$.  Then observe that $A_i^k(T) \cap A_j^k(T) = \varnothing$ for any $i \neq j$.  Indeed, if some $T' \in A_i^k(T) \cap A_j^k(T)$ then $T'_{i+1} \neq T_{i+1}$ and $T'_{j+1} \neq T_{j+1}$.  Since $T' \in A_j^k(T)$ and the vertex $T'_{i+1}$ is modified, we get $j < i + 1 \le j + k$.  At the same time, since $T' \in A_i^k(T)$ and the vertex $T'_{j+1}$ is modified, $i < j + 1 \le i + k$.  This is only possible if $i = j$.

Observe that 
$$
\bigcup_{i=1}^m \Phi_i^k(T) \subseteq \{ T \} \cup \bigcup_{i=1}^m A_i^k(T) \,.
$$
Indeed, if $T' \in \Phi_i^k(T)$ for some $i$, then either $T' = T$ or there exists $i < j \le i + k$ such that $T'_j \neq T_j$ and $T'_p = T_p$ for every $p = i + 1, i + 2, \ldots, j - 1$.  In the latter case $T' \in A_j^k(T)$.  At the same time, 
$$
\{ T \} \cup \bigcup_{i=1}^m A_i^k(T) \subseteq \bigcup_{i=1}^m \Phi_i^k(T)
$$
since $A_i^k(T) \subset \Phi_i^k(T)$ and $T \in \Phi_i^k(T)$ for any $i$.  Hence, 
$$
\{ T \} \cup \bigcup_{i=1}^m A_i^k(T) = \bigcup_{i=1}^m \Phi_i^k(T) = N_\text{$k$-FO}(T) \,.
$$  
Recall that $A_i^k(T) \cap A_j^k(T) = \varnothing$ and observe that $|A_i^k(T)| = O\left((ks - 1) s^{k-1} (k - 1)! \right)$.  Hence, $|N_\text{$k$-FO}(T)| = O\left(m (ks - 1) s^{k-1} (k - 1)! \right)$.

Compare it to the number $O(m s^k k!)$ of candidates considered by $m$ runs of the algorithm $\mathcal{F}_r$, $r \in \{ 1, 2 \}$, which is $O\left(\dfrac{ks}{ks - 1}\right)$ times larger than $|N_\text{$k$-FO}(T)|$.  We conclude that this relatively small overhead is not worth complicating the algorithm.

Let \FO{k} be the local search with the $N_\text{$k$-FO}(T)$ neighborhood.  Then, depending on the implementation, its time complexity is either $O(m k! s^2)$ or $O(m k^2 2^k s^2)$.

\section{Computational Experiments}
\label{sec:gtsp_ls_experiments}

Theoretical analysis is a useful tool in the algorithm design.  However, empirical analysis is absolutely necessary to select the most efficient neighborhoods and algorithms.

\subsection{Experiments Prerequisites}
\label{sec:gtsp_experiments_prerequisites}

In this section we use the GTSP test bed discussed in Section~\ref{sec:gtsp_testbed}.  In order to save some space, we usually include only every fifth instance in our tables.

In order to generate the starting tour for the local search procedures, we use a simplified Nearest Neighbor~\citep{Noon1988} construction heuristic (\NN)\@.  Unlike that proposed by Noon, our algorithm tries only one starting vertex.  Trying every vertex as a starting one significantly slows down the heuristic and almost does not influence the quality of the solutions obtained after applying a local search.  Note that in what follows the running time of a local search includes the running time of the construction heuristic.

Observe that performance of a local search depends on the initial solution and the performance of a stochastic algorithm vary from time to time.  To smooth out the results, we repeat every experiment 10 times.  It requires some special measures to ensure that an algorithm $H$ proceeds differently for different runs, i.e., $H^i(I) \neq H^j(I)$ in general case, where $i$ and $j$ are the run numbers.  For \GK$^r$ (a memetic algorithm which will be discussed later) the run number $r$ is the random generator seed value.  In \NN$^{r}$, we start the tour construction from the vertex $C_{r,1}$, i.e., from the first vertex of the $r$th cluster of the instance.  This also affects all the local searches since they start from the \NN$^r$ solutions.

All the algorithms are implemented in Visual C++; the evaluation platform is based on an Intel Core i7 2.67~GHz processor.

\subsection{Local Search Strategy}

Until now, we considered only the first improvement strategy which applies an improvement as soon as it is found.  Alternatively, one can use the best improvement strategy which first explores the whole neighborhood and then applies the best found improvement.  Note that the first improvement strategy is normally faster while the best improvement strategy gives better solution quality.

We implemented both strategies for most of the algorithms discussed above.  The experimental analysis clearly shows that the difference in solution quality between the strategies is negligible while the running time is very different.  Thus, we decided to use only the first improvement strategy in all further experiments.

\subsection{Implementations Evaluation}
\label{sec:gtsp_ls_experiments_implementations}

It is interesting to find out what are the advantages of introducing all the improvements discussed above.  In this section we compare different implementations of the \CO{}, \twoopt{}{} and \FO{} algorithms.

\bigskip

The results of experiments with the \CO{} algorithms are provided in Tables~\ref{tab:gtsp_co_variations_one} and~\ref{tab:gtsp_co_variations_more}.  Table~\ref{tab:gtsp_co_variations_one} includes only the instances with $\gamma = 1$ while Table~\ref{tab:gtsp_co_variations_more} includes all the instances with $\gamma > 1$.

The modifications of \CO{} are as follows:
\begin{description}
	\item[\CO{}$_1$] is a pure implementation of the algorithm.
	\item[\CO{}$_2$] uses an optimized order of the shortest paths.
	\item[\CO{}$_3$] tries to reduce the size $\gamma$ of the smallest cluster.
	\item[\CO{}$_4$] applies both improvements.
\end{description}

In spite of the fact that all the instances in the test bed had small $\gamma$ (the largest $\gamma$ in the test bed is 3), the experiments clearly show that the cluster reduction technique is very efficient (see the results for \CO$_3$ and \CO$_4$).  It was able to significantly improve the running times for almost every instance in Table~\ref{tab:gtsp_co_variations_more}, while, obviously, it does not slow down the algorithm if $\gamma = 1$.  Note that the differences between $\CO{}_1$ and $\CO{}_3$, and between $\CO{}_2$ and $\CO{}_4$ in Table~\ref{tab:gtsp_co_variations_one} are because of the measurement errors.

The optimized calculations order is also beneficial, but not by much.  It is more efficient when $\gamma > 1$.  Indeed, it is easy to see that if $\gamma = 1$ and the condition (\ref{eq:co_order_condition}) is met, then either $|\T{2}| = 1$ or $|\T{4}| = 1$.  Hence, this improvement can be applied quite rarely and only if at least two clusters involved in the calculation are of size 1, i.e., the whole calculation in relatively fast itself.

We conclude that the proposed refinements are usually insignificant if $\gamma = 1$ but they are very efficient if $\gamma > 1$.  However, for some other test bed both cluster reduction technique and optimized calculations order may become crucial for the algorithm's efficiency.

\bigskip

Table~\ref{tab:gtsp_implementations_two_opt} reports the running times of two Basic and two Global adaptations of \twoopt{}{}.  For details of the \twoopt{B}{} and \twoopt{B}{adv} implementations see Section~\ref{sec:gtsp_two_opt}.  \twoopt{G}{} is a fully optimized implementation which applies all the improvements discussed in Section~\ref{sec:gtsp_global_refinements}.  In contrast, \twoopt{G}{simple} is a simplified variation of the algorithm which uses the dynamic programming approach to calculate the shortest cycle for every \twoopt{}{} move but it does not introduce any supporting clusters or lower bounds.

One can see that \twoopt{B}{adv} is usually inefficient for GTSP\@.  Observe that the time required to generate the lists $l(v)$ is $O(m^2 \log{m})$ while it takes only $O(m^2)$ operations to explore the whole neighborhood $N_\text{2-opt}(T)$.  Moreover, in order to use the lists several times, i.e., for different GTSP tours, one should include clusters $C$, not vertices, in the lists $l(v)$ and use the maximum distance $w_\text{max}(v \to C)$ when ordering the lists.  An obvious speed-up heuristic is to include in $l(v)$ only the vertices (clusters) which are close to $v$.  However, this still does not improve the algorithm enough to compete with the first 2-opt implementation, see Algorithm~\ref{alg:two_opt_basic}.  We assume that \twoopt{B}{adv} may be useful as a part of a powerful metaheuristic which applies \twoopt{}{} many times for one instance; then the lists $l(v)$ are repeatedly reused.

As regards the Global implementations, it follows from our experiments that, on average, \twoopt{G}{} is more than 10 times faster than \twoopt{G}{simple}.  Note that the speed-up is higher for large instances.  This is because the lower bound is more accurate for large values of $m$.

\bigskip

We have two algorithms to explore the $N_\text{$k$-FO}(T)$ neighborhood; the first algorithm is faster for small values of $k$ while the second one is faster for large $k$.  Table~\ref{tab:gtsp_fo_implementations} compares the running times of both algorithms.  It is clearly visible that the first implementation is faster for $k \le 4$ while for $k > 4$ the second implementation is preferable.  Observe that already for $k = 7$ the difference in performances of the algorithms is significant and it is approximately the same for all the instances in the test bed.

\subsection{Simple Local Search Evaluation}
\label{sec:gtsp_simple_evaluation}

In this Section we provide and discuss the results of all the simple local search algorithms discussed above.

The results for \twoopt{}{} adaptations are provided in Table~\ref{tab:gtsp_two_opt}.  The Basic adaptation \twoopt{B}{} is the fastest and the weakest one.  It takes only 1~ms to proceed even for the largest instances, however, it is not able to change the vertex selection which makes its solution quality noncompetitive.  The \twoopt{B}{co} modification, thus, is significantly better with respect to solution quality and it is not much slower.  Other adaptations further improve the results at the cost of noticeably larger running times.  However, though the neighborhood of \twoopt{G}{} is much larger than all other \twoopt{}{} neighborhoods, it is only about 3 times slower than \twoopt{L}{} or \twoopt{L}{co}.  This shows again the efficiency of the refinements proposed in Section~\ref{sec:gtsp_global_refinements}.

Similar results are obtained for the \Insertion{}{} adaptations, see Table~\ref{tab:gtsp_insertion}.  The Basic adaptation is extremely fast, though it is not as fast as \twoopt{B}{} is.  The cluster optimization in \Insertion{B}{co} significantly improves the solution quality of the heuristic.  All other adaptations are significantly slower.  \Insertion{G}{} is again about 3 times slower than \Insertion{L}{} or \Insertion{L}{co}.  Note that \Insertion{L}{co} and especially \Insertion{G}{} find the optimal solutions for some small instances.

\bigskip

Finally we provide the results for the \FO{} algorithm, see Table~\ref{tab:gtsp_fo}.  Note that we use the first algorithm for $k \le 4$ and the second algorithm for $k > 4$, see Sections~\ref{sec:gtsp_fragment_optimization} and~\ref{sec:gtsp_ls_experiments_implementations}.  Obviously, the heuristic yields very good solutions when $k$ is close to $m$, but it is quite slow for large values of $k$.  We conclude that the most reasonable values of $k$ are 2, 3, 4, 5 and sometimes 6.

\FO{} neighborhood is rather interesting in combination with some other neighborhoods than as a stand-alone heuristic.  However, combination of several neighborhoods is a subject for a separate research.

\subsection{Fair Competition}
\label{sec:gtsp_fair_competition}

Observe that the number of \LK{} variations presented in Section~\ref{sec:lin_kernighan} is really huge and one cannot report and compare the results of these heuristics like we did in Section~\ref{sec:gtsp_simple_evaluation}.  In fact, the problem of a fair comparison of several heuristics is quite common.  Indeed, every experiment result consist of at least two parameters: solution error and running time.  It is a trade-off between speed and quality, and both quick (and low-quality) and slow (and high-quality) heuristics are of interest.  A heuristic should only be considered as useless if it is \emph{dominated} by another heuristic, i.e., it is both slower and yields lower quality solutions.

Hence, one can clearly separate a set of successful from a set of dominated heuristics.  However, this only works for a single experiment.  If the experiment is conducted for several test instances, the comparison becomes unobvious.  Indeed, a heuristic may be successful in one experiment and unsuccessful in another one.  A natural solution of this problem is to use average values but it is often incorrect to compare solution qualities and running times for different instances.  E.g., if some algorithm takes $t_1 = 1$~second to solve some test instance and it takes $t_2 = 100$~seconds to solve another test instance, the average running time $t_\text{avg} = (t_1 + t_2) / 2$ for this algorithm would not actually include the results of the first experiment: $t_\text{avg} \approx t_2 / 2$. 

In a fair competition, one should compare heuristics which have similar running times.  For every time $\tau_i \in $ \{0.02 s, 0.05 s, 0.1 s, 0.2 s, \ldots, 50 s\} we compare solution quality of all the heuristics which were able to solve an instance in less than $\tau_i$.  In order to further reduce the size of the table and to smooth out the experimental results, we additionally group similar instances together and report only the average values for each group.  Recall also that we additionally repeat every experiment 10 times, see Section~\ref{sec:gtsp_experiments_prerequisites}.

The following heuristics were included in the experiments:
\begin{enumerate}
	\item The Basic variations of \LKtsp{}, i.e., \LKb{$x$}{$\alpha$}{} and \LKb{$x$}{$\alpha$}{\,co}, where $\alpha \in \{ 2, 3, 4, 5 \}$ and $x \in \{ 1, 2, 3, 4 \}$ define the backtracking depth and gain acceptance strategy, respectively.
	
	\item The Closest variations of \LKtsp{}, i.e., \LKc{$x$}{$\alpha$}{} and \LKc{$x$}{$\alpha$}{\,co}, where $\alpha \in \{ 2, 3, 4 \}$ and $x \in \{ 1, 2, 3, 4, 5 \}$.

	\item The Shortest variations of \LKtsp{}, i.e., \LKs{$x$}{$\alpha$}{} and \LKs{$x$}{$\alpha$}{\,co}, where $\alpha \in \{ 2, 3, 4 \}$ and $x \in \{ 1, 2, 3, 4, 5 \}$.

	\item The Global variations of \LKtsp{}, i.e., \LKe{$x$}{$\alpha$}, where $\alpha \in \{ 1, 2, 3 \}$ and $x \in \{ 1, 2, 3, 4, 5 \}$.

	\item \FO{$k$} for $k = 2, 3, \ldots, 10$ and adaptations of 2-opt (\optshort{2}{}{}), 3-opt (\optshort{3}{}{}) and Insertion (\Insertion{}{}) local searches according to Section~\ref{sec:gtsp_ls_adaptation}.  Note that the adaptations used in this comparison include most but not all the improvements described above.
	
	\item The state-of-the-art memetic algorithm \GK{} which will be discussed in detail in Section~\ref{sec:gtsp_ma}.
\end{enumerate}

Finally we get Table~\ref{tab:gtsp_ls_competition}.  Roughly speaking, every cell of this table reports the most successful heuristics for a given range of instances and being given some limited time.  More formally, let $\tau = \{ \tau_1, \tau_2, \ldots \}$ be a set of predefined time limits.  Let $\mathcal{I} = \{ \mathcal{I}_1, \mathcal{I}_2, \ldots \}$ be a set of predefined instance groups such that all instances in every $\mathcal{I}_j$ have similar difficulty.  Let $\mathcal{H}$ be a set of all heuristics included in the competition.  $H(I)_\text{time}$ and $H(I)_\text{error}$ are the running time and the relative solution error, respectively, of the heuristic $H \in \mathcal{H}$ for the instance $I \in \mathcal{I}$:
$$
H(I)_\text{error} = \frac{w(H(I)) - w(I_\text{best})}{w(I_\text{best})} \,,
$$
where $I_\text{best}$ is the optimal or best known solution for the instance $I$.  $H(\mathcal{I}_j)_\text{time}$ and $H(\mathcal{I}_j)_\text{error}$ denote the corresponding values averaged for all instances $I \in \mathcal{I}_j$ and all $r \in \{ 1, 2, \ldots, 10 \}$.

For every cell $i,j$ we define a winner heuristic $\var{Winner}_{i,j} \in \mathcal{H}$ as follows:
\begin{enumerate}
	\item $\var{Winner}_{i,j}^r(I)_\text{time} \le \tau_i$ for every instance $I \in \mathcal{I}_j$ and every $r \in \{ 1, 2, \ldots, 10 \}$.
	\item $\var{Winner}_{i,j}(\mathcal{I}_j)_\text{error} < \var{Winner}_{i-1,j}(\mathcal{I}_j)_\text{error}$ (it is only applicable if $i > 1$).
	\item If several heuristics meet the conditions above, we choose the one with the smallest $H_{i,j}(\mathcal{I}_j)_\text{error}$. 
	\item If several heuristics meet the conditions above and have the same solution quality, we choose the one with the smallest $H_{i,j}(\mathcal{I}_j)_\text{time}$.
\end{enumerate}

Apart from the winner, every cell contains all the heuristics $H \in \mathcal{H}$ which meet the following conditions:
\begin{enumerate}
	\item $H^r(I)_\text{time} \le \tau_i$ for every instance $I \in \mathcal{I}_j$ and every $r \in \{ 1, 2, \ldots, 10 \}$.
	\item $H(\mathcal{I}_j)_\text{error} < \var{Winner}_{i-1,j}(\mathcal{I}_j)_\text{error}$ (it is only applicable if $i > 1$).
	\item $H(\mathcal{I}_j)_\text{error} \le 1.1 \cdot Winner_{i,j}(\mathcal{I}_j)_\text{error}$.
	\item $H(\mathcal{I}_j)_\text{time} \le 1.2 \cdot Winner_{i,j}(\mathcal{I}_j)_\text{time}$.
\end{enumerate}

Since \LK{} is a powerful heuristic, we did not consider any instances with less than 30 clusters in this competition.  Note that all the smaller instances are relatively easy to solve, e.g., \GK{} was able to solve of them to optimality in our experiments, and it took only about 30~ms on average, and for \LKs{5}{2}{co} it takes less than a half millisecond to get 0.3\% error, on average, see Table~\ref{tab:gtsp_lk_detailed_small}.  

We use the following groups $\mathcal{I}_j$ of instances:
\begin{description}
	\item[Tiniest] includes 
	\instance{30ch150},
	\instance{30kroA150},
	\instance{30kroB150},
	\instance{31pr152},
	\instance{32u159} and
	\instance{39rat195}.

	\item[Tiny] includes 
	\instance{40kroa200},
	\instance{40krob200},
	\instance{41gr202},
	\instance{45ts225},
	\instance{45tsp225} and
	\instance{46pr226}.
	
	\item[Small] includes
	\instance{46gr229},
	\instance{53gil262},
	\instance{56a280},
	\instance{60pr299} and
	\instance{64lin318}.
	
	\item[Moderate] includes
	\instance{80rd400},
	\instance{84fl417},
	\instance{87gr431},
	\instance{88pr439} and
	\instance{89pcb442}.
	
	\item[Large] includes
	\instance{99d493},
	\instance{107att532},
	\instance{107ali535},
	\instance{113pa561},
	\instance{115u574} and
	\instance{115rat575}.

	\item[Huge] includes
	\instance{132d657},
	\instance{134gr666},
	\instance{145u724} and
	\instance{157rat783}.

	\item[Giant] includes
	\instance{200dsj1000},
	\instance{201pr1002},
	\instance{212u1060} and
	\instance{217vm1084}.
\end{description}

Note that the instances \instance{35si175}, \instance{36brg180}, \instance{40d198}, \instance{53pr264}, \instance{107si535}, \instance{131p654} and \instance{207si1032} are excluded from this competition since they are significantly harder to solve than the other instances of the corresponding groups.  This is discussed in Section~\ref{sec:detailed_results} and the results for these instances are included in Tables~\ref{tab:gtsp_lk_quality} and~\ref{tab:gtsp_lk_time}.

One can see from Table~\ref{tab:gtsp_ls_competition} that there is a clear tendency: the proposed Lin-Kernighan adaptation outperforms all other heuristics in a wide range of trade-offs between solution quality and running time.  Only the state-of-the-art memetic algorithm \GK{} is able to beat \LK{} being given large time.  There are several occurrences of \opt{2}{}{} in the upper right corner (i.e., for Huge and Giant instances and less than 5~ms time) but this is because this given time is too small for even the most basic variations of \LK{}\@.

Clearly, the most important parameter of \LK{} is its variation, and each of the four variations (\heuristic{basic}, \heuristic{closest}, \heuristic{shortest} and \heuristic{global}) is successful in a certain running time range.  \LKb{}{}{} wins the competition for small running times.  For the middle range of running times one should choose \LKc{}{}{} or \LKs{}{}{}.  The \LKe{}{} variation wins only in a small range of times; having more time, one should choose the memetic algorithm \GK{}~\citep{GK_GTSP_GA_2008}.

Here are some tendencies with regards to the rest of the \LK{} parameters:
\begin{itemize}
	\item It is usually beneficial to apply \CO{} every time a tour improvement is found.
	\item The most successful gain acceptance options are~\ref{item:gain4} and~\ref{item:gain5} (see Section~\ref{sec:gain}).
	\item The larger the backtracking depth $\alpha$, the better the solutions.  However, it is an expensive way to improve the solutions; one should normally keep $\alpha \in \{ 2, 3, 4 \}$.
\end{itemize}

Table~\ref{tab:gtsp_ls_competition}, however, does not make it clear what parameters one should use in practice.  In order to give some recommendations, we found the distance $d(H)$ between each of the heuristics $H \in \mathcal{H}$ to the winner algorithms.  For every column $j$ of Table~\ref{tab:gtsp_ls_competition} we calculated $d_j(H)$:
$$
d_j(H) = \frac{H(\mathcal{I}_j)_\text{error} - \var{Winner}_{i,j}(\mathcal{I}_j)_\text{error}}{\var{Winner}_{i,j}(\mathcal{I}_j)_\text{error}} \,,
$$
where $i$ is minimized such that $H^r(I)_\text{time} \le \tau_i$ for every $I \in \mathcal{I}_j$ and $r \in \{ 1, 2, \ldots, 10 \}$.  Then $d_j(H)$ were averaged for all $j$ to get the required distance: $d(H) = \overline{d_j(H)}$\@.  The list of the heuristics $H$ with the smallest distances $d(H)$ is presented in Table~\ref{tab:bestheuristics}.  In fact, we added \optshort{2}{b}{co}, \LKb{2}{2}{co} and \LKe{4}{2}{} to this list only to fill the gaps.  Every heuristic $H$ in Table~\ref{tab:bestheuristics} is also provided with the average running time $T(H)$, in \% of \GK{} running time:
\begin{multline*}
T(H) = \overline{T(H, I, r)} \text{ averaged for all instances $I \in \mathcal{I}$ and all $r \in \{1, 2, \ldots, 10 \}$} \,, \\
\text{where } T(H, I, r) = \frac{H^r(I)_\text{time}}{\func{MA}(I)_\text{time}} \cdot 100\% \\
\text{and } \func{MA}(I)_\text{time} = \overline{\func{MA}^r(I)_\text{time}} \text{ averaged for all $r \in \{1, 2, \ldots, 10 \}$} \,.
\end{multline*}

\begin{table}[htb]
\caption[List of the most successful GTSP heuristics.]{List of the most successful GTSP heuristics.  The heuristics $H$ are ordered according to their running times, from the fastest to the slowest ones.}
\label{tab:bestheuristics}
\begin{center}
\begin{tabular}{l @{\qquad} r @{\qquad} r}
\toprule
$H$ & $d(H)$, \% & Time, \% of \GK{} time \\
\cmidrule{1-3}

\optshort{2}{B}{co} & 44 & 0.04\\
\LKb{2}{2}{co} & 34 & 0.10\\
\LKc{5}{2}{co} & 12 & 0.40\\
\LKs{5}{2}{co} & 19 & 0.97\\
\LKs{5}{3}{co} & 19 & 2.53\\
\LKs{5}{4}{co} & 35 & 8.70\\
\LKs{4}{4}{co} & 32 & 15.34\\
\LKe{4}{2}{} & 56 & 43.62\\
\GK & 0 & 100.00\\

\bottomrule
\end{tabular}
\end{center}
\end{table}

\subsection{Experiment Details for Selected Heuristics}
\label{sec:detailed_results}

In this section we provide detailed information on the experimental results for the most successful heuristics, see Section~\ref{sec:gtsp_fair_competition}.  Tables~\ref{tab:gtsp_lk_detailed_small}, \ref{tab:gtsp_lk_quality} and~\ref{tab:gtsp_lk_time} include the following information:
\begin{itemize}
	\item The `Instance' column contains the instance name as described above.

	\item The `Best' column contains the best known or optimal objective values of the test instances.
	
	\item The rest of the columns correspond to different heuristics and report either relative solution error or running time in milliseconds.  Since the results of the local searches depend on the initial tour and the memetic algorithm is non-deterministic, we averaged every value in these tables for ten runs, see Section~\ref{sec:gtsp_fair_competition} for details.
	
	\item The `Average' row reports the averages for all the instances in a table.

	\item The `Light avg' row reports the averages for all the instances used in Section~\ref{sec:gtsp_fair_competition}.

	\item Similarly, the `Heavy avg' row reports the averages for all the instances ($m \ge 30$) excluded from the competition in Section~\ref{sec:gtsp_fair_competition}.
\end{itemize}

All the small instances ($m < 30$) are separated from the rest of the test bed to Table~\ref{tab:gtsp_lk_detailed_small}.  One can see that all these instances are relatively easy to solve; in fact several heuristics are able to solve all or almost all of them to optimality in every run and it takes only a small fraction of second.  A useful observation is that \LKe{4}{2} solves all the instances with up to 20 clusters to optimality, and in this range \LKe{4}{2} is significantly faster than \GK{}.

As regards the larger instances ($m \ge 30$), it is worth noting that there exist several `heavy' instances among them: \instance{35si175}, \instance{36brg180}, \instance{40d198}, \instance{53pr264}, \instance{107si535}, \instance{131p654} and \instance{207si1032}\@.  Some heuristics perform extremely slowly for these instances: the running time of \LKs{5}{3}{co}, \LKs{5}{4}{co}, \LKs{4}{4}{co} and \LKe{4}{2} is 3 to 500 times larger for every `heavy` instance than it is for the other similar size instances.  Other \LK{} variations are also affected, though, this mostly relates to the ones which use the `optimistic' gain acceptance functions (Options~\ref{item:gain4} and~\ref{item:gain5}), see Section~\ref{sec:gain}.

Our analysis has shown that all of these instances have an unusual weight distribution.  In particular, all these instances have enormous number of `heavy' edges, i.e., the the weights which are close to the maximum weight in the instance, prevail over the smaller weights.  Recall that \LK{} bases on the assumption that a randomly selected edge will probably have a `good' weight.  Then we can optimize a path in the hope to find a good option to close it up later.  However, the probability to find a `good' edge is low in the `heavy' instances.  Hence, the termination condition \func{GainIsAcceptable} does not usually stop the search though a few tour improvements can be found.  This, obviously, slows down the algorithm.

Observe that such `unfortunate' instances can be easily detected before the algorithm's run.  Observe also that even the fast heuristics yield relatively good solutions for these instances (see Tables~\ref{tab:gtsp_lk_quality} and~\ref{tab:gtsp_lk_time}).  Hence, one can use a lighter heuristic to get a reasonable solution quality in a reasonable time in this case.

\section{Conclusion}

Several GTSP neighborhoods are discussed and evaluated in this chapter.  Special attention is paid to adaptation of a TSP neighborhood for GTSP and to exploration of the adapted neighborhood efficiently.  Theoretical analysis shows that these algorithms are significantly faster than any algorithms known from the literature, and the experiments confirm the success of our approaches.

Important theoretical results were obtained for the Cluster Optimization procedure which is widely used in the literature because of its efficiency, simplicity and perfect theoretical properties.  Our tuned implementation of the algorithm showed a noticeable improvement over the original one in our experiments.  We have also proven that the best implementation of Cluster Optimization cannot be significantly faster than the proposed algorithm.

A new class of GTSP neighborhoods, namely Fragment Optimization, is introduced in the work.  It is an exponential neighborhood, and we proposed two efficient algorithms to explore it.  

A number of adaptations of the Lin-Kernighan heuristic for GTSP is presented.  The experimental evaluation confirms the success of these adaptations.

Based on the experimental results, we selected the most successful heuristics for different solution quality/running time requirements.  Note that this list mostly consists of \LK{} adaptations.  However, this does not mean that the simple local searches have no practical application.  These neighborhoods are not intended to be used alone but they are expected to be very efficient in combination with some other local search.  This is a subject of our future research.
\chapter{Local Search Algorithms for MAP}
\label{sec:map_ls}

In this chapter we collect and generalize all MAP local search heuristics known from the literature, propose some new neighborhoods and evaluate them all both theoretically and experimentally.

Recall that we consider only the general case of MAP and, thus, all the heuristics relying on any special structures of the weight matrices are not included in our comparison.

\section{Dimensionwise Variations Heuristics}
\label{sec:map_dv}

The heuristics of this group were first introduced by Bandelt et al.~\cite{Bandelt2004} for MAP with decomposable costs.  However, having a very large neighborhood (see below), they are very efficient even in the general case.  The fact that this approach was also used by Huang and Lim as a local search procedure for their memetic algorithm~\cite{Huang2006} confirms its efficiency.

The idea of the dimensionwise variation heuristics is as follows.  Consider the initial assignment $A$ in the permutation form $A = \pi_1 \pi_2 \ldots \pi_s$ (see Section~\ref{sec:map}).  Let $p(A, \rho_1, \rho_2, \ldots, \rho_s)$ be an assignment obtained from $A$ by applying the permutations $\rho_1, \rho_2, \ldots, \rho_s$ to $\pi_1, \pi_2, \ldots, \pi_s$ respectively:
\begin{equation}
\label{eq:p}
p(A, \rho_1, \rho_2, \ldots, \rho_s) = \rho_1(\pi_1) \rho_2(\pi_2) \ldots \rho_s(\pi_s) \,.
\end{equation}
Let $p_D(A, \rho)$ be an assignment $p(A, \rho_1, \rho_2, \ldots, \rho_s)$, where $\rho_j = \rho$ if $j \in D$ and $\rho_j = 1_n$ otherwise ($1_n$ is the identity permutation of size $n$): 
\begin{equation}
\label{eq:p_D}
p_D(A, \rho) = p\left(A, \left\{ \begin{array}{@{}l@{~}l@{}}\rho & \text{if } 1 \in D\\ 1_n & \text{otherwise}\end{array} \right., \left\{ \begin{array}{@{}l@{~}l@{}}\rho & \text{if } 2 \in D\\ 1_n & \text{otherwise}\end{array} \right., \ldots, \left\{ \begin{array}{@{}l@{~}l@{}}\rho & \text{if } s \in D\\ 1_n & \text{otherwise}\end{array} \right.\right) \,.
\end{equation}
On every iteration, the heuristic selects some nonempty set $D \subsetneq \{ 1, 2, \ldots, s \}$ of dimensions and searches for a permutation $\rho$ such that $w(p_D(A, \rho))$ is minimized.
\nomenclature{Dimensionwise heuristic}{is a MAP local search.  On every iteration, it fixes several dimensions of the problem and groups together the rest of the dimensions; then it construct the corresponding AP, solves it and corrects the assignment accordingly}

For every subset of dimensions $D$, there are $n!$ different permutations $\rho$ but the optimal one can be found in the polynomial time.  Let $swap(u, v, D)$ be a vector which is equal to vector $u$ in all dimensions $j \in \{ 1, 2, \ldots, s \} \setminus D$ and equal to vector $v$ in all dimensions $j \in D$:
\begin{equation}
\label{eq:swap}
swap(u, v, D)_j = \left\{ \begin{array}{l @{\quad} l}
u_j & \text{if } j \notin D \\
v_j & \text{if } j \in D \\
\end{array} \right. \text{\qquad for } j = 1, 2, \ldots, s.
\end{equation}
Let matrix $[M_{i,j}]_{n \times n}$ be constructed as
\begin{equation}
\label{eq:M}
M_{i,j} = w(swap(A^i, A^j, D)) \,.
\end{equation}
It is clear that the solution of the corresponding 2-AP is exactly the required permutation $\rho$.  Indeed, assume there exists some permutation $\rho'$ such that $w(p_D(A, \rho')) < w(p_D(A, \rho))$.  Observe that $p_D(A, \rho) = \{ swap(A^i, A^{\rho(i)}, D) :\ i \in \{ 1, 2, \ldots, n \} \}$.  Then we have
$$
\sum_{i = 1}^n w(swap(A^i, A^{\rho'(i)}, D)) < \sum_{i = 1}^n w(swap(A^i, A^{\rho(i)}, D)) \,.
$$
Since $w(swap(A^i, A^{\rho(i)}, D)) = M_{i, \rho(i)}$, the sum $\sum_{i = 1}^n w(swap(A^i, A^{\rho(i)}, D))$ is already minimized to the optimum and no $\rho'$ can exist.

\bigskip

The neighborhood of a dimensionwise heuristic is as follows:
\begin{equation}
\label{eq:dv_neighborhood}
N_\text{DV}(A) = \big\{ p_D(A, \rho) :\ D \in \mathcal{D} \text{ and $\rho$ is a permutation} \big\} \,,
\end{equation}
where $\mathcal{D}$ includes all dimension subsets acceptable by a certain heuristic.  Observe that
\begin{equation}
\label{eq:p_D_symmetry}
p_D(A, \rho) = p_{\overline{D}}(A, \rho^{-1}) \,,
\end{equation}
where $\rho^{-1}(\rho) = \rho(\rho^{-1}) = 1_s$ and $\overline{D} = \{ 1, 2, \ldots, s \} \setminus D$, and, hence,
\begin{equation}
\label{eq:P_D_symmetry}
\big\{ p_D(A, \rho) :\ \text{$\rho$ is a permutation} \big\} = \big\{ p_{\overline{D}}(A, \rho) :\ \text{$\rho$ is a permutation} \big\}
\end{equation}
for any $D$.  From (\ref{eq:P_D_symmetry}) and the obvious fact that $p_\varnothing(A, \rho) = p_{\{ 1, 2, \ldots, s \}}(A, \rho) = A$ for any $\rho$ we introduce the following restrictions for $\mathcal{D}$:
\begin{equation}
\label{eq:D_restrictions}
D \in \mathcal{D} \Rightarrow\overline{D} \notin \mathcal{D} \text{\quad\ and \quad} \varnothing, \{ 1, 2, \ldots, s \} \notin \mathcal{D} \,.
\end{equation}
With these restrictions, one can see that for any pair of distinct sets $D_1, D_2 \in \mathcal{D}$ the equation $p_{D_1}(A, \rho_1) = p_{D_2}(A, \rho_2)$ holds if and only if $\rho_1 = \rho_2 = 1_n$.  Hence, the size of the neighborhood $N_\text{DV}(A)$ is
\begin{equation}
\label{eq:dv_neighborhood_size}
|N_\text{DV}(A)| = |\mathcal{D}| \cdot (n! - 1) + 1 \,.
\end{equation}

In~\cite{Bandelt2004} it is decided that the number of iterations should not be exponential with regards to neither $n$ nor $s$ while the size of the maximum $\mathcal{D}$ is $|\mathcal{D}| = 2^{s-1} - 1$.  Therefore two heuristics, LS1 and LS2, are evaluated in~\cite{Bandelt2004}.  LS1 includes only singleton values of $D$, i.e., $\mathcal{D} = \{ D :\ |D| = 1 \}$; LS2 includes only doubleton values of $D$, i.e., $\mathcal{D} = \{ D :\ |D| = 2 \}$.  It is surprising but according to both~\cite{Bandelt2004} and our computational experience, the heuristic LS2 produces worse solutions than LS1 though it obviously has larger neighborhood and larger running times.  We improve the heuristic by allowing $|D| \le 2$, i.e., $\mathcal{D} = \{ D :\ |D| \le 2 \}$.  This does not change the theoretical time complexity of the algorithm but improves its performance.  The heuristic LS1 is called \OneDV{} in our research; LS2 with $|D| \le 2$ is called \TwoDV.  We also assume (see Section~\ref{sec:map}) that the value of $s$ is a small fixed constant and, thus, introduce a heuristic \MDV{} which enumerates all feasible (recall (\ref{eq:D_restrictions})) $D \subset \{ 1, 2, \ldots, s \}$.
\nomenclature[1DV]{\OneDV}{is a MAP dimensionwise local search with $"|\mathcal{D}"| = 1$}
\nomenclature[2DV]{\TwoDV}{is a MAP dimensionwise local search with $"|\mathcal{D}"| \le 2$}
\nomenclature[sDV]{\MDV}{is a MAP dimensionwise local search with $"|\mathcal{D}"| \le s$}

The order in which the heuristics take the values $D \in \mathcal{D}$ in our implementations is as follows.  For \OneDV\ it is $\{ 1 \}$, $\{ 2 \}$, \ldots, $\{ s \}$.  \TwoDV\ begins as \OneDV\ and then takes all pairs of dimensions: $\{ 1, 2 \}$, $\{ 1, 3 \}$, \ldots, $\{ 1, s \}$, $\{ 2, 3 \}$, \ldots, $\{ s - 1, s \}$.  Note that because of~(\ref{eq:D_restrictions}) it enumerates no pairs of vectors for $s = 3$, and for $s = 4$ it only takes the following pairs: $\{ 2, 3 \}$, $\{ 2, 4 \}$ and $\{ 3, 4 \}$.  \MDV\ takes first all sets $D$ of size 1, then all sets $D$ of size 2 and so on up to $|D| = \lfloor s / 2 \rfloor$.  If $s$ is even then we should take only half of the sets $D$ of size $s / 2$ (recall (\ref{eq:P_D_symmetry})); for this purpose we take all the subsets of $D \subset \{ 2, 3, \ldots, s \}$, $|D| = s / 2$ in the similar order as before.

It is obvious that $N_\text{\OneDVPlain}(A) \subseteq N_\text{\TwoDVPlain}(A) \subseteq N_\text{\MDVPlain}(A)$ for any $s$; however for $s = 3$ all the neighborhoods are equal and for $s = 4$ \TwoDV\ and \MDV\ also coincide.

According to (\ref{eq:D_restrictions}) and (\ref{eq:dv_neighborhood_size}), the neighborhood size of \OneDV\ is 
$$
|N_\text{\OneDVPlain}(A)| = s \cdot (n! - 1) + 1 \,,
$$
of \TwoDV\ is 
$$
|N_\text{\TwoDVPlain}(A)| = \left\{ \begin{array}{ll}
(2^{s-1} - 1) \cdot (n! - 1) + 1 & \text{if } s \in \{ 3, 4 \} \\
\left(\binom{s}{2} + s\right) \cdot (n! - 1) + 1 & \text{if } s \ge 5
\end{array} \right. \,,
$$
and of \MDV\ is 
$$
|N_\text{\MDVPlain}(A)| = (2^{s-1} - 1) \cdot (n! - 1) + 1 \,.
$$

The time complexity of every run of DV is $O(|\mathcal{D}| \cdot n^3)$ as every 2-AP takes $O(n^3)$ and, hence, the time complexity of \OneDV\ is $O(s \cdot n^3)$, of \TwoDV\ is $O(s^2 \cdot n^3)$ and of MDV is $O(2^{s-1} \cdot n^3)$.

\section{\texorpdfstring{$k$-opt}{k-opt}}
\label{sec:map_kopt}

The \Kopt\ heuristic for 3-AP for $k = 2$ and $k = 3$ was first introduced by Balas and Saltzman~\cite{Balas1991} as a \emph{pairwise} and \emph{triple interchange heuristic}.  \Twoopt\ as well as its variations were also discussed in~\cite{Aiex2005,Clemons2004,Murphey1998,Oliveira2004,Pasiliao2005,Robertson2001} and some other papers.  We generalize the heuristic for arbitrary values of $k$ and $s$.

The heuristic proceeds as follows.  For every subset of $k$ vectors taken in the assignment $A$ it removes all these vectors from $A$ and inserts some new $k$ vectors such that the assignment feasibility is preserved and its weight is minimized.  Another definition is as follows: for every set of distinct vectors $e^1, e^2, \ldots, e^k \in A$ let $X'_j = \{ e_j^1, e_j^2, \ldots, e_j^k \}$ for $j = 1, 2, \ldots, s$.  Let $A' = \{ e'^1, e'^2, \ldots, e'^k \}$ be a solution of this $s$-AP of size $k$.  Replace the vectors $e^1, e^2, \ldots, e^k$ in the initial assignment $A$ with $e'^1, e'^2, \ldots, e'^k$.

The time complexity of \Kopt\ is obviously $O\left( \binom{n}{k} \cdot k!^{s-1} \right)$; for $k \ll n$ it can be replaced with $O(n^k \cdot k!^{s-1})$.  It is a natural question if one can use some faster solver on every iteration.  Indeed, according to Section~\ref{sec:map} it is possible to solve $s$-AP of size $k$ in $O(k!^{s-2} \cdot k^3)$.  However, it is easy to see that $k!^{s-1} < k!^{s-2} \cdot k^3$ for every $k$ up to 5, i.e., it is better to use the exhaustive search for any reasonable $k$.  One can doubt that the exact algorithm actually takes $k!^{s-2} \cdot k^3$ operations but even for the lower bound $\Omega(k!^{s-2} \cdot k^2)$ the inequality $k!^{s-1} < k!^{s-2} \cdot k^2$ holds for any $k \le 3$, i.e., for all the values of $k$ we actually consider.

Now let us find the neighborhood of the heuristic.  For some set $\mathcal{I}$ and a subset $I \subset \mathcal{I}$ let a permutation $\rho$ of elements in $\mathcal{I}$ be an \emph{$I$-permutation} if $\rho(i) = i$ for every $i \in \mathcal{I} \setminus I$, i.e., if $\rho$ does not move any elements except elements from $I$\@.  Let $E = \{ e^1, e^2, \ldots, e^k \} \subset A$ be a set of $k$ distinct vectors in $A$\@.  For $j = 2, 3, \ldots, s$ let $\rho_j$ be an $E_j$-permutation, where $E_j = \{ e_j^1, e_j^2, \ldots, e_j^k \}$.  Then a set $W(A, E)$ of all assignments which can be obtained from $A$ by swapping coordinates of vectors $E$ can be described as follows:
$$
W(A, E) = \big\{ p(A, 1_n, \rho_2, \rho_3, \ldots, \rho_s) :\ \rho_j \text{ is an $E_j$-permutation for } j = 2, 3, \ldots, s \big\} \,.
$$
Recall that $1_n$ is the identity permutation of size $n$ and $p(A, \rho_1, \rho_2, \ldots, \rho_s)$ is defined by (\ref{eq:p}).



The neighborhood $N_{k\text{-opt}}(A)$ is defined as follows:
\begin{equation}
\label{eq:KoptNeighborhood}
N_{k\text{-opt}}(A) = \bigcup_{E \subset A, |E| = k} W(A, E) \,.
\end{equation}

Let $Y, Z \subset A$ such that $|Y| = |Z| = k$.  Observe that $W(A, Y) \cap W(A, Z)$ is nonempty and apart from the initial assignment $A$ this intersection may contain assignments which are modified only in the common vectors $Y \cap Z$.  To calculate the size of the neighborhood of \Kopt{} let us introduce $W'(A, E)$ as a set of all assignments in $W(A, E)$ such that every vector in $E$ is modified in at least one dimension, where $E \subset A$ is the set of $k$ selected vectors in the assignment $A$:
$$
W'(A, E) = \big\{ A' \in W(A, E) :\ |A \cap A'| = n - k \big\} \,.
$$
Then the neighborhood $N_{k\text{-opt}}(A)$ of \Kopt{} is
\begin{equation}
N_{k\text{-opt}}(A) = \bigcup_{E \subset A, |E| \le k} W'(A, E)
\end{equation}
and since $W(A, Y) \cap W(A, Z) = \varnothing$ if $Y \neq Z$ we have
\begin{equation}
\label{eq:kopt_neighborhood_size}
|N_{k\text{-opt}}(A)| = \sum_{E \subset A, |E| \le k} |W'(A, E)| = \sum_{i = 0}^k \binom{n}{i} N^i \,,
\end{equation}
where $N^i = |W(A, E)|$ for any $E$ with $|E| = i$.  Observe that
$$
W'(A, E) = W(A, E) \setminus \bigcup_{E' \subsetneq E} W'(A, E')
$$
and $|W(A, E)| = k!^{s-1}$ for $|E| = k$ and, hence,
\begin{equation}
\label{eq:kopt_n_k}
N^k = k!^{s-1} - \sum_{i = 0}^{k-1}\binom{k}{i} N^{i} \,.
\end{equation}
It is obvious that $N^0 = 1$ since one can obtain exactly one assignment (the given one) by changing no vectors.  From this and~(\ref{eq:kopt_n_k}) we have $N^1 = 0$, $N^2 = 2^{s-1} - 1$ and $N^3 = 6^{s-1} - 3 \cdot 2^{s-1} + 2$.  From this and~(\ref{eq:kopt_neighborhood_size}) follows
\begin{equation}
\label{eq:twoopt_neighborhood_size}
|N_\text{2-opt}(A)| = 1 + \binom{n}{2} (2^{s-1} - 1) \,,
\end{equation}
\begin{equation}
|N_\text{3-opt}(A)| = 1 + \binom{n}{2} (2^{s-1} - 1) + \binom{n}{3} (6^{s-1} - 3 \cdot 2^{s-1} + 2) \,.
\end{equation}

\bigskip

In our implementation, we skip an iteration if the corresponding set of vectors $E$ either consists of the vectors of the minimal weight ($w(e) = \min_{e \in X} w(e)$ for every $e \in E$) or all these vectors have remained unchanged during the previous run of $k$-opt.

\bigskip

It is assumed in the literature~\cite{Balas1991,Pasiliao2005,Robertson2001} that \Kopt\ for $k > 2$ is too slow to be applied in practice.  However, the neighborhood $N_{k\text{-opt}}$ do not only includes the neighborhood $N_{(k-1)\text{-opt}}$ but also grows exponentially with the growth of $k$ and, thus, becomes very powerful.  We decided to include \Twoopt\ and \Threeopt\ in our research.  Greater values of $k$ are not considered in this research because of nonpractical time complexity (observe that the time complexity of \Opt{4} is $O(n^4 \cdot 24^{s-1})$) and even \Threeopt\ with all the improvements described above still takes a lot of time to proceed.  However, \Threeopt\ is more robust when used in a combination with some other heuristic (see Section~\ref{sec:map_vnd}).

\bigskip

It is worth noting that our extension of the pairwise (triple) interchange heuristic~\cite{Balas1991} is not typical.  Many papers~\cite{Aiex2005,Clemons2004,Murphey1998,Pasiliao2005,Robertson2001} consider another neighborhood:
$$
N_\text{\OptStarPlain{$k$}}(A) = \big\{ p_D(A, \rho) :\ D \subset \{ 1, 2, \ldots, s \}, |D| = 1 \text{ and $\rho$ moves at most $k$ elements} \big\} \,,
$$
where $p_D$ is defined in~(\ref{eq:p_D}).  The size of such neighborhood is $|N_\text{\OptStarPlain{$k$}}(A)| = s \cdot \binom{n}{k} \cdot (k! - 1) + 1$ and the time complexity of one run of \OptStar{$k$} in the assumption $k \ll n$ is $O(s \cdot n^k \cdot k!)$, i.e., unlike \Kopt, it is not exponential with respect to the number of dimensions $s$ which is considered to be important by many researchers.  However, as it is stated in Section~\ref{sec:map}, we assume that $s$ is a small fixed constant and, thus, the time complexity of \Kopt\ is still reasonable.  At the same time, observe that $N_{k\text{-opt*}}(A) \subset N_\text{1-DV}(A)$ for any $k \le n$, i.e., \OneDV\ performs as good as \OptStar{$n$} with the time complexity of \OptStar{3}.  Only in the case of $k = 2$ the heuristic \OptStar{2} is faster in theory however it is known~\cite{Burkard1999} that the expected time complexity of AP is significantly less than $O(n^3)$ and, thus, the running times of \OptStar{2} and \OneDV\ are similar while \OneDV\ is definitely more powerful.  Because of this we do not consider \OptStar{2} in our comparison.



\section{Lin-Kernighan}
\label{sec:map_vopt}

The \emph{Variable Depth Interchange} (\VDI) was first introduced by Balas and Saltzman for 3-AP as a local search based on the well-known Lin-Kernighan heuristic for the Traveling Salesman Problem~\cite{Balas1991}.  We provide here a natural extension \Vopt{} of the \VDI{} local search for the $s$-dimensional case, $s \ge 3$, and then improve this extension.  Our computational experiments show that the improved version of \Vopt{} is superior to the natural extension of \VDI{} with respect to solution quality at the cost of a reasonable increase in running time.  In what follows, \Vopt{} refers to the improved version of the heuristic unless otherwise specified.
\nomenclature{Vectorwise heuristic}{is a class of MAP local searches.  On every iteration, it removes several vectors from the solution and inserts some new ones such that the assignment preserves its feasibility}
\nomenclature[v-opt]{\Vopt}{is a MAP vectorwise local search based on the idea of the Lin-Kernighan heuristic}
\nomenclature[VDI]{\VDI}{stand for the Variable Depth Interchange.  It is a MAP vectorwise local search based on the idea of the Lin-Kernighan heuristic.  In this work it is extended and called \Vopt{}}

In~\cite{Balas1991}, \VDI{} is described quite briefly.  Our contribution is not only in extending, improving and analyzing it but also in a more detailed and, we believe, clearer explanation of it.  We describe the heuristic in a different way to the description provided in~\cite{Balas1991}, however, both versions of our algorithm are equal to \VDI{} in case of $s = 3$.  This fact was also checked by reproducing experimental results reported in~\cite{Balas1991}.

%

\bigskip

In what follows we will use function $U(u, v)$ which returns a set of swaps between the vectors $u$ and $v$.  The difference between two versions of \Vopt{} is only in the $U(u, v)$ definition.  For the natural extension of \VDI{}, let $U(u, v)$ be a set of all the possible swaps (see~(\ref{eq:swap})) in at most one dimension between the vectors $u$ and $v$, where the coordinates in at most one dimension are swapped:
$$
U(u, v) = \big\{ swap(u, v, D) :\ D \subset \{ 1, 2, \ldots, s \} \text{ and } |D| \le 1 \big\} \,.
$$

For the improved version of \Vopt{}, let $U(u, v)$ be a set of all the possible swaps in at most $\lfloor s / 2 \rfloor$ dimensions between the vectors $v$ and $w$:
$$
U(u, v) = \big\{ swap(u, v, D) :\ D \subset \{ 1, 2, \ldots, s \} \text{ and } |D| \le s / 2 \big\} \,.
$$
The constraint $|D| \le s / 2$ guarantees that at least half of the coordinates of every swap are equal to the first vector coordinates.  The computational experiments show that removing this constraint increases the running time and decreases the average solution quality.

Let vector $\mu(u, v)$ be the minimum weight swap between vectors $u$ and $v$:
$$
\mu(u, v) = \argmin_{e \in U(u, v)} w(e) \,.
$$
Let $A$ be an initial assignment.

\begin{enumerate}
\item \label{item:vopt_mainloop} For every vector $c \in A$ do the rest of the algorithm.

\item \label{item:vopt_loopbegin} Initialize the \emph{total gain} $G = 0$, the \emph{best assignment} $A_\text{best} = A$, and a set of available vectors $L = A \setminus \{ c \}$.

\item \label{item:vopt_search} Find vector $m \in L$ such that $w(\mu(c, m))$ is minimized.  Set $v = \mu(c, m)$ and $\overline{v}_j = \{ c_j, m_j\} \setminus \{ v_j \}$ for every $1 \le j \le s$.
Now $v \in U(c, m)$ is the minimum weight swap of $c$ with some other vector $m$ in the assignment, and $\overline{v}$ is the complementary vector.

\item \label{item:vopt_stop_condition} Set $G = G + w(c) - w(v)$.  If now $G \le 0$, set $A = A_\text{best}$ and go to the next iteration (Step~\ref{item:vopt_mainloop}).

\item Mark $m$ as an unavailable for the further swaps: $L = L \setminus \{ m \}$.  Note that $c$ is already marked unavailable: $c \notin L$.

\item Replace $m$ and $c$ with $v$ and $\overline{v}$.  Set $c = \overline{v}$.

\item If $w(A) < w(A_\text{best})$, save the new assignment as the best one: $A_\text{best} = A$.

\item Repeat from Step~\ref{item:vopt_search} while the total gain is positive (see Step \ref{item:vopt_stop_condition}) and $L \neq \varnothing$.

\end{enumerate}

The heuristic repeats until no improvement is found during a run.  The time complexity of one run of \Vopt{} is $O(n^3 \cdot 2^{s-1})$.  The time complexity of the natural extension of \VDI{} is $O(n^3 \cdot s)$, and the computation experiments also show a significant difference between the running times of the improved and the natural extensions.  However, the solution quality of the natural extension for $s \ge 7$ is quite poor, while for the smaller values of $s$ it produces solutions similar to or even worse than \MDV{} solutions at the cost of much larger running times.

The neighborhood $N_\text{\VoptPlain}(A)$ is not fixed and depends on the MAP instance and initial assignment $A$.  The number of iterations (runs of Step~\ref{item:vopt_search}) of the algorithm can vary from $n$ to $n^2$.  Moreover, there is no guarantee that the algorithm selects a better assignment even if the corresponding swap is in $U(c, m)$.  Thus, we do not provide any results for the neighborhood of \Vopt{}.

\section{Variable Neighborhood Descend}
\label{sec:map_vnd}

Above we have presented two types of neighborhoods, let us say \emph{dimensionwise} (Section~\ref{sec:map_dv}) and \emph{vectorwise} (Sections~\ref{sec:map_kopt} and~\ref{sec:map_vopt}).  The idea of our combined heuristic is to use the dimensionwise and the vectorwise neighborhoods together, combining them into so called Variable Neighborhood Descent~\cite{Talbi2009}.  The combined heuristic improves the assignment by moving it into a local optimum with respect to the dimensionwise neighborhood, then it improves it by moving it to a local minimum with respect to the vectorwise neighborhood.  The procedure is repeated until the assignment occurs in a local minimum with respect to both the dimensionwise and the vectorwise neighborhoods.
\nomenclature[DV]{\Comb{DV}{opt}}{is a combination of two MAP local searches: a dimensionwise ($\text{DV}$) and a vectorwise ($\text{opt}$)}

More formally, the combined heuristic \Comb{DV}{opt} consists of a dimensionwise heuristic $DV$ (either \OneDV, \TwoDV\ or \MDV) and a vectorwise heuristic $opt$ (either \Twoopt, \Threeopt\ or \Vopt).  \Comb{DV}{opt} proceeds as follows:
\begin{enumerate}
	\item \label{item:combined_first} Apply the dimensionwise heuristic $A = DV(A)$.
	\item Repeat:
	\begin{enumerate}
		\item Save the assignment weight $x = w(A)$ and apply the vectorwise heuristic $A = opt(A)$.
		\item If $w(A) = x$ stop the algorithm.
		\item Save the assignment weight $x = w(A)$ and apply the dimensionwise heuristic $A = DV(A)$.
		\item If $w(A) = x$ stop the algorithm.
	\end{enumerate}
\end{enumerate}
\nomenclature[1DV2]{\OneDVtwo}{is a combination of \OneDV{} and \Twoopt{} MAP local searches}
\nomenclature[2DV2]{\TwoDVtwo}{is a combination of \TwoDV{} and \Twoopt{} MAP local searches}
\nomenclature[1DV3]{\OneDVthree}{is a combination of \OneDV{} and \Threeopt{} MAP local searches}
\nomenclature[2DV3]{\TwoDVthree}{is a combination of \TwoDV{} and \Threeopt{} MAP local searches}
\nomenclature[sDV3]{\MDVthree}{is a combination of \MDV{} and \Threeopt{} MAP local searches}
\nomenclature[1DVV]{\OneDVV}{is a combination of \OneDV{} and \Vopt{} MAP local searches}
\nomenclature[2DVV]{\TwoDVV}{is a combination of \TwoDV{} and \Vopt{} MAP local searches}
\nomenclature[sDVV]{\MDVV}{is a combination of \MDV{} and \Vopt{} MAP local searches}




Step~\ref{item:combined_first} of the combined heuristic is the hardest one.  Indeed, it is typical that it takes a lot of iterations to move a bad solution to a local minimum while for a good solution it takes just a few iterations.  Hence, the first of the two heuristics should be the most efficient one, i.e., it should perform quickly and produce a good solution.  In this case the dimensionwise heuristics are more efficient because, having approximately the same as vectorwise heuristics time complexity, they search much larger neighborhoods.  The fact that the dimensionwise heuristics are more efficient than the vectorwise ones is also confirmed by experimental evaluation (see Section~\ref{sec:map_ls_experiments}).

It is clear that the neighborhood of a combined heuristic is defined as follows:
\begin{equation}
\label{eq:combined_neighborhood}
N_\text{\CombPlain{DV}{opt}}(A) = N_\text{DV}(A) \cup N_\text{opt}(A) \,,
\end{equation}
where $N_\text{DV}(A)$ and $N_\text{opt}(A)$ are neighborhoods of the corresponding dimensionwise and vectorwise heuristics respectively.  To calculate the size of the neighborhood $N_\text{\CombPlain{DV}{opt}}(A)$ we need to find the size of the intersection of these neighborhoods.  Observe that
\begin{equation}
\label{eq:combined_intersection}
N_\text{DV}(A) \cap N_{k\text{-opt}}(A) = \big\{ p_D(A, \rho) :\ D \in \mathcal{D} \text{ and $\rho$ moves at most $k$ elements} \big\} \,,
\end{equation}
where $p_D(A, \rho)$ is defined by~(\ref{eq:p_D}).  This means that, if $r_k$ is the number of permutations on $n$ elements which move at most $k$ elements, the intersection (\ref{eq:combined_intersection}) has size
\begin{equation}
\label{eq:combined_intersection_size}
|N_\text{DV}(A) \cap N_{k\text{-opt}}(A)| = |\mathcal{D}| \cdot (r_k - 1) + 1 \,.
\end{equation}
The number $r_k$ can be calculated as 
\begin{equation}
\label{eq:combined_r_k}
r_k = \sum_{i = 0}^k \binom{n}{i} \cdot d_i \,,
\end{equation}
where $d_i$ is the number of derangements on $i$ elements, i.e., permutations on $i$ elements such that none of the elements appear on their places; $d_i = i! \cdot \sum_{m = 0}^i (-1)^m / m!$~\cite{Harris2008}.  For $k = 2$, $r_2 = 1 + \binom{n}{2}$; for $k = 3$, $r_3 = 1 + \binom{n}{2} + 2 \binom{n}{3}$.  From (\ref{eq:dv_neighborhood_size}), (\ref{eq:kopt_neighborhood_size}), (\ref{eq:combined_neighborhood}) and (\ref{eq:combined_intersection_size}) we immediately have
\begin{equation}
\left|N_\text{\CombPlain{DV}{\KoptPlain}}(A)\right| = 1 + |\mathcal{D}| \cdot (n! - 1) + \left[\sum_{i = 2}^k \binom{n}{i} N^i\right] - |\mathcal{D}| \cdot (r_k - 1) \,,
\end{equation}
where $N^i$ and $r_k$ are calculated according to (\ref{eq:kopt_n_k}) and (\ref{eq:combined_r_k}) respectively.  Substituting the value of $k$, we have:
\begin{equation}
\label{eq:combined_neighborhood_size_2}
\left|N_\text{\CombPlain{DV}{\TwooptPlain}}(A)\right| = 1 + |\mathcal{D}| \cdot (n! - 1) + \binom{n}{2} (2^{s-1} - 1) - |\mathcal{D}| \cdot \binom{n}{2} \qquad \text{and}
\end{equation}
\begin{multline}
\label{eq:combined_neighborhood_size_3}
\left|N_\text{\CombPlain{DV}{\ThreeoptPlain}}(A)\right| = 1 + |\mathcal{D}| \cdot (n! - 1) + \binom{n}{2} (2^{s-1} - 1) \\
+ \binom{n}{3} (6^{s-1} - 3 \cdot 2^{s-1} + 2) - |\mathcal{D}| \cdot \left[\binom{n}{2} + 2 \binom{n}{3}\right]
\end{multline}

One can easily substitute $|\mathcal{D}| = s$, $|\mathcal{D}| = \binom{s}{2}$ and $|\mathcal{D}| = 2^{s-1} - 1$ to (\ref{eq:combined_neighborhood_size_2}) or (\ref{eq:combined_neighborhood_size_3}) to get the neighborhood sizes of \OneDVtwo, \TwoDVtwo, \MDVtwo, \OneDVthree, \TwoDVthree\ and \MDVthree.  We will only show the results for \MDVtwo:
%
\begin{multline}
\left|N_\text{\MDVtwoPlain}(A)\right| = 1 + (2^{s-1} - 1) \cdot (n! - 1) + \binom{n}{2} (2^{s-1} - 1) - (2^{s-1} - 1) \cdot \binom{n}{2} \\
= 1 + (2^{s-1} - 1) \cdot (n! - 1) \,,
\end{multline}
i.e., $\left|N_\text{\MDVtwoPlain}(A)\right| = \left|N_\text{\MDVPlain}(A)\right|$\@.  Since $N_\text{\MDVPlain}(A) \subseteq N_\text{\MDVtwoPlain}(A)$ (see~(\ref{eq:combined_neighborhood})), we can conclude that $N_\text{\MDVtwoPlain}(A) = N_\text{\MDVPlain}(A)$.  Indeed, the neighborhood of \Twoopt\ can be defined as follows: 
$$
N_\text{\TwooptPlain} = \big\{ p_D(A, \rho) :\ D \subset \{ 2, 3, \ldots, s \} \text{ and } \rho \text{ swaps at most two elements} \big\} \,,
$$
which is obviously a subset of $N_\text{\MDVPlain}(A)$ (see (\ref{eq:dv_neighborhood})).  Hence, the combined heuristic \MDVtwo{} is of no interest.

For other combinations the intersection (\ref{eq:combined_intersection}) is significantly smaller than both neighborhoods $N_\text{DV}(A)$ and $N_{\KoptPlain}(A)$ (recall that the neighborhood $N_\text{\VoptPlain}$ has a variable structure).  Indeed, $|N_\text{DV}(A)| \gg |N_\text{DV}(A) \cap N_{k\text{-opt}}(A)|$ because $|\mathcal{D}| \cdot (n! - 1) \gg |\mathcal{D}| \cdot (r_k - 1)$ for $k \ll n$.  Similarly, $|N_\text{\TwooptPlain}(A)| \gg |N_\text{DV}(A) \cap N_{k\text{-opt}}(A)|$ because $\binom{n}{2} (2^{s-1} - 1) \gg |\mathcal{D}| \cdot \binom{n}{2}$ if $|\mathcal{D}| \ll 2^{s-1}$, which is the case for \OneDV\ and \TwoDV\ if $s$ is large enough.  Finally, $|N_\text{\ThreeoptPlain}(A)| \gg |N_\text{DV}(A) \cap N_{k\text{-opt}}(A)|$ because $\binom{n}{2} (2^{s-1} - 1) + \binom{n}{3} (6^{s-1} - 3 \cdot 2^{s-1} + 2) \gg |\mathcal{D}| \cdot \left[ \binom{n}{2} + 2 \binom{n}{3} \right]$, which is true even for $|\mathcal{D}| = 2^{s-1}$, i.e., for \MDV.

The time complexity of the combined heuristic is $O(n^k \cdot k!^{s-1} + |\mathcal{D}| \cdot n^3)$ in case of $opt = \text{\KoptPlain}$ and $O(n^3 \cdot (2^{s-1} + |\mathcal{D}|))$ if $opt = \text{\VoptPlain}$.  The particular formulas are provided in the following table:

\begin{center}
\begin{tabular}{llll}
\toprule
 & \TwooptPlain & \ThreeoptPlain & \VoptPlain \\
\cmidrule(){1-4}
\OneDVPlain & $O(2^{s-1} \cdot n^2 + s \cdot n^3)$ & $O(6^{s-1} \cdot n^3)$ & $O(2^s \cdot n^3)$ \\
\TwoDVPlain & $O(2^{s-1} \cdot n^2 + s^2 \cdot n^3)$ & $O(6^{s-1} \cdot n^3)$ & $O(2^s \cdot n^3)$ \\
\MDVPlain & (no interest) & $O(6^{s-1} \cdot n^3)$ & $O(2^s \cdot n^3)$ \\
\bottomrule
\end{tabular}
\end{center}

Note that all the combinations with \Threeopt\ and \Vopt\ have equal time complexities; this is because the time complexities of \Threeopt\ and \Vopt\ are dominant.  Our experiments show that the actual running times of \Threeopt\ and \Vopt\ are really much higher then even the \MDV\ running time.  This means that the combinations of these heuristics with \MDV{} are approximately as fast as the combinations of these heuristics with light dimensionwise heuristics \OneDV\ and \TwoDV\@.  Moreover, as it was noticed above in this section, the dimensionwise heuristic, being executed first, simplifies the job for the vectorwise heuristic and, hence, the increase of the dimensionwise heuristic power may decrease the running time of the whole combined heuristic.  At the same time, the neighborhoods of the combinations with \MDV\ are significantly larger than the neighborhoods of the combinations with \OneDV\ and \TwoDV\@.  We can conclude that the `light' heuristics \OneDVthree, \TwoDVthree, \OneDVV\ and \TwoDVV\ are of no interest because the `heavy' heuristics \MDVthree\ and \MDVV, having the same theoretical time complexity, are more powerful and, moreover, outperformed the `light' heuristics in our experiments with respect to both solution quality and running time on average and in most of single experiments.




%
%
%


\section{Other Algorithms}

Here we provide a list of some other MAP algorithms presented in the literature.

\begin{itemize}
\item A host of local search procedures and construction heuristics which often have some approximation guarantee (\cite{Bandelt2004,Burkard1996,Crama1992,Isler2005,Kuroki2007,Murphey1998} and some others) are proposed for several special cases of MAP (usually with decomposable weights, see Section~\ref{sec:decomposable}) and exploit the specifics of these instances.  However, as it was stated in Section~\ref{sec:map}, we consider only the general case of MAP, i.e., all the algorithms included in this research do not rely on any special structure of the weight matrix.

\item A number of construction heuristics are intended to generate solutions for the general case of MAP, see Section~\ref{sec:map_construction_heuristics}.  While some of them are fast and low quality, like \Greedy, some, like \MaxRegret, are significantly slower but produce much better solutions.  A special class of construction heuristics, Greedy Randomized Adaptive Search Procedure (GRASP), was also investigated by many researchers~\cite{Aiex2005,Murphey1998,Oliveira2004,Robertson2001}.

\item Several metaheuristics, including a simulated annealing procedure~\cite{Clemons2004} and a memetic algorithm~\cite{Huang2006}, were proposed in the literature.  Metaheuristics are sophisticated algorithms intended to search for the near optimal solutions in a reasonably large time.  Proceeding for much longer than local search and being hard for theoretical analysis of the running time or the neighborhood, metaheuristics cannot be compared straightforwardly to local search procedures.

\item Some weak variations of \Twoopt\ are considered in~\cite{Aiex2005,Murphey1998,Pasiliao2005,Robertson2001}.  While our heuristic \Twoopt\ tries all possible recombinations of a pair of assignment vectors, i.e., $2^{s-1}$ combinations, these variations only try the swaps in one dimension at a time, i.e., $s$ combinations for every pair of vectors.  We have already decided that these variations have no practical interest, for details see Section~\ref{sec:map_kopt}.
\end{itemize}

\section{Experiment Results}
\label{sec:map_ls_experiments}

In this section, the results of empirical evaluation are reported and discussed.  In our experiments we use the instance families discussed in Section~\ref{sec:map_testbed}.  The sizes of all but \GP{} instances are selected such that an algorithm could process them all in approximately the same time.  The \GP{} instances are included in order to examine the behavior of the heuristics on smaller instances (recall that \GP{} is the only instance set for which we know the exact solutions for small instances).


All the heuristics are implemented in Visual C++.  The evaluation platform is based on AMD Athlon 64 X2 3.0~GHz processor.

We present the results of three different types of experiments:
\begin{itemize}
\item In Subsection~\ref{sec:map_ls_pure}, the local search heuristics are applied to the assignments generated by some construction heuristic.  These experiments allow us to exclude several local searches from the rest of the experiments.  However, the comparison of the results is complicated because of the significant difference in both the solution quality and the running time.

\item In Subsection~\ref{sec:map_ls_metaheuristics}, two simple metaheuristics are used to equate the running times of different heuristics.  This is done by varying of number of iterations of the metaheuristics.

\item In Subsection~\ref{sec:map_ls_heuristics_comparison}, the results of all the discussed approaches are gathered in two tables to find the most successful solvers for the instances with independent and decomposable weights for every particular running time.
\end{itemize}

\subsection{Pure Local Search Experiments}
\label{sec:map_ls_pure}

In this section, we run every local search heuristic for every instance exactly once.  The local search is applied to a solution generated with one of the following construction heuristics (for details see Section~\ref{sec:map_construction_heuristics}):
\begin{enumerate}
	\item \Trivial, which was first mentioned in~\cite{Balas1991} as \emph{Diagonal}.  \Trivial\ construction heuristic simply assigns $A_j^i = i$ for every $i = 1, 2, \ldots, n$ and $j = 1, 2, \ldots, s$.

	\item \Greedy{} was discussed in many papers, see, e.g.~\cite{Balas1991,Burkard1996,Gutin2008,GK_Greedy_Chapter,GK_Some_Theory}.  It was proven~\cite{Gutin2008} that in the worst case \Greedy\ produces the unique worst solution; however, it was shown~\cite{GK_Greedy_Chapter} that in some cases \Greedy\ may be a good selection as a fast and simple heuristic.

	\item \MaxRegret{} was discussed in a number of papers, see, e.g.,~\cite{Balas1991,Burkard1996,Gutin2008,Robertson2001}.  As for \Greedy, it is proven~\cite{Gutin2008} that in the worst case \MaxRegret\ produces the unique worst solution however many researchers~\cite{Balas1991} noted that \MaxRegret\ is quite powerful in practice.

	\item \ROM{} was first introduced in~\cite{Gutin2008} as a heuristic of a large domination number.

\end{enumerate}

We use the improved versions of the construction heuristics, see Section~\ref{sec:map_construction_heuristics_improved}.

We will begin our discussion from the experiments started from trivial assignments.  Recall that our test bed includes 10 instances of every size and type and, hence, every result reported in Tables~\ref{tab:ls_solutions_from_trivial} and~\ref{tab:ls_times_from_trivial} is an average for 10 experiments, one experiment for each instance.  Each table is split into two parts; the first part contains only the instances with independent weights (\GP\ and \random) while the second part contains only the instances with decomposable weights (\clique, \geometric, \product\ and \squareroot).  The average values for different instance families and numbers of dimensions are provided at the bottom of each part of each table.  The tables are also split vertically according to the classes of heuristics.  The winner in every row and every class of heuristics is underlined.

The value of the solution error is calculated as $\left(\dfrac{w(A)}{w(A_\text{best})} - 1\right) \cdot 100\%$, where $A$ is the obtained assignment and $A_\text{best}$ is the optimal assignment (or the best known one, see Section~\ref{sec:map_testbed}).

\bigskip

In the group of the vectorwise heuristics the most powerful one is definitely \Threeopt.  \Vopt{} outperforms it only in a few experiments, mostly three dimensional ones (recall that the neighborhood of \Kopt\ increases exponentially with the increase of the number of dimensions $s$).  As it was expected, \Twoopt\ never outperforms \Threeopt\ since $N_\text{2-opt} \subset N_\text{3-opt}$ (see Section~\ref{sec:map_kopt}).  The tendencies for the independent weight instances and for the decomposable weight instances are similar; the only difference which is worth to note is that all but \Vopt\ heuristics of this group solve the \product\ instances very well.  Note that the dispersion of the weights in \product\ instances is really high and, thus, \Vopt, which minimizes the weight of only one vector in every pair of vectors while the weight of the complementary vector may increase arbitrary, cannot be efficient for them.

As one can expect, \MDV\ is more successful than \TwoDV\ and \TwoDV\ is more successful than \OneDV\ with respect to the solution quality (obviously, all the heuristics of this group perform equally for 3-AP and \TwoDV\ and \MDV\ are also equal for 4-AP, see Section~\ref{sec:map_dv}).  However, for the instances with decomposable weights all the dimensionwise heuristics perform very similarly and even for the large $s$, \MDV\ is not significantly more powerful than \OneDV\ or \TwoDV\ which means that in case of decomposable instances the most efficient iterations are when $|D| = 1$.  We can assume that if $c$ is the number of edges connecting the fixed and unfixed parts of the clique, then an iteration of a dimensionwise heuristic is rather efficient when $c$ is small.  Observe that, e.g., for \clique\ the diversity of values in the weight matrix $[M_{i, j}]_{n \times n}$ (see~(\ref{eq:M})) decreases with the increase of the number $c$ and, hence, the space for optimization on every iteration is decreasing.  Observe also that in the case $c = 1$ the iteration leads to the optimal match between the fixed and unfixed parts of the assignment vectors.

All the combined heuristics show improvements in the solution quality over each of their components, i.e., over both corresponding vectorswise and dimensionwise local searches.  In particular, \OneDVtwo\ outperforms both \Twoopt\ and \OneDV, \TwoDVtwo\ outperforms both \Twoopt\ and \TwoDV, \MDVthree\ outperforms both \Threeopt\ and \MDV\ and \MDVV\ outperforms both \Vopt\ and \MDV\@.  Moreover, \MDVthree\ is significantly faster than \Threeopt\ and \MDVV\ is significantly faster than \Vopt.  Hence, we will not discuss the single heuristics \Threeopt\ and \Vopt\ in the rest of the chapter.  The heuristics \OneDVtwo\ and \TwoDVtwo, obviously, perform equally for 3-AP instances.

While for the instances with independent weights the combination of the dimensionwise heuristics with the vectorwise ones significantly improves the solution quality, it is not the case for the instances with decomposable weights (observe that \OneDV\ performs almost as well as the most powerful heuristic \MDVthree) which shows the importance of the instances division.  We conclude that the vectorwise neighborhoods are not efficient for the instances with decomposable weights.

\bigskip

Next we conducted the experiments starting from the other construction heuristics.  But first we compared the construction heuristics themselves, see Table~\ref{tab:construction}.  It is not surprising that \Trivial\ produces the worst solutions.  However, one can see that \Trivial\ outperforms \Greedy\ and \MaxRegret\ for every \product\ instance.  The reason is in the extremely high dispersion of the weights in \product.  Both \Greedy\ and \MaxRegret\ construct the assignments by adding new vectors to it.  The decision which vector should be added does not depend (or does not depend enough in case of \MaxRegret) on the rest of the vectors and, thus, at the end of the procedure only the vectors with huge weights are available.  For other instance families, \Greedy, \MaxRegret\ and \ROM\ perform similarly though the running time of the heuristics is very different.  \MaxRegret\ is definitely the slowest construction heuristic; \Greedy\ is very fast for the \random\ instances (this is because of the large number of vectors of the weight $a$ and the implementation features, see Section~\ref{sec:map_construction_heuristics_improved} for details) and relatively slow for the rest of the instances; \ROM's running time almost does not depend on the instance and is constantly moderate.

Starting from \Greedy\ (Table~\ref{tab:greedy}) significantly improves the solution quality.  This mostly influenced the weakest heuristics, e.g., \Twoopt\ average error decreased in our experiments from 59\% and 20\% to 15\% and 6\% for independent and decomposable weights respectively, though, e.g., the most powerful heuristic \MDVthree\ error also noticeably decreased (from 2.8\% and 5.8\% to 2.0\% and 2.5\%).  As regards the running time, \Greedy\ is slower than most of the local search heuristics and, thus, the running times of all but \MDVthree\ and \MDVV\ heuristics are very similar.  The best of the rest of the heuristics in this experiment is \MDV\ though \OneDVtwo\ and \TwoDVtwo\ perform similarly.

Starting from \MaxRegret\ improves the solution quality even more but at the cost of very large running times.  In this case the difference in the running time of the local search heuristics almost disappears and \MDVthree, the best one, reaches the average error values 1.3\% and 2.2\% for independent and decomposable weights respectively.  Starting from \ROM\ improves the quality only for the worst heuristics.  This is probably because all the best heuristics contain \MDV\ which does a good vectorwise optimization (recall that \ROM\ exploits a similar to the dimensionwise neighborhood idea).  At the same time, starting from \ROM\ increases the running time of the heuristics significantly; the results for both \MaxRegret\ and \ROM\ are excluded from this work; one can find them on the web~\cite{Karapetyan}.

It is clear that the construction heuristics are quite slow comparing to the local search\footnote{Note that a basic construction heuristic is normally faster that a local search procedure.  However, a typical MAP instance has a huge weight matrix ($n^s$ elements), and every of the considered construction heuristics scans this matrix at least once which takes $O(n^s)$ operations.  In contrast, exploration of a local search neighborhood may be relatively quick like $O(n^s 2^s)$ or $O(n^3 s)$ operations for the \Twoopt{} and \OneDV{} heuristics, respectively.} and we should answer the following question: is it worth to spend so much time on the initial solution construction or there is some way to apply local search several times in order to improve the assignments iteratively?  It is known that the algorithms which apply local search several times are called metaheuristics.  There is a number of different metaheuristic approaches such as tabu search or memetic algorithms, but this is not the subject of this research.  In what follows, we are going to use two simple metaheuristics, \Chain\ and \Multichain.

\subsection{Experiments With Metaheuristics}
\label{sec:map_ls_metaheuristics}

It is obvious that there is no sense in applying a local search procedure to one solution several times because a local search moves the solution to a local minimum with respect to its neighborhood, i.e., the second exploration of this neighborhood is useless.  In order to apply a local search several times, one should perturb the solution obtained on the previous iteration.  This idea immediately brings us to the first metaheuristic, let us say \Chain:
\begin{enumerate}
\item Initialize an assignment $A$;
\item Set $A_\text{best} = A$;
\item Repeat:
\begin{enumerate}
	\item Apply local search $A = LS(A)$;
	\item If $w(A) < w(A_\text{best})$ set $A_\text{best} = A$;
	\item Perturb the assignment $A = Perturb(A)$.
\end{enumerate}
\end{enumerate}

In this algorithm we use two subroutines, $LS(A)$ and $Perturb(A)$.  The first one is some local search procedure and the second one is an algorithm which moves the given assignment away from the local minimum by a random perturbation of it.  The perturbation should be strong enough such that the assignment will not come back to the previous position on the next iteration every time though it should not be too strong such that the results of the previous search would be totally destroyed.  Our perturbation procedure selects $p = \lceil n / 25 \rceil + 1$ vectors in the assignment and perturbs them randomly.  In other words, $Perturb(A)$ is just a random move of the \Opt{p} heuristic.  The parameters of the procedure are obtained empirically.

One can doubt if \Chain\ is good enough for large running times and, thus, we introduce a little bit more sophisticated metaheuristic, \Multichain.  Unlike \Chain, \Multichain\ maintains several assignments on every iteration:
\begin{enumerate}
\item Initialize assignment $A_\text{best}$;
\item Set $P = \varnothing$ and repeat the following $c (c + 1) / 2$ times:\\
$P = P \cup \{ LS(Perturb(A_\text{best})) \}$\\
(recall that $Perturb(A)$ produces a different assignment every time);
\item Repeat:
\begin{enumerate}
	\item Save the best $c$ assignments from $P$ into $C_1, C_2, \ldots, C_c$ such that $w(C_i) \le w(C_{i+1})$;
	\item If $w(C_1) < w(C_\text{best})$ set $A_\text{best} = C_1$.
	\item Set $P = \varnothing$ and for every $i = 1, 2, \ldots, c$ repeat the following $c - i + 1$ times: $P = P \cup \{ LS(Perturb(C_i)) \}$.
\end{enumerate}
\end{enumerate}

The parameter $c$ is responsible for the power of \Multichain; we use $c = 5$ and, thus, the algorithm performs $c (c + 1) / 2 = 15$ local searches on every iteration.

\bigskip

The results of the experiments with \Chain\ running for 5 and 10 seconds are provided in Tables~\ref{tab:chain5} and~\ref{tab:chain10} respectively.  The experiments are repeated for three construction heuristics, \Trivial, \Greedy\ and \ROM\@.  It was not possible to include \MaxRegret\ in the comparison because it takes much more than 10 seconds for some of the instances.

The diversity in solution quality of the heuristics decreased with the usage of a metaheuristic.  This is because the fast heuristics are able to repeat more times than the slow ones.  Note that \MDVthree, which is the most powerful single heuristic, is now outperformed by other heuristics.  The most successful heuristics for the instances with independent and decomposable weights are \MDVV\ and \OneDV\ respectfully, though \OneDVtwo\ and \TwoDVtwo\ are slightly more successful than \MDVV\ for the \GP\ instances.  This result also holds for \Multichain, see Tables~\ref{tab:multichain5} and~\ref{tab:multichain10}.  The success of \OneDV\ confirms again that a dimensionwise heuristic is most successful when $|D| = 1$ if the weights are decomposable and that it is more efficient to repeat these iterations many times rather than try $|D| > 1$.  For the explanation of this phenomenon see Section~\ref{sec:map_ls_pure}.  The success of \OneDVtwo\ and \TwoDVtwo\ for \GP\ means existence of a certain structure in the weight matrices of these instances.

One can see that the initialization of the assignment is not crucial for the final solution quality.  However, using \Greedy\ instead of \Trivial\ clearly improves the solutions for almost every instance and local search heuristic.  In contrast to \Greedy, using of \ROM\ usually does not improve the solution quality.  It only influences \Twoopt\ which is the only pure vectorwise local search in the comparison (recall that \ROM\ has a dimensionwise structure and, thus, it is good in combination with vectorwise heuristics).

The \Multichain\ metaheuristic, given the same time, obtains better results than \Chain.  However, \Multichain\ fails for some combinations of slow local search and hard instance because it is not able to complete even the first iteration in the given time.  \Chain, having much easier iterations, do not have this disadvantage.

Giving more time to a metaheuristic also improves the solution quality.  Therefore, one is able to obtain high quality solutions using metaheuristics with large running times.

\subsection{Solvers Comparison}
\label{sec:map_ls_heuristics_comparison}

To compare all the heuristics and metaheuristics discussed in this research we use the same technique as in Section~\ref{sec:gtsp_fair_competition}.  The results are presented in Tables~\ref{tab:heuristics_independent} and~\ref{tab:heuristics_decomposable}.  These tables indicate which heuristics should be chosen to solve particular instances in the given time limitations.  Several best heuristics are selected for every combination of the instance and the given time.  A heuristic is included in the table if it was able to solve the problem in the given time, and if its solution quality is not worse than $1.1 \cdot w(A_\text{best})$ and its running time is not larger than $1.1 \cdot t_\text{best}$, where $A_\text{best}$ is the best assignment produced by the considered heuristics and $t_\text{best}$ is the time spent to produce $A_\text{best}$.

The following information is provided for every solver in Tables~\ref{tab:heuristics_independent} and~\ref{tab:heuristics_decomposable}:
\begin{itemize}
	\item Metaheuristic type (\textbf{C} for \Chain{}, \textbf{MC} for \Multichain{} or empty if the experiment is single).
	\item Local search procedure (\Twoopt, \OneDV, \TwoDV, \MDV, \OneDVtwo, \TwoDVtwo, \MDVthree\, \MDVV\ or empty if no local search was applied to the initial solution).
	\item Construction heuristic the experiment was started with (\heuristic{Gr}, \heuristic{M-R} or empty if the assignment was initialized by \Trivial).
	\item The solution error in percent.
\end{itemize}

The following solvers were included in this experiment:
\begin{itemize}
	\item Construction heuristics \Greedy, \MaxRegret\ and \ROM.

	\item Single heuristics \Twoopt, \OneDV, \TwoDV, \MDV, \OneDVtwo, \TwoDVtwo, \MDVthree\ and \MDVV started from either \Trivial, \Greedy, \MaxRegret\ or \ROM.

	\item \Chain\ and \Multichain\ metaheuristics for either \Twoopt, \OneDV, \TwoDV, \MDV, \OneDVtwo, \TwoDVtwo, \MDVthree\ or \MDVV\ and started from either \Trivial, \Greedy, \MaxRegret\ or \ROM\@.  The metaheuristics proceeded until the given time limitations.
\end{itemize}

Note that for certain instances we exclude duplicating solvers (recall that all the dimensionwise heuristics perform equally for 3-AP as well as \TwoDV\ and \MDV\ perform equally for 4-AP, see Section~\ref{sec:map_dv}).  The common rule is that we leave \MDV\ rather than \TwoDV\ and \TwoDV\ rather than \OneDV\@.  For example, if the list of successful solvers for some 3-AP instance contains \textbf{C}~\OneDVPlain~Gr, \textbf{C}~\TwoDVPlain~Gr and \textbf{C}~\MDVPlain~Gr, then only \textbf{C}~\MDVPlain~Gr will be included in the table.  This is also applicable to the combined heuristics, e.g, having \OneDVtwoPlain~R and \TwoDVtwoPlain~R for a 3-AP instance, we include only \TwoDVtwoPlain~R in the final results.

The last row in every table indicates the heuristics which are the most successful on average, i.e., the heuristics which can solve all the instances with the best average results.

\bigskip

Single construction heuristics are not presented in the tables; single local search procedures appear only for the small allowed times when all other heuristics take more time to run; the most of the best solvers are the metaheuristics.  \Multichain\ seems to be more suitable than \Chain\ for large running times; however, \Multichain\ does not appear for the instances with small $n$.  This is probably because the power of the perturbation degree increases with the decrease of the instance size (note that $perturb(A)$ perturbs at least two vectors in spite of $n$).

The most successful heuristics for the assignment initialization are \Trivial\ and \Greedy; \Trivial\ is useful rather for small running times.  \MaxRegret\ and \ROM\ appear only a few times in the tables. 

The success of a local search depends on the instance type.  The most successful local search heuristic for the instances with independent weights is definitely \MDVV.  The \MDV\ heuristic also appears several times in Table~\ref{tab:heuristics_independent}, especially for small running times.  For the instances with decomposable weights, the most successful are the dimensionwise heuristics and, in particular, \OneDV.

\section{Conclusion}

Several neighborhoods are generalized and discussed in this chapter.  An efficient approach of joining different neighborhoods is successfully applied; the yielded heuristics showed that they combine the strengths of their components.  The experimental evaluation for a set of instances of different types show that there are several superior heuristic approaches suitable for different kinds of instances and running times.  Two kinds of instances are selected: instances with independent weights and instances with decomposable weights.  The first ones are better solvable by a combined heuristic \MDVV; the second ones are better solvable by \OneDV\@.  In both cases, it is good to initialize the assignment with the \Greedy\ construction heuristic if there is enough time; otherwise one should use a trivial assignment as the initial one.  The results can also be significantly improved by applying metaheuristic approaches for as long as possible.

Thereby, it is shown in this chapter that the metaheuristic approach, being based on the fast local search, dominates slow heuristics and, thus, further research of some more sophisticated metaheuristics such as memetic algorithms is of interest.
\chapter{Memetic Algorithms}
\label{sec:memetic}

A memetic algorithm (AM) is a combination of an evolutionary algorithm with a local search procedure \citep{Hart2005,Krasnogor2005}.  The memetic approach is a template for an algorithm rather than a set of rules for designing a powerful heuristic.  A typical frame of MA is presented in Figure~\ref{fig:ma} (for a formal definition of a MA main loop see, e.g., \cite{Krasnogor2008}).
\begin{figure}[!hbtp]
\frame{\parbox{\textwidth}{
\begin{enumerate}
\item \label{item:first_generation} Produce the first generation, i.e., a set of feasible solutions.
\item \label{item:first_generation_ls} Apply a local search procedure to every solution in the first generation.
\item \label{item:termination} Repeat the following while a termination criterion is not met:
	\begin{enumerate}
	\item \label{item:next_generation} Produce a set of new solutions by applying so-called genetic operators to solutions from the previous generation.
	\item \label{item:next_generation_ls} Improve every solution in this set with the local search procedure.
	\item \label{item:selection} Select several best solutions from this set to the next generation.
	\end{enumerate}
\end{enumerate}
}}
\caption{A typical memetic algorithm frame.}
\label{fig:ma}
\end{figure}

After a thorough research of GTSP and MAP local search, we can use our knowledge to design MAs for these problems.  However, the development of the GTSP local search preceded our research of GTSP MA and, thus, in Section~\ref{sec:gtsp_ma} we do not use some of the improvements proposed in Section~\ref{sec:gtsp_ls}.  It is in our future plans to update our GTSP MA according to the most recent results.  Nevertheless, the algorithm presented in Section~\ref{sec:gtsp_ma} is the state-of-the-art GTSP metaheuristic.  It clearly dominates all other metaheuristics known from the literature.  Moreover, it outperforms all the algorithms~\cite{Bontoux2009,Tasgetiren2010} appeared after our results were published.

\section{GTSP Memetic Algorithm}
\label{sec:gtsp_ma}

We start with a general scheme of our heuristic, which is similar to the general schemes of many memetic algorithms, see Figure~\ref{fig:ma}.

\begin{itemize}
	\item[Step 1] \emph{Initialize}.  Construct the first generation of solutions.  To produce a solution we use a semirandom construction heuristic (see Section~\ref{sec:gtsp_first_generation}).

	\item[Step 2] \emph{Improve}.  Use a local search procedure to replace each of the first generation solutions by a local optimum.  Eliminate duplicate solutions.

	\item[Step 3] \emph{Produce next generation}.  Use reproduction, crossover, and mutation genetic operators to produce the non-optimized next generation.  Each of the genetic operators selects parent solutions from the previous generation.  The weight of a solution is used as the evaluation function.

	\item[Step 4] \emph{Improve next generation}.  Use a local search procedure to replace each of the current generation solutions except the reproduced ones by the local optimum.  Eliminate duplicate solutions.

	\item[Step 5] \emph{Evolute}.  Repeat Steps 3--4 until a termination condition is reached.
\end{itemize}

\subsection{Coding}

MA requires each solution to be coded in a \emph{chromosome}, i.e., to be represented by a sequence of \emph{genes}.  Unlike~\cite{Snyder2000,Tasgetiren2007} we use a natural coding of the solutions as in~\cite{Silberholz2007}.  The coded solution is a sequence of numbers ($T_1$ $T_2$ \ldots $T_m$) such that $T_i$ is the vertex at the position $i$ of the solution.  For example (2~5~9~4) represents the cycle visiting vertex 2, then vertex 5, then vertex 9, then vertex 4, and then returning to vertex 2.  Note that not any sequence corresponds to a feasible solution as the feasible solution should contain exactly one vertex from each cluster, i.e., $\Cluster(T_i) \neq \Cluster(T_j)$ for any $i \neq j$.

Note that, using natural coding, each solution can be represented by $m$ different chromosomes: the sequence can be `rotated', i.e., the first gene can be moved to the end of the chromosome or the last gene can be inserted before the first one and these operations will preserve the cycle.  For example, chromosomes (2~5~9~4) and (5~9~4~2) represent the same solution.  We need to take this into account when considering several solutions together, i.e., in exactly two cases: when we compare two solutions, and when we apply crossover operator.  In these cases we `normalise' the chromosomes by rotating each of them such that the vertex $v \in V_1$ (the vertex that represents the cluster 1) takes the first place in the chromosome.  For example, if we had a chromosome (2~5~9~4) and the vertex 5 belongs to the cluster 1, then we rotate the chromosome in the following way: (5~9~4~2).

In the case of the symmetric problem the chromosome can also be `reflected' while preserving the solution.  However, our heuristic is designed for both symmetric and asymmetric instances and, thus, the chromosomes (1~5~9~4) and (4~9~5~1) are considered as the chromosomes corresponding to distinct solutions.

The main advantage of the natural coding is its efficiency in local search implementations\footnote{Recall that this result was obtained before our thorough research of GTSP local search; also observe that the natural coding is efficient for the Basic adaptations of TSP neighborhoods, see Section~\ref{sec:gtsp_ls}.}.  Since the local search is the most time consuming part of our heuristic, the coding should be optimized for it.

\subsection{First Generation}
\label{sec:gtsp_first_generation}

We produce $2 m$ solutions for the first generation, where $m$ is the number of clusters.  The solutions are generated by a \emph{semirandom construction heuristic}.  The semirandom construction heuristic generates a random cluster permutation and then finds the best vertex in each cluster by applying the Cluster Optimization, see Section~\ref{sec:gtsp_cluster_optimization}).

The advantages of the semirandom construction heuristic are that it is fast and its cycles have no regularity.  The latter is important as each completely deterministic heuristic can cause solutions uniformity and as a result some areas of solution space may not be explored.

\subsection{Next Generations}
\label{sec:gtsp_next_generation}

Each generation except the first one is based on the previous generation.  To produce the next generation one uses genetic operators, which are algorithms that construct a solution or two from one or two so-called parent solutions.  Parent solutions are chosen from the previous generation using some \emph{selection strategy}.  We perform $r$ runs of \emph{reproduction}, $8 r$ runs of \emph{crossover}, and $2 r$ runs of \emph{mutation} operator.  The value $r$ is calculated as 
\begin{equation}
\label{eq:gtsp_ma_r}
r = 0.2 g + 0.05 m + 10 \,,
\end{equation}
where $g$ is the number of generations produced before the current one.  As a result, we obtain at most $11 r$ solutions in each generation but the first one (note that we remove duplicate solutions from the population and, hence, the number of solutions in each generation may vary).

One may expect the number of local minima found by the algorithm to increase from generation to generation.  This number may also be expected to grow when the number of clusters $m$ grows.  Thus, in the formula above $r$ depends on both $g$ and $m$.  All the coefficients in this section were obtained empirically.  Note that slight variations in selection of these coefficients do not significantly influence the results of the algorithm.

\subsection{Reproduction}
\label{sec:gtsp_reproduction}

Reproduction is a process of copying solutions from the previous generation.  Reproduction operator requires a selection strategy.  In our algorithm, we select $r$ (see~(\ref{eq:gtsp_ma_r})) lightest solutions from the previous generation to copy them to the current generation.

\subsection{Crossover}

A \emph{crossover} operator is a genetic operator that combines two different solutions from the previous generation.  We use a modification of the two-point crossover introduced by Silberholz and Golden~\cite{Silberholz2007} as an extension of an \emph{Ordered Crossover}~\cite{Davis1985}.  Our crossover operator produces just one child solution $(r_1~r_2~\ldots~r_m)$ from the parent solutions $(p_1~p_2~\ldots~p_m)$ and $(q_1~q_2~\ldots~q_m)$.  At first, it selects a random position $a$ and a random fragment length $1 \le l < m$ and copies the fragment $[a, a + l)$ of the first parent to the beginning of the child solution: $r_i = p_{i+a}$ for each $i = 0, 1, \ldots, l - 1$\footnote{We assume that $T_{i+m} = T_i$ for the solution $(T_1~T_2~\ldots~T_m)$ and for any $1 \le i \le m$.}.  To produce the rest of the child solution, we introduce a sequence $q'$ as follows: $q'_i = q_{i+a+l-1}$, where $i = 1, 2, \ldots, m$.  Then, for every $i$, we remove $q'_i$ if the cluster $\Cluster(q'_i)$ is already visited by the child solution $r$.  As a result, $l$ vertices are removed: $|q'| = m - l$.  We extend the child solution $r$ by the sequence $q'$: $r = (r_1~r_2~\ldots~r_l~q'_1~q'_2~\ldots~q'_{m-l})$.

The main advantage of this crossover is that it tends to preserve the vertex order of both parents.

\bigskip

\noindent
\emph{Crossover example}. Let the first parent $p$ be $(1~2~3~4~5~6~7)$ and the second parent $q = (3~2~5~7~6~1~4)$ (for simplicity, we assume that every cluster contains exactly one vertex: $V_i = \{ i \}$).  First of all, we rotate the parent solutions such that $\Cluster(p_1) = \Cluster(q_1) = 1$:\\
$p = (1~2~3~4~5~6~7)$ (remains the same) and\\
$q = (1~4~3~2~5~7~6)$. \\
Then we choose a random fragment in the parent solutions:\\
$p = (1~2~\underline{3~4}~5~6~7)$\\
$q = (1~4~\underline{3~2}~5~7~6)$\\
and copy this fragment from the first parent $p$ to the child solution: $r = (3~4)$.  Next we produce the sequence $q' = (5~7~6~1~4~3~2)$ and remove vertices 3 and 4 from it as the corresponding clusters are already visited by $r$: $q' = (5~7~6~1~2)$.  Finally, we extend the child solution $r$ by $q'$:\\
$r = (3~4~5~7~6~1~2)$.

\bigskip

The crossover operator requires some strategy to select two parent solutions from the previous generation.  In our algorithm an elitist strategy is used; the parents are chosen randomly between the best $33 \%$ of all the solutions in the previous generation.

\subsection{Mutation}

A \emph{mutation} operator partially modifies a solution from the previous generation.  The modification should be stochastic and usually worsens the solution.  The goal of mutation is to increase the solution diversity.

Our mutation operator removes a random fragment of the solution and inserts it at some random position.  The size of the fragment is selected between $0.05 m$ and $0.3 m$.  An elitist strategy is used in our algorithm; the parent is selected randomly among $75 \%$ best solutions in the previous generation.

\bigskip

\noindent
\emph{Mutation example}. Let the parent solution be (1~2~3~4~5~6~7).  Let the random fragment start at the position 2 and be of length 3.  The new fragment position is 3, for example.  After removing the fragment, we get (1~5~6~7).  Now insert the fragment (2~3~4) at the position 3: (1~5~2~3~4~6~7).

\subsection{Termination condition}
\label{sec:gtsp_ma_termination}

For the termination condition we use the concept of idle generations.  We call a generation \emph{idle} if its best solution has the same weight as the weight of the best solution in the preceding generation.  In other words, if some generation has not improved the solution, it is called idle.  Our MA stops if a certain number of idle generations are produced sequentially.

In particular, we implemented the following new condition.  Let $I(l)$ be the number of sequential idle generations with the best solution of the weight $l$.  Let $I_{cur} = I(l_{cur})$, where $l_{cur}$ is the weight of the best solution obtained so far.  Let $\displaystyle{I_{max} = \max_{l > l_{cur}} I(l)}$.  Then our heuristic stops if $I_{cur} \ge \max\{1.5 I_{max}, 0.05 m + 5\}$.  This formula means that we are ready to wait for the next improvement 1.5 times more generations than we have ever waited previously.  The constant $0.05 m + 5$ is the minimum boundary for the number of generations we are ready to wait for improvement.  

All the coefficients used in this section are obtained empirically.

\subsection{Asymmetric instances}

Our algorithm is designed to process both symmetric and asymmetric instances, however, some parameters should take different values for these types of instances for the purpose of high efficiency.  In particular, we double the size of the first generation ($4 m$ instead of $2 m$, see Section~\ref{sec:gtsp_first_generation}) and increase the minimum number of idle generations by 5 (i.e., $I_{cur} \ge \max\{1.5 I_{max}, 0.05 m + 10\}$, see Section~\ref{sec:gtsp_ma_termination}).  The local improvement procedure (see below) has also some differences for symmetric and asymmetric instances.

\subsection{Local Improvement Part}
\label{sec:gtsp_local_improvement}

We apply a \emph{local improvement procedure} for each solution added to the current generation.  The local improvement procedure runs several local search heuristics sequentially.  The following local search heuristics are used in our algorithm:

\begin{itemize}
	\item \emph{Swap} tries to swap every non-neighboring pair of vertices, see Section~\ref{sec:gtsp_swap}.  The heuristic applies all the improvements found during one cycle of swaps.

	\item \emph{$k$-Neighbor Swap} is a naive implementation of the Fragment Optimization local search (see Section~\ref{sec:gtsp_fragment_optimization}).  It tries different permutations of every solution subsequence $(T_{i}~T_{i+1}~\ldots~T_{i+k-1})$.  In particular, it tries all the non-trivial permutations which are not covered by any of $k'$-Neighbor Swap, $k' = 2, 3, \ldots, k-1$.  For each permutation, the best vertex selection within the considered cluster subsequence is calculated.  The best permutation is accepted if it improves the solution.  The heuristic applies all the improvements found during one cycle.

	\item \emph{2-opt} (see the Basic adaptation of 2-opt TSP neighborhood, Section~\ref{sec:gtsp_two_opt}) tries to replace every non-adjacent pair of edges $T_i \to T_{i+1}$ and $T_j \to T_{j+1}$ in the solution by the edges $T_i \to T_j$ and $T_{i+1} \to T_{j+1}$ if the new edges are lighter, i.e., the sum of their weights is smaller than the sum of the weights of old edges.  The heuristic applies all the improvements found during one cycle.

	\item \emph{Direct 2-opt} is a modification of the 2-opt heuristic.  It only considers long edges in the tour, i.e., it selects the edges $T_i \to T_{i+1}$ and $T_j \to T_{j+1}$ from a list of the the heaviest edges in the tour.

	\item \emph{Insertion} (see the Local adaptation \Insertion{L}{} in Section~\ref{sec:gtsp_insertion}) tries to remove a vertex from the solution and to insert it in a different position.  The best vertex in the inserted cluster is selected after the insertion.  The heuristic tries every combination of the old and the new positions except the neighboring positions, and applies all the improvements found.

	\item \label{item:cluster_optimization} \emph{Cluster Optimization} (CO) is exactly the same algorithm as described in Section~\ref{sec:gtsp_cluster_optimization}.  However, this implementation does not use all the improvements proposed in Section~\ref{sec:gtsp_co_refinements}. 
\end{itemize}

For each local search algorithm with some local cluster optimization, i.e., for $k$-Neighbor Swap and Insert, we use a speed-up heuristic.  We calculate a lower bound $l_\text{new}$ of the new solution weight and compare it to the previous weight $l_\text{prev}$ of the solution before the move.  If $l_\text{new} \ge l_\text{prev}$, the solution modification is immediately declined.  The lower bound $l_\text{new}$ is calculated in the assumption that every new edge $x \to y$ in the tour has the weight $w_\text{min}(X \to Y)$ of the shortest edge between the corresponding clusters $X = \Cluster(x)$ and $Y = \Cluster(y)$.

Some of these heuristics form a heuristic-vector $\mathcal{H}$ as follows:

\bigskip
\noindent
\begin{tabular}{ll}
Symmetric instances								& Asymmetric instances \\
\cmidrule(){1-2}
Insert												& Swap \\
Direct 2-opt for $m / 4$ longest edges		& Insert \\
2-opt								& Direct 2-opt for $m / 4$ longest edges \\
2-Neighbor Swap				& 2-opt \\
3-Neighbor Swap				& 2-Neighbor Swap \\
4-Neighbor Swap				& 3-Neighbor Swap \\
\cmidrule(){1-2}
\end{tabular}
\bigskip

The improvement procedure applies all the local search heuristics from $\mathcal{H}$ cyclically.  Once some heuristic fails to improve the tour, it is excluded from $\mathcal{H}$.  If 2-opt heuristic fails, we also exclude Direct 2-opt from $\mathcal{H}$.  Once $\mathcal{H}$ is empty, the CO heuristic is applied to the solution and the improvement procedure stops.

\subsection{Modification of the Algorithm for Preprocessing}
\label{sec:gtsp_ma_modified}

Recall that in Section~\ref{sec:gtsp_preprocessing} we proposed two reduction algorithms intended to reduce the size of a GTSP instance.  The first algorithm removes some vertices and, hence, does not significantly change the structure of the instance.  In contrast, the second algorithm removes some edges or, to be precise, sets some weights to a very large number.  Observe that, if an edge of a very large weight is occasionally included in a tour, this may totally change the algorithm's behavior.

For the purpose of evaluating the reduction algorithms, we modified our GTSP memetic algorithm (\GK{}) as follows:
\begin{itemize}
\item The 2-opt heuristic was extended with the cluster optimization (see \twoopt{L}{} adaptation in Section~\ref{sec:gtsp_two_opt}).

\item Direct 2-opt heuristic was excluded from the Local Search Procedure.

\item Every time before starting the Cluster Optimization, we remove all the vertices that cannot be included in the solution.  In other words, a vertex $y \in Y$ is excluded if $w_\text{min}(X \to y) = \infty$ or $w_\text{min}(y \to Z) = \infty$, where $X = \Cluster(T_i)$, $Y = \Cluster(T_{i+1})$ and $Z = \Cluster(T_{i+2})$.

\item Since the modified Local Search Procedure is more powerful than the previous one, we reduced the number of solutions in a generation (see Section~\ref{sec:gtsp_next_generation}): $r = 0.2g + 0.03m + 8$.  We have also changed the termination condition (see Section~\ref{sec:gtsp_ma_termination}): $I_\text{cur} \ge \max(1.5 I_\text{max}, 0.025m + 2)$.
\end{itemize}

\section{Experimental Evaluation of GTSP Memetic Algorithm}

Our test bed includes all the instances described in Section~\ref{sec:gtsp_testbed} with $40 \le m \le 217$.  Unlike in~\cite{Silberholz2007,Snyder2000,Tasgetiren2007}, smaller instances are not considered.

Tables~\ref{tab:gtsp_ma_quality}, \ref{tab:gtsp_ma_time} and~\ref{tab:gtsp_ma_details} report the experiments results.  We compare the following heuristics:
\begin{description}
	\item[GK] is our GTSP memetic algorithm.
	\item[SG] is the heuristic by Silberholz and Golden~\cite{Silberholz2007}.
	\item[SD] is the heuristic by Snyder and Daskin~\cite{Snyder2000}.
	\item[TSP] is the heuristic by Tasgetiren, Suganthan, and Pan~\cite{Tasgetiren2007}.
\end{description}

The results for \GK{} (our memetic algorithm) and \SD{} were obtained in our own experiments.  Other results are taken from the corresponding papers.  Each result for \GK{} and \SD{} is averaged for ten algorithm runs (recall that all these heuristics are non-deterministic and, hence, their results may very from time to time).  The results for \SG{} and \TSPma{} were collected after five runs.
\nomenclature[GK]{\GK{} (GTSP)}{is our memetic algorithm}

To compare the running times of all the considered heuristics we need to convert the running times of \SG{} and \TSPma{} obtained from the corresponding papers to the running times on our evaluation platform.  Let us assume that the running time of some Java implemented algorithm on the \SG{} evaluation platform is $t_\text{SG} = k_\text{SG} \cdot t_\text{GK}$, where $k_\text{SG}$ is some constant and $t_\text{GK}$ is the running time of the same but \texttt{C++} implemented algorithm on our evaluation platform.  Similarly, let us assume that the running time of some algorithm on the \TSPma{} evaluation platform is $t_\text{TSP} = k_\text{TSP} \cdot t_\text{GK}$.

The computer used for \GK{} and \SD{} evaluation is based on AMD Athlon 64 X2 3.0~GHz processor.  The evaluation platforms used for \SG{} and \TSPma{} are based on Intel Pentium 4 3.0~GHz and Intel Centrino Duo 1.83~GHz processors, respectively.  \GK{}, \SD{}, and \TSPma{} are implemented in \texttt{C++} (\GK{} is implemented in \texttt{C\#} but the most time critical fragments are implemented in \texttt{C++}).  \SG{} is implemented in \texttt{Java}.  Some rough estimation of \texttt{Java} performance in the combinatorial optimization applications shows that \texttt{C++} implementation could be approximately two times faster than the \texttt{Java} implementation.  hence, we assume that $k_\text{SG} \approx 3$ and $k_\text{TSP} \approx 2$.

In order to assess our estimation, we can compare the results of \SD{} reported in different papers (note that \SD{} was reimplemented in \texttt{Java} in~\cite{Silberholz2007}).  The time ratio between the \SD{} running times from~\cite{Silberholz2007} and our own results vary significantly for different instances, but for some moderate size instances the ratio is about 2.5 to 3.  These results correlate well with the previous estimation.  The suggested value $k_\text{TSP} \approx 2$ is also confirmed by this method.

\bigskip

The headers of the tables in this section are as follows:
\begin{description}
	\item[Name] is the instance name as described in Section~\ref{sec:gtsp_testbed}.

	\item[Error, \%] is the average solution error.  It is calculated as $\frac{\var{value} - \var{opt}}{\var{opt}} \times 100 \%$, where $\var{value}$ is the obtained solution weight and $\var{opt}$ is the optimal solution weight.  Recall that the exact solutions are known from~\cite{Ben-Arieh2003} and from~\cite{Fischetti1997} for only 17 of the considered instances.  For the rest of the instances we use the best solutions ever obtained in our experiments instead or in the literature.

	\item[Time, sec] is the average running time for the considered heuristic.  The running times for \SG{} and for \TSPma{} are obtained from the corresponding papers and, thus, these values have to be adjusted before comparison, see above.

	\item[Quality impr., \%] is the improvement of the average solution quality of the \GK{} with respect to some other heuristic.  The improvement is calculated as $E_H - E_\text{GK}$ where $E_H$ is the average error of the considered heuristic $H$ and $E_\text{GK}$ is the average error of our heuristic.

	\item[Time impr.] is the improvement of the \GK{} average running time with respect to some other heuristic running time.  The improvement is calculated as $T_H / T_\text{GK}$ where $T_H$ is the average running time of the considered heuristic $H$ and $T_\text{GK}$ is the average running time of our heuristic.

	\item[Opt., \%] is the number of tests, in percent, in which the optimal solution was reached.  This value is displayed for only three heuristics since we do not have this information for \SG{}.

	\item[Opt.] is the weight of the best known solution for the given instance.

	\item[Value] is the average solution weight.

	\item[\# gen.] is the average number of generations produced by our heuristic.
\end{description}

The results of the experiments presented in Table~\ref{tab:gtsp_ma_quality} show that our heuristic (\GK{}) clearly outperforms all other heuristics with respect to solution quality.  For each of the considered instances the average solution reached by our heuristic is always not worse than the average solution reached by any other heuristic and the percent of the runs in which the optimal solution was reached is not less than for any other considered heuristic (note that we are not able to compare our heuristic with \SG{} with respect to this parameter).

The average values are calculated for four \emph{instance sets} (IS).  The \emph{Full IS} includes all the instances in our test bed, both symmetric and asymmetric.  The \emph{Sym.~IS} includes all the symmetric instances in our test bed.  The \emph{SG~IS} includes all the instances considered in~\cite{Silberholz2007} such that $m \ge 40$.  The \emph{TSP~IS} includes all the instances considered in~\cite{Tasgetiren2007} such that $m \ge 40$.

One can see that the average quality of our \GK{} heuristic is approximately 10 times better than that of \SG{} heuristic, approximately 30 times better than that of \SD{}, and for TSP~IS our heuristic reaches the optimal solution in every run and for every instance, in contrast to \TSPma{} that has 0.44\% average error.  The maximum error of \GK{} is 0.27\% while the maximum error of \SG{} is 2.25\% and the maximum error of \SD{} is 3.84\%.

The running times of the considered heuristics are presented in Table~\ref{tab:gtsp_ma_time}.  in every experiment, the running time of \GK{} is not worse than the running time of any other heuristic: the minimum time improvement with respect to \SG{} is 6.6 that is greater than 3 (recall that 3 is the adjusting coefficient for \SG{} evaluation platform, see above), the time improvement with respect to \SD{} is never less than 1.0 (recall that both heuristics were tested on the same platform), and the minimum time improvement with respect to \TSPma{} is 4.6 that is greater than 2 (recall that 2 is the adjusting coefficient for \TSPma{} evaluation platform, see above).  The average time improvement is approximately 12~times for \SG{} (or 4~times if we take into account the platforms difference), 3~times for \SD{}, and 11~times for \TSPma{} (or 5~times if we take into account the platforms difference).

The stability of \GK{} is high. E.g., we ran \GK{} for the \texttt{89pcb442} instance 100 times, and the optimal solution was obtained in every run.  The time standard deviation was 0.27~sec; the minimum and the maximum running times were 1.29~s and 2.45~s, respectively, and the average was 1.88~s.  For 100 runs of \texttt{217vm1084}, the average running time was 65.32~s, the minimum and the maximum times were 44.30~s and the standard deviation was~13.57 s.  The average solution was 130994 (0.22\% above the best known), the minimum and the maximum were 130704 (exactly the best known) and 131845 (0.87\% above best known), and the standard deviation was 331.

Some details on the \GK{} experiments are presented in Table~\ref{tab:gtsp_ma_details}.  The table includes the average number of generations produced by the heuristic.  One can see that this number is relatively small: note that the \SD{} and \TSPma{} limit the number of generation to 100 while they consider only the instances with $m < 90$; \SG{} terminates the algorithm after 150 idle generations.  This shows that our heuristic does not require a lot of generations because of the powerful local search procedure and large population sizes.

All the materials required to reproduce our experiments is available online~\cite{Karapetyan}.

\section{Population Sizing}
When implementing a memetic algorithm, one faces a lot of questions.  Some of these questions, like selecting the most appropriate local search or crossover operators, were widely discussed in the literature while others are still not investigated enough.  In this research we focus our attention on the population sizing in memetic algorithms.

Population size is the number of solutions (chromosomes) maintained at a time by a memetic algorithm.  Many researchers indicate the importance of selecting proper population sizes \citep{Glover2003,Harik1999,Hart2005}.  However, the most usual way to define the population size is to fix it to some constant at the design time \citep{Cotta2008,Grefenstette1986,Hart2005,Huang2006}.  Several more sophisticated models based on statistical analysis of the problem or self-adaptive techniques are proposed for genetic, particle swarm optimization and some other evolutionary algorithms \citep{Cotta2008,Eiben2004,Goldberg1991,Harik1999,Hart2005,Kaveh2007,Lee1993} but they all are not suitable for memetic algorithms because of the totally different algorithm dynamics.

It is known \citep{Hart2005} that in memetic algorithms the population size, the solution quality and the running time are mutually dependent.  Often the population size is fixed at the design time which, for a given algorithm with a certain termination criterion, determines the solution quality and the running time.  However, in many applications it is the running time which has to be fixed.  This leads to a problem of finding the most appropriate population size $m$ for a fixed running time $\tau$ such that the solution quality is optimized.  However, the population size $m$ depends not only on the given time $\tau$ but also on the instance type and size, on the local search performance and on the computational platform.  The fact that the optimal population size depends on the particular instance, forces researchers to use parameter control to adapt dynamically the population size for all the factors during the run (see, e.g., \cite{Coelho2008,Eiben2004,Kaveh2007}).  However, none of these approaches consider the running time of the whole algorithm and, hence, are poorly suitable for a strict time limitation.

Instead of it, we have found a parameter encapsulating all these factors, i.e, a parameter which reflects on the relation between the instance, the local search procedure and the computation platform.  It is the average running time $t$ of the local search procedure applied to some solutions of the given instance.  Definitely this time depends on the particular solutions but later we will show that $t$ can be measured at any point of the memetic algorithm run with a good enough precision.

Now we can find a near-optimal population size $m_\text{opt}$ as a function of $\tau$ and $t$.  In particular, it can be calculated as
\begin{equation}
\label{eq:mo}
m_\text{opt}(\tau, t) = a \cdot \frac{\tau^b}{t^c}\,,
\end{equation}
where $a$, $b$ and $c$ are some tuned \citep{Eiben1999} constants which reflect on the specifics of the other algorithm factors.

Observe that this is not a pure parameter tuning.  Indeed, the population size depends on the average local search running time $t$ which is obtained during the algorithm run.  Thus, our approach is a combination of the parameter tuning and control.

In our previous attempt (see Section~\ref{sec:gtsp_next_generation}) to adjust the population size we assumed that it depends on the instance size $n$ only (i.e., $m = m(n)$) but an obvious disadvantage of this approach is that it does not differentiate between instance types.

\subsection{Managing Solution Quality and Population Sizing}
\label{sec:general}

Having some fixed procedures for production of the first generation (Step~\ref{item:first_generation} in Figure~\ref{fig:ma}), improving a solution (Steps~\ref{item:first_generation_ls} and \ref{item:next_generation_ls}) and obtaining the next generation from the previous one (Steps~\ref{item:next_generation} and \ref{item:selection}), the algorithm designer is able to manage the solution quality and the running time of the algorithm by varying the termination criterion (Step~\ref{item:termination}) and the population size, i.e., the number of maintained solutions in Steps~\ref{item:first_generation} and \ref{item:selection}.

Usually, a termination condition in a memetic algorithm tries to predict the point after which any further effort is useless or, at least, not efficient.  A typical approach is to count the number $I_\text{idle}$ of running generations which did not improve the best result and to stop the algorithm when this number reaches some predefined value.  A slightly more advanced prediction method is applied in our state-of-the-art algorithm for GTSP (see Section~\ref{sec:gtsp_ma_termination}).  It stops the algorithm when $I_\text{idle}$ reaches $k \cdot I_\text{prev}$, where $k > 1$ is a constant and $I_\text{prev}$ is the maximum $I_\text{idle}$ obtained before the current solution was found.

In case of such termination conditions, the running time of the algorithm is unpredictable and, hence, cannot be adjusted for one's needs.  Observe that many applications (like real-time systems) in fact have strict time limitations.  To satisfy these limitations, we bound our algorithm within some fixed running time and aim to use this time with the most possible efficiency.  Below we discuss how the parameters of the algorithm should be adjusted for this purpose.

\subsection{Population Size}
\label{sec:populationsize}

Population size is the number of solutions maintained by a memetic algorithm at the same time.  This number may vary from generation to generation but we decided to keep the population size constant during the algorithm run in order to simplify the research.

\begin{figure}[t]
\centering
\includegraphics{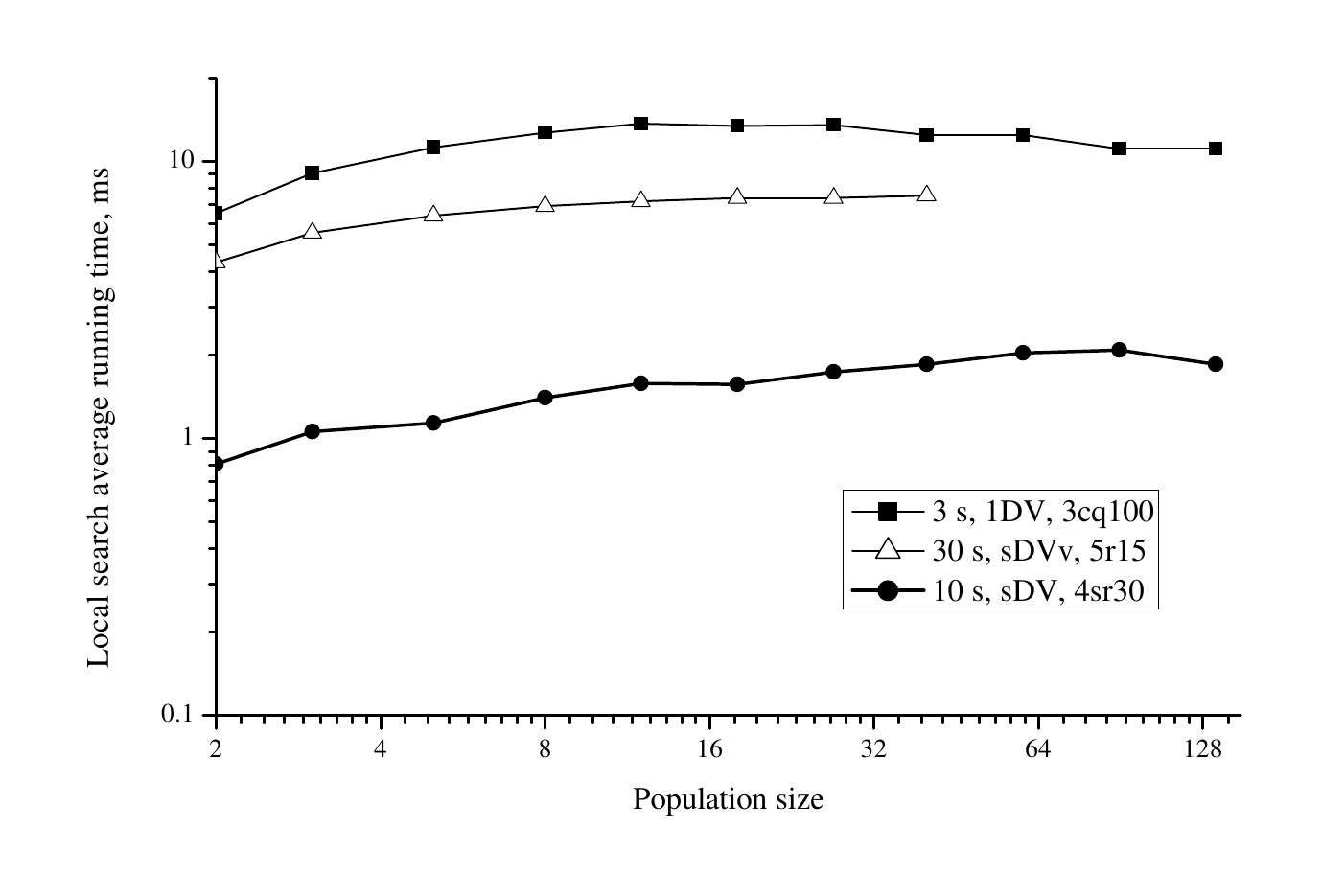}
\caption[The average time required for one local search run depends only marginally on the proportion between the population size and the number of generations.]{The average time required for one local search run depends only marginally on the proportion between the population size and the number of generations.  These three lines correspond to three runs of our memetic algorithm.  In every run we used different local search procedures (\OneDV{}, \MDV{} and \MDVV{}, for details see Section~\ref{sec:local_search}) and different given times $\tau$ (3 s, 10 s and 30 s).}
\label{fig:lstime}
\end{figure}

Let $I$ be the total number of generations during the algorithm run and $m$ be the population size.  Then the running time of the whole algorithm is proportional to $I \cdot m$.  Indeed, the most time consuming part of a memetic algorithm is local search.  The number of times the local search procedure is applied is proportional to $I \cdot m$, and we have shown empirically (see Figure~\ref{fig:lstime}) that the average running time of a local search depends only marginally on the population size.  Since we fix the running time of the whole algorithm, we get:
$$
I \cdot m \approx \mathrm{const} \,.
$$

In other words, we claim that inversely proportional change of $I$ and $m$ preserves the running time of the whole algorithm; our experiments confirm it.

\begin{figure}[t]
\centering
\includegraphics{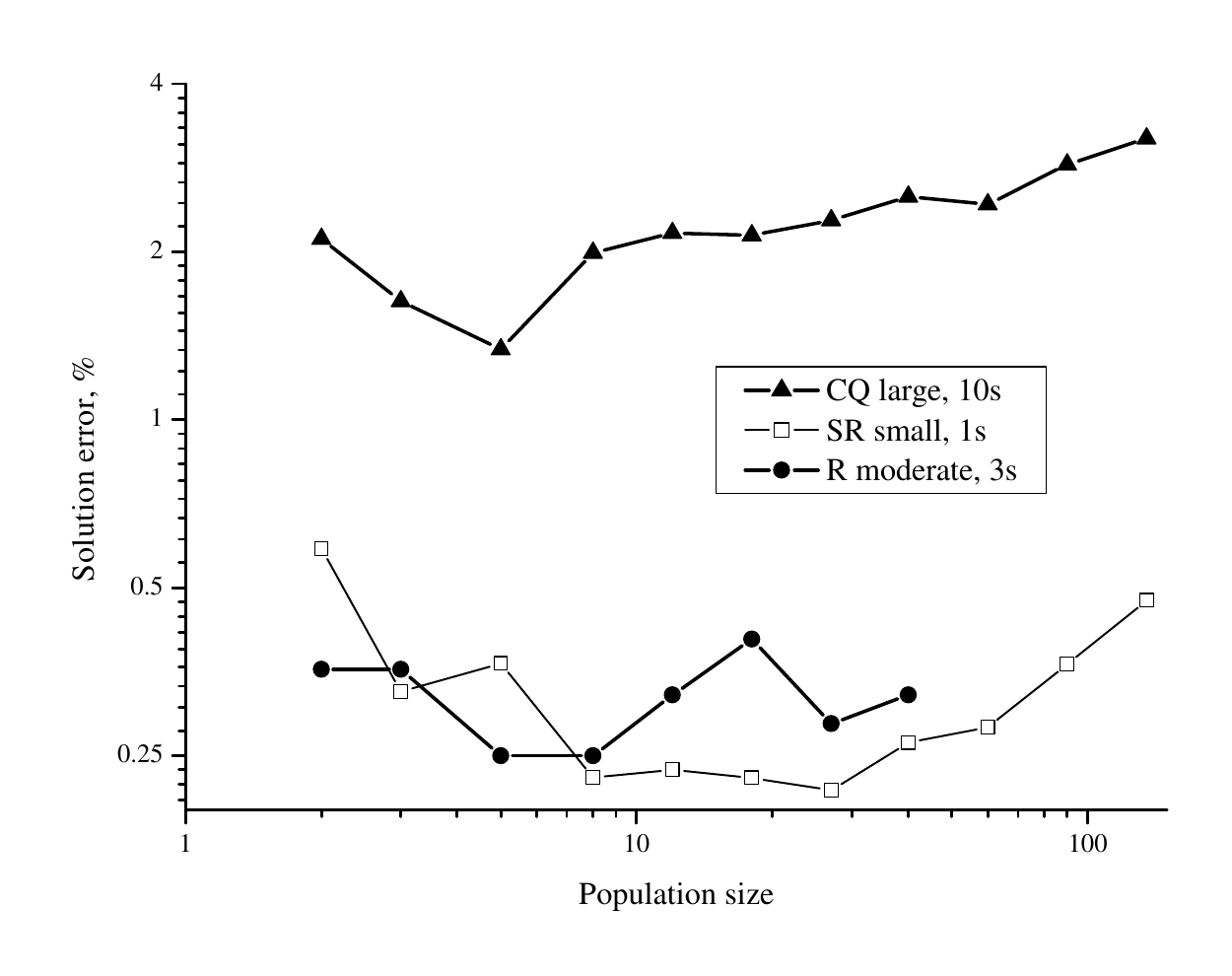}
\caption[The solution quality significantly depends on the population size.]{The solution quality significantly depends on the population size.  For every instance, local search and given time, there exists some optimal population size.  On this plot we show how the relative solution error depends on the population size for different types and sizes of instances (for detailed descriptions of the particular instance types, see Section~\ref{sec:map_testbed}).}
\label{fig:error_of_m}
\end{figure}

Since $I \cdot m = \text{const}$, we need to find the optimal ratio between $I$ and $m$.  Our experimental analysis shows that this ratio is crucial for the algorithm performance: for a wrongly selected ratio between $I$ and $m$, the relative solution error, i.e., the percentage above the optimal objective value, may be twice as big as the relative solution error for a well fitted ratio, see Figure~\ref{fig:error_of_m}.

Observe that the optimal ratio between $I$ and $m$ depends on the following factors:
\begin{itemize}
	\item Given time $\tau$;
	\item Instance type and size;
	\item Computational platform;
	\item Local search procedure;
	\item Genetic operators and selection strategies.
\end{itemize}
Note that all factors but the first one are hard to formalize.  Next we will discuss relations between these factors.


Since we assume that almost only the local search consumes the processor time (see above), the computational platform affects only the local search procedure.  Another parameter which greatly influences the local search performance is the problem instance; it is incorrect to discuss a local search performance without considering a particular instance.  

Let $t$ be the average running time of the local search procedure applied to some solution of the given instance being run on the given computational platform.  (Recall that this time stays almost constant during the algorithm run, see Figure~\ref{fig:lstime}\@.)  Our idea is to use $t$ as the value which encapsulates the specifics of the instance, of the computational platform and of the local search procedure.

Definitely the local search and the instance are also related to the genetic operators and selection strategies, but we assume that this relation is not that important; our computational experience confirms this.

Hence, we can calculate the near-optimal population size $m_\text{opt} = f(t, \tau)$, and the rest of the factors are indirectly included into the function $f$ definition.  Obviously $m_\text{opt}$ grows with the growth of $\tau$ and reduces with the growth of $t$.  We propose a flexible function (\ref{eq:mo}) for $m_\text{opt}$.  The constants $a$, $b$ and $c$ are intended to reflect on the specifics of genetic operators and selection strategies.  Observe that variation of $a$, $b$ and $c$ may significantly change the behavior of $m_\text{opt}$.

Since $a$, $b$ and $c$ are only related to the fixed parts of the algorithm, they should be adjusted before the algorithm's run, i.e., these parameters should be tuned \citep{Eiben1999}.  However, the whole approach should be considered as a combination of parameter tuning and control since the time $t$ is obtained during the algorithm's run.


\subsection{\texorpdfstring{Choosing Constants $a$, $b$ and $c$}{Choosing Constants a, b and c}}

Our approach has two stages: tuning the constants $a$, $b$ and $c$ according to the algorithm structure, and finding the average running time $t$ of the local search procedure.  Having all these values, we can calculate the near-optimal population size $m_\text{opt}$ according to (\ref{eq:mo}) and run the algorithm.

This section discusses the first stage of our approach, i.e., tuning the constants $a$, $b$ and $c$.  The next section discusses finding the value $t$.

The constants $a$, $b$ and $c$ in (\ref{eq:mo}) should be selected to minimize the solution error for all combinations of local searches $\lambda$, instances $\phi$ and given times $\tau$ which are of interest.  In practice this means that one should select a representative instance set $\Phi$, assign the most appropriate local search $\lambda = \lambda(\phi)$ for every instance $\phi \in \Phi$ and define several given times $\tau \in T$ which will be used in practice.  Note that if $|T| = 1$, i.e., only one given time is required, then the number of constants in (\ref{eq:mo}) can be reduced: $m_\text{opt}(t) = a / t^c$.

Let $A_\text{MA}(m, \lambda, \phi, \tau)$ be a solution obtained by the memetic algorithm for the population size $m$, local search $\lambda$, instance $\phi$ and given time $\tau$.  
Let $w(A)$ be the objective value of a solution $A$.  

We need some measure of the memetic algorithm quality which reflects on the success of choosing a particular population size.  This measure should not depend on the rest of the algorithm parameters, i.e., it should have similar values for all the solutions obtained for the best chosen population sizes whatever is the instance, the local search or the given time.  Clearly one cannot use the relative solution error since its value hugely depends on the given time and other factors.  

We propose using \emph{scaled\footnote{Sometimes in the literature it is also called \emph{differential}.} solution errors} as follows.  Let $w_\text{min}(\lambda, \phi, \tau)$ and $w_\text{max}(\lambda, \phi, \tau)$ be the minimum and the maximum objective values obtained for the given $\lambda$, $\phi$ and $\tau$:
$$
w_\text{min}(\lambda, \phi, \tau) = \min_m w(A_\text{MA}(m, \lambda, \phi, \tau)) 
\text{\quad and}
$$
$$
w_\text{max}(\lambda, \phi, \tau) = \max_m w(A_\text{MA}(m, \lambda, \phi, \tau)) \ . \quad
$$
Then the scaled error $\epsilon(m, \lambda, \phi, \tau)$ of the solution $A_\text{MA}(m, \lambda, \phi, \tau)$ is calculated as follows:
$$
\epsilon(m, \lambda, \phi, \tau) = \frac{w(A_\text{MA}(m, \lambda, \phi, \tau)) - w_\text{min}(\lambda, \phi, \tau)}{w_\text{max}(\lambda, \phi, \tau) - w_\text{min}(\lambda, \phi, \tau)} \cdot 100\% \ .
$$
In other words, the scaled solution error shows the position of the solution obtained for the given population size between the solutions obtained for the best and for the worst values of $m$.  The scaled solution error is varied in $[0\%, 100\%]$; the smaller $\epsilon$, the better the solution.  Note that this scaled error has some useful theoretical properties \citep{Zemel1981}.

Since all the scaled solution errors have comparable values, we can use the average for every combination of $\tau \in T$ and $\phi \in \Phi$ as an indicator of $m_\text{opt}$ function success:
\begin{equation}
\label{eq:gamma}
\gamma = \overline{\epsilon\big(m_\text{opt}(\tau, t(\lambda, \phi)), \lambda, \phi, \tau\big)} \ .
\end{equation}
(Note that we use $t(\lambda, \phi)$ because the average local search running time $t$ depends on the local search procedure $\lambda$ and the instance $\phi$; recall $\lambda = \lambda(\phi)$.)  Obviously, $0\% \le \gamma \le 100\%$, and the smaller $\gamma$, the better $m_\text{opt}$.

The number of runs of the memetic algorithm required to find the best values of $a$, $b$ and $c$ can be huge\footnote{Note that since memetic algorithms are stochastic, one should run every experiment several times in order to get a better precision.} which makes the approach proposed in this research unaffordable.  For the purpose of decreasing the computation time we suggest the following dynamic programming technique.

\begin{enumerate}
	\item Let $\Phi$ be the test bed and $T$ be the set of the given times we are going to use for our algorithm.
	\item For every instance $\phi \in \Phi$ set the most appropriate local search $\lambda = \lambda(\phi)$.
	\item Let $M$ be the set of reasonable population sizes.  One can even reduce it by removing, e.g., all odd values from $M$, or leaving only certain values, e.g., $M = \{ 2, 4, 8, 16, \ldots \}$.
	\item Calculate and save $e(m, \lambda(\phi), \phi, \tau)$ for every $m \in M$, $\phi \in \Phi$ and $\tau \in T$.  
	\item Measure and save $t(\lambda(\phi), \phi)$ for every $\phi$.  For this purpose run the local search $\lambda(\phi)$ after a construction heuristic.
	\item Now for every combination of $a$, $b$ and $c$ compute $\gamma$ according to (\ref{eq:gamma}); every time the relative solution error $e(m, \lambda(\phi), \phi, \tau)$ is required, find $m' \in M$ which is the closest one to $m$ and use the corresponding precalculated value.  The discretization of $a$, $b$ and $c$ should be chosen according to available resources.
	\item Fix the combination of $a$, $b$ and $c$ which minimizes $\gamma$.  This finishes the tuning process.
\end{enumerate}

\subsection{\texorpdfstring{Finding Local Search Average Running Time $t$}{Finding Local Search Average Running Time t}}
\label{sec:ls_time}

In order to calculate the near-optimal population size $m_\text{opt}$ according to (\ref{eq:mo}), we need to find $t$ at the beginning of the memetic algorithm run.  Recall that the value $t$ is the average running time of the local search procedure applied to some solutions of the given instance.  Definitely this value significantly depends on the particular solutions.  However, the solutions in a memetic algorithm are permanently perturbed and, thus, they are always moved out from the local minima before the local search is applied.  This guaranties some uniformity in the improvement process during the whole algorithm.  Hence, we are able to measure the time $t$ at any point.

Our algorithm produces and immediately improves the solutions for the first generation until $m_1 \le m_\text{opt}(\tau, t_\text{cur} / m_1)$, where $m_1$ is the number of already produced solutions, $\tau$ is the time given to the whole memetic algorithm, $t_\text{cur}$ is the time already spent to generate solutions for the first generation and $m_\text{opt}(\tau, t)$ is the population size calculated according to (\ref{eq:mo}).  When the first generation is produced, the size of the population for all further generations is set to $m = m_\text{opt}(\tau, t_\text{cur} / m_1)$.

\section{Other Details of MAP Memetic Algorithm}
\label{sec:map_ma}

As a case study for the population sizing proposed in Section~\ref{sec:general} we decided to use the Multidimensional Assignment Problem.

\subsection{Main Algorithm Scheme}
\label{sec:map_ma_main_scheme}

While the general scheme of a typical memetic algorithm (see Figure~\ref{fig:ma}) is quite common for all memetic algorithms, the set of genetic operators and the way they are applied can vary significantly.  In this research we use quite a typical (see, e.g., \cite{Krasnogor2008}) procedure to obtain the next generation:
\begin{equation}
\label{eq:next_generation}
g^{i+1} = \selection \Big( \{ g_1^i \} \cup \mutation \big( g^i \setminus \{ g_1^i \} \big) \cup \crossover \big( g^i \big) \Big) \ ,
\end{equation}
where $g^k$ is the $k$th generation and $g_1^k$ is the best assignment in the $k$th generation.  For a set of assignments $G$ the function $\selection(G)$ simply returns $m_{i+1}$ best distinct assignments among them, where $m_k$ is the size of the $k$th generation (if the number of distinct assignments in $G$ is less than $m_{i+1}$, $\selection$ returns all the distinct assignments and updates the value of $m_{i+1}$ accordingly).  Note that the assignment $g_1^i$ avoids the $\mutation$ thus preserving the currently best result.  The function $\mutation(G)$ is defined as follows:
\begin{equation}
\label{eq:mutation}
\mutation(G) = \bigcup_{g \in G} 
	\left\{
	\begin{array}{ll}
		\LocalSearch(\perturb(g, \mu_m))	& \text{if } r < p_m \\
		g						& \text{otherwise}
	\end{array}
	\right.
\end{equation}
where $r \in [0, 1]$ is chosen randomly every time and the constants $p_m = 0.5$ and $\mu_m = 0.1$ define the probability and the strength of mutation operator respectively.  The function $crossover(G)$ is calculated as follows:
\begin{equation}
\label{eq:crossover}
\crossover(G) = \bigcup_{j = 1}^{(l \cdot m_{i+1} - m_i) / 2} \LocalSearch(\crossover(u_j, v_j))
\end{equation}
where $u_j$ and $v_j$ are assignments from $G$ randomly selected for every $j = 1, 2, \ldots, (l \cdot m_{i+1} - m_i) / 2$ and $l = 3$ defines ratio between the produced and selected for the next generation solutions.  The functions $\crossover(x, y)$, $\perturb(x, \mu)$ and $\LocalSearch(x)$ are discussed below.

\subsection{Coding}
\label{sec:coding}

Coding is a way of representing a solution as a sequence of atom values such as boolean values or numbers; genetic operators are applied to such sequences.  Good coding should meet the following requirements:
\begin{itemize}
\item Coding $\code(x)$ should be invertible, i.e., there should exist a decoding procedure $\decode$ such that $\decode(\code(x)) = x$ for any feasible solution $x$.

\item Evaluation of the quality (fitness function) of a coded solution should be fast.

\item Every fragment of the coded solution should refer to just a part of the whole solution, so that a small change in the coded sequence should not change the whole solution.

\item It should be relatively easy to design algorithms for random modification of a solution (mutation) and for combination of two solutions (crossover) which produce feasible solutions.
\end{itemize}

Huang and Lin~\cite{Huang2006} use a local search procedure which, given first two dimensions of an assignment, determines the third dimension (recall that the algorithm in \cite{Huang2006} is designed only for 3-AP)\@.  Since the first dimension can always be fixed without any loss of generality (see Section~\ref{sec:map}), one needs to store only the second dimension of an assignment.  Unfortunately, this coding requires a specific local search and is robust for 3-AP only.  We use a different coding; a vector of an assignment is considered as an atom in our algorithm and, thus, a coded assignment is just a list of its vectors.  The vectors are always stored in the first coordinate ascending order, e.g., an assignment consisting of vectors $(2, 1, 1)$, $(4, 4, 2)$, $(3, 2, 3)$ and $(1, 3, 4)$ (see Fig.~\ref{fig:map_example}) would be represented as 
$$
(1, 3, 4), (2, 1, 1), (3, 2, 3), (4, 4, 2) \ .
$$

Two assignments are considered equal if they have equal codes.

\subsection{First Generation}
\label{sec:first_generation}

As it was shown is Chapter~\ref{sec:map_ls} (and we also confirmed it empirically by testing our memetic algorithm with construction heuristics described in Section~\ref{sec:map_construction_heuristics}), it is beneficial to start any MAP local search or metaheuristic from a Greedy construction heuristic.  Thus, we start from running Greedy (we use the improved implementation, see Section~\ref{sec:map_construction_heuristics_improved}) and then perturb it using our $perturb$ procedure (see Section~\ref{sec:perturb}) to obtain every item of the first generation:
$$
g_j^1 = \LocalSearch(\perturb(\mathit{greedy}, \mu_f)),
$$
where $\mathit{greedy}$ is an assignment constructed by Greedy and $\mu_f = 0.2$ is the perturbation strength coefficient.  Since $\perturb$ performs a random modification, it guarantees some diversity in the first generation.

The number of assignments to be produced for the first generation is discussed in Section~\ref{sec:ls_time}.

\subsection{Crossover}
\label{sec:crossover}

A typical crossover operator combines two solutions, parents, to produce two new solutions, children.  Crossover is the main genetic operator, i.e., it is the source of a genetic algorithm strength.  Due to the selection operator, solutions consisting of `successful' fragments are spread wider than others and that is why, if both parents have some similar fragments, these fragments are assumed to be `successful' and should be copied without any change to the children solutions.  Other parts of the solution can be randomly mixed and modified though they should not be totally destroyed.

The one-point crossover is the simplest example of a crossover; it produces two children $x'$ and $y'$ from two parents $x$ and $y$ as follows: $x'_i = x_i$ and $y'_i = y_i$ for every $i = 1, 2, \ldots, k$, and $x'_i = y_i$ and $y'_i = x_i$ for every $i = k + 1, k + 2, \ldots, n$, where $k \in \{ 1, 2, \ldots, n - 1 \}$ is chosen randomly.  One can see that if $x_i = y_i$ for some $i$, then the corresponding values in the children sequences will be preserved: $x'_i = y'_i = x_i = y_i$.

However, the one-point and some other standard crossovers do not preserve feasibility of MAP assignments since not every sequence of vectors can be decoded into a feasible assignment.  We propose a special crossover operator.  Let $x$ and $y$ be the parent assignments and $x'$ and $y'$ be the child assignments.  First, we retrieve equal vectors in the parent assignments and initialize both children with this set of vectors:
$$
x' = y' = x \cap y \,.
$$
Let $k = |x \cap y|$, i.e., the number of equal vectors in the parent assignments, $p = x \setminus x'$ and $q = y \setminus y'$, where $p$ and $q$ are ordered sets.  Let $\pi$ and $\omega$ be random permutations of size $n - k$.  Let $r$ be an ordered set of random values uniformly distributed in $[0, 1]$.  For every $j = 1, 2, \ldots, n - k$ the crossover sets
$$
x' = x' \cup \left\{ \begin{array}{ll}
p_{\pi(j)} & \text{if } r_j < 0.8 \\
q_{\omega(j)} & \text{otherwise}
\end{array} \right. \mbox{\quad and \quad}
y' = y' \cup \left\{ \begin{array}{ll}
q_{\omega(j)} & \text{if } r_j < 0.8 \\
p_{\pi(j)} & \text{otherwise}
\end{array} \right. \,.
$$

Since this procedure can yield infeasible assignments, it requires additional correction of the child solutions.  For this purpose, the following is performed for every dimension $d = 1, 2, \ldots, s$ and for every child assignment $c$.  For every $i$ such that $\exists j < i: c_d^j = c_d^i$ set $c_d^i = r$ where $r \in \{ 1, 2, \ldots, n \} \setminus \{ c_d^1, c_d^2, \ldots, c_d^n \}$ is chosen randomly.  In the end of the correction procedure, sort the assignment vectors in the ascending order of the first coordinates (see Section~\ref{sec:coding}).

In other words, our crossover copies all equal vectors from the parent assignments to the child ones.  Then it copies the rest of the vectors; every time it chooses randomly a pair of vectors, one from the first parent and one from the second one.  Then it adds this pair of vectors either to the first and to the second child respectively (probability 80\%) or to the second and to the first child respectively (probability 20\%).  Since the obtained child assignments can be infeasible, the crossover corrects each one; for every dimension of every child it replaces all duplicate coordinates with randomly chosen correct ones, i.e., with the coordinates which are not currently used for that dimension.

Note that (\ref{eq:crossover}) requires $l \cdot m_{i+1} - m_i$ to be even.  If $m_{i+1} = m_i = m_o(\tau, t)$ then $l \cdot m_{i+1} - m_i$ is always even (recall that $l = 3$).  However, the size of the population is not guaranteed and, hence, $l \cdot m_{i+1} - m_i = (l - 1) \cdot m$ may take odd values.  To resolve this issue, we remove the worst solution from the $i$th generation if $l \cdot m_{i+1} - m_i$ appears to be odd.

We also tried the crossover operator used in \citep{Huang2006} but it appeared to be less efficient than the one proposed here.

\subsection{Perturbation Algorithm}
\label{sec:perturb}

The perturbation procedure $\perturb(x, \mu)$ is intended to modify randomly an assignment $x$, where the parameter $\mu$ defines how strong is the perturbation.  In our memetic algorithm, perturbation is used to produce the first generation and to mutate assignments from the previous generation when producing the next generation.  

Our perturbation procedure $\perturb(x, \mu)$ performs $\lceil n \mu / 2 \rceil$ random swaps.  In particular, each swap randomly selects two vectors and some dimension and then swaps the corresponding coordinates: swap $x_u^d$ and $x_v^d$, where $u, v \in \{ 1, 2, \ldots, n \}$ and $d \in \{ 1, 2, \ldots, s \}$ are chosen randomly; repeat the procedure $\lceil n \mu / 2 \rceil$ times.  For example, if $\mu = 1$, our perturbation procedure modifies up to $n$ vectors in the given assignment.

\subsection{Local Search Procedure}
\label{sec:local_search}

An extensive study of a number of local search heuristics for MAP is presented in Chapter~\ref{sec:map_ls}; it includes both fast and slow but powerful algorithms.  It also shows that a combination of two heuristics can yield a heuristic superior to the original ones.

Here we omit the results for \Threeopt{} and \Vopt{}.  The experiments show that our memetic algorithm is unsuccessful with these local searches.  Note that similar conclusions were indicated in Chapter~\ref{sec:map_ls}.

\bigskip

Recall that we proposed a division of instances into two groups: instances with independent weights and instances with decomposable weights.  The weight matrices of the instances with independent weights have no structure, i.e., there is no correlation between weights $w(u)$ and $w(v)$ even if the vectors $u$ and $v$ are different in only one coordinate.  In contrast, the weights of the instances with decomposable weights are defined using the graph formulation of MAP (see Section~\ref{sec:map}) and have the structure of (\ref{eq:decomposable_w}).  Most of the instances which have some practical interest and which do not belong to the group of independent weight instances can be represented as instances with decomposable weights, see, e.g., \clique{} and \squareroot{} instance families in Section~\ref{sec:map_testbed}.

It is known that even for a fixed optimization problem there is no local search procedure which would be the best choice for all types of instances \citep{Krasnogor2001,Krasnogor2005}.  Splitting all the MAP instances into two groups, namely instances with independent and decomposable weights, gives us a formal way to use appropriate local searches for every instance.  In particular, it was shown that the instances with independent weights are better solvable by \MDVV{} while the dimensionwise heuristics are the best choice for the instances with decomposable weights.

Table~\ref{tab:ls_comparison} presents a comparison of the results of our memetic algorithm based on the local search procedures discussed above.  The time given for every run of the algorithm is 3 seconds.  The table reports the relative solution error for every instance and every considered algorithm.  The column `best' shows the best known solution for each instance.  

One can see that the outcomes of Chapter~\ref{sec:map_ls} are repeated here, i.e., for the \random{} instances (see Section~\ref{sec:map_testbed}) \MDVV{} provides clearly the best performance; for the instances with decomposable weights, i.e., for the \clique{} and \squareroot{} instances, the fast heuristics \OneDV{}, \TwoDV{}, \MDV{}, \OneDVtwo{} and \TwoDVtwo{} perform better than others in almost every experiment, and \MDV{} shows the best average result among them (though in Table~\ref{tab:ls_comparison} \TwoDV{} slightly outperforms it, for other given times \MDV{} shows the best results).  

Thereby, in what follows we use \MDVV{} as a local search for the instances with independent weights and \MDV{} for the instances with decomposable weights.

\subsection{Population Size Adjustment}

The constants $a$, $b$ and $c$ were selected to minimize $\gamma$ (see Section~\ref{sec:populationsize}); as an instance set $\Phi$ we used the full test bed (see Section~\ref{sec:map_testbed}), the given times were $T = \{ 1~\text{s}, 3~\text{s}, 10~\text{s}, 30~\text{s}, 100~\text{s}\}$, the generation sizes were $M = \{ 2, 3, 5, 8, 12, 18, 27, 40, 60, 90, 135 \}$ and local search $\lambda(\phi)$ was selected according to Section~\ref{sec:local_search}.  The best value of $\gamma = 13\%$ was obtained for $a = 0.08$, $b = 0.35$ and $c = 0.85$ (see (\ref{eq:mo})).  Note that these values are not a compromize and present a minimum for every separate instance set and given time.  Observe also that fixing $m$ to some value leads to $\gamma > 19\%$ for the same set of instances, local searches and given times.

Slight variations of the constants $a$, $b$ and $c$ do not influence the performance of the algorithm significantly.  Moreover, there exist some other values for these parameters which also yield good results.  The values of the constants should not be adjusted for every computational platform.

\section{Experimental Evaluation of MAP Memetic Algorithm}

Three metaheuristics were compared in our experiments:
\begin{itemize}
	\item An extended version of the memetic algorithm in \cite{Huang2006} (\HL{}).
	\nomenclature[HL]{\HL}{is a 3-AP memetic algorithm by Huang and Lim}

	\item An extended version of the simulated annealing algorithm in \cite{Clemons2004} (\SA{}).
	\nomenclature[SA]{\SA}{is a 3-AP Simulated Annealing algorithm by Clemons et al.}

	\item Our memetic algorithm (\GK{}).
	\nomenclature[GK]{\GK{} (MAP)}{is our memetic algorithm}
\end{itemize}

All the heuristics are implemented in Visual C++ and evaluated on a platform based on AMD Athlon 64 X2 3.0~GHz processor.  Our implementations as well as the test bed generator and the best known assignments are available on the web \citep{Karapetyan}.

\subsection{HL Heuristic}

For the purpose of comparison, the Huang and Lim's memetic algorithm was extended as follows:
\begin{itemize}
\item The coded assignment contains not only the second dimension but it stores sequentially all the dimensions except the first and the last ones, i.e., an assignment $\{ e^1, e^2, \ldots, e^s \}$ is represented as $e^1_2$, $e^2_2$, \ldots, $e^n_2$, $e^1_3$, $e^2_3$, \ldots, $e^n_3$, \ldots, $e^1_{s-1}$, $e^2_{s-1}$, \ldots, $e^n_{s-1}$ ($e_1^i = i$ for each $i$ and $e_s^i$ can be chosen in an optimal way by solving an AP, see Section~\ref{sec:coding}).

\item The local search heuristic, that was initially designed for 3-AP, is extended to \OneDV{} as described in Section~\ref{sec:map_dv}.

\item The crossover, proposed in \citep{Huang2006}, is applied separately to every dimension (except the first and the last ones) since it was designed for one dimension only (recall that the memetic algorithm from \citep{Huang2006} stores only the second dimension of an assignment, see Section~\ref{sec:coding}).

\item The termination criterion is replaced with a time check; the algorithm terminates when the given time is elapsed.
\end{itemize}

Our computational experience show that the solution quality of our implementation of the Huang and Lim's heuristic is similar to the results reported in \citep{Huang2006} and the running time is reasonably larger because of the extension for $s > 3$.

\subsection{SA Heuristic}

The Simulated Annealing heuristic in \cite{Clemons2004} was originally proposed for an arbitrary number of dimensions.  We reimplemented it and our computational experience show that both the solution quality and the running times\footnote{In our experiments, the running times of the heuristic were always approximately 20 times smaller than the results reported in \citep{Clemons2004} which can be explained by a difference in the computational platforms.} of our implementation of the Simulated Annealing heuristic are similar to the results reported in \citep{Clemons2004}.

For the purpose of comparison to other heuristics we needed to fit \SA{} for using a predefined running time.  We tried two strategies:
\begin{itemize}
\item An adaptive cooling ratio $R$ (see \cite{Clemons2004}).  The value $R$ is updated before each change of the temperature as follows:
$$
R = \sqrt[m]{\frac{0.1}{T}} \text{\quad and\quad} f = (\tau - t_\text{e}) \cdot \frac{i}{t_\text{e}} \ ,
$$
where $T$ is the current temperature (see \cite{Clemons2004}), $t_\text{e}$ is the elapsed time, $\tau$ is the given time and $f$ is the expected number of further iterations which is calculated according to the number $i$ of already finished iterations.

\item An adaptive number of local search iterations $\mathit{NUM}_\text{max}$ (see \cite{Clemons2004}).  The value $\mathit{NUM}_\text{max}$ is updated before each change of the temperature as follows:
$$
\mathit{NUM}_\text{max} = (\tau - t_\text{e}) \cdot \frac{c}{t_\text{e}} \cdot \frac{1}{I - i} \ ,
$$
where $t_\text{e}$ is the elapsed time, $\tau$ is the given time, $c$ is the total number of local search iterations already performed, $i$ is the number of the algorithm iterations already performed and $I$ is the number of algorithm iterations to be performed.  Since the cooling ratio $R$ as well as the initial and the final temperatures $T_\text{start}$ and $T_\text{final}$ are fixed, the number $I$ of iterations of the algorithm is also fixed:
$$
I = \log_R \frac{T_\text{final}}{T_\text{initial}} \ .
$$
\end{itemize}

In both adaptations, the algorithm terminates if the given time is elapsed: $t \ge \tau$.

Both adaptations yielded competitive algorithms though according to our experimental evaluation the second adaption which varies the number of local search iterations appears to be more efficient.  One can assume that the best adaptation should vary both the cooling ratio and the number of local search iterations but this is a subject for another research.  Hence, in what follows the \SA{} algorithm refers to the extension with the adaptive number of local search iterations.

\subsection{Experiment Results}

The main results are reported in Tables~\ref{tab:metaheuristics1} and~\ref{tab:metaheuristics2}; in these tables, we compare our algorithm (\GK) to the Simulated Annealing heuristic (\SA{}) and the memetic algorithm by Huang and Lim (\HL{})\@.  The comparison is performed for the following given times $\tau$: 0.3 s, 1 s, 3 s, 10 s, 30 s, 100 s and 300 s.  Every entry of these tables contains the relative solution error averaged for 10 instances of some fixed type and size but of different seed values (see Section~\ref{sec:map_testbed} for details); we did not repeat every experiment several times which is typical for stochastic algorithms.  The value of the relative solution error $e(A)$ is calculated as follows
\begin{equation}
\label{eq:solution_error}
e(A) = \left(\frac{w(A)}{w(A_\text{best})} - 1\right) \cdot 100\% \ .
\end{equation}
where $A$ is the obtained solution and $A_\text{best}$ is the best known solution\footnote{The best known solutions were obtained during our experiments with different heuristics and the corresponding weights can be found in Table~\ref{tab:ls_comparison}.  For the \random{} instances we actually know the optimal objective values; it is proven for large values of $n$ that a \random{} instance has a solution of the minimal possible weight (see Section~\ref{sec:map_random_estimation}); since we obtained the minimal possible solutions for every \random{} instance in our experiments, we can extend the results of Section~\ref{sec:map_random_estimation} to all the \random{} instances in our test bed.  Hence, we do not need the \GP{} instances anymore.}.  

The results for \product{} and \geometric{} instances were excluded from Tables~\ref{tab:ls_comparison}, \ref{tab:metaheuristics1} and~\ref{tab:metaheuristics2} because even the stand alone local searches used in our memetic algorithm are able to solve \geometric{} instances to optimality and \product{} instances to less than 0.04\% over optimality\footnote{We believe that the best known solutions for both \geometric{} and \product{} instances are optimal but we are not able to verify it.}.  Similar result were reported in Chapter~\ref{sec:map_ls}.

The average values for different instance families, numbers of dimensions and instance sizes are provided at the bottom of each table.  The best among \HL{}, \SA{} and \GK{} results are underlined in every row for every particular given time.

One can see that \GK{} clearly outperforms both \SA{} and \HL{} for all the given times.  Moreover, \GK{} is not worse than the other heuristics in every experiment which proves its flexibility and robustness.  A two-sided paired $t$-test confirms statistical difference even between \GK{} with $\tau = 1$ s and \HL{} with $\tau = 100$ s because the $p$-value in this case was less than $0.0001$ for both instances with independent and decomposable weights.  This shows that \HL{} is not able to use large time efficiently.

The solution quality of \GK{} significantly depends on the given time: for the instances with both independent and decomposable weights a three times increase of the running time improves the solution quality in approximately 1.2 to 2 times.  Recall that the approach proposed in this research to select the most appropriate population size reduces $\gamma$ more than 1.5 times (see Section~\ref{sec:populationsize}) and, hence, it would take roughly 1.5 to 10 times more time to get the same solution quality for a memetic algorithm with a fixed population size\footnote{Note that $\gamma$ is not just the average for the solution errors and, thus, these calculations are very approximate.}.

It is worth noting that we experimented with different values of the \GK{} algorithm parameters such as $\mu_f$, $\mu_m$, $p_m$, $l$, etc.\ and concluded that small variations of these values do not significantly influence the algorithm performance.

For the instances with independent weights all the algorithms perform better for the large instances rather than for the small ones.  One can explain it by showing that the number of vectors of the minimal weight in \random{} is proportional to $n^s$ while the number of vectors in an assignment is $n$ and, thus, the number of global minima increases with the increase of $n$ (see Section~\ref{sec:map_random_estimation}.  In contrast, the instances with decomposable weights become harder with the growth of $n$.

Since the \HL{} heuristic uses \OneDV{} local search, it performs quite well for the instances with decomposable weights and yields solutions of poor quality for the instances with independent weights.  Due to the fixed population size, it does not manage to solve some large instances in short times which results in huge solution errors reported in Table~\ref{tab:metaheuristics1} for the instances \heuristic{3cq70}, \heuristic{3sr70}, \heuristic{3cq100} and \heuristic{3sr100}.  \HL{} was initially designed for 3-AP and tested on small instances \citep{Huang2006} and, hence, it performs better for the instances with small $s$ and $n$.

The \SA{} heuristic is less successful than the others; for both instances with independent and decomposable weights it is worse than both \HL{} and \GK{} in almost every experiment.  The solution quality of \SA{} improves quite slowly with the increase of the running time; it seems that \SA{} would not be able to significantly improve the solution quality even if it is given much larger time.

\section{Conclusion}

In this chapter, we proposed two memetic algorithms, namely for the Generalized Traveling Salesman Problem and the Multidimensional Assignment Problem.  The first algorithm is featured with a powerful local search procedure, variable population size, well-fitted genetic operators and new efficient termination condition.  Unlike other heuristics in the literature, our algorithm is able to process both symmetric and asymmetric GTSP instances.  Experimental analysis shows that the proposed heuristic dominates all other evolutionary algorithm for GTSP known from the literature.  Moreover, it was able to significantly improve the best known solutions for a number of standard instances.

Based on our experience in GTSP algorithm design, we have developed the idea of the variable population size and replaced the termination condition with a predefined running time.  In this case the goal was to adjust the population size such that the given time would be used with the maximum efficiency.  In our approach, this is achieved by adjusting the population size according to the given time, the local search procedure and the problem instance.  As a case study problem, we used MAP.  

Our experiments have confirmed that the proposed population sizing leads to an outstanding flexibility of the algorithm.  Indeed, it was able to perform efficiently for a wide range of instances, being given from 0.3 to 300 seconds of the running time and with totally different local search procedures.  As an evidence of its efficiency, we compared it to two other MAP metaheuristics proposed in the literature and concluded that our algorithm clearly outperforms the other heuristics with no exception.  Moreover, the difference in the solution quality of our memetic algorithm (\GK{}) and the previous state-of-the-art memetic algorithm (\HL{}) continuously grows with the increase of the given time which confirms that \GK{} is much more flexible than \HL{}.

The main factors influencing the performance of a memetic algorithm are running time, computational platform, problem instance, local search procedure, population size and genetic operators.  We did not focus on the genetic operators investigation in this research; however we believe that the operators used in our MAP MA are well fitted since our attempts to improve the algorithm results by changing the operators have failed.  The local search procedure and the population size are varied according to the problem instance; after an extensive study of the local searches, we show that there are two totally different cases of MAP, and for these cases one should use different local search procedures.  Since these local searches have very different running times, the memetic algorithm has to adapt for them.  In our approach this is achieved by using the adjustable population size which is a function of the average running time of the local search.  Thereby, the average running time of the local search encapsulates not only the local search specifics but also the specifics of the instance and the computational platform performance.  Since the algorithm is self-adjustable, the running time can be used as a parameter responsible for the `solution quality'/`running time' balance and, thus, the population size should also depend on the given time.

The adjustable population size requires several constants to be tuned prior to using the algorithm; we proposed a procedure to find the optimal values of these constants.

In conclusion we note that choosing the most appropriate population size is crucial for the performance of a memetic algorithm.  Our approach to calculate the population size according to the average running time of the local search and the time given to the whole algorithm, used to perform well for a large variation of the instances and given times and for two totally different local searches.  Observe, however, that the whole discussion of the population sizing does not involve any MAP specifics and, hence, we can conclude that the obtained results can be extended to any hard optimization problem.

Further research is required to evaluate the proposed approach in application to other hard combinatorial optimization problems.  It is also an interesting question if changing the population size during the algorithm's run can further improve the results.

\chapter{Conclusion}
\label{conclusion}

\markright{\thechapter.\ Conclusion}

In this research we proposed a number of algorithms and approaches for combinatorial optimization problems.  We focused on two combinatorial optimization problems, namely, the Generalized Traveling Salesman Problem and the Multidimensional Assignment Problem, and significantly developed the knowledge on these problems.  Both GTSP and MAP have a lot of important applications but were not studied enough in the literature.

\bigskip

One of the most important questions in heuristic design is experimental evaluation.  In order to conduct a proper experiment, one needs a representative test bed.  In Chapter~\ref{sec:heuristics_design} we provide an example of successful adaptation of a well-developed TSP test bed for GTSP\@.  However, this approach seems to be inapplicable for MAP and, thus, we generalize the existing instance families, propose some new ones and introduce a classification; our division of the MAP instances into the instances with independent weights and the instances with decomposable weights turns out to be essential in our further research.

It is often important to know the optimal objective value for a test instance.  However, it may be impossible to solve a large instance to optimality in any reasonable time.  In this work, we propose a probabilistic analysis in application to one of the most used MAP instance family, namely the Random instance family.  For a large enough instance, it is possible to estimate the optimal objective value with a very high precision.  This approach may also be applied to the randomly generated instances for some other problems.

Another technique which is applicable to a wide range of problems is preprocessing.  In our example, preprocessing is applied to GTSP\@.  Observe that the shortest cycle for a GTSP instance visits only $m$ vertices and, hence, we may remove up to $n - m$ vertices from the instance in advance and this does not influence the optimal solution.  Moreover, the problem remains a typical GTSP but of a smaller size and, hence, any ordinary solvers may be applied to the reduced instances.  Our computational experiments showed that this reduction may significantly speed up slow GTSP solvers.

While many researchers focus only on the theoretical properties of the algorithms, the implementation issues may also be crucial for the algorithm's performance.  We considered four construction heuristics for MAP and proposed some simple transformations for each of them in order to optimize the implementations with regards to the computer architecture.  The improved implementations appear to be several times faster than the original ones.

We also discussed data structures for GTSP and proposed some new approaches which are easier and faster for implementation of algorithms.

\bigskip

In Chapter~\ref{sec:gtsp_ls} we thoroughly discussed the local search for GTSP\@.  Unfortunately, previously there was no stand-alone research in this area and, thus, we collected all the existing approaches, classified, extended and improved them and also proposed some new algorithms.

One of the most important GTSP local searches is Cluster Optimization.  It finds the best vertex selection for a fixed cluster order and it takes only $O(s \gamma n)$ time.  The algorithm was widely discussed in the literature, however, we proposed two new refinements which can significantly speed it up in certain circumstances.

Another class of GTSP local searches consists of adaptations of TSP local searches.  We provided some theoretical discussion of possible adaptations and proposed a unified framework for this approach.  Moreover, we significantly improved the existing results for the most powerful `Global' adaptation.  Apart from being powerful, this adaptation has some nice theoretical properties but previously it was too slow.  Our approach makes the `Global' adaptation applicable in practice.

One of the most successful TSP local searches is the Lin-Kernighan heuristic.  We proposed several adaptations of this algorithm for GTSP\@.  Note that it is not a typical local search and, thus, it cannot be adapted straightforwardly.  Moreover, the original heuristic is very complicated for understanding and implementation and so we propose a new explanation of a simplified Lin-Kernighan heuristic.  We claim that our simplified version preserves the main features of the algorithm.  This statement is supported by the success of our Lin-Kernighan adaptations.

We also proposed a new class of GTSP local searches, namely Fragment Optimization.  These local searches, being quite natural for GTSP, are relatively slow when implemented naively.  We proposed two algorithms to explore the corresponding neighborhoods efficiently.

\bigskip

Chapter~\ref{sec:map_ls} is devoted to MAP local search.  Similarly to the GTSP, there was no stand-alone research in this area before.  We collected all existing MAP local searches, extended them when necessary and proposed some new ones.  We also introduced a division of all the MAP local searches into two classes: dimensionwise and vectorwise.  Having this classification, we proposed to combine the local searches of different types together.  After a thorough experimental evaluation we selected the most efficient heuristics, and the combined ones turn out to be very successful.

\bigskip

A significant part of our research is devoted to memetic algorithms.  This type of evolutionary algorithms intensively applies local search and, thus, is essentially interesting in this work.

Our memetic algorithm for GTSP is featured by a number of new approaches like an efficient termination condition or a variable population size which is selected according to the problem size and varies during the algorithm's run.  It is worth noting that the local search procedure in our memetic algorithm is a combination of several local searches and, thus, it is relatively powerful.  All these and some other features of the algorithm make it extremely successful and, in particular, it dominates all the existing GTSP metaheuristics known from the literature.

When designing a memetic algorithm for MAP, we have further improved our population sizing.  In our approach, the population size as one of the most important parameters of a memetic algorithm is intended to adjust the whole algorithm according to the particular instance, local search procedure and time requirements.  Being applied to MAP, this approach yielded an extremely flexible memetic algorithm capable to work efficiently in a wide range of these parameters.  We compared our memetic algorithm with two other MAP metaheuristics known from the literature and concluded that our algorithm, being given the same time, significantly outperforms the other heuristics.

\bigskip

Our research leaves some questions open.  In particular, one can doubt that the proposed simplified Lin-Kernighan heuristic preserves the main features of the original algorithm and, thus, a formal or experimental proof is of interest.  It is also interesting to consider combinations of different GTSP neighborhoods in the so-called Variable Neighborhood Descend.  According to our expectations, this may yield several very successful heuristics.  In order to make our experimental evaluation more representative, it may be beneficial to consider some additional instance families and to test the existing approaches on significantly larger instances.  We are also going to apply our population sizing approach to the GTSP memetic algorithm in order to prove its efficiency on different optimization problems.

\clearpage
\phantomsection
\addcontentsline{toc}{chapter}{Glossary}
\renewcommand{\nomname}{Glossary}
\markboth{\nomname}{\nomname}
\printnomenclature[1em]

\clearpage
\phantomsection
\addcontentsline{toc}{chapter}{Bibliography}
\bibliographystyle{plain}
\bibliography{../gtsp}{}

\singlespacing
\setlength{\topmargin}{-0.25in}
\setlength{\textwidth}{6in}
\setlength{\textheight}{9in}
\setlength{\oddsidemargin}{0.5in}
\setlength{\headwidth}{1\textwidth}

\appendix
\chapter{Tables}
\label{sec:tables}

\markright{\thechapter.\ Tables}

\footnotesize


\enlargethispage{10\baselineskip}

\begin{table}[!ht] \centering
	\footnotesize
	\caption{Time benefit of the GTSP Reduction for the \ExactSolver{} heuristic.} 
	\label{tab:reduction_exact}
\begin{tabular}{l r c r r c r r c r r r }
\toprule{}
&
    && \multicolumn{2}{c}{Vertices Red.}
    && \multicolumn{2}{c}{Edge Red.}
    && \multicolumn{3}{c}{Combined Reduction} \\

\cmidrule{4-5}  \cmidrule{7-8}  \cmidrule{10-12}

Instance & $T_0$, sec
    &&  $R_v$, \% & $B$, \%
    &&  $R_e$, \% & $B$, \%
    &&  $R_v$, \% & $R_e$, \% & $B$, \% \\
\cmidrule{1-12}
5gr21     	&	0.8	&&	38.1	&	40	&&	48.7	&	52	&&	38.0	&	45.0	&	56	\\
5ulysses22	&	1.7	&&	50.0	&	60	&&	44.3	&	48	&&	50.0	&	39.5	&	79	\\
5gr24     	&	0.2	&&	54.2	&	74	&&	33.1	&	53	&&	54.1	&	10.4	&	81	\\
6fri26    	&	0.9	&&	50.0	&	67	&&	28.7	&	18	&&	50.0	&	20.3	&	74	\\
6bayg29	&	6.0	&&	41.4	&	19	&&	0.0	&	59	&&	41.3	&	33.6	&	70	\\
10gr48    	&	16.1	&&	37.5	&	57	&&	27.0	&	2	&&	37.5	&	25.5	&	55	\\
10hk48    	&	52.7	&&	12.5	&	16	&&	34.2	&	6	&&	12.5	&	32.3	&	22	\\
11eil51	&	32.8	&&	17.6	&	37	&&	32.6	&	17	&&	17.6	&	28.8	&	42	\\
14st70 	&	150.4	&&	17.1	&	43	&&	36.5	&	17	&&	17.1	&	24.6	&	50	\\
\cmidrule{1-12}																	
Average	&	&&	35.4	&	45.9	&&	31.7	&	30.2	&&	35.3	&	28.9	&	58.8	\\
\bottomrule{}
\end{tabular}
\end{table}

\begin{table}[!ht] 
	\centering
	\footnotesize
	\caption{Time benefit of the GTSP Reduction for the \SD{} heuristic.}
	\label{tab:reduction_sd}
\begin{tabular}{l r c r r c r r c r r r }
\toprule{}
&
    && \multicolumn{2}{c}{Vertices Red.}
    && \multicolumn{2}{c}{Edge Red.}
    && \multicolumn{3}{c}{Combined Reduction} \\

\cmidrule{4-5}  \cmidrule{7-8}  \cmidrule{10-12}

Instance & $T_0$, sec
    &&  $R_v$, \% & $B$, \%
    &&  $R_e$, \% & $B$, \%
    &&  $R_v$, \% & $R_e$, \% & $B$, \% \\
\cmidrule{1-12}
157rat783 	&	23.6	&&	3.2	&	11	&&	9.8	&	5	&&	3.1	&	8.4	&	36	\\
200dsj1000	&	100.3	&&	0.8	&	47	&&	9.6	&	36	&&	0.8	&	9.4	&	42	\\
201pr1002	&	54.9	&&	1.9	&	12	&&	9.2	&	22	&&	1.9	&	8.7	&	43	\\
207si1032	&	21.3	&&	8.2	&	3	&&	12.1	&	-1	&&	8.2	&	10.2	&	24	\\
212u1060 	&	88.8	&&	3.3	&	8	&&	14.4	&	35	&&	3.3	&	11.2	&	42	\\
217vm1084 	&	78.1	&&	22.2	&	49	&&	24.0	&	-2	&&	22.2	&	8.9	&	57	\\
235pcb1173	&	107.9	&&	0.9	&	5	&&	8.2	&	30	&&	0.9	&	8.2	&	32	\\
259d1291	&	169.4	&&	3.7	&	9	&&	12.4	&	25	&&	3.7	&	9.8	&	26	\\
261rl1304 	&	140.4	&&	1.5	&	9	&&	7.9	&	47	&&	1.4	&	7.2	&	66	\\
265rl1323 	&	132.6	&&	1.8	&	20	&&	7.8	&	20	&&	1.7	&	7.0	&	32	\\
276nrw1379	&	111.5	&&	0.8	&	4	&&	7.4	&	22	&&	0.7	&	7.1	&	46	\\
\cmidrule{1-12}
Average	&		&&	4.4	&	16.1	&&	11.2	&	21.7	&&	4.4	&	8.7	&	40.5	\\
\bottomrule{}
\end{tabular}
\end{table}

\begin{table}[!ht] 
	\centering
	\footnotesize
	\caption{Time benefit of the GTSP Reduction for the \SG{} heuristic.}
	\label{tab:reduction_sg}
\begin{tabular}{l r c r r}
\toprule{}
&
    && \multicolumn{2}{c}{Vertices Red.}\\

\cmidrule{4-5}

Instance & $T_0$, sec
    &&  $R_v$, \% & $B$, \%\\
\cmidrule{1-5}
84fl417.gtsp	&	4.5	&&	10.3	&	12	\\
87gr431.gtsp	&	8.3	&&	0.0	&	6	\\
88pr439.gtsp	&	10.2	&&	2.3	&	-3	\\
89pcb442.gtsp	&	11.5	&&	5.4	&	0	\\
99d493.gtsp	&	20.0	&&	0.8	&	7	\\
107att532.gtsp	&	25.1	&&	3.9	&	11	\\
107si535.gtsp	&	16.9	&&	17.9	&	34	\\
107ali535.gtsp	&	29.1	&&	5.4	&	20	\\
113pa561.gtsp	&	14.5	&&	26.2	&	31	\\
\cmidrule{1-5}
Average	&		&&	8.0	&	13	\\
\bottomrule{}
\end{tabular}
\end{table}

\begin{table}[!ht] 
	\centering
	\footnotesize
	\caption{Time benefit of the GTSP Reduction for the \GK{} heuristic.}
	\label{tab:reduction_gk}
\begin{tabular}{l r c r r c r r c r r r }
\toprule{}
&
    && \multicolumn{2}{c}{Vertices Red.}
    && \multicolumn{2}{c}{Edge Red.}
    && \multicolumn{3}{c}{Combined Reduction} \\

\cmidrule{4-5}  \cmidrule{7-8}  \cmidrule{10-12}

Instance & $T_0$, sec
    &&  $R_v$, \% & $B$, \%
    &&  $R_e$, \% & $B$, \%
    &&  $R_v$, \% & $R_e$, \% & $B$, \% \\
\cmidrule{1-12}
89pcb442.gtsp	&	3.43	&&	5.4	&	16	&&	12.0	&	-2	&&	5.4	&	9.8	&	7	\\
99d493.gtsp	&	6.36	&&	0.8	&	2	&&	17.9	&	0	&&	0.8	&	19.4	&	2	\\
107att532.gtsp	&	5.96	&&	3.9	&	7	&&	20.6	&	10	&&	3.9	&	18.1	&	11	\\
107si535.gtsp	&	4.52	&&	17.9	&	14	&&	26.5	&	8	&&	17.9	&	18.0	&	15	\\
107ali535.gtsp	&	8.91	&&	5.4	&	17	&&	16.6	&	19	&&	5.4	&	14.3	&	25	\\
113pa561.gtsp	&	6.86	&&	26.2	&	20	&&	31.3	&	6	&&	26.2	&	22.6	&	23	\\
115u574.gtsp	&	7.43	&&	1.9	&	-2	&&	14.4	&	-6	&&	1.9	&	14.0	&	-1	\\
115rat575.gtsp	&	7.29	&&	3.1	&	0	&&	11.3	&	0	&&	3.1	&	10.9	&	2	\\
131p654.gtsp	&	5.47	&&	13.5	&	11	&&	32.7	&	2	&&	13.5	&	28.3	&	13	\\
\cmidrule{1-12}																	
	&		&&	8.7	&	9	&&	20.4	&	4	&&	8.7	&	17.3	&	11	\\
\bottomrule{}
\end{tabular}
\end{table}

\clearpage

\begin{longtable}{l c r r r c r r r c r r r }
	\caption{Test results of GTSP reduction algorithms.} 
	\label{tab:gtsp_reduction_results} \\
\toprule
    && \multicolumn{3}{c}{Vertex reduction}
    && \multicolumn{3}{c}{Edge reduction}
    && \multicolumn{3}{c}{Combined reduction} \\

\cmidrule{3-5}  \cmidrule{7-9}  \cmidrule{11-13}

Instance
    && $R_v$		& $R_v$, \%	& $T$
    && $R_e$, \%	& $R_v$		& $T$
    && $R_v$, \%,	& $R_e$, \%	& $T$ \\
\cmidrule{1-13}
\endfirsthead
\toprule
    && \multicolumn{3}{c}{Vertex reduction}
    && \multicolumn{3}{c}{Edge reduction}
    && \multicolumn{3}{c}{Combined reduction} \\

\cmidrule{3-5}  \cmidrule{7-9}  \cmidrule{11-13}

Instance
    && $R_v$		& $R_v$, \%	& $T$
    && $R_e$, \%	& $R_v$		& $T$
    && $R_v$, \%,	& $R_e$, \%	& $T$ \\
\cmidrule{1-13}
\endhead

\multicolumn{13}{l}{{Continued on next page\ldots}} \\
\bottomrule
\endfoot

\bottomrule
\endlastfoot

4ulysses16	&&	9	&	56.3	&	0.0	&&	62.0	&	4	&	0.0	&&	56.3	&	23.5	&	0.0	\\
4gr17     	&&	11	&	64.7	&	0.0	&&	35.8	&	3	&	0.0	&&	64.7	&	23.0	&	0.0	\\
5gr21     	&&	8	&	38.1	&	0.0	&&	48.7	&	3	&	0.0	&&	38.1	&	45.0	&	0.0	\\
5ulysses22	&&	11	&	50.0	&	0.0	&&	44.3	&	2	&	0.0	&&	50.0	&	39.5	&	0.0	\\
5gr24     	&&	13	&	54.2	&	0.0	&&	33.1	&	3	&	0.0	&&	54.2	&	10.4	&	0.0	\\
6fri26    	&&	13	&	50.0	&	0.0	&&	28.7	&	3	&	0.0	&&	50.0	&	20.3	&	0.0	\\
6bayg29   	&&	12	&	41.4	&	0.0	&&	37.9	&	5	&	0.0	&&	41.4	&	33.6	&	0.0	\\
9dantzig42	&&	6	&	14.3	&	0.0	&&	36.2	&	0	&	0.0	&&	14.3	&	24.9	&	0.0	\\
\end{longtable}


\begin{table}[!ht] \centering
	\footnotesize
	\caption[Experiments with the different variations of the \CO{} algorithm for GTSP\@.  $\gamma = 1$.]{Experiments with the different variations of the \CO{} algorithm for GTSP.  The test bed includes only the instances with $\gamma = 1$.}
	\label{tab:gtsp_co_variations_one}
\begin{tabular}{lrrcrrrr}
\toprule

&&&&\multicolumn{4}{c}{Running time, $\mu$s}\\
\cmidrule(r){5-8}
Instance&$\gamma$&$s$&&\CO{}$_1$&\CO{}$_2$&\CO{}$_3$&\CO{}$_4$\\
\cmidrule(){1-8}

12brazil58&1&16&&1.0&0.9&0.9&0.9\\
20kroa100&1&8&&1.6&1.6&1.6&1.6\\
26bier127&1&27&&2.8&2.7&2.9&2.7\\
32u159&1&16&&2.4&2.1&2.4&2.1\\
41gr202&1&17&&4.4&3.8&4.4&3.8\\
53pr264&1&12&&5.1&5.2&5.2&5.2\\
87gr431&1&58&&12.0&11.3&12.0&11.2\\
107att532&1&20&&12.2&11.5&12.1&11.5\\
131p654&1&25&&20.9&20.1&20.9&20.1\\
200dsj1000&1&19&&25.8&25.2&25.9&25.2\\
\cmidrule(){1-8}

Average&1.0&21.8&&8.8&8.4&8.8&8.4\\
\bottomrule

\end{tabular}
\end{table}

\begin{table}[!ht] \centering
	\footnotesize
	\caption[Experiments with the different variations of the \CO{} algorithm for GTSP\@.  $\gamma > 1$.]{Experiments with the different variations of the \CO{} algorithm for GTSP\@.  The test bed includes only the instances with $\gamma > 1$.}
	\label{tab:gtsp_co_variations_more}
\begin{tabular}{lrrcrrrr}
\toprule

&&&&\multicolumn{4}{c}{Running time, $\mu$s}\\
\cmidrule(r){5-8}
Instance&$\gamma$&$s$&&\CO{}$_1$&\CO{}$_2$&\CO{}$_3$&\CO{}$_4$\\
\cmidrule(){1-8}

10gr48&2&10&&1.1&1.1&0.8&0.8\\
11eil51&2&7&&1.6&1.4&0.9&0.8\\
20rat99&2&11&&3.0&2.7&1.9&1.8\\
20kroc100&2&13&&2.9&2.7&2.3&2.1\\
20krod100&2&9&&3.6&2.9&2.4&2.3\\
20rd100&2&8&&2.9&2.5&3.3&2.8\\
21lin105&2&12&&3.0&2.3&3.1&2.4\\
22pr107&3&7&&5.4&4.9&2.2&2.0\\
25pr124&2&13&&3.8&3.6&2.2&2.1\\
26ch130&2&10&&4.3&3.7&2.5&2.5\\
29pr144&2&10&&4.4&4.2&2.7&2.6\\
30ch150&2&15&&5.7&5.0&3.3&3.2\\
30kroa150&2&11&&5.5&4.7&5.8&5.0\\
36brg180&2&110&&2.7&2.8&4.1&4.3\\
39rat195&2&9&&6.8&6.0&4.0&3.7\\
45ts225&3&9&&8.1&6.0&8.5&6.3\\
56a280&2&10&&9.8&8.8&5.3&5.2\\
207si1032&2&15&&48.3&44.6&27.6&26.8\\
\cmidrule(){1-8}

Average&2.1&16.1&&6.8&6.1&4.6&4.3\\
\bottomrule

\end{tabular}
\end{table}

\begin{table}[!ht] \centering
	\footnotesize
	\caption[Comparison of GTSP \twoopt{}{} implementations.]{Comparison of GTSP \twoopt{}{} implementations.  The reported values are running times, in ms.}
	\label{tab:gtsp_implementations_two_opt}
\begin{tabular}{lrrc@{\qquad}rr}
\toprule

&\multicolumn{2}{c}{Basic}&&\multicolumn{2}{c}{Global}\\
\cmidrule(r){2-3}
\cmidrule(r){5-6}
Instance&\optshort{2}{B}{}&\optshort{2}{B}{adv}&&\optshort{2}{G}{}&\optshort{2}{E}{simple}\\
\cmidrule(){1-6}

10att48&0.3&0.2&&0.3&0.5\\
12brazil58&0.0&0.0&&0.1&0.3\\
20rat99&0.0&0.1&&0.4&1.6\\
20kroe100&0.0&0.1&&0.2&1.1\\
24gr120&0.0&0.1&&0.3&2.0\\
28gr137&0.0&0.1&&0.6&3.3\\
31pr152&0.0&0.1&&0.4&2.7\\
40d198&0.0&0.3&&1.9&12.9\\
45tsp225&0.0&0.3&&2.1&12.6\\
56a280&0.1&0.5&&3.2&24.5\\
87gr431&0.1&1.1&&4.9&57.7\\
107att532&0.2&1.8&&7.0&76.7\\
131p654&0.3&2.5&&6.5&179.6\\
200dsj1000&1.0&6.8&&51.8&780.4\\
\cmidrule(){1-6}

Average&0.2&1.0&&5.7&82.6\\
\bottomrule

\end{tabular}
\end{table}

\begin{table}[!ht] \centering
	\footnotesize
	\caption[GTSP \FO{} implementations comparison.]{GTSP \FO{} implementations comparison.  The reported values are running times, in ms.}
	\label{tab:gtsp_fo_implementations}
\begin{tabular}{lrrrrrcrrrrr}
\toprule

&\multicolumn{5}{c}{Algorithm 1}&&\multicolumn{5}{c}{Algorithm 2}\\
\cmidrule(){2-6}
\cmidrule(){8-12}
Instance \quad $k$: &3&4&5&6&7&&3&4&5&6&7\\
\cmidrule(){1-12}

10att48&0.3&0.4&0.9&4.1&24.4&&0.3&0.4&0.7&1.4&3.0\\
12brazil58&0.1&0.2&1.0&5.8&37.0&&0.1&0.3&0.7&1.8&4.3\\
20rat99&0.1&0.4&2.2&10.5&70.7&&0.1&0.5&1.4&3.2&8.3\\
20kroe100&0.1&0.3&1.6&12.3&87.4&&0.2&0.4&1.1&3.7&9.9\\
24gr120&0.1&0.6&2.2&13.0&88.5&&0.2&0.5&1.4&3.9&10.0\\
28gr137&0.1&0.4&2.6&16.1&121.7&&0.1&0.5&1.7&5.2&14.9\\
31pr152&0.1&0.5&2.7&13.5&93.0&&0.2&0.6&1.8&4.1&10.7\\
40d198&0.2&0.7&3.5&21.3&210.6&&0.3&0.9&2.5&6.9&26.0\\
45tsp225&0.2&0.9&4.5&26.2&265.2&&0.4&1.0&3.0&8.2&29.7\\
56a280&0.3&0.9&4.7&28.3&224.7&&0.4&1.1&3.2&8.9&27.6\\
87gr431&0.6&2.2&11.2&64.6&426.0&&0.8&2.6&8.2&22.7&59.8\\
107att532&0.6&2.2&10.7&62.2&497.8&&0.8&2.6&7.7&21.4&67.3\\
131p654&1.0&3.2&15.9&90.9&578.9&&1.4&3.9&11.9&35.1&90.0\\
200dsj1000&1.5&4.8&21.0&120.5&838.1&&1.9&5.3&14.2&38.5&107.0\\
\cmidrule(){1-12}

Average&0.4&1.3&6.1&34.9&254.6&&0.5&1.5&4.2&11.8&33.5\\
\bottomrule
\end{tabular}
\end{table}

\begin{table}[!ht] \centering
	\footnotesize
	\caption{\twoopt{}{} adaptations for GTSP comparison.}
	\label{tab:gtsp_two_opt}
\begin{tabular}{lcrrrrrcrrrrr}
\toprule

&&\multicolumn{5}{c}{Solution error, \%}&&\multicolumn{5}{c}{Running time, ms}\\
\cmidrule(){3-7}
\cmidrule(){9-13}
Instance&&\optshort{2}{B}{}&\optshort{2}{B}{co}&\optshort{2}{L}{}&\optshort{2}{L}{co}&\optshort{2}{G}{}&&\optshort{2}{B}{}&\optshort{2}{B}{co}&\optshort{2}{L}{}&\optshort{2}{L}{co}&\optshort{2}{G}{}\\
\cmidrule(){1-13}

10att48&&8.5&6.3&2.3&2.3&2.3&&0.2&0.2&0.2&0.2&0.3\\
12brazil58&&14.0&2.1&1.5&1.5&1.1&&0.0&0.0&0.0&0.0&0.1\\
20rat99&&22.1&17.1&14.5&13.7&0.8&&0.0&0.0&0.1&0.1&0.4\\
20kroe100&&15.2&1.3&2.7&2.7&0.0&&0.0&0.0&0.1&0.1&0.2\\
24gr120&&30.2&16.8&8.4&10.3&15.2&&0.0&0.0&0.2&0.3&0.4\\
28gr137&&9.6&2.5&3.7&2.7&1.9&&0.0&0.1&0.2&0.2&0.7\\
31pr152&&9.8&4.1&2.4&2.4&1.3&&0.0&0.0&0.2&0.2&0.4\\
40d198&&7.3&8.7&3.4&5.0&1.5&&0.1&0.2&0.6&0.8&2.0\\
45tsp225&&20.8&14.0&9.7&9.4&6.8&&0.0&0.1&0.6&0.8&2.2\\
56a280&&26.9&13.3&13.3&10.8&14.6&&0.1&0.2&1.0&1.1&3.4\\
87gr431&&10.3&4.8&7.0&6.9&4.2&&0.2&0.9&1.3&1.6&5.2\\
107att532&&16.8&9.2&14.7&14.2&7.9&&0.2&0.7&2.4&4.1&8.1\\
131p654&&4.1&6.9&7.7&7.7&4.0&&0.4&1.7&3.7&5.6&7.9\\
200dsj1000&&23.3&15.7&16.5&16.1&12.9&&1.0&3.2&16.5&20.4&53.5\\
\cmidrule(){1-13}

Average&&15.6&8.8&7.7&7.5&5.3&&0.2&0.5&1.9&2.6&6.1\\
\bottomrule
\end{tabular}
\end{table}

\begin{table}[!ht] \centering
	\footnotesize
	\caption{\Insertion{}{} adaptations for GTSP comparison.}
	\label{tab:gtsp_insertion}
\begin{tabular}{lcrrrrrcrrrrr}
\toprule

&&\multicolumn{5}{c}{Solution error, \%}&&\multicolumn{5}{c}{Running time, ms}\\
\cmidrule(){3-7}
\cmidrule(){9-13}
Instance&&\Insertion{B}{}&\Insertion{B}{co}&\Insertion{L}{}&\Insertion{L}{co}&\Insertion{G}{}&&\Insertion{B}{}&\Insertion{B}{co}&\Insertion{L}{}&\Insertion{L}{co}&\Insertion{G}{}\\
\cmidrule(){1-13}

10att48&&4.7&2.4&0.0&0.0&0.0&&0.2&0.2&0.2&0.2&0.3\\
12brazil58&&15.5&2.1&0.6&0.0&0.0&&0.0&0.0&0.0&0.0&0.1\\
20rat99&&32.0&16.5&4.0&0.0&0.0&&0.0&0.0&0.1&0.2&0.8\\
20kroe100&&18.5&7.7&6.6&6.6&6.6&&0.0&0.0&0.1&0.1&0.7\\
24gr120&&35.1&20.7&8.0&5.7&2.0&&0.0&0.0&0.1&0.2&0.6\\
28gr137&&9.6&8.2&1.9&7.4&0.0&&0.0&0.0&0.2&0.3&1.0\\
31pr152&&12.4&8.2&6.1&5.9&5.9&&0.0&0.1&0.3&0.3&0.6\\
40d198&&24.6&21.9&19.3&21.6&15.8&&0.1&0.1&0.5&0.4&2.9\\
45tsp225&&43.4&24.8&25.1&24.6&15.2&&0.1&0.2&0.9&1.0&3.2\\
56a280&&31.9&22.3&23.0&21.4&20.9&&0.1&0.1&1.0&0.9&5.0\\
87gr431&&11.0&7.8&6.9&7.1&6.7&&0.2&0.4&2.4&2.7&10.2\\
107att532&&20.8&16.7&13.1&12.0&11.6&&0.6&0.9&6.3&7.7&26.5\\
131p654&&23.0&24.0&21.7&21.4&19.0&&1.5&2.6&22.4&17.1&58.0\\
200dsj1000&&38.7&30.9&28.1&28.4&27.4&&4.6&4.5&53.9&66.3&206.3\\
\cmidrule(){1-13}

Average&&22.9&15.3&11.7&11.6&9.4&&0.5&0.7&6.3&7.0&22.6\\\bottomrule
\end{tabular}
\end{table}

\begin{table}[!ht] \centering
	\footnotesize
	\caption{GTSP \FO{$k$} for different $k$ comparison.}
	\label{tab:gtsp_fo}
\begin{tabular}{lcrrrrrcrrrrr}
\toprule

&&\multicolumn{5}{c}{Solution error, \%}&&\multicolumn{5}{c}{Running time, ms}\\
\cmidrule(){3-7}
\cmidrule(){9-13}
Instance&&\FO{2}&\FO{4}&\FO{6}&\FO{8}&\FO{10}&&\FO{2}&\FO{4}&\FO{6}&\FO{8}&\FO{10}\\
\cmidrule(){1-13}

10att48&&8.2&0.0&0.0&0.0&---&&0.2&0.4&1.4&7.8&---\\
12brazil58&&2.1&0.0&0.0&0.0&0.0&&0.0&0.2&1.8&10.5&44.8\\
20rat99&&18.5&17.9&6.8&0.8&0.0&&0.0&0.4&3.2&54.8&177.8\\
20kroe100&&24.3&24.3&23.4&23.4&0.0&&0.0&0.3&3.7&26.3&156.0\\
24gr120&&34.6&11.9&12.4&0.0&0.0&&0.0&0.6&3.9&48.9&187.2\\
28gr137&&15.0&12.8&2.4&2.1&6.3&&0.0&0.4&5.2&38.6&193.5\\
31pr152&&11.0&7.2&7.2&0.7&0.7&&0.0&0.5&4.1&38.5&218.4\\
40d198&&29.6&24.7&23.9&15.9&4.5&&0.1&0.7&6.9&72.7&508.6\\
45tsp225&&43.8&39.5&31.5&24.4&12.3&&0.1&0.9&8.2&56.0&468.0\\
56a280&&25.4&25.2&21.6&18.1&18.1&&0.1&0.9&8.9&80.7&464.9\\
87gr431&&11.1&7.6&6.6&6.0&5.9&&0.2&2.2&22.7&151.4&892.4\\
107att532&&24.0&22.4&21.7&20.6&13.5&&0.3&2.2&21.4&148.3&1202.9\\
131p654&&33.4&31.3&29.4&27.3&26.6&&0.4&3.2&35.1&227.9&1426.0\\
200dsj1000&&43.7&39.1&37.4&35.9&34.3&&0.7&4.8&38.5&267.1&1822.5\\
\cmidrule(){1-13}

Average&&23.2&18.9&16.0&12.5&9.4&&0.2&1.3&11.8&87.8&597.2\\
\bottomrule
\end{tabular}
\end{table}

\begin{table}[!ht] \centering
	\footnotesize
	\caption[GTSP local search fair competition.]{GTSP local search fair competition.  Every cell reports the most successful heuristics being given some limited time (see the first column) for a given range of instances (see the header).  Every heuristic is provided with the average relative solution error in percent.  To make the table easier to read, all the \LKb{}{}{} and \LKe{}{} adaptations of \LK{} are selected with bold font.  All the cells where the dominating heuristic is \LKc{}{}{} or \LKs{}{}{} are highlighted with gray background.}
	\label{tab:gtsp_ls_competition}

\end{table}

\begin{table}[!ht] \centering
	\footnotesize
	\caption{Detailed experiment results for the most successful GTSP heuristics ($m < 30$).}
	\label{tab:gtsp_lk_detailed_small}
\begin{tabular}{l @{\quad} r @{\quad} c *{5}{@{\hspace{1em}} r} @{\quad} c *{5}{@{\hspace{1em}} r}}
\toprule
&&& \multicolumn{5}{c}{Solution error, \%} && \multicolumn{5}{c}{Running time, ms} \\
\cmidrule(r){4-8} 
\cmidrule(r){10-14}
Instance&Best&&\optshort{2}{b}{co}&\LKc{5}{2}{co}&\LKs{5}{2}{co}&\LKe{4}{2}{}&\GK&&\optshort{2}{b}{co}&\LKc{5}{2}{co}&\LKs{5}{2}{co}&\LKe{4}{2}{}&\GK\\
\cmidrule(){1-14}

10att48&5394&&6.3&0.0&0.0&0.0&0.0&&0.24&0.25&0.28&2.53&18.72\\
10gr48&1834&&4.9&0.0&0.0&0.0&0.0&&0.01&0.03&0.06&1.45&12.48\\
10hk48&6386&&0.0&0.3&0.0&0.0&0.0&&0.01&0.08&0.17&1.36&18.72\\
11eil51&174&&4.0&0.6&0.0&0.0&0.0&&0.01&0.04&0.13&1.23&17.16\\
11berlin52&4040&&0.0&0.0&0.4&0.0&0.0&&0.01&0.06&0.12&1.17&12.48\\
12brazil58&15332&&2.1&0.0&0.0&0.0&0.0&&0.01&0.06&0.11&1.67&12.48\\
\bottomrule

\end{tabular}
\end{table}

\begin{table}[!ht] \centering
	\footnotesize
	\caption[Detailed experiment results for the most successful GTSP heuristics ($m \ge 30$).  Solution errors.]{Detailed experiment results for the most successful GTSP heuristics ($m \ge 30$).  The reported values are relative solution errors, \%.}
	\label{tab:gtsp_lk_quality}
\begin{tabular}{l@{\hspace{1.5em}}r@{\hspace{1.5em}}c@{\hspace{1.5em}}r@{\hspace{1.5em}}r@{\hspace{1.5em}}r@{\hspace{1.5em}}r@{\hspace{1.5em}}r@{\hspace{1.5em}}r@{\hspace{1.5em}}r@{\hspace{1.5em}}r@{\hspace{1.5em}}r}
\toprule

Instance&Best&&\optshort{2}{b}{co}&\LKb{2}{2}{co}&\LKc{5}{2}{co}&\LKs{5}{2}{co}&\LKs{5}{3}{co}&\LKs{5}{4}{co}&\LKs{4}{4}{co}&\LKe{4}{2}{}&\GK\\
\cmidrule(){1-12}

30ch150&2750&&6.5&7.1&1.7&1.1&0.0&0.0&0.3&1.1&0.0\\
30kroa150&11018&&16.2&8.2&0.1&1.6&0.0&0.0&0.0&0.0&0.0\\
30krob150&12196&&5.4&5.4&0.0&1.0&0.0&0.5&0.0&0.0&0.0\\
31pr152&51576&&4.1&3.3&3.9&1.9&0.0&0.0&0.0&1.2&0.0\\
32u159&22664&&24.9&10.2&0.8&0.4&0.0&0.0&0.0&1.1&0.0\\
35si175&5564&&2.6&3.8&0.0&0.0&0.0&0.0&0.0&0.0&0.0\\
36brg180&4420&&314.5&314.5&0.5&78.3&0.0&0.0&0.0&0.0&0.0\\
\bottomrule

\end{tabular}
\end{table}

\begin{table}[!ht] \centering
	\footnotesize
	\caption[Detailed experiment results for moderate and large instances ($m \ge 30$).  Running times.]{Detailed experiment results for moderate and large instances ($m \ge 30$). The reported values are running times, ms.}
	\label{tab:gtsp_lk_time}
\begin{tabular}{l@{\hspace{1em}}c@{\hspace{1em}}r@{\hspace{1em}}r@{\hspace{1em}}r@{\hspace{1em}}r@{\hspace{1em}}r@{\hspace{1em}}r@{\hspace{1em}}r@{\hspace{1em}}r@{\hspace{1em}}r}
\toprule

Instance&&\optshort{2}{b}{co}&\LKb{2}{2}{co}&\LKc{5}{2}{co}&\LKs{5}{2}{co}&\LKs{5}{3}{co}&\LKs{5}{4}{co}&\LKs{4}{4}{co}&\LKe{4}{2}{}&\GK\\
\cmidrule(){1-11}

30ch150&&0.1&0.1&0.5&1.4&2.8&2.7&8.0&46.7&56.2\\
30kroa150&&0.0&0.1&0.4&1.0&1.8&4.0&4.2&32.3&57.7\\
30krob150&&0.0&0.1&0.4&1.2&1.5&2.5&7.0&50.6&65.5\\
31pr152&&0.0&0.2&0.4&1.4&4.5&25.3&33.4&38.8&39.0\\
32u159&&0.1&0.1&0.3&0.9&2.7&4.2&25.7&31.9&62.4\\
35si175&&0.1&0.2&1.8&3.6&10.0&23.5&358.8&232.5&64.0\\
36brg180&&0.0&0.2&0.4&0.4&1.2&2.3&279.3&46.4&53.0\\
\bottomrule

\end{tabular}
\end{table}

\clearpage


\begin{table}[!ht] \centering
	\scriptsize
	\caption{MAP construction heuristics comparison.}
	\label{tab:construction}
\begin{tabular}{@{} l @{\hspace{0.5em}} r @{\hspace{0.5em}} c @{\hspace{2em}} *{3}{r @{\hspace{0.5em}}} r @{} c @{\hspace{2em}} *{3}{r @{\hspace{0.5em}}} r @{}}
\toprule
 &  &  & \multicolumn{4}{c}{Solution error, \%} &  & \multicolumn{4}{c}{Running times, ms}\\
\cmidrule(){4-7}
\cmidrule(){9-12}
Inst. & Best &  & Trivial & Greedy & Max-Regret & ROM &  & Trivial & Greedy & Max-Regret & ROM\\
\cmidrule(){1-12}
3gp100 & 504.4& & 157 & \underline{6} & 6 & 10 & & \underline{0} & 40 & 799 & 9\\
3r150 & 150.0& & 4\,997 & 54 & \underline{29} & 34 & & \underline{0} & 14 & 4\,253 & 26\\
4gp30 & 145.2& & 158 & 9 & 9 & \underline{2} & & \underline{0} & 35 & 206 & 7\\
4r80 & 80.0& & 4\,985 & 74 & \underline{49} & 76 & & \underline{0} & 12 & 27\,285 & 278\\
5gp12 & 66.2& & 147 & 13 & \underline{9} & 9 & & \underline{0} & 6 & 36 & 2\\
5r40 & 40.0& & 4\,911 & 159 & \underline{116} & 169 & & \underline{0} & 6 & 37\,214 & 686\\
\bottomrule
\end{tabular}
\end{table}

\begin{table}[!ht] \centering
	\scriptsize
	\caption{Solution errors of MAP local search heuristics started from Trivial.}
	\label{tab:ls_solutions_from_trivial}
\begin{tabular}{@{} l@{\hspace{0.5em}}r@{\hspace{0.5em}} c @{\hspace{2em}} *{2}{r @{\hspace{0.5em}}} r @{} c @{\hspace{2em}} *{2}{r @{\hspace{0.5em}}} r @{} c @{\hspace{2em}} *{3}{r @{\hspace{0.5em}}} r @{}}
\toprule
 &  &  & \multicolumn{12}{c}{Solution error, \%}\\
\cmidrule(){4-15}
Inst. & Best &  & \TwooptPlain & \ThreeoptPlain & \VoptPlain &  & \OneDVPlain & \TwoDVPlain & \MDVPlain &  & \OneDVtwoPlain & \TwoDVtwoPlain & \MDVthreePlain & \MDVVPlain\\
\cmidrule(){1-15}
3gp100 & 504.4& & 19.6 & \underline{10.0} & 19.8 & & \underline{4.9} & \underline{4.9} & \underline{4.9} & & 4.9 & 4.9 & \underline{4.6} & 4.9\\
3r150 & 150.0& & 134.5 & 16.0 & \underline{1.5} & & \underline{2.4} & \underline{2.4} & \underline{2.4} & & 2.4 & 2.4 & 2.1 & \underline{0.7}\\
4gp30 & 145.2& & 17.4 & \underline{4.2} & 13.4 & & 11.1 & \underline{7.9} & \underline{7.9} & & 10.7 & 7.9 & \underline{4.2} & 7.5\\
4r80 & 80.0& & 115.0 & 7.3 & \underline{2.0} & & 20.5 & \underline{11.5} & \underline{11.5} & & 18.9 & 11.5 & 4.1 & \underline{1.6}\\
5gp12 & 66.2& & 10.6 & \underline{2.1} & 8.5 & & 12.5 & \underline{6.9} & \underline{6.9} & & 10.1 & 6.9 & \underline{1.8} & 6.9\\
5r40 & 40.0& & 104.5 & 4.3 & \underline{3.8} & & 63.0 & \underline{34.3} & \underline{34.3} & & 47.3 & 34.3 & \underline{3.5} & 5.3\\
\bottomrule
\end{tabular}
\end{table}

\begin{table}[!ht] \centering
	\scriptsize
	\caption{Running times of MAP local search heuristics started from Trivial.}
	\label{tab:ls_times_from_trivial}
\begin{tabular}{@{} l@{\hspace{0.5em}} c @{\hspace{1em}} *{2}{r @{\hspace{0.5em}}} r @{} c @{\hspace{2em}} *{2}{r @{\hspace{0.5em}}} r @{} c @{\hspace{2em}} *{3}{r @{\hspace{0.5em}}} r @{}}
\toprule
 &  & \multicolumn{12}{c}{Running time, ms}\\
\cmidrule(){3-14}
Inst. &  & \TwooptPlain & \ThreeoptPlain & \VoptPlain &  & \OneDVPlain & \TwoDVPlain & \MDVPlain &  & \OneDVtwoPlain & \TwoDVtwoPlain & \MDVthreePlain & \MDVVPlain\\
\cmidrule(){1-14}
3gp100& & \underline{6.2} & 820.6 & 181.8 & & 14.3 & \underline{14.0} & 16.5 & & 18.4 & \underline{16.7} & 430.6 & 79.0\\
3r150& & \underline{19.8} & 1\,737.9 & 65.7 & & 17.6 & 18.7 & \underline{17.1} & & 22.7 & \underline{18.9} & 147.8 & 45.4\\
4gp30& & \underline{1.5} & 150.3 & 45.0 & & \underline{0.7} & 1.2 & 1.1 & & \underline{1.4} & 1.4 & 116.9 & 17.5\\
4r80& & \underline{10.5} & 987.5 & 64.5 & & \underline{7.9} & 18.0 & 15.3 & & \underline{11.2} & 18.4 & 344.8 & 98.2\\
5gp12& & \underline{0.3} & 38.5 & 3.6 & & \underline{0.2} & 0.4 & 0.5 & & \underline{0.5} & 0.5 & 30.6 & 1.6\\
5r40& & \underline{16.9} & 425.9 & 34.3 & & \underline{2.3} & 7.2 & 6.3 & & \underline{4.6} & 8.6 & 386.9 & 35.3\\
\bottomrule
\end{tabular}
\end{table}

\begin{table}[!ht] \centering
	\scriptsize
	\caption{Solution errors of MAP local search heuristics started from Greedy.}
	\label{tab:greedy}
\begin{tabular}{@{} l @{\hspace{0.5em}}  c @{\hspace{1em}} *{7}{r @{\hspace{0.3em}}} r @{} c @{\hspace{2em}} *{7}{r @{\hspace{0.3em}}} r @{}}
\toprule
 &  & \multicolumn{8}{c}{Solution error, \%} &  & \multicolumn{8}{c}{Running times, ms}\\
\cmidrule(){3-10}
\cmidrule(){12-19}
Inst. &  & \TwooptPlain & \OneDVPlain & \TwoDVPlain & \MDVPlain & \OneDVtwoPlain & \TwoDVtwoPlain & \MDVthreePlain & \MDVVPlain &  & \TwooptPlain & \OneDVPlain & \TwoDVPlain & \MDVPlain & \OneDVtwoPlain & \TwoDVtwoPlain & \MDVthreePlain & \MDVVPlain\\
\cmidrule(){1-19}
3gp100& & 4.3 & 3.4 & 3.4 & 3.4 & 3.4 & 3.4 & \underline{3.3} & 3.4 & & \underline{0.04} & 0.04 & 0.04 & 0.04 & 0.05 & 0.05 & 0.36 & 0.09\\
3r150& & 16.7 & 1.2 & 1.2 & 1.2 & 1.2 & 1.2 & 0.8 & \underline{0.7} & & \underline{0.02} & 0.02 & 0.02 & 0.03 & 0.02 & 0.03 & 0.11 & 0.05\\
4gp30& & 4.5 & 3.7 & 3.6 & 3.6 & 3.6 & 3.6 & \underline{2.6} & 3.6 & & 0.04 & \underline{0.03} & 0.04 & 0.04 & 0.04 & 0.04 & 0.11 & 0.05\\
4r80& & 15.8 & 7.9 & 6.1 & 6.1 & 7.9 & 6.1 & 2.6 & \underline{1.5} & & \underline{0.01} & 0.02 & 0.02 & 0.02 & 0.02 & 0.02 & 0.21 & 0.08\\
5gp12& & 5.4 & 6.3 & 4.5 & 4.5 & 5.3 & 4.5 & \underline{1.8} & 4.5 & & \underline{0.01} & 0.01 & 0.01 & 0.01 & 0.01 & 0.01 & 0.03 & 0.01\\
5r40& & 18.5 & 19.8 & 13.5 & 13.5 & 15.0 & 13.5 & \underline{2.3} & 3.5 & & \underline{0.01} & 0.01 & 0.01 & 0.01 & 0.01 & 0.01 & 0.18 & 0.04\\
\bottomrule
\end{tabular}
\end{table}

\begin{table}[!ht] \centering
	\tiny
	\caption[Chain metaheuristic for MAP started from Trivial, Greedy and ROM.]{Chain metaheuristic for MAP started from Trivial, Greedy and ROM\@.  5 seconds given.  1 --- \TwooptPlain, 2 --- \OneDVPlain, 3 --- \TwoDVPlain, 4 --- \MDVPlain, 5 --- \OneDVtwoPlain, 6 --- \TwoDVtwoPlain, 7 --- \MDVthreePlain, 8 --- \MDVVPlain.}
	\label{tab:chain5}
\begin{tabular}{@{} l@{\hspace{0.5em}} c @{\hspace{0.5em}} *{7}{r @{\hspace{0.5em}}} r @{} c @{\hspace{1.5em}} *{7}{r @{\hspace{0.5em}}} r @{} c @{\hspace{1.5em}} *{7}{r @{\hspace{0.5em}}} r @{}}
\toprule
 &  & \multicolumn{26}{c}{Solution error, \%}\\
\cmidrule(){3-28}
 &  & \multicolumn{8}{c}{Trivial} &  & \multicolumn{8}{c}{Greedy} &  & \multicolumn{8}{c}{ROM}\\
\cmidrule(){3-10}
\cmidrule(){12-19}
\cmidrule(){21-28}
Inst. &  & 1 & 2 & 3 & 4 & 5 & 6 & 7 & 8 &  & 1 & 2 & 3 & 4 & 5 & 6 & 7 & 8 &  & 1 & 2 & 3 & 4 & 5 & 6 & 7 & 8\\
\cmidrule(){1-28}
3gp100& & 15.3 & \underline{1.8} & \underline{1.8} & \underline{1.8} & 1.8 & 1.8 & 2.8 & 2.5 & & 5.3 & 1.7 & \underline{1.7} & \underline{1.7} & 1.8 & 1.8 & 2.9 & 2.3 & & 9.8 & \underline{1.9} & \underline{1.9} & \underline{1.9} & 1.9 & 1.9 & 2.6 & 2.3\\
3r150& & 77.7 & \underline{0.0} & \underline{0.0} & \underline{0.0} & \underline{0.0} & \underline{0.0} & 0.1 & \underline{0.0} & & 41.4 & \underline{0.0} & \underline{0.0} & \underline{0.0} & \underline{0.0} & \underline{0.0} & \underline{0.0} & \underline{0.0} & & 33.6 & \underline{0.0} & \underline{0.0} & \underline{0.0} & \underline{0.0} & \underline{0.0} & \underline{0.0} & \underline{0.0}\\
4gp30& & 7.0 & 1.6 & 1.1 & 1.1 & 1.8 & 1.1 & \underline{0.8} & 1.4 & & 6.3 & 1.9 & 0.8 & \underline{0.8} & 1.8 & 0.9 & \underline{0.8} & 1.4 & & 2.2 & 1.7 & 0.9 & 0.9 & 1.7 & 1.0 & \underline{0.8} & 1.4\\
4r80& & 55.0 & 4.4 & 1.9 & 1.9 & 4.1 & 2.3 & 0.4 & \underline{0.0} & & 41.6 & 4.6 & 1.6 & 1.6 & 4.5 & 1.8 & 0.8 & \underline{0.0} & & 57.0 & 4.3 & 2.0 & 2.0 & 4.3 & 2.0 & 0.9 & \underline{0.0}\\
\bottomrule
\end{tabular}
\end{table}

\begin{table}[!ht] \centering
	\tiny
	\caption[Chain metaheuristic for MAP started from Trivial, Greedy and ROM.]{Chain metaheuristic for MAP started from Trivial, Greedy and ROM\@.  10 seconds given.  1 --- \TwooptPlain, 2 --- \OneDVPlain, 3 --- \TwoDVPlain, 4 --- \MDVPlain, 5 --- \OneDVtwoPlain, 6 --- \TwoDVtwoPlain, 7 --- \MDVthreePlain, 8 --- \MDVVPlain.}
\label{tab:chain10}
\begin{tabular}{@{} l@{\hspace{0.5em}} c @{\hspace{0.5em}} *{7}{r @{\hspace{0.5em}}} r @{} c @{\hspace{2em}} *{7}{r @{\hspace{0.5em}}} r @{} c @{\hspace{2em}} *{7}{r @{\hspace{0.5em}}} r @{}}
\toprule
 &  & \multicolumn{26}{c}{Solution error, \%}\\
\cmidrule(){3-28}
 &  & \multicolumn{8}{c}{Trivial} &  & \multicolumn{8}{c}{Greedy} &  & \multicolumn{8}{c}{ROM}\\
\cmidrule(){3-10}
\cmidrule(){12-19}
\cmidrule(){21-28}
Inst. &  & 1 & 2 & 3 & 4 & 5 & 6 & 7 & 8 &  & 1 & 2 & 3 & 4 & 5 & 6 & 7 & 8 &  & 1 & 2 & 3 & 4 & 5 & 6 & 7 & 8\\
\cmidrule(){1-28}
3gp100& & 15.1 & \underline{1.6} & \underline{1.6} & \underline{1.6} & 1.7 & 1.7 & 2.3 & 2.2 & & 5.3 & 1.6 & 1.6 & \underline{1.6} & 1.6 & 1.6 & 2.5 & 2.1 & & 9.8 & \underline{1.6} & \underline{1.6} & \underline{1.6} & 1.8 & 1.7 & 2.2 & 2.1\\
3r150& & 75.3 & \underline{0.0} & \underline{0.0} & \underline{0.0} & \underline{0.0} & \underline{0.0} & \underline{0.0} & \underline{0.0} & & 41.4 & \underline{0.0} & \underline{0.0} & \underline{0.0} & \underline{0.0} & \underline{0.0} & \underline{0.0} & \underline{0.0} & & 33.6 & \underline{0.0} & \underline{0.0} & \underline{0.0} & \underline{0.0} & \underline{0.0} & \underline{0.0} & \underline{0.0}\\
4gp30& & 6.5 & 1.4 & 0.8 & 0.8 & 1.3 & 1.0 & \underline{0.7} & 1.3 & & 6.2 & 1.7 & 0.8 & 0.8 & 1.5 & 0.8 & \underline{0.7} & 1.1 & & 2.2 & 1.4 & 0.8 & 0.8 & 1.4 & 0.8 & \underline{0.7} & 1.2\\
4r80& & 52.1 & 3.9 & 1.1 & 1.0 & 3.6 & 1.1 & 0.1 & \underline{0.0} & & 41.4 & 3.9 & 1.0 & 1.0 & 4.3 & 1.1 & 0.4 & \underline{0.0} & & 55.0 & 4.0 & 1.1 & 1.1 & 3.4 & 1.3 & 0.4 & \underline{0.0}\\
\bottomrule
\end{tabular}
\end{table}

\begin{table}[!ht] \centering
	\tiny
	\caption[Multichain metaheuristic for MAP started from Trivial, Greedy and ROM.]{Multichain metaheuristic for MAP started from Trivial, Greedy and ROM\@.  5 seconds given.  1 --- \TwooptPlain, 2 --- \OneDVPlain, 3 --- \TwoDVPlain, 4 --- \MDVPlain, 5 --- \OneDVtwoPlain, 6 --- \TwoDVtwoPlain, 7 --- \MDVthreePlain, 8 --- \MDVVPlain.}
	\label{tab:multichain5}
\begin{tabular}{@{} l@{\hspace{0.5em}} c @{\hspace{0.5em}} *{7}{r @{\hspace{0.4em}}} r @{} c @{\hspace{1.5em}} *{7}{r @{\hspace{0.4em}}} r @{} c @{\hspace{1.5em}} *{7}{r @{\hspace{0.4em}}} r @{}}
\toprule
 &  & \multicolumn{26}{c}{Solution error, \%}\\
\cmidrule(){3-28}
 &  & \multicolumn{8}{c}{Trivial} &  & \multicolumn{8}{c}{Greedy} &  & \multicolumn{8}{c}{ROM}\\
\cmidrule(){3-10}
\cmidrule(){12-19}
\cmidrule(){21-28}
Inst. &  & 1 & 2 & 3 & 4 & 5 & 6 & 7 & 8 &  & 1 & 2 & 3 & 4 & 5 & 6 & 7 & 8 &  & 1 & 2 & 3 & 4 & 5 & 6 & 7 & 8\\
\cmidrule(){1-28}
3gp100& & 11.8 & \underline{1.1} & 1.2 & \underline{1.1} & 1.3 & 1.3 & 156.9 & 2.0 & & 5.3 & 1.2 & 1.2 & \underline{1.1} & 1.4 & 1.3 & 5.6 & 2.2 & & 9.7 & \underline{1.2} & \underline{1.2} & \underline{1.2} & 1.3 & 1.3 & 9.8 & 2.0\\
3r150& & 68.1 & \underline{0.0} & \underline{0.0} & \underline{0.0} & \underline{0.0} & \underline{0.0} & \underline{0.0} & \underline{0.0} & & 41.4 & \underline{0.0} & \underline{0.0} & \underline{0.0} & \underline{0.0} & \underline{0.0} & \underline{0.0} & \underline{0.0} & & 33.6 & \underline{0.0} & \underline{0.0} & \underline{0.0} & \underline{0.0} & \underline{0.0} & \underline{0.0} & \underline{0.0}\\
4gp30& & 3.0 & 0.9 & \underline{0.7} & \underline{0.7} & 0.8 & \underline{0.7} & \underline{0.7} & 1.1 & & 3.1 & 0.8 & \underline{0.7} & \underline{0.7} & 1.0 & \underline{0.7} & \underline{0.7} & 1.0 & & 2.1 & \underline{0.7} & \underline{0.7} & \underline{0.7} & \underline{0.7} & \underline{0.7} & 0.8 & 1.0\\
4r80& & 45.3 & 3.4 & 1.5 & 1.4 & 2.3 & 1.5 & 0.4 & \underline{0.0} & & 38.9 & 3.3 & 0.9 & 0.9 & 3.1 & 0.9 & 0.8 & \underline{0.0} & & 44.6 & 2.4 & 1.0 & 1.0 & 2.6 & 1.1 & 45.3 & \underline{0.0}\\
\bottomrule
\end{tabular}
\end{table}

\begin{table}[!ht] \centering
	\tiny
	\caption[Multichain metaheuristic for MAP started from Trivial, Greedy and ROM.]{Multichain metaheuristic for MAP started from Trivial, Greedy and ROM\@.  10 seconds given.  1 --- \TwooptPlain, 2 --- \OneDVPlain, 3 --- \TwoDVPlain, 4 --- \MDVPlain, 5 --- \OneDVtwoPlain, 6 --- \TwoDVtwoPlain, 7 --- \MDVthreePlain, 8 --- \MDVVPlain.}
	\label{tab:multichain10}
\begin{tabular}{@{} l@{\hspace{0.5em}} c @{\hspace{0.5em}} *{7}{r @{\hspace{0.5em}}} r @{} c @{\hspace{2em}} *{7}{r @{\hspace{0.5em}}} r @{} c @{\hspace{2em}} *{7}{r @{\hspace{0.5em}}} r @{}}
\toprule
 &  & \multicolumn{26}{c}{Solution error, \%}\\
\cmidrule(){3-28}
 &  & \multicolumn{8}{c}{Trivial} &  & \multicolumn{8}{c}{Greedy} &  & \multicolumn{8}{c}{ROM}\\
\cmidrule(){3-10}
\cmidrule(){12-19}
\cmidrule(){21-28}
Inst. &  & 1 & 2 & 3 & 4 & 5 & 6 & 7 & 8 &  & 1 & 2 & 3 & 4 & 5 & 6 & 7 & 8 &  & 1 & 2 & 3 & 4 & 5 & 6 & 7 & 8\\
\cmidrule(){1-28}
3gp100& & 11.2 & \underline{1.0} & \underline{1.0} & \underline{1.0} & 1.1 & 1.1 & 2.5 & 1.7 & & 5.3 & \underline{1.0} & \underline{1.0} & \underline{1.0} & 1.1 & 1.1 & 2.8 & 1.8 & & 9.7 & 1.0 & \underline{0.9} & 1.0 & 1.2 & 1.1 & 2.3 & 1.7\\
3r150& & 65.3 & \underline{0.0} & \underline{0.0} & \underline{0.0} & \underline{0.0} & \underline{0.0} & \underline{0.0} & \underline{0.0} & & 41.4 & \underline{0.0} & \underline{0.0} & \underline{0.0} & \underline{0.0} & \underline{0.0} & \underline{0.0} & \underline{0.0} & & 33.6 & \underline{0.0} & \underline{0.0} & \underline{0.0} & \underline{0.0} & \underline{0.0} & \underline{0.0} & \underline{0.0}\\
4gp30& & 2.9 & \underline{0.7} & \underline{0.7} & \underline{0.7} & \underline{0.7} & \underline{0.7} & \underline{0.7} & 1.0 & & 2.6 & 0.8 & \underline{0.7} & \underline{0.7} & \underline{0.7} & \underline{0.7} & \underline{0.7} & 0.9 & & 2.1 & \underline{0.7} & \underline{0.7} & \underline{0.7} & \underline{0.7} & \underline{0.7} & \underline{0.7} & 0.8\\
4r80& & 43.8 & 2.5 & 0.9 & 0.9 & 2.1 & 1.0 & 0.2 & \underline{0.0} & & 38.1 & 2.3 & 0.6 & 0.6 & 2.3 & 0.6 & 0.5 & \underline{0.0} & & 42.4 & 2.0 & 0.8 & 0.8 & 2.0 & 0.8 & 0.5 & \underline{0.0}\\
\bottomrule
\end{tabular}
\end{table}

\begin{table}[!ht] \centering
	\scriptsize
	\caption{All MAP heuristics comparison for the instances with independent weights.}
	\label{tab:heuristics_independent}
%
\end{table}


\begin{table}[!ht] \centering
	\scriptsize
	\caption{All MAP heuristics comparison for the instances with decomposable weights.}
	\label{tab:heuristics_decomposable}
%
\end{table}


\begin{table}
	\footnotesize
	\caption{GTSP metaheuristics quality comparison.}
	\label{tab:gtsp_ma_quality}
\begin{tabular}{l @{\qquad} rrrr @{\qquad} rrr @{\qquad} rrr}
\toprule
\multirow{2}{*}{Name}		&	\multicolumn{4}{c}{Error, \%}							&	\multicolumn{3}{c}{Quality impr., \%}					&	\multicolumn{3}{c}{Opt., \%}					 \\
\cmidrule(r){2-5} \cmidrule(r){6-8} \cmidrule{9-11}
		&	GK	&	SG	&	SD	&	TSP	&	SG	&	SD	&	TSP	&	GK	&	SD	&	TSP	\\
\cmidrule{1-11}																						
40d198		&	0.00	&	0.00	&	0.00	&	0.00	&	0.00	&	0.00	&	0.00	&	100	&	100	&	100	\\
40kroa200		&	0.00	&	0.00	&	0.00	&	0.00	&	0.00	&	0.00	&	0.00	&	100	&	100	&	100	\\
40krob200		&	0.00	&	0.05	&	0.01	&	0.00	&	0.05	&	0.01	&	0.00	&	100	&	70	&	100	\\
41gr202		&	0.00	&		&	0.00	&		&		&	0.00	&		&	100	&	100	&		\\
45ts225		&	0.00	&	0.14	&	0.09	&	0.04	&	0.14	&	0.09	&	0.04	&	100	&	0	&	60	\\
45tsp225		&	0.00	&		&	0.01	&		&		&	0.01	&		&	100	&	90	&		\\
\bottomrule
\end{tabular}
\end{table}

\begin{table}
	\footnotesize
	\caption{GTSP metaheuristics running time comparison.}
	\label{tab:gtsp_ma_time}
\begin{tabular}{ l @{\qquad} rrrr @{\qquad} rrr }
\toprule

\multirow{2}{*}{Name}		&	\multicolumn{4}{c}{Time, sec}							&	\multicolumn{3}{c}{Time impr., \%}					\\
\cmidrule(r){2-5} \cmidrule(r){6-8}
		&	GK	&	SG	&	SD	&	TSP	&	SG	&	SD	&	TSP	\\
\cmidrule{1-8}
40d198		&	0.14	&	1.63	&	1.18	&	1.22	&	11.6	&	8.4	&	8.7	\\
40kroa200		&	0.14	&	1.66	&	0.26	&	0.79	&	12.1	&	1.9	&	5.8	\\
40krob200		&	0.16	&	1.63	&	0.80	&	2.70	&	10.2	&	5.0	&	16.8	\\
41gr202		&	0.21	&		&	0.65	&		&		&	3.2	&		\\
45ts225		&	0.24	&	1.71	&	0.46	&	1.42	&	7.0	&	1.9	&	5.8	\\
45tsp225		&	0.19	&		&	0.55	&		&		&	2.9	&		\\
\bottomrule
\end{tabular}
\end{table}

\begin{table}
	\footnotesize
	\caption{\texttt{GK} experiments details.}
	\label{tab:gtsp_ma_details}
\begin{tabular}{lrrrrrr}
\toprule
Name		&	Opt.	&	Value	&	Error, \%	&	Opt., \%	&	Time, sec	&	\# gen.	\\
\cmidrule{1-7}

40d198		&	10557	&	10557.0	&	0.00	&	100	&	0.14	&	9.1	\\
40kroa200		&	13406	&	13406.0	&	0.00	&	100	&	0.14	&	9.0	\\
40krob200		&	13111	&	13111.0	&	0.00	&	100	&	0.16	&	10.3	\\
41gr202		&	23301	&	23301.0	&	0.00	&	100	&	0.21	&	9.8	\\
45ts225		&	68340	&	68340.0	&	0.00	&	100	&	0.24	&	12.7	\\
45tsp225		&	1612	&	1612.0	&	0.00	&	100	&	0.19	&	10.4	\\
46pr226		&	64007	&	64007.0	&	0.00	&	100	&	0.10	&	9.0	\\
\bottomrule
\end{tabular}
\end{table}


\begin{table}[!ht] \centering
	\scriptsize
	\caption{Comparison of the MAP Memetic Algorithm based on different local search procedures.  The given time is 3 s.}
	\label{tab:ls_comparison}
\begin{tabular}{@{} l@{\quad}r@{\quad} c  *{7}{r @{\quad}} r @{}}
\toprule
 &  &  & \multicolumn{8}{c}{Relative solution error, \%}\\
\cmidrule(){4-11}
Inst. & Best &  & \TwooptPlain & \OneDVPlain & \TwoDVPlain & \MDVPlain & \OneDVtwoPlain & \TwoDVtwoPlain & \MDVthreePlain & \MDVVPlain\\
\cmidrule(){1-11}
3r40 & 40.0& & 122.00 & 26.75 & 30.00 & 27.25 & 32.25 & 32.50 & 32.00 & \underline{6.25}\\
3r70 & 70.0& & 102.71 & 11.43 & 11.14 & 11.57 & 11.71 & 11.57 & 15.00 & \underline{0.71}\\
3r100 & 100.0& & 83.90 & 3.00 & 3.20 & 3.10 & 3.30 & 3.10 & 5.80 & \underline{0.00}\\
4r20 & 20.0& & 68.00 & 46.00 & 28.00 & 29.50 & 39.50 & 32.00 & 17.50 & \underline{0.00}\\
4r30 & 30.0& & 73.00 & 31.00 & 23.67 & 23.67 & 27.00 & 21.67 & 14.33 & \underline{0.00}\\
4r40 & 40.0& & 73.50 & 24.00 & 15.25 & 15.00 & 23.00 & 15.75 & 11.25 & \underline{0.00}\\
\bottomrule
\end{tabular}
\end{table}

\begin{table}[!ht] \centering
	\scriptsize
	\caption{MAP metaheuristics comparison for small running times.}
	\label{tab:metaheuristics1}
\begin{tabular}{@{} l@{\hspace{0.5em}} c @{\hspace{2em}} *{2}{r @{\hspace{0.5em}}} r @{} c @{\hspace{2em}} *{2}{r @{\hspace{0.5em}}} r @{} c @{\hspace{2em}} *{2}{r @{\hspace{0.5em}}} r @{}}
\toprule
 &  & \multicolumn{11}{c}{Relative solution error, \%}\\
\cmidrule(){3-13}
 &  & \multicolumn{3}{c}{0.3 sec.} &  & \multicolumn{3}{c}{1 sec.} &  & \multicolumn{3}{c}{3 sec.}\\
\cmidrule(){3-5}
\cmidrule(){7-9}
\cmidrule(){11-13}
Instance &  & HL & SA & GK &  & HL & SA & GK &  & HL & SA & GK\\
\cmidrule(){1-13}
3r40& & 49.75 & 120.00 & \underline{10.75} & & 44.25 & 99.00 & \underline{9.75} & & 41.50 & 84.50 & \underline{6.25}\\
3r70& & 512.86 & 102.86 & \underline{3.29} & & 18.14 & 82.86 & \underline{1.71} & & 16.86 & 72.71 & \underline{0.71}\\
3r100& & 5051.50 & 100.30 & \underline{1.10} & & 15.40 & 70.10 & \underline{0.20} & & 4.90 & 59.20 & \underline{0.00}\\
4r20& & 73.50 & 153.50 & \underline{6.00} & & 71.00 & 133.00 & \underline{0.50} & & 59.00 & 100.50 & \underline{0.00}\\
4r30& & 56.67 & 126.33 & \underline{2.00} & & 50.33 & 114.00 & \underline{0.00} & & 45.00 & 94.00 & \underline{0.00}\\
\bottomrule
\end{tabular}
\end{table}

\begin{table}[!ht] \centering
	\scriptsize
	\caption{MAP metaheuristics comparison for large running times.}
	\label{tab:metaheuristics2}
\begin{tabular}{@{} l@{\hspace{0.5em}} c @{\hspace{2em}} *{2}{r @{\hspace{0.5em}}} r @{} c @{\hspace{2em}} *{2}{r @{\hspace{0.5em}}} r @{} c @{\hspace{2em}} *{2}{r @{\hspace{0.5em}}} r @{} c @{\hspace{2em}} *{2}{r @{\hspace{0.5em}}} r @{}}
\toprule
 &  & \multicolumn{15}{c}{Relative solution error, \%}\\
\cmidrule(){3-17}
 &  & \multicolumn{3}{c}{10 sec.} &  & \multicolumn{3}{c}{30 sec.} &  & \multicolumn{3}{c}{100 sec.} &  & \multicolumn{3}{c}{300 sec.}\\
\cmidrule(){3-5}
\cmidrule(){7-9}
\cmidrule(){11-13}
\cmidrule(){15-17}
Instance &  & HL & SA & GK &  & HL & SA & GK &  & HL & SA & GK &  & HL & SA & GK\\
\cmidrule(){1-17}
3r40& & 38.25 & 63.50 & \underline{4.50} & & 32.50 & 60.75 & \underline{4.75} & & 28.75 & 51.75 & \underline{2.50} & & 27.25 & 47.00 & \underline{1.75}\\
3r70& & 14.00 & 55.00 & \underline{0.57} & & 13.29 & 45.14 & \underline{0.00} & & 11.43 & 37.71 & \underline{0.00} & & 10.71 & 34.29 & \underline{0.00}\\
3r100& & 4.10 & 45.60 & \underline{0.00} & & 3.50 & 36.60 & \underline{0.00} & & 3.00 & 30.80 & \underline{0.00} & & 2.40 & 24.80 & \underline{0.00}\\
4r20& & 49.50 & 94.50 & \underline{0.00} & & 44.00 & 80.00 & \underline{0.00} & & 38.50 & 63.00 & \underline{0.00} & & 34.00 & 52.00 & \underline{0.00}\\
4r30& & 37.33 & 83.00 & \underline{0.00} & & 33.67 & 68.00 & \underline{0.00} & & 31.00 & 58.00 & \underline{0.00} & & 28.00 & 45.00 & \underline{0.00}\\
\bottomrule
\end{tabular}
\end{table}

\end{document}